\documentclass{aa}

\usepackage{txfonts}
\usepackage{graphicx}
\begin{document}

\title{MHD wave propagation in the neighbourhood of two dipoles}
\author{J.~A. McLaughlin \and A.~W. Hood}

\offprints{J.~A. McLaughlin, \email{james@mcs.st-and.ac.uk}}

\institute{School of Mathematics and Statistics, University of St
Andrews, KY16 9SS, UK}
%\offprints{J.~A. McLaughlin,
%\email{james@mcs.st-and.ac.uk}
\date{Received 23 November 2005 / Accepted 22 February 2006}

\authorrunning{McLaughlin \& Hood}

% ----------------------------------------------------------------

\abstract
{This paper is the third in a series of investigations by the authors.}
{The nature of fast magnetoacoustic and Alfv\'en waves is investigated in a 2D $\beta=0$ plasma in the neighbourhood of two dipoles.}
{We use both numerical simulations (two-step Lax-Wendroff scheme) and analytical techniques (WKB approximation).}
{It is found that  the propagation of the  linear fast wave is dictated by the  Alfv\'en speed profile and that close to the null, the  wave is attracted to the neutral point. However, it is also found that  in this magnetic configuration some of the wave can  escape the refraction effect; this had not  been seen in previous investigations  by the authors. The wave split occurs  near the regions of very high  Alfv\'en speed (found near the loci of the two dipoles). Also, for the set-up investigated it was found that  $40\%$ of the wave energy accumulates at the null.  Ohmic dissipation will then extract the wave energy at this point.  The Alfv\'en wave behaves in a different manner in that part of the wave accumulates along the separatrices and part escapes. Hence, the current density will accumulate at this part of the topology and this is where wave heating will occur.}
{The phenomenon of wave accumulation at a specific place is a feature of both wave types, as is the result that a fraction of the wave can now escape the numerical box when propagating in this magnetic configuration.}

\keywords{Magnetohydrodynamics (MHD) -- Waves -- Sun:~corona -- Sun:~ magnetic fields -- Sun:~oscillations}

\maketitle

% ----------------------------------------------------------------

\section{Introduction}\label{section1}

The solar corona is a complicated, inhomogenous magnetised environment. It is also know that there are MHD wave motions in the corona and that the environment plays a key role in their propagation. Wave motions have been observed with the SOHO and TRACE satellites, for example slow MHD waves (Berghmans \& Clette, 1999; De Moortel \emph{et al.}, 2000) and fast MHD waves (Nakariakov \emph{et al.}, 1999). To begin to understand this magnetic environment, it is useful to look at the  structure (topology) of the magnetic field itself. Potential field extrapolations of the coronal magnetic field can be made from photospheric magnetograms. Such extrapolations show the existence of an important feature of the topology; \emph{null points}. Null points are points in the field where the Alfv\'en speed is zero. Detailed investigations of the coronal magnetic field, using such potential field calculations, can be found in \cite{Beveridge2002} and \cite{Brown2001}.

%The solar corona is dominated by the magnetic field. In order to understand the myriad of phenomena that occur on the Sun, it is important to understand the structure (topology) of the magnetic field itself. 

%Potential field extrapolations of the coronal magnetic field can be made from photospheric magnetograms. Such extrapolations show the existence of an important feature of the topology; \emph{null points}.

% Null points are points in the field where the Alfv\'en speed is zero. Detailed investigations of the coronal magnetic field, using such potential field calculations, can be found in \cite{Beveridge2002} and \cite{Brown2001}.

McLaughlin \& Hood (2004) (hereafter referred to as {{Paper I}}) found that for a single 2D null point, the fast magnetoacoustic wave was attracted to the null and the wave energy accumulated there. In addition, they found that the Alfv\'en wave energy accumulated along the separatrices of the topology. McLaughlin \& Hood (2005) (hereafter Paper II) found that the key results from Paper I carried over from the simple 2D single magnetic null point to the more complicated magnetic configuration of two null points. The aim of this paper (effectively the third in a series) is to see if  their ideas  carry through to a more realistic magnetic configuration (that still involves a magnetic null).

Waves in the neighbourhood of a single 2D null point have been investigated
by various authors. \cite{Bulanov1980} provided a
detailed discussion of the propagation of fast and Alfv\'en waves
using cylindrical symmetry. In their paper, harmonic fast waves are generated and
these propagate towards the null point. However, the assumed
cylindrical symmetry means that the disturbances can only propagate
either towards or away from the null point. \cite{CraigWatson1992}
mainly consider the radial propagation of the $m=0$ mode (where $m$ is the
azimuthal wavenumber) using a mixture of analytical and numerical
solutions. In their investigation, the outer radial boundary is held fixed so that any
outgoing waves will be reflected back towards the null point. This
means that all the energy in the wave motions is contained within a
fixed region. They show that the propagation of the $m=0$ wave
towards the null point generates an exponentially large increase in
the current density and that magnetic resistivity dissipates this current in
a time related to $\log { \eta }$. Their initial disturbance is given as a
function of radius. In this paper, we are interested in generating
the disturbances at the boundary rather than internally. Craig and
McClymont (1991, 1993) and Hassam (1992) investigate the normal mode solutions for both
$m=0$ and $m\ne 0$ modes with resistivity included. Again they
emphasise that the current builds up as the inverse square of the
radial distance from the null point. However, attention was restricted to a circular reflecting boundary.

%2D analysis in this paper. Investigate basic process. Investigate fast and Alfv\'en uncoupled. Same behaviour in 3D but waves coupled together.
The propagation of fast magnetoacoustic waves in an inhomogeneous
coronal plasma has been investigated by \cite{Nakariakov1995},
who showed how the waves are refracted into regions of low Alfv\'en
speed. In the case of null points, it is the aim of this paper to
see how this refraction proceeds when the Alfv\'en speed actually
drops to zero.

The propagation of fast waves in 2D coronal arcades has been investigated by \cite{oliver1}. When the background Alfv\'en speed is either constant or decreases monotonically with height, the waves are leaky (\cite{toni}). However, when the density is chosen so that there exists a local minimum in the Alfv\'en speed, the fast waves can be trapped (\cite{erwin}, \cite{brady}). The present paper investigates the nature of fast wave propagation when there is a null point at which the Alfv\'en speed actually drops to zero at  a local minimum. The same idea of trapping at least a fraction of the wave is found. Thus, fast waves in 2D coronal arcades can be trapped by either a density enhancement or the presence of a null point.

%The paper has the following outline. In Section \ref{sec:1}, the basic equations are described. The results for an uncoupled fast magnetoacoustic wave are presented in Section \ref{sec:2}. This section discusses fast wave propagation with a pulse coming in from the top and side boundaries, and numerical and anaytical results are presented. Section \ref{sec:3} discusses the propagation of Alfv\'en waves and the conclusions are given in Section \ref{sec:4}

The paper has the following outline. In Section \ref{sec:1}, the basic equations are described. The results for an uncoupled fast magnetoacoustic wave are presented in Section \ref{sec:2}. This section discusses fast wave propagation with a pulse generated on the lower boundary, and numerical and anaytical results for three scenarios are presented. Section \ref{sec:3} discusses the propagation of Alfv\'en waves and the conclusions are given in Section \ref{sec:4}.

\section{Basic Equations}\label{sec:1}

The usual MHD equations for an ideal, zero $\beta$ plasma appropriate to
the solar corona are used. Hence,
\begin{eqnarray}
\qquad  \rho \left( {\partial {\bf{v}}\over \partial t} + \left( {\bf{v}}\cdot\nabla \right) {\bf{v}} \right) &=& {1\over \mu}\left(\nabla \times {\bf{B}}\right)\times {\bf{B}}\; ,\label{eq:2.1a} \\
  {\partial {\bf{B}}\over \partial t} &=& \nabla \times \left ({\bf{v}}\times {\bf{B}}\right ) \; ,\label{eq:2.1b} \\
{\partial \rho\over \partial t} + \nabla \cdot \left (\rho {\bf{v}}\right ) &=& 0\; , \label{eq:2.1c}
\end{eqnarray}
where $\rho$ is the mass density, ${\bf{v}}$ is the plasma
velocity, ${\bf{B}}$ the magnetic induction (usually called the
magnetic field) and  $ \mu = 4 \pi \times 10^{-7} \/\mathrm{Hm^{-1}}$  the magnetic
permeability. The gas pressure and the
adiabatic energy equation are neglected in the zero $\beta$ approximation. The magnetic diffusivity is neglected in the ideal approximation.

\subsection{Basic equilibrium}\label{sec:1.1}

\begin{figure}
\begin{center}
\includegraphics[width=2.2in]{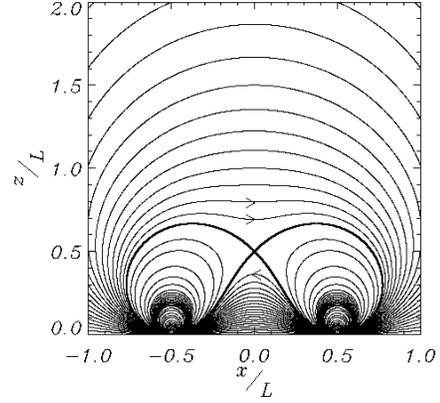}
\caption{Our choice of equilibrium magnetic field.}
\label{fig:dipolemagneticfield}
\end{center}
\end{figure}

Paper II showed that the key results from Paper I carried over from the simple 2D single magnetic null point to the more complicated magnetic configuration of two null points. However, the choice of magnetic fields used in both these papers are only valid  in the neighbourhood of the null point(s), because as $x$ and $z$ get very large, ${\bf{B}}_0$ also gets large, which is unphysical. Thus, in this paper we now want to investigate the behaviour of MHD waves near a more realistic magnetic configuration (and one that still involves a magnetic null). One such choice is a magnetic field created by two dipoles, of the form $  {\bf{B}}_0 = {B_0}{L^2}    \left( B_x,0,B_z \right)$, where:
\begin{eqnarray}
 B_x =  \frac{ - \left( x+a\right)^2 + z^2 }{ \left[\left( x+a\right)^2 +z^2\right]^2} +   \frac{ - \left( x-a\right)^2 + z^2 }{ \left[\left( x-a\right)^2 +z^2\right]^2}\; ,\nonumber\\
 B_z=-\frac { 2\left( x+a\right)z}{ \left[\left( x+a\right)^2 +z^2\right]^2} -  \frac { 2\left( x-a\right)z}{ \left[\left( x-a\right)^2 +z^2\right]^2} \label{twodipoles}
\end{eqnarray}
where $B_0$ is a characteristic field strength, $L$ is the length scale for magnetic field variations and $2a$ is the separation of the dipoles. We choose $a=0.5\:L$ in our simulations. This magnetic field can be seen in Figure \ref{fig:dipolemagneticfield}. It comprises of four separatrices and an X-point located at $(x,z)=(0,a)$. Note that as $x$ or $z$ gets very large, the field strength becomes small (this is a more physical field than those previously investigated in Papers I and II).

\subsection{Linearised equations}

As in Papers I and II, the linearised MHD equations are used to study the nature of wave propagation near the two dipoles. Using subscripts of $0$ for equilibrium quantities and $1$ for
perturbed quantities, the linearised equation of motion becomes
\begin{equation}\label{eq:2.3}
\qquad  \rho_0 \frac{\partial \mathbf{v}_1}{\partial t} = \left(\frac{ \nabla \times \mathbf{B}_1}{\mu} \right) \times \mathbf{B}_0 \; ,
\end{equation}
the linearised induction equation
\begin{equation}\label{eq:2.5}
\qquad    {\partial {\bf{B}}_1\over \partial t} = \nabla \times
    ({\bf{v}}_1 \times {\bf{B}}_0) \; ,
\end{equation}
and the linearised equation of mass continuity
\begin{equation}\label{eq:2.4}
\qquad \frac{\partial \rho_1} {\partial t} + \nabla\cdot\left( \rho_0 \mathbf{v} _1 \right) =0 \; .
\end{equation}
We will not discuss equation (\ref{eq:2.4}) further as it can be solved once we know $\mathbf{v} _1$. In fact, it has no influence on the momentum equation (in the zero $\beta$ approximation) and so in effect the plasma is arbitrarily compressible (Craig \& Watson 1992). We assume the background gas density is uniform and label it as $\rho_0$. A spatial variation in $\rho _0$ can cause phase mixing (\cite{Heyvaerts1983}, \cite{DeMoortel1999}, \cite{Hood2002}).

%\subsection{Non-dimensionalise}

We now consider a change of scale to non-dimensionalise; let ${\mathbf{\mathrm{v}}}_1 = \bar{\rm{v}} {\mathbf{v}}_1^*$, ${\mathbf{B}}_0 = B_0 {\mathbf{B}}_0^*$, ${\mathbf{B}}_1 = B_0 {\mathbf{B}}_1^*$, $x = L x^*$, $z=Lz^*$, $\nabla = \frac{1}{L}\nabla^*$ and $t=\bar{t}t^*$, where we let * denote a dimensionless quantity and $\bar{\rm{v}}$, $B_0$, $L$ and $\bar{t}$ are constants with the dimensions of the variable they are scaling. We then set $\frac {B_0}{\sqrt{\mu \rho _0 } } =\bar{\rm{v}}$ and $\bar{\rm{v}} =  {L} / {\bar{t}}$ (this sets $\bar{\rm{v}}$ as a sort of constant background Alfv\'{e}n speed). This process non-dimensionalises equations (\ref{eq:2.3}) and (\ref{eq:2.5}), and under these scalings, $t^*=1$ (for example) refers to $t=\bar{t}=  {L} / {\bar{\rm{v}}}$; i.e. the time taken to travel a distance $L$ at the background Alfv\'en speed. For the rest of this paper, we drop the star indices; the fact that they are now non-dimensionalised is understood.

The ideal linearised MHD equations naturally decouple into two equations for the fast
MHD wave and the Alfv\'en wave. The slow MHD wave is absent in
this limit and there is no velocity component along the background
magnetic field, as can be seen by taking the scalar product of equation (\ref{eq:2.3}) with ${\mathbf{B}} _0$.

The linearised equations for the fast magnetoacoustic wave are:
\begin{eqnarray}
\qquad \frac{\partial V}{\partial t} &=& v_A^2 \left( x,z \right) \left( \frac{\partial b_z}{\partial x} - \frac{\partial b_x}{\partial z}  \right) \nonumber \\
\frac{\partial b_x}{\partial t} &=& -\frac{\partial V}{\partial z} \; , \; \frac{\partial b_z}{\partial t} =   \frac{\partial V}{\partial x} \; \label{fastalpha},
\end{eqnarray}
where  ${\mathbf{B}} _1 = \left( b_x,0,b_z \right)$, the Alfv\'{e}n speed $v_A \left( x,z \right)$ is equal to $ \sqrt{B_x^2+B_z^2}$ and the variable $ V $ is related to the perpendicular velocity; $ V = \left[ \left( \mathbf{v} _1 \times { \mathbf{B} } _0 \right) \cdot {\hat{\mathbf{e}} }_y \right] $.
These equations can be combined to form a single wave equation:
\begin{eqnarray}
\qquad \frac{\partial ^2 V}{\partial t^2} = v_A^2 \left( x,z \right) \left( \frac{\partial^2 V}{\partial x^2} + \frac{\partial ^2 V}{\partial z^2}  \right) \; \label{fastbeta}.
\end{eqnarray}

The linearised equations for the Alfv\'en wave, with ${\mathbf{v}} _1 = \left( 0,v_y,0 \right)$ and ${\mathbf{B}} _1 = \left( 0,b_y,0 \right)$ are:
\begin{eqnarray}
\qquad \frac {\partial v_y }{\partial t} = B_x \frac {\partial b_y }{\partial x} + B_z\frac {\partial b_y }{\partial z} \; , \quad \frac {\partial b_y }{\partial t} = B_x \frac {\partial v_y }{\partial x} + B_z\frac {\partial v_y }{\partial z} \; \label{alfvenalpha} ,
\end{eqnarray}
which can be combined to form a single wave equation:
\begin{eqnarray}
\qquad \frac {\partial^2 v_y }{\partial t^2} = \left(B_x \frac {\partial }{\partial x} +B_z\frac {\partial }{\partial z} \right) ^2 v_y \; \label{alfvenbeta}.
\end{eqnarray}

It is worth noting that the zero $\beta$ approximation will be invalid near the null points, where the pressure terms become important. In fact, a finite $\beta$ model would introduce slow waves, which would be driven by the excited fast waves (\cite{McLaughlin2006}). Also, the model is linear but the inhomogenity of the medium would lead to nonlinear coupling of the modes (\cite{Nakariakov1997}).

\subsection{Alfv\'en speed profile}\label{jeangrey}

\begin{figure*}[t]
\begin{center}
\includegraphics[width=2.0in]{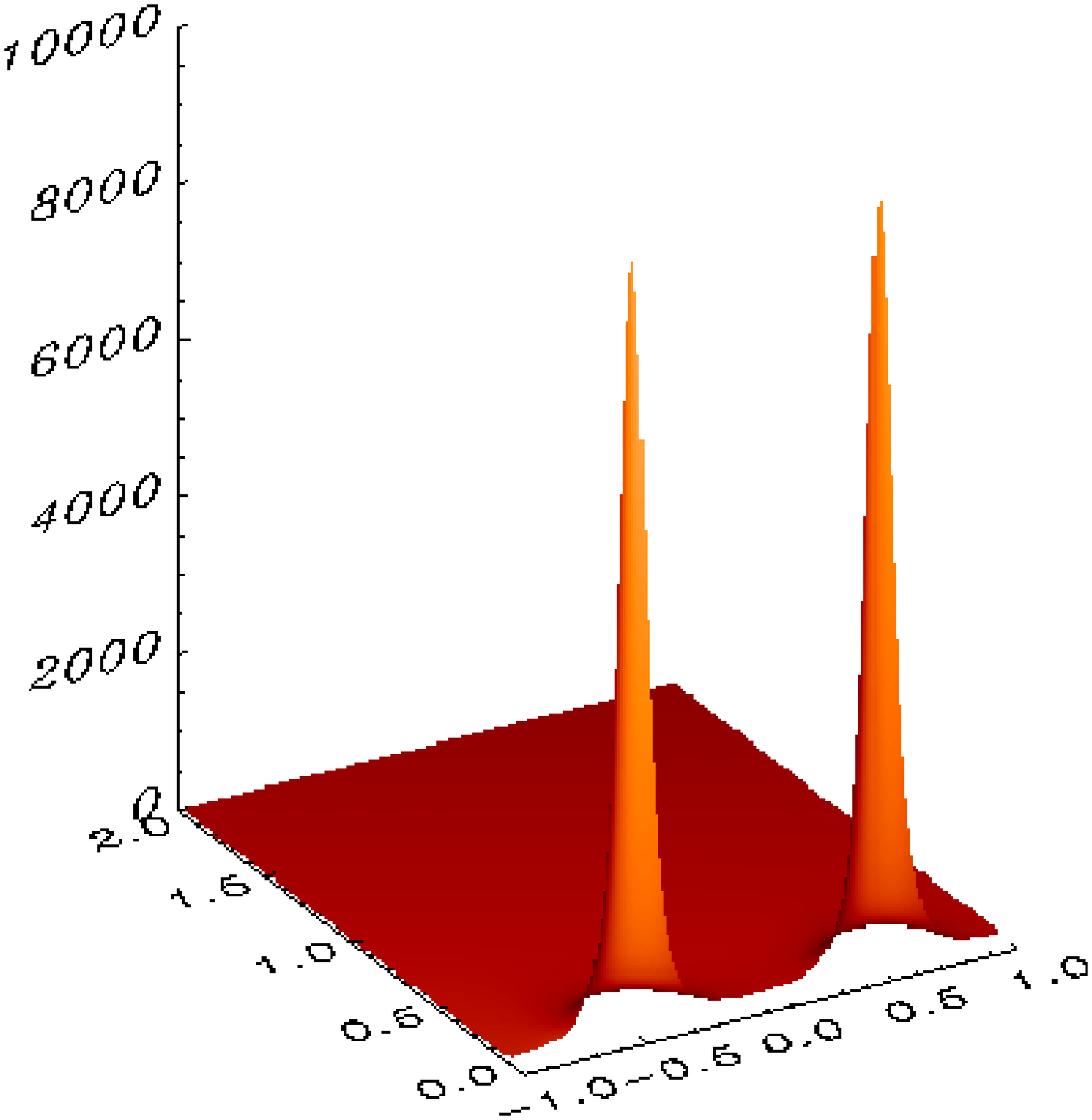}
\hspace{0.2in}
\includegraphics[width=2.0in]{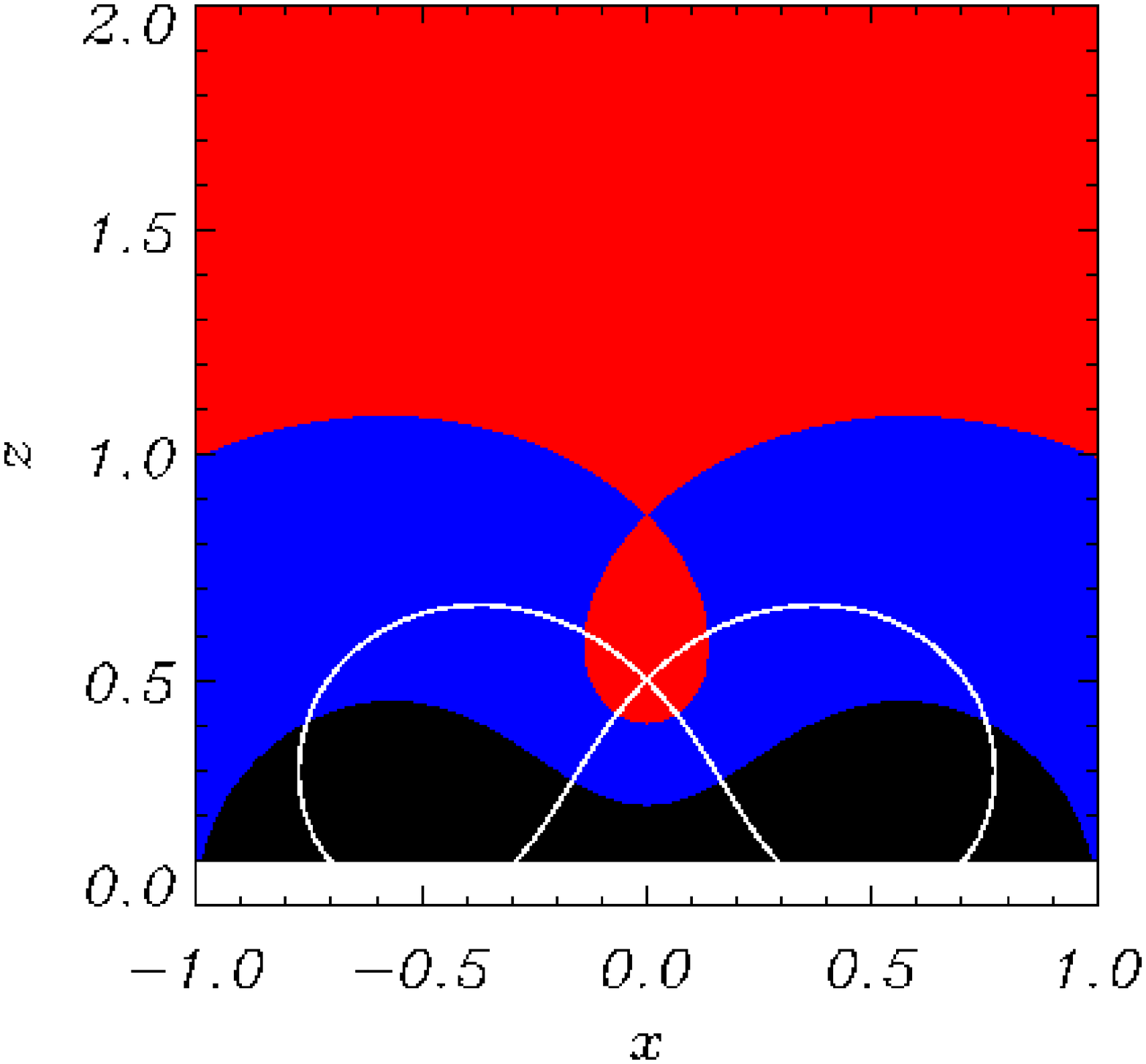}
\caption{{\emph{Left}} shows a shaded surface of  $v_A^2$. {\emph{Right}} shows a colour contour of $v_A^2$; red represents values  $0 \le v_A^2 \le 1$, blue  $1 \le v_A^2 \le 20$ and black  $v_A^2 \ge 20$. The white curve shows  the magnetic skeleton. Both representations are cut-off at $z=z_0=0.1$.}
\label{VA2}
\end{center}
\end{figure*}

We shall be solving equation (\ref{fastbeta}) numerically in the following sections and sending  a planar wave into the system from the lower boundary of our numerical box. However, due to our choice of equilibrium  magnetic field, there are two singularities in the field at $(x,z) =(\pm a,0)$ and hence  $v_A^2 \rightarrow \infty$ at these points. If we were to generate a wave along $z=0$, our numerical scheme would be  unable to deal with this extreme speed differential. Thus, in this chapter we will generate our waves not along $z=0$ but along $z=z_0$, where $z_0$ is small (and so close to the  $z=0$ line). This choice still starts the waves in a region of varying Alfv\'en speed, but reduces the massive differential to something easier to work with numerically. Thus, starting the waves at $z=z_0$ results in very little loss of information about the system. In addition, the CFL  condition dictates that our timestep has to be very small (because our speed is so large in some places). Thus, to reduce the computing time and with very little loss of understanding, $z_0=0.1$ was set in the experiments  (recall $a=0.5$ in our simulations).

Figure \ref{VA2} shows a  shaded surface (left) and colour contour (right) of $v_A^2(x,z) = B_x^2(x,z) + B_z^2(x,z)$. The shaded surface clearly shows that the velocity profile changes massively across the magnetic region and reaches a maximum at  $(x,z)=(\pm a,0)$. The contour (right) is  colour coded; red represents values  $0 \le v_A^2 \le 1$, blue represents  $1 \le v_A^2 \le 20$ and black represents  $v_A^2 \ge 20$. Thus, we can see that around the X-point there is a small island of low  Alfv\'en speed. The white curve shows  the skeleton of the magnetic field (equation \ref{twodipoles}).

\section{Fast waves}\label{sec:2}

We now look at the behaviour of fast MHD waves in the neighbourhood of two dipoles. There are three scenarios which we shall visit in order to fully understand the system. Firstly, the two dipoles create a natural X-point and we want to know if the fast wave wraps around this null point in a similar way to that described in Paper I. Thus, Simulation One describes a localised, linear fast wave approaching this X-point. This magnetic configuration also contains regions of very high  Alfv\'en speed along the lower boundary. Thus, Simulation Two describes a linear fast wave generated in this area, but localised to one side of the (symmetric) system. Simulation Three describes a scenario similar to Simulation Two, but now with a pulse generated along the entire lower boundary. Simulations One and Two will build towards an understanding of the complete system of Simulation Three.

All three simulations generate a linear fast wave pulse on the lower boundary, which then propagates up into the magnetic configuration. The lower boundary was set at $z_0$ in all three experiments, where $z_0=0.1$, $a=0.5$, $\omega=8\pi$ and we focus our attention in the region  $-1 \leq x \leq 1$ and $z_0 \leq z \leq 2$.

%The areas inside contain areas of very high (background) Alfv\'en speed.  
%Initially, $z_0=0$ was set, but  the CFL condition (Section \ref{CFL Condition}) dictates that our timestep has to be very small. Thus, to reduce the computing time and with very little loss of understanding, $z_0=0.1$ was set in the experiments where $a=0.5$.

\subsection{Simulation One}

We solve the linearised MHD equations for the fast wave (equations \ref{fastalpha}), numerically using a two-step Lax-Wendroff scheme. The numerical scheme is run in a box with $-0.6 \leq x \leq 0.6$ and $z_0 \leq z \leq 1$, using the magnetic field shown in Figure \ref{fig:dipolemagneticfield}. We initially consider a single wave pulse coming in from the bottom boundary, localised along $-0.2 \leq x \leq 0.2$, $z=z_0=0.1$.  The boundary conditions were set such that:
\begin{eqnarray*}
V(x, z_0,t) = \left\{  \begin{array}{cl} 
\sin { \omega t } \;\sin{\left[\frac{5\pi}{2}\left( {x +1}\right)\right]}  & {\mathrm{for}}\; \left\{ \begin{array}{c} 
 {-0.2 \leq x \leq 0.2}\\
 {0 \leq t \leq \frac {\pi}{\omega}}\end{array}\right. \\
0 & { \mathrm{otherwise} }\end{array} \right. \\
\left.\frac {\partial V} {\partial x }  \right| _{x=-0.6} = 0 \; , \quad \left.\frac {\partial V} {\partial x }  \right| _{x=0.6} = 0 \; , \quad \left.\frac {\partial V} {\partial z }   \right| _{z=1}  = 0 \; .
\end{eqnarray*}
Tests show that the central behaviour is largely unaffected by these choices of side and top boundary conditions. The other boundary conditions on the perturbed magnetic field follow from the remaining equations and the solenodial condition.

\begin{figure*}[t]
\hspace{0in}
\vspace{0.1in}
%\hspace{0.2in}
\includegraphics[width=1.2in]{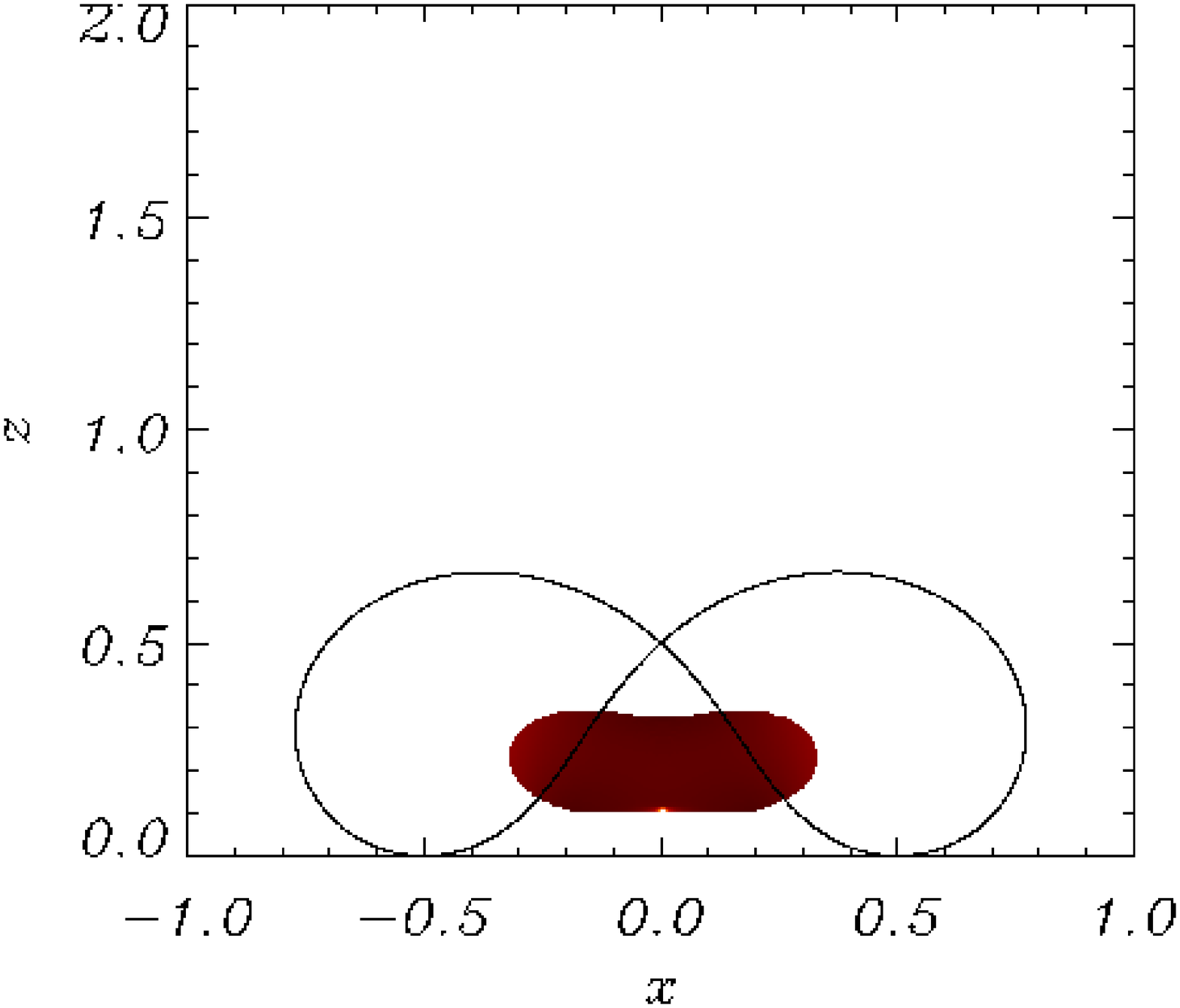}
\hspace{0.0in}
\includegraphics[width=1.2in]{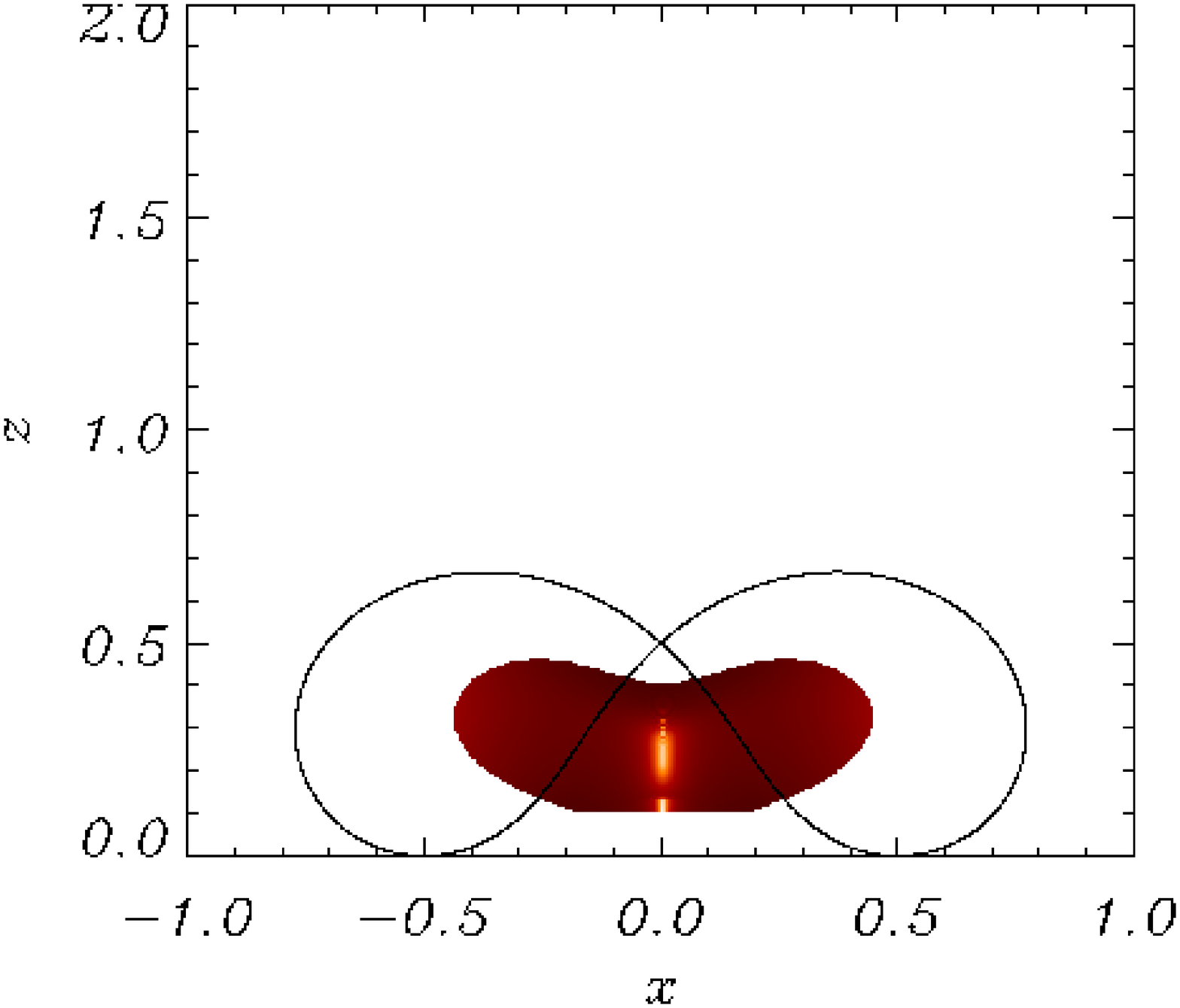}
\hspace{0.0in}
\includegraphics[width=1.2in]{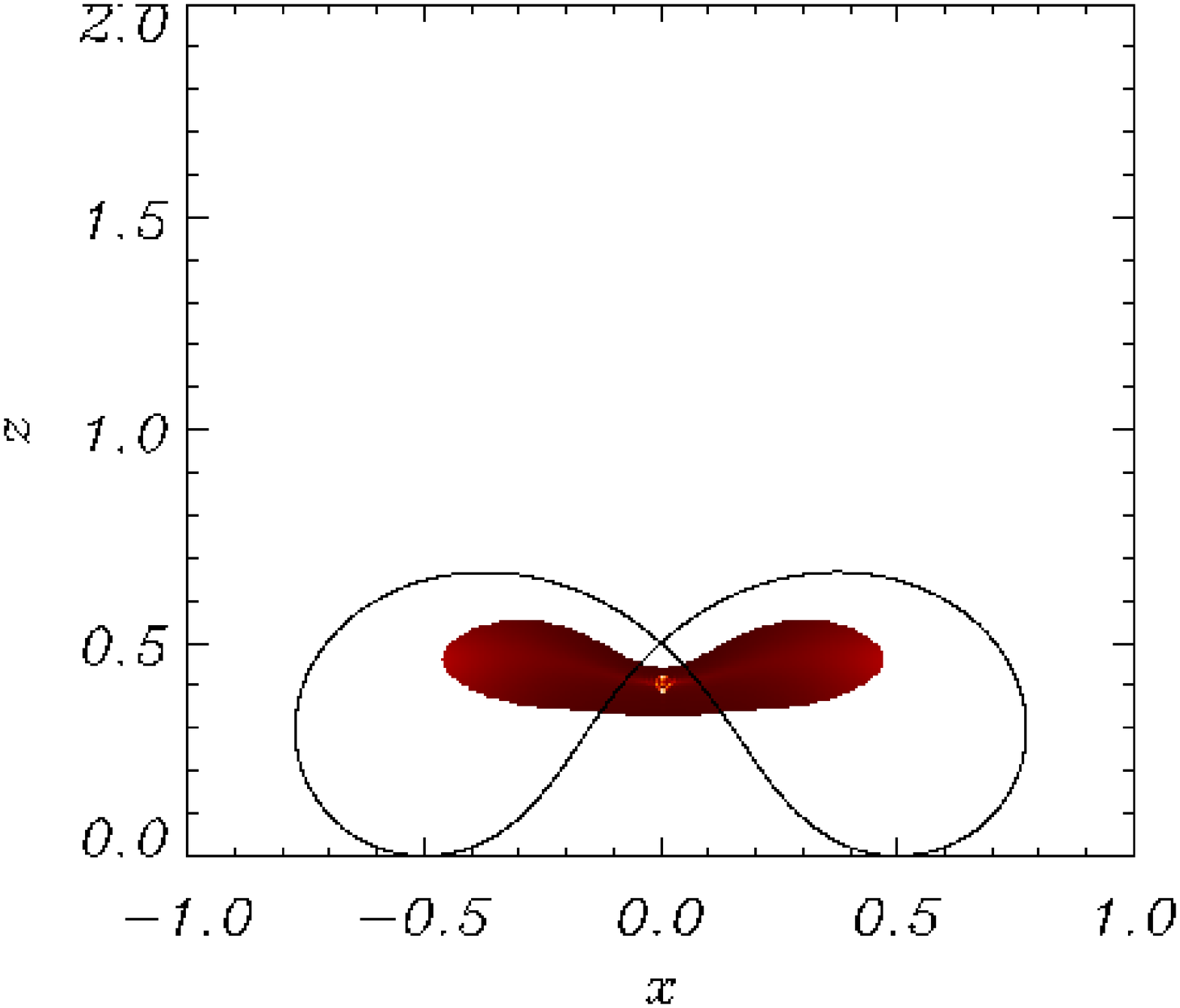}
\hspace{0.0in}
\includegraphics[width=1.2in]{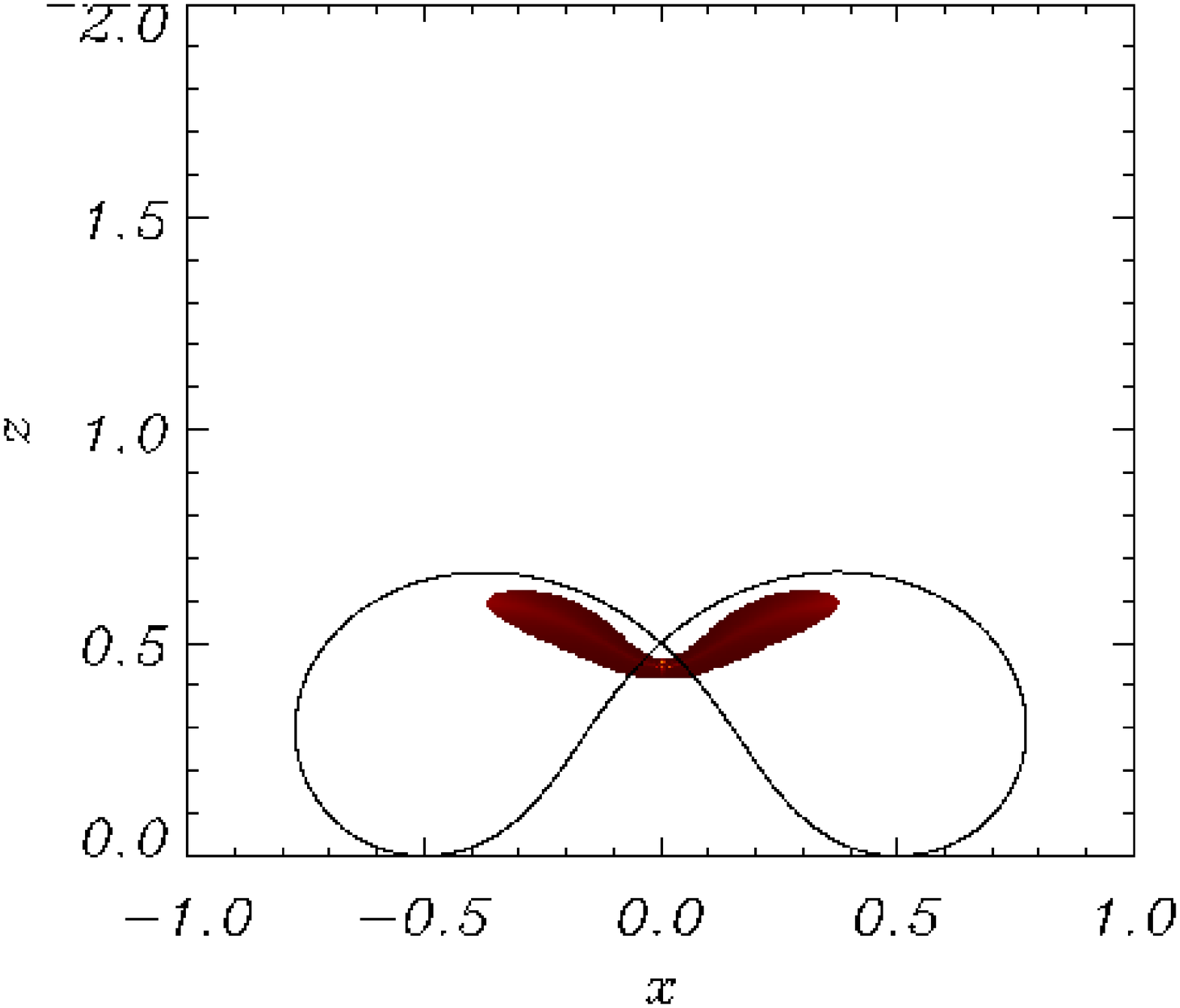}\\
\hspace{0in}
\vspace{0.1in}
%\hspace{0.2in}
\includegraphics[width=1.2in]{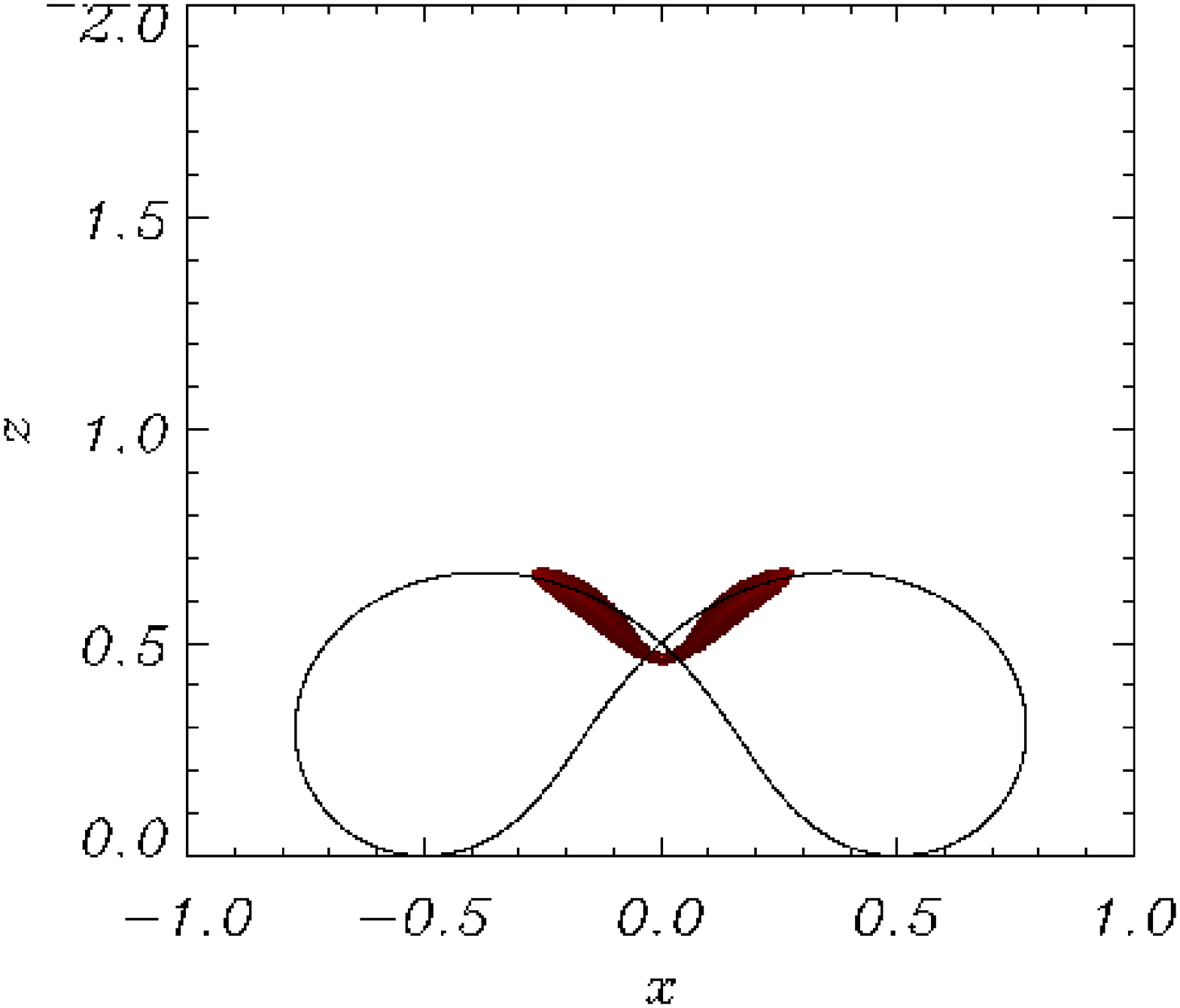}
\hspace{0.0in}
\includegraphics[width=1.2in]{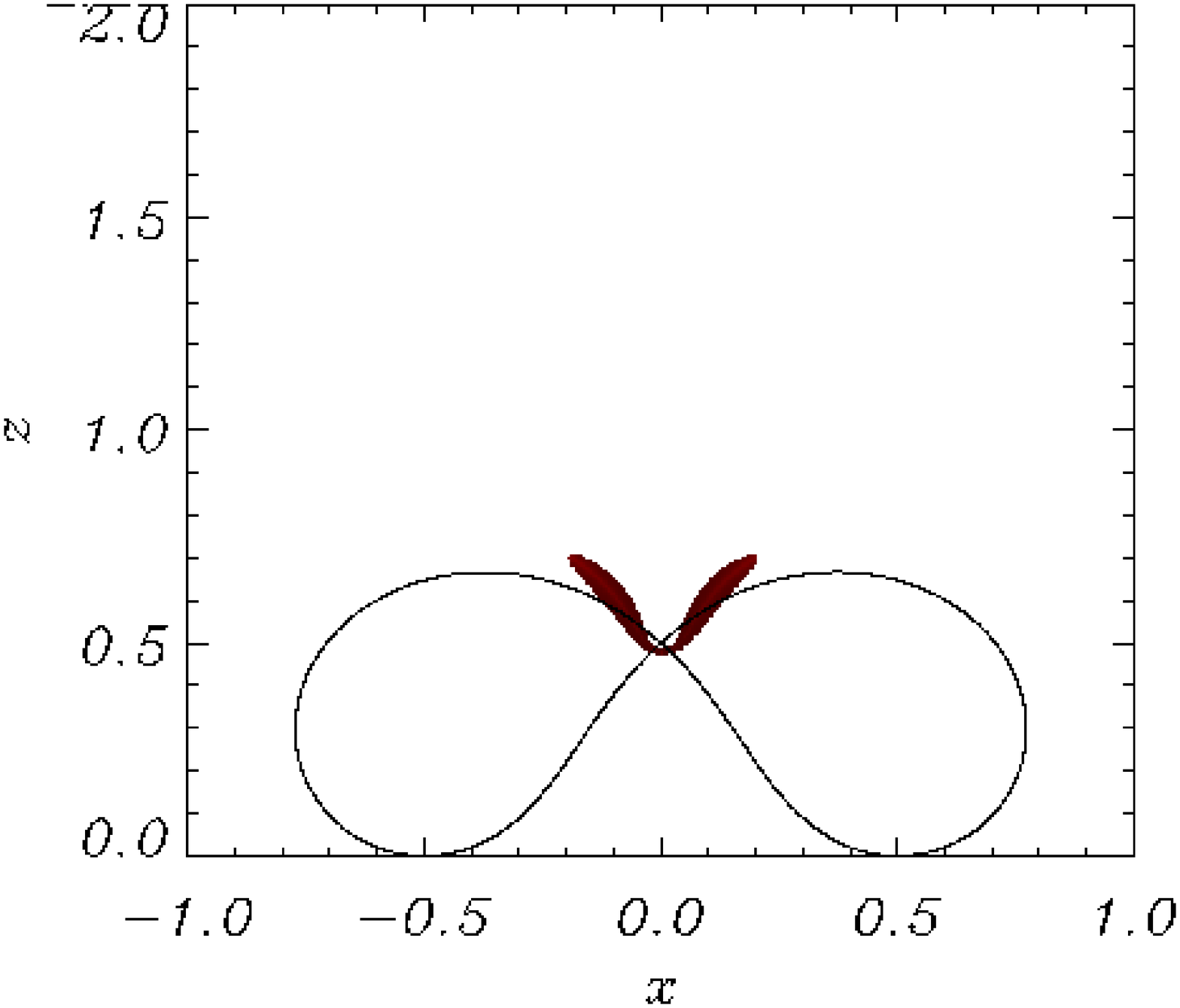}
\hspace{0.0in}
\includegraphics[width=1.2in]{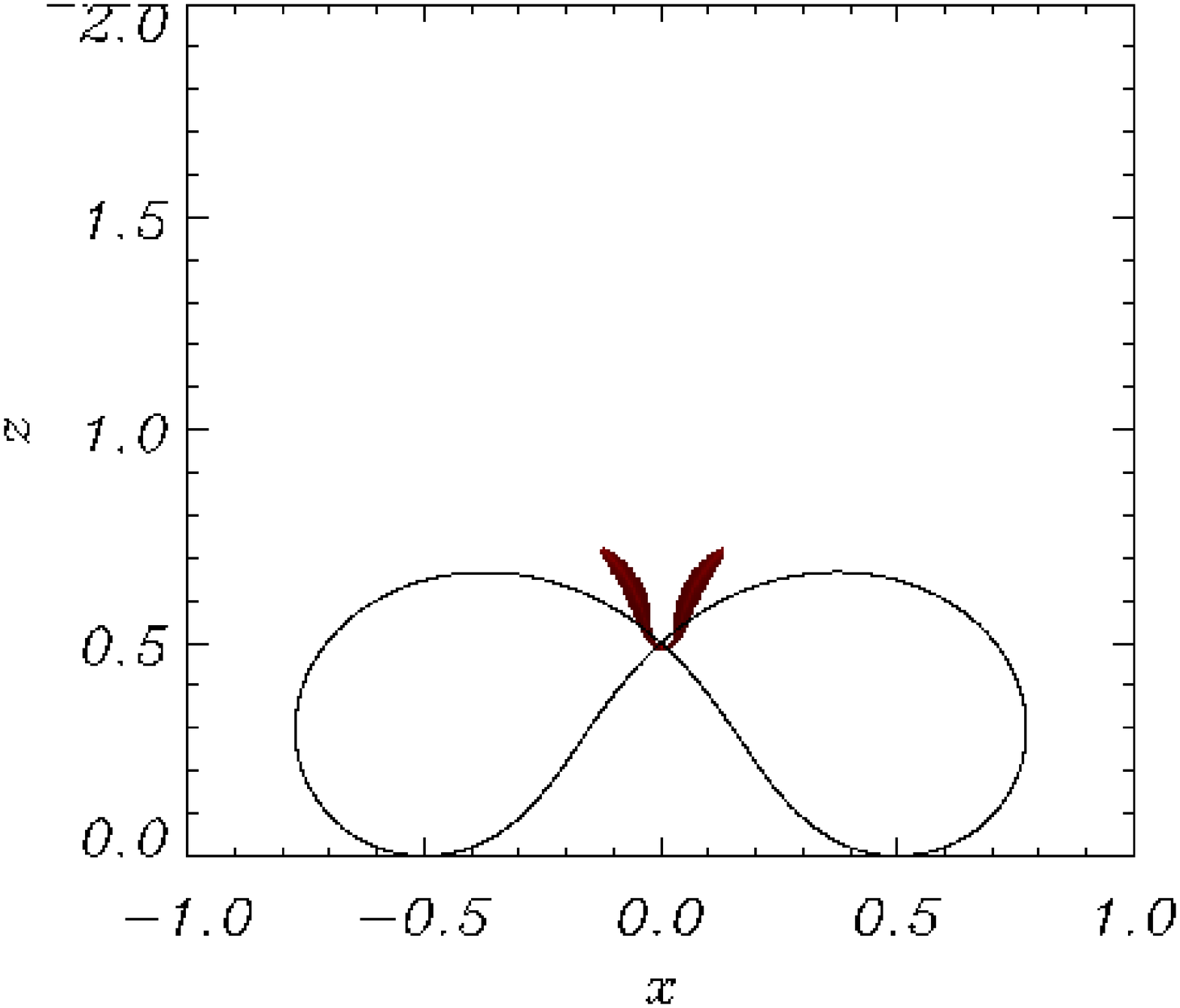}
\hspace{0.0in}
\includegraphics[width=1.2in]{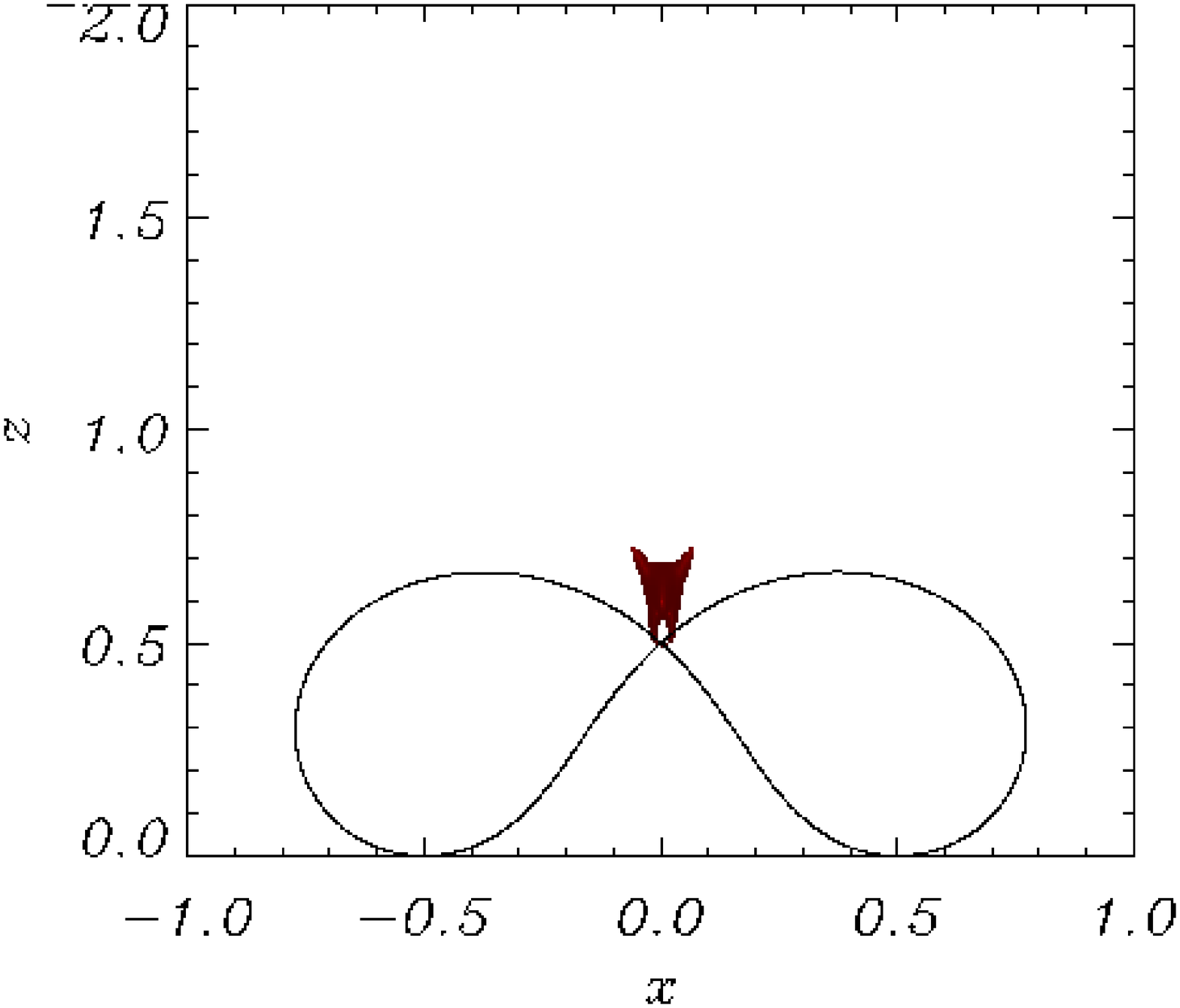}\\
\hspace{0in}
\vspace{0.1in}
%\hspace{0.2in}
\includegraphics[width=1.2in]{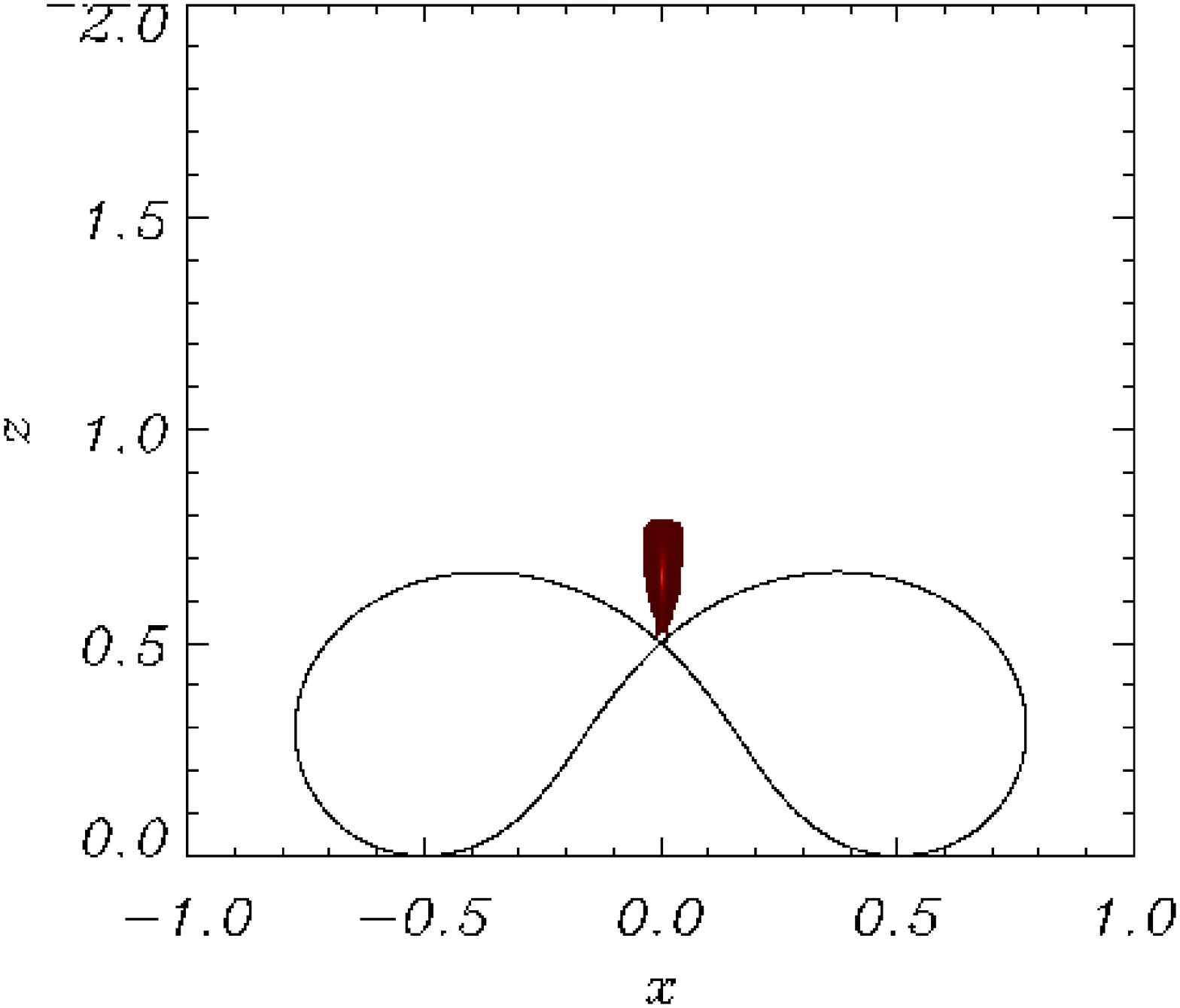}
\hspace{0.0in}
\includegraphics[width=1.2in]{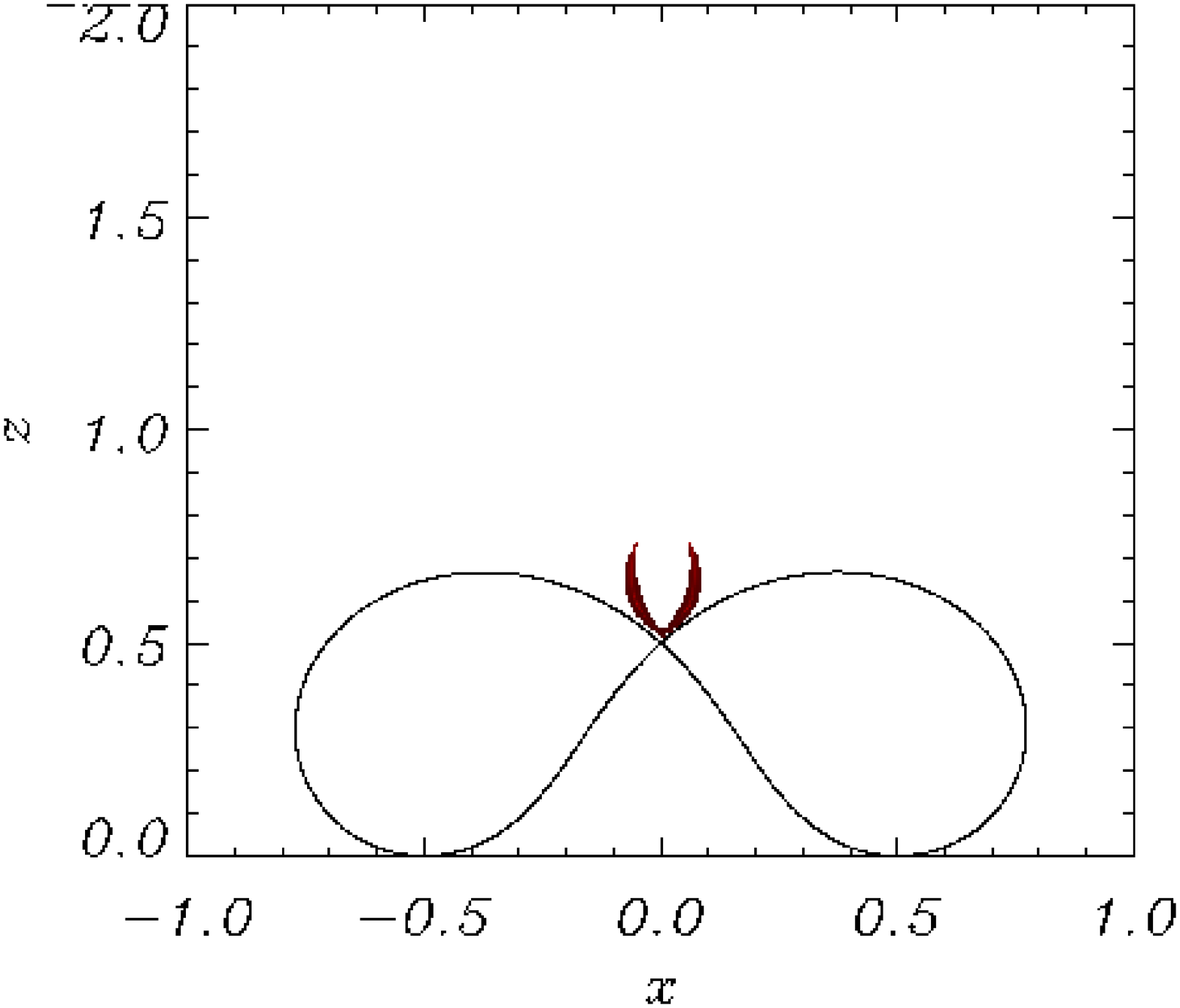}
\hspace{0.0in}
\includegraphics[width=1.2in]{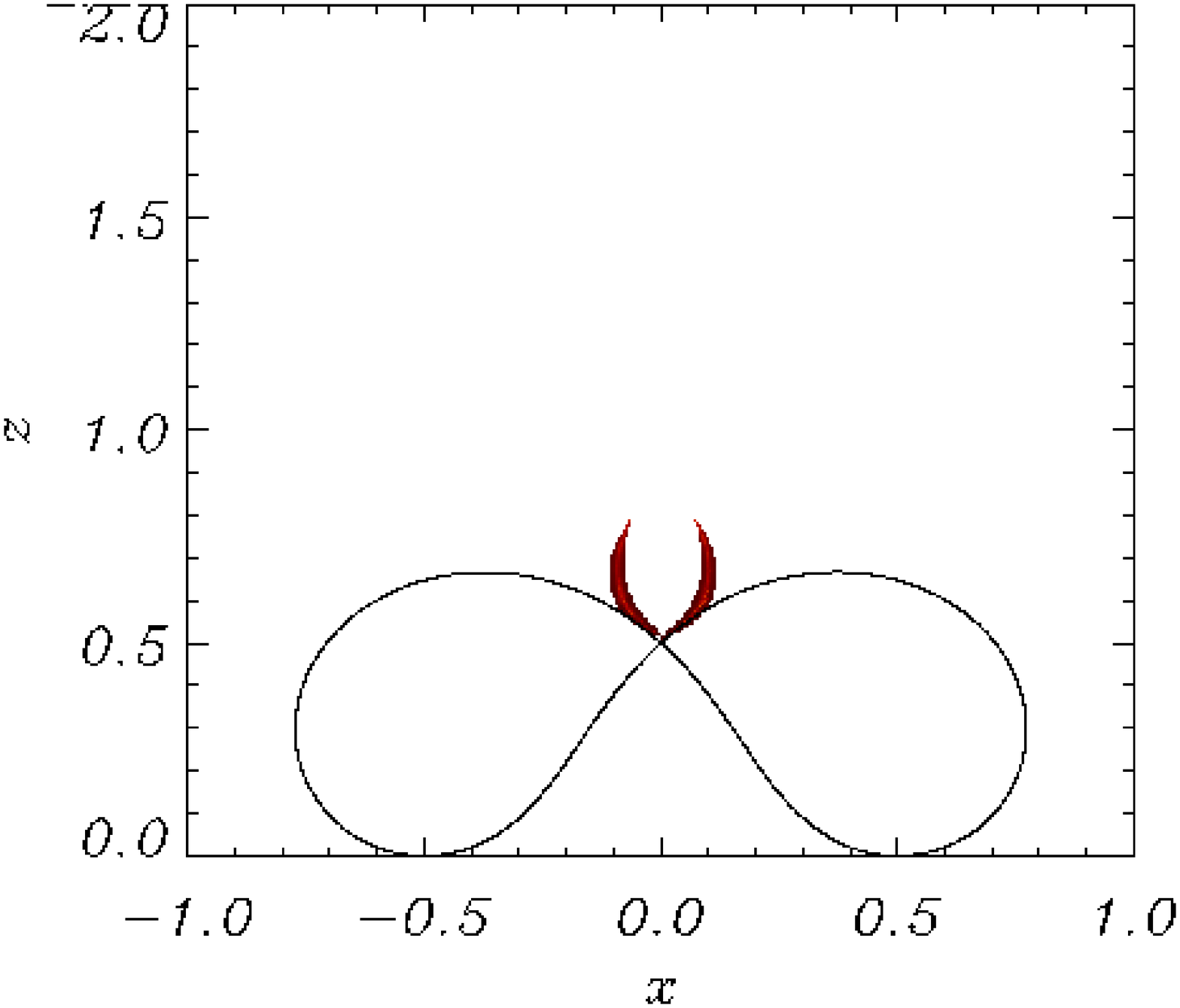}
\hspace{0.0in}
\includegraphics[width=1.55in]{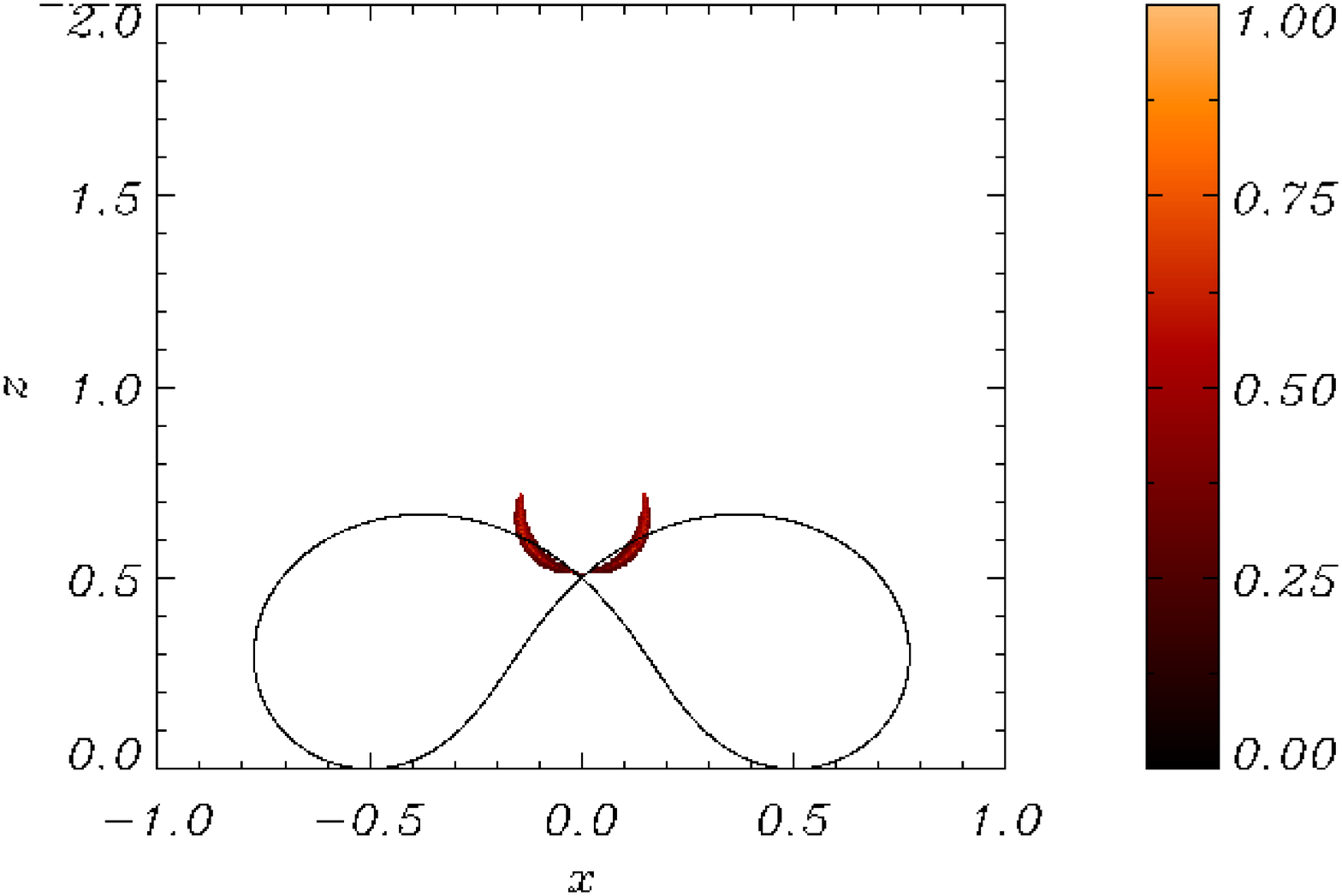}
\caption{Contours of numerical simulation $V$ for a fast wave sent in from lower boundary for $-0.2 \leq x \leq 0.2 $, $z=0.1$ and its resultant propagation at times $(a)$ $t$=0.05, $(b)$ $t$=0.1, $(c)$ $t$=0.15, $(d)$ $t=$0.2, $(e)$ $t$=0.25 and $(f)$ $t$=0.3, $(g)$ $t$=0.35, $(h)$ $t$=0.4, $(i)$ $t$=0.45, $(j)$ $t=$0.5, $(k)$ $t$=0.55 and $(l)$ $t$=0.6, labelling from top left to bottom right. The black curve shows  the magnetic skeleton with the null point located at $(x,z)=(0,0.5)$.}
\label{fig:4.5.2.1}
\end{figure*}

We find that the linear, fast magnetoacoustic wave propagates up from the lower boundary (from $z_0$) and begins to wrap around the null point. This can be seen in  Figure \ref{fig:4.5.2.1}. In the early subfigures, we can see that the pulse travels faster at the ends (wings) than in its centre, due to the variation in Alfv\'en speed. The wave then continues to \emph{refract} around the X-point, and wraps itself around it. The  Alfv\'en speed is zero at $(x,z)=(0,0.5)$ (the null)  and so the fast wave cannot cross this point. Consequently, the null acts as an immovable barrier to the propagation of the wave and thus is a focus for the wrapping around effect. The wave continues to wrap around the null and becomes thinner. The wave (eventually) accumulates very close to the X-point. This is a similar effect to that described in Paper I. This refraction effect is a key feature of fast wave propagation.

Since the Alfv\'{e}n speed drops to zero at the null point, the wave never reaches there, but the length scales (this can be thought of as the distance between the leading and trailing edges of the wave pulse) rapidly decrease, indicating that the current (and all other gradients) will increase.

% As a simple illustration, consider the wavefront as it propagates up the $z$ axis at $x=0$. Here the vertical velocity is $v_z = \frac {d z} {d t} = -z $. 

%Here the vertical velocity is $v_z = \frac {d z} {d t} = -z $. Thus, the start of the wave is located at a position $z_s=6 e^{-t}$, when the wave is initally at $z=6$. If the end of the wave leaves $z=6$ at $t=t_0$ then the location of the end of the wave is $z_e=6 e^{t_0 - t}$. Thus, the distance between the leading and trailing edge of the wave is $\delta z = 6 \left( e^{t_0} -1 \right) e^{-t} $ and this decreases with time, suggesting that all gradients will increase exponentially.

%\clearpage

%\newpage
%\newpage

\subsection{Simulation Two}

%\subsubsection{Numerical simulation}

In this second simulation, we again solve the linearised MHD equations for the fast wave numerically (using the same two-step Lax-Wendroff scheme). We again  consider a single wave pulse coming in from the lower  boundary, but now localised along $-1 \leq x \leq 0$, $z=z_0=0.1$. This allows us to explore a different topological area than that of Simulation One. The numerical scheme is run in a box with $-1.4 \leq x \leq 1.4$ and $z_0 \leq z \leq 1.7$. The boundary conditions were set such that:
\begin{eqnarray*}
V(x, z_0,t) = \left\{  \begin{array}{cl} 
 \sin { \omega t } \;\sin{\left[\pi\left( {x +1}\right)\right]}  & {\mathrm{for}} \; \left\{  \begin{array}{c}
{-1 \leq x \leq 0} \\
{0 \leq t \leq \frac {\pi}{\omega}}\end{array} \right. \\
0 & { \mathrm{otherwise} }\end{array} \right. \\
\left.\frac {\partial V } {\partial x }  \right| _{x=-1.4} = 0 \; , \quad \left.\frac {\partial V} {\partial x }   \right| _{x=1.4} = 0 \; , \quad \left.\frac {\partial  V } {\partial z }  \right| _{z=1.7}  = 0 \; .
\end{eqnarray*}

\begin{figure*}[t]
\hspace{0in}
\vspace{0.1in}
%\hspace{0.2in}
\includegraphics[width=1.2in]{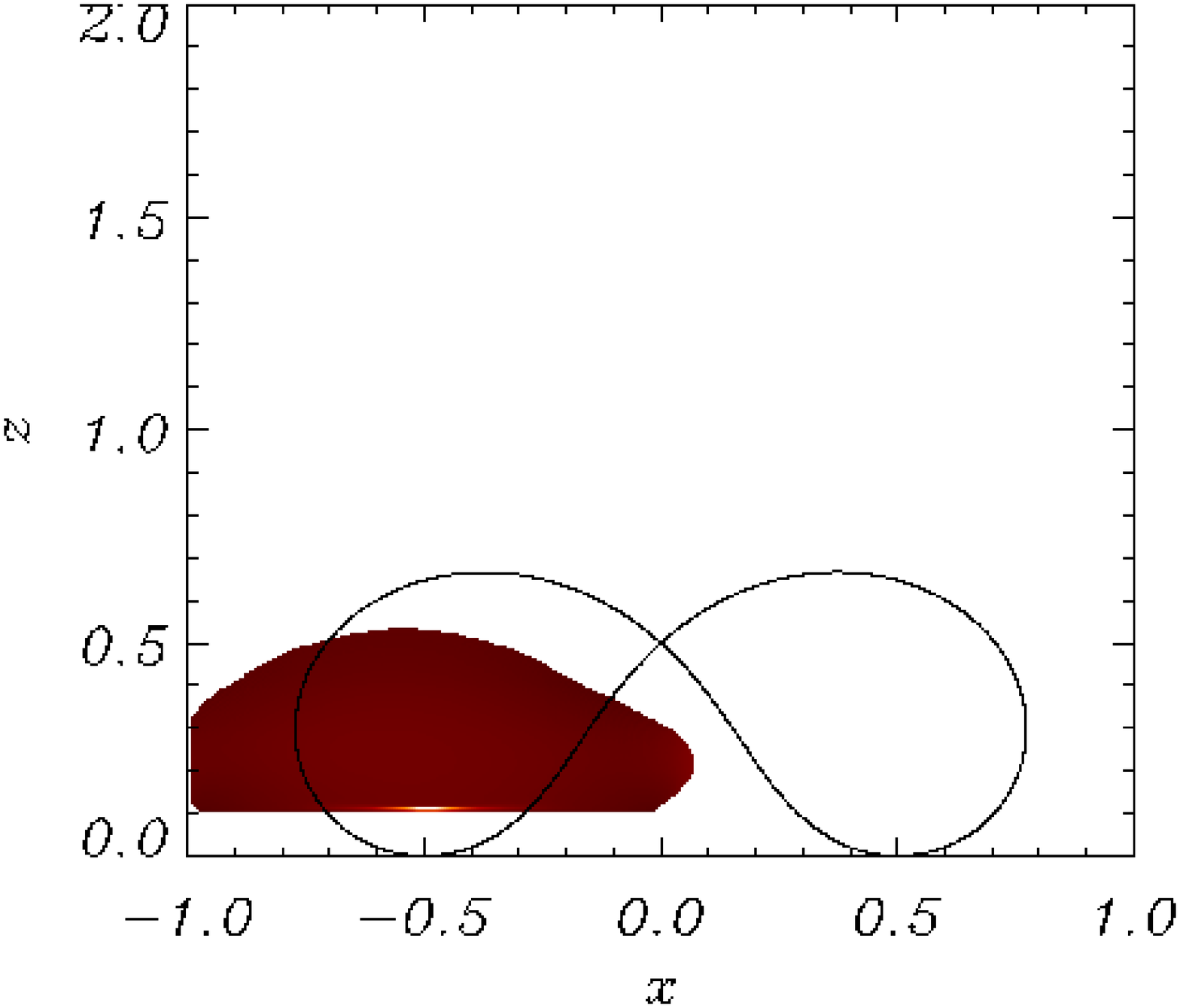}
\hspace{0.0in}
\includegraphics[width=1.2in]{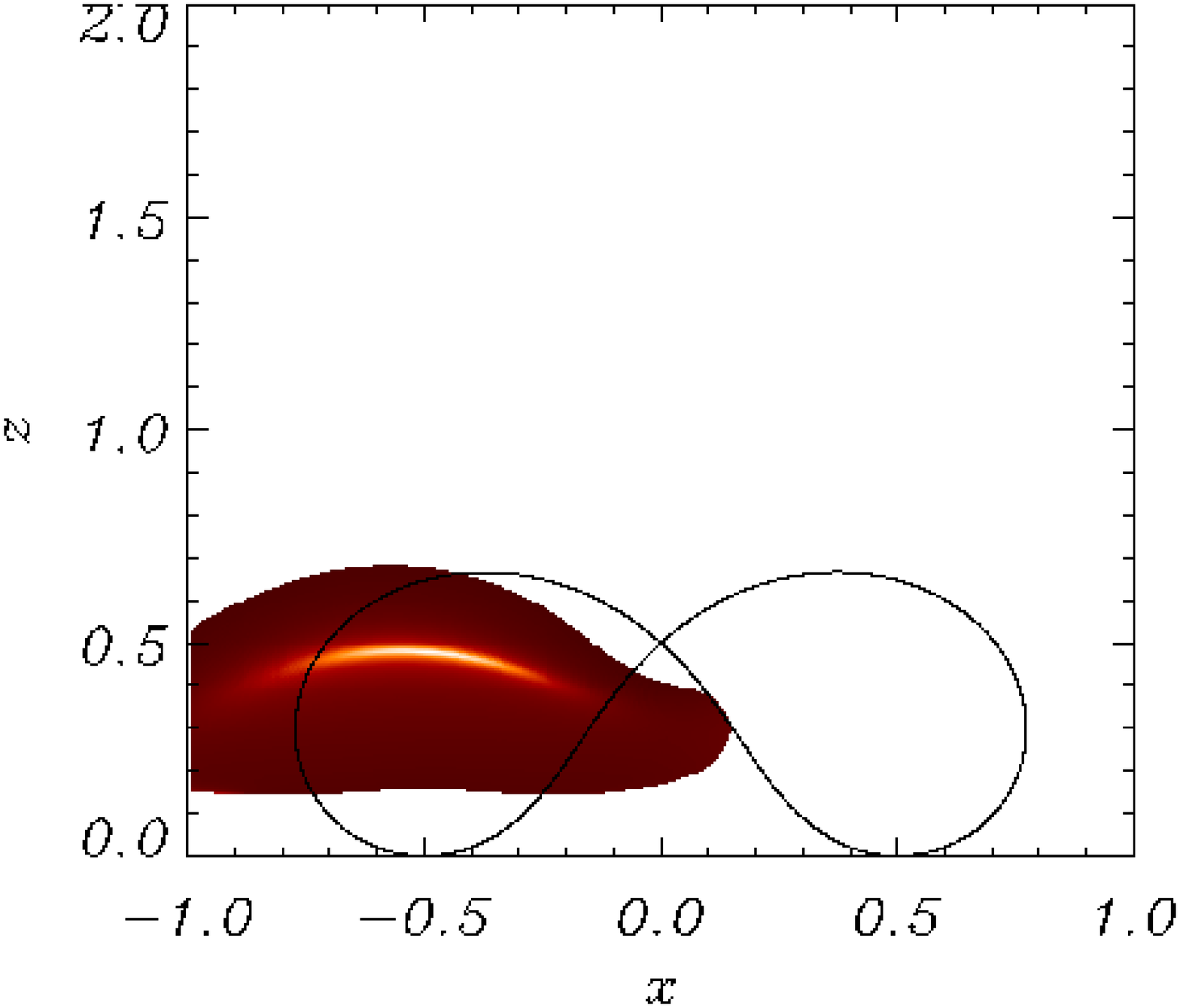}
\hspace{0.0in}
\includegraphics[width=1.2in]{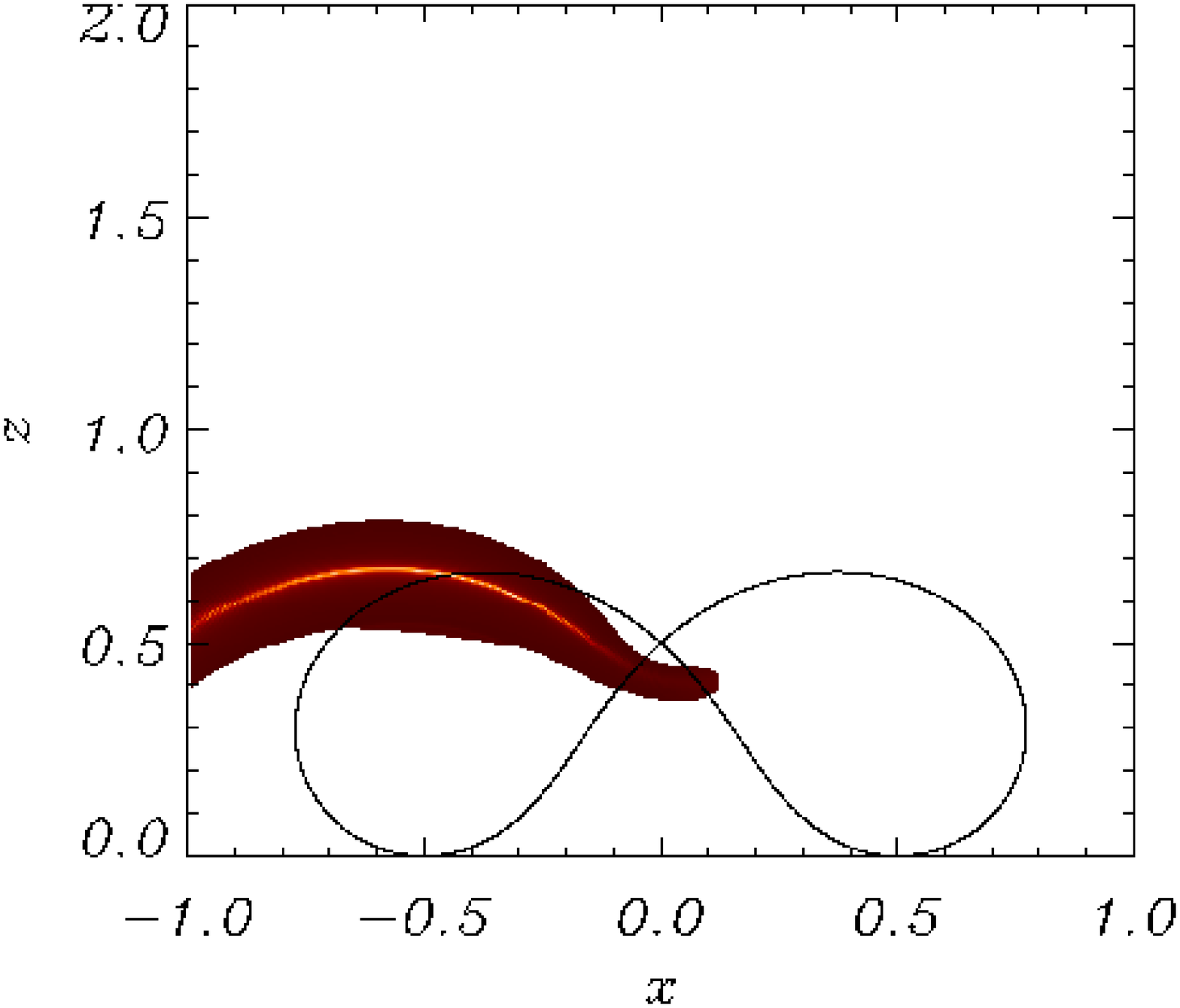}
\hspace{0.0in}
\includegraphics[width=1.2in]{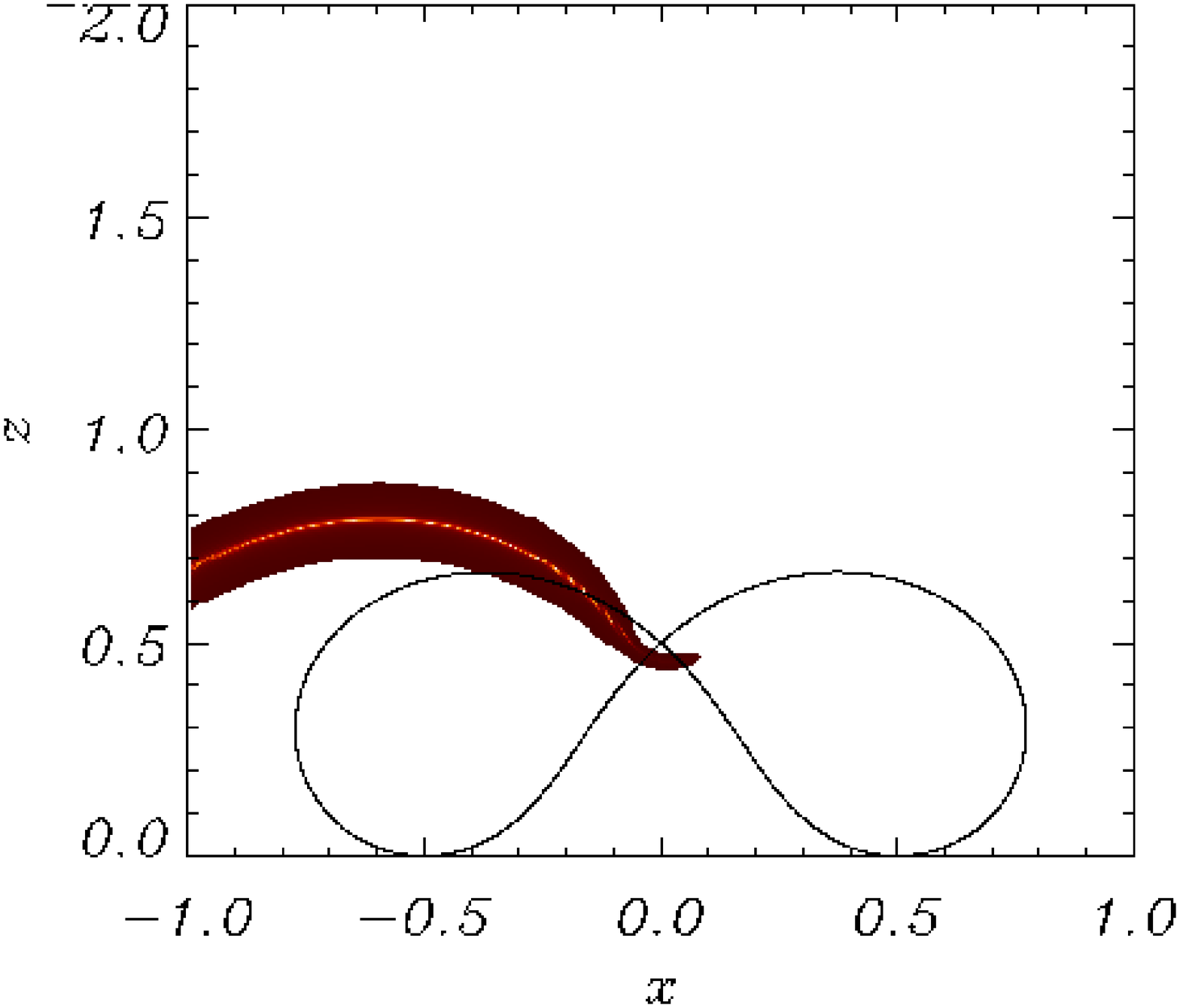}\\
\hspace{0in}
\vspace{0.1in}
%\hspace{0.2in}
\includegraphics[width=1.2in]{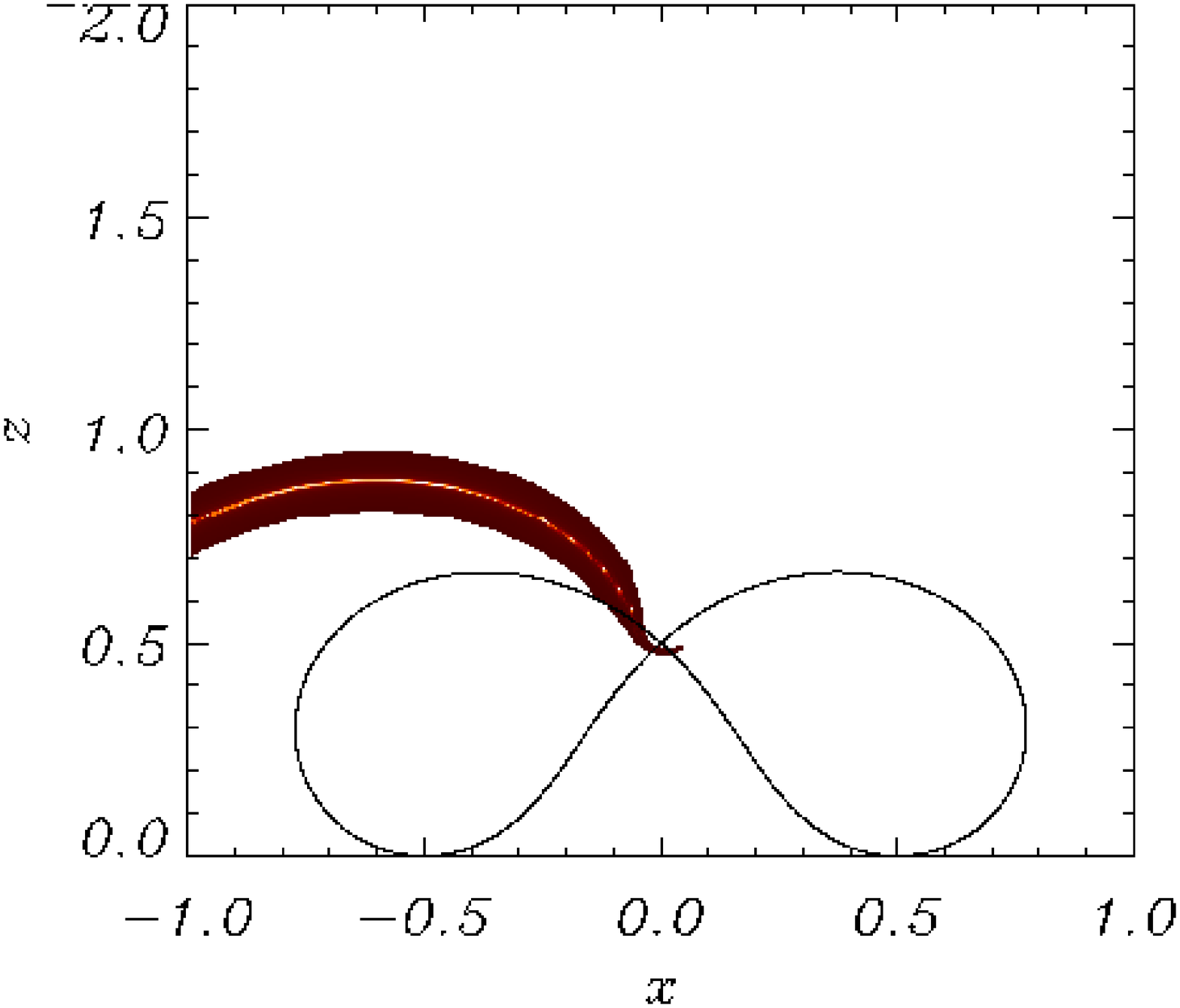}
\hspace{0.0in}
\includegraphics[width=1.2in]{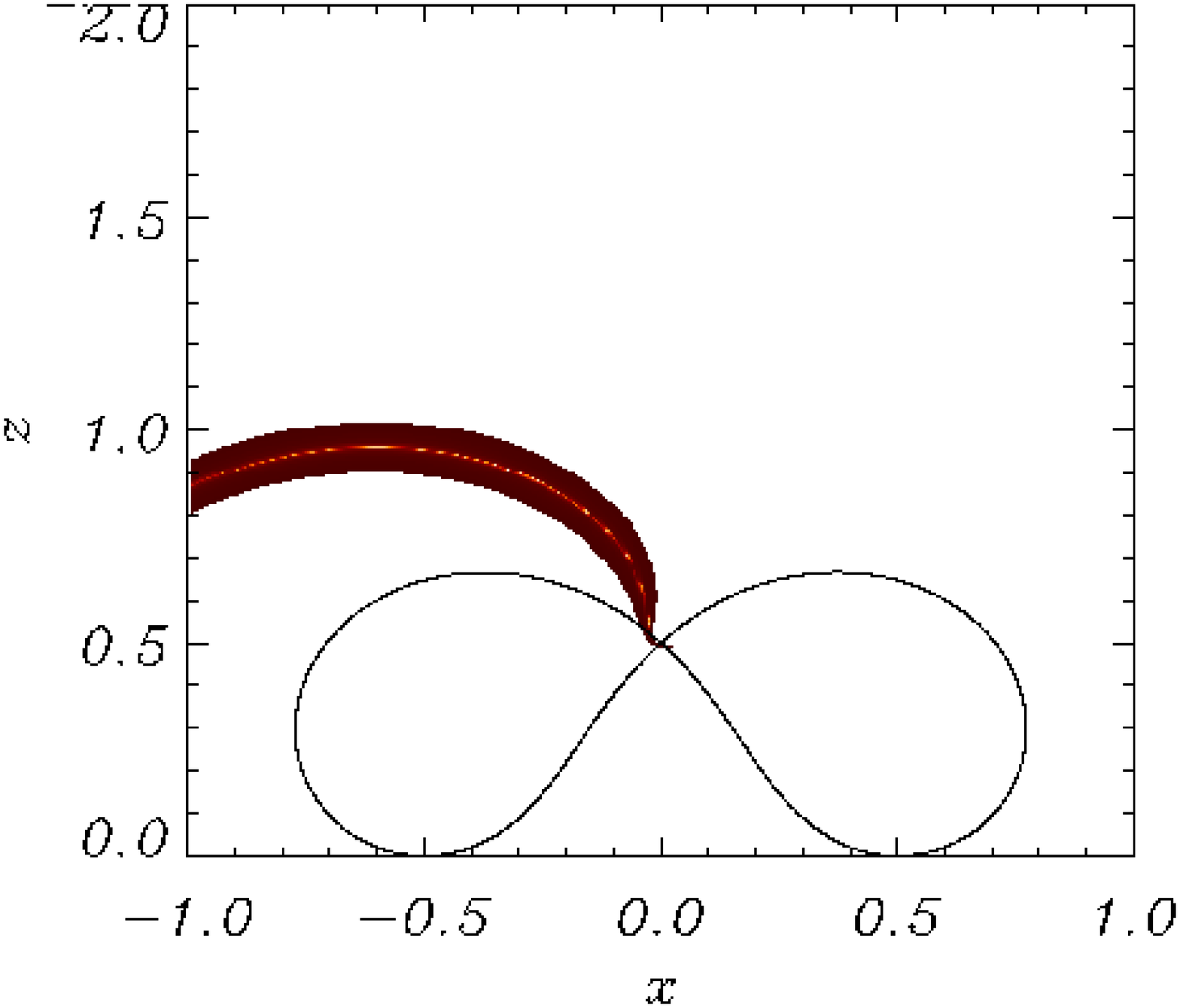}
\hspace{0.0in}
\includegraphics[width=1.2in]{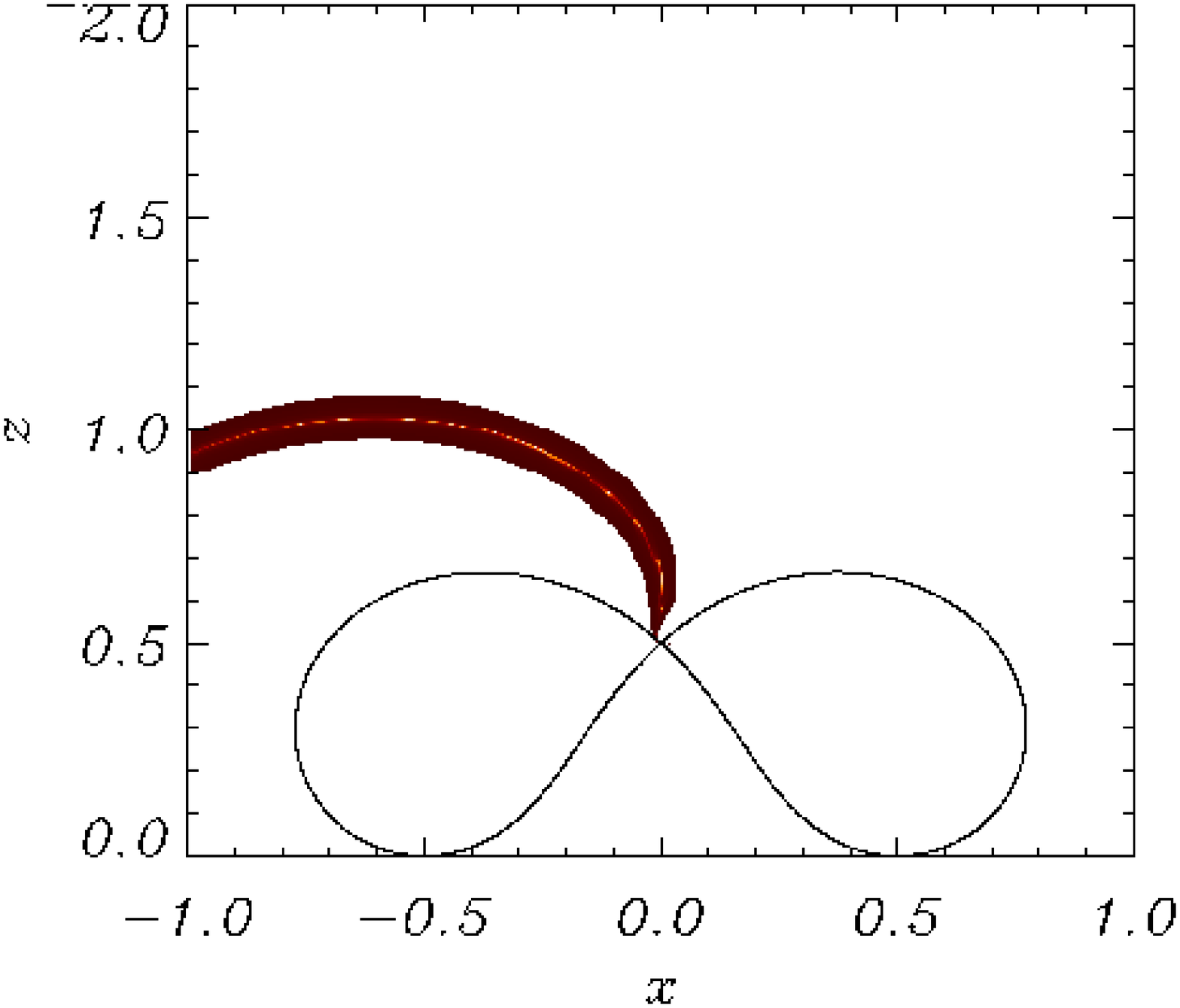}
\hspace{0.0in}
\includegraphics[width=1.2in]{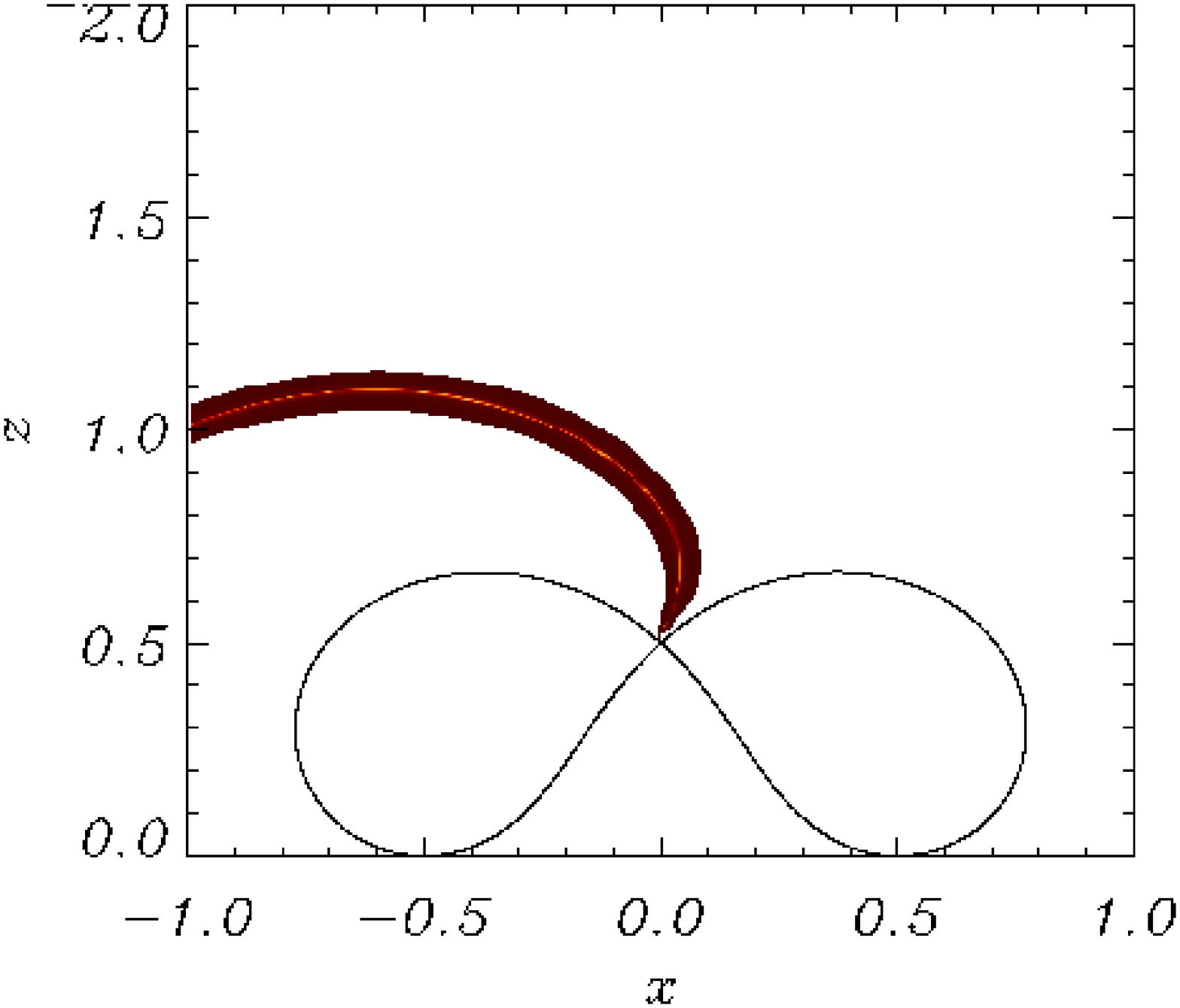}\\
\hspace{0in}
\vspace{0.1in}
%\hspace{0.2125in}
\includegraphics[width=1.2in]{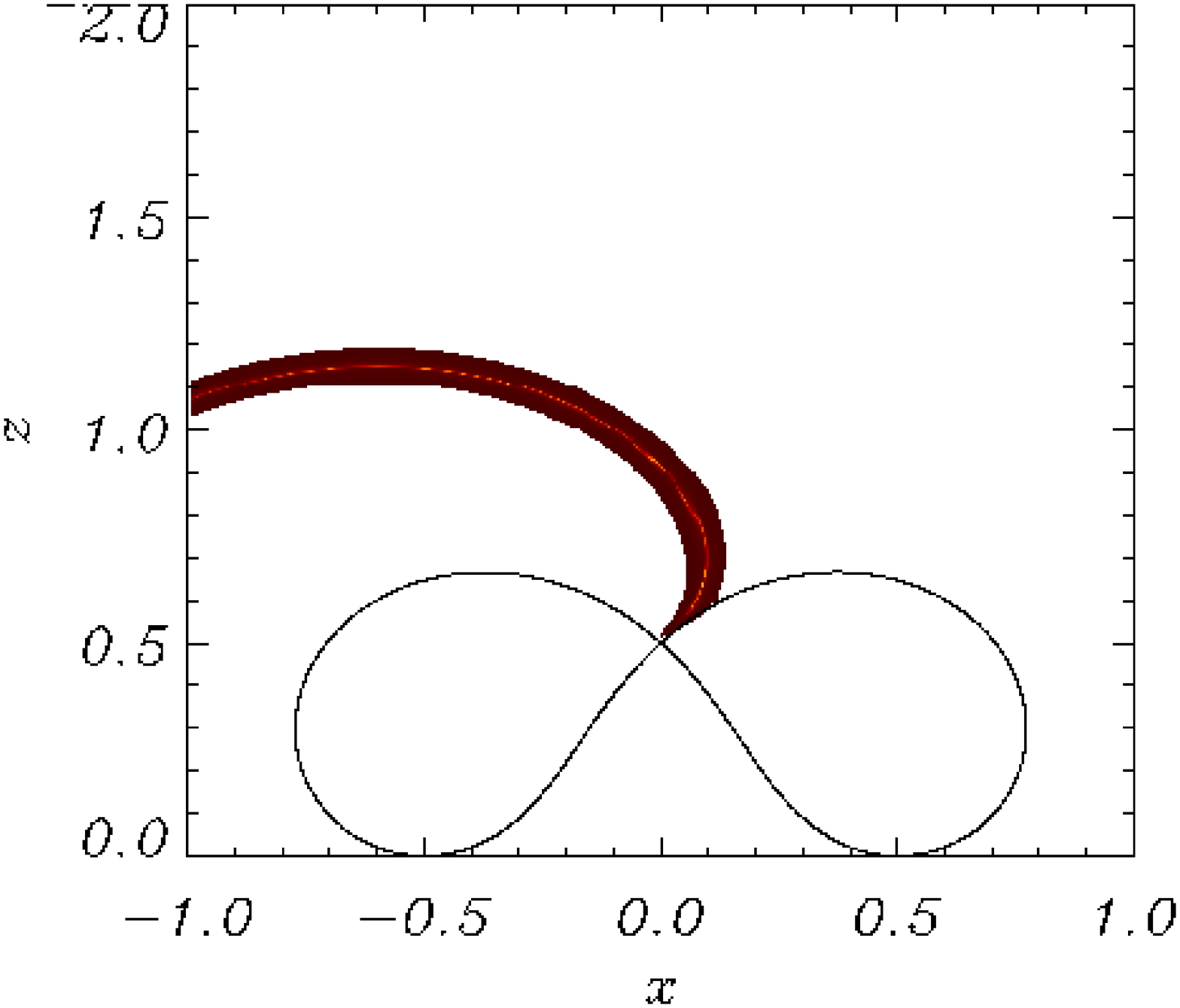}
\hspace{0.0in}
\includegraphics[width=1.2in]{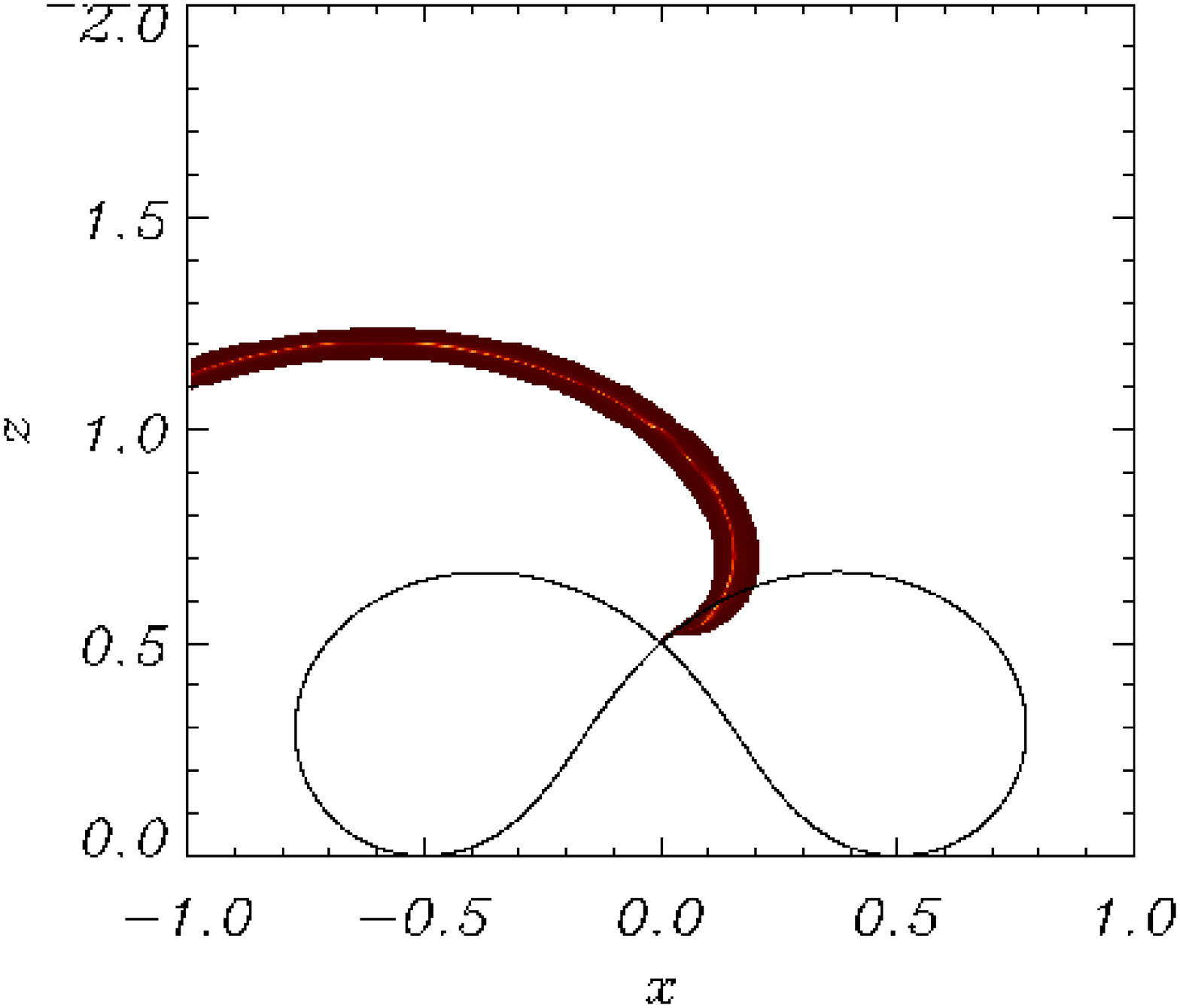}
\hspace{0.0in}
\includegraphics[width=1.2in]{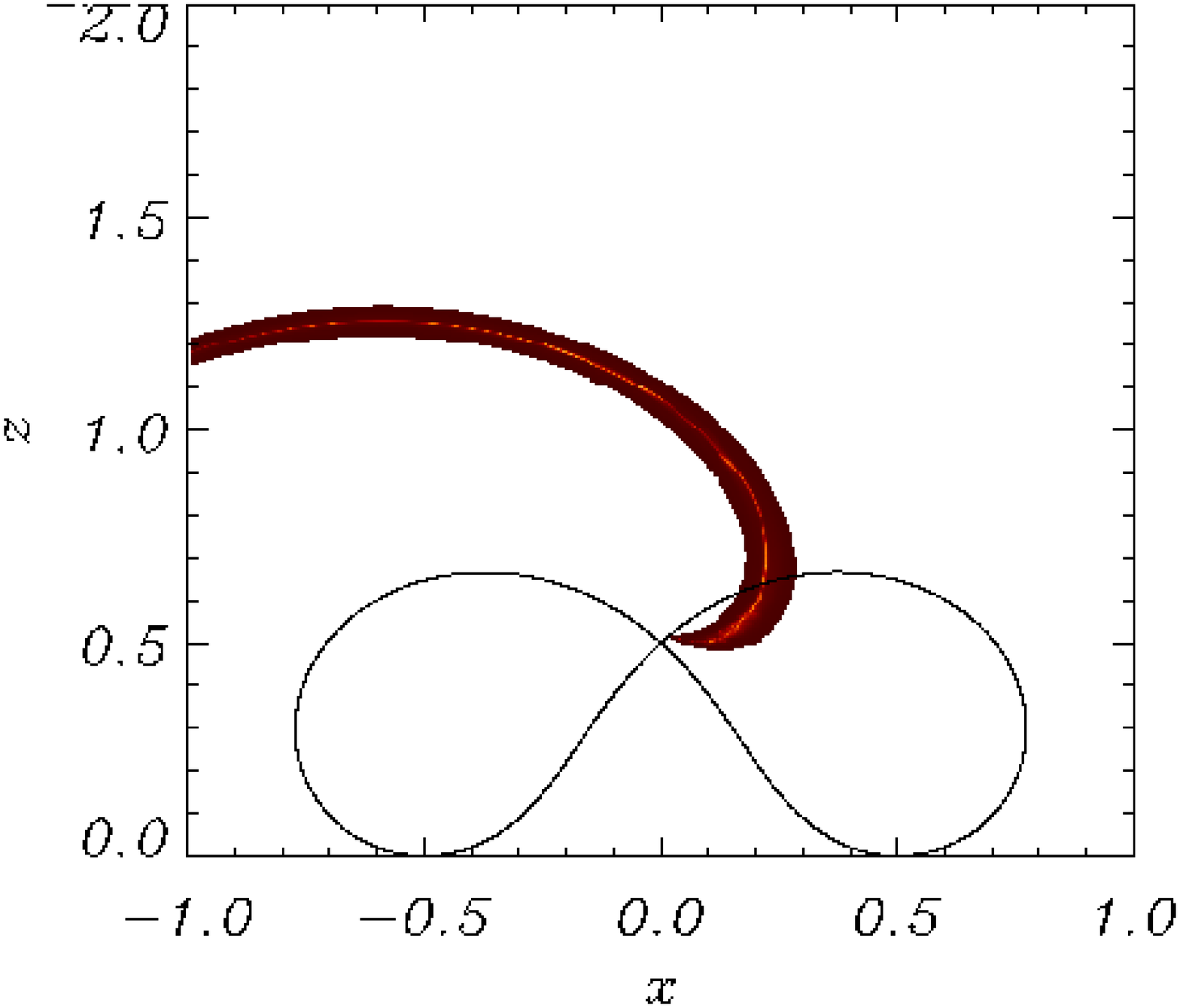}
\hspace{0.0in}
\includegraphics[width=1.2in]{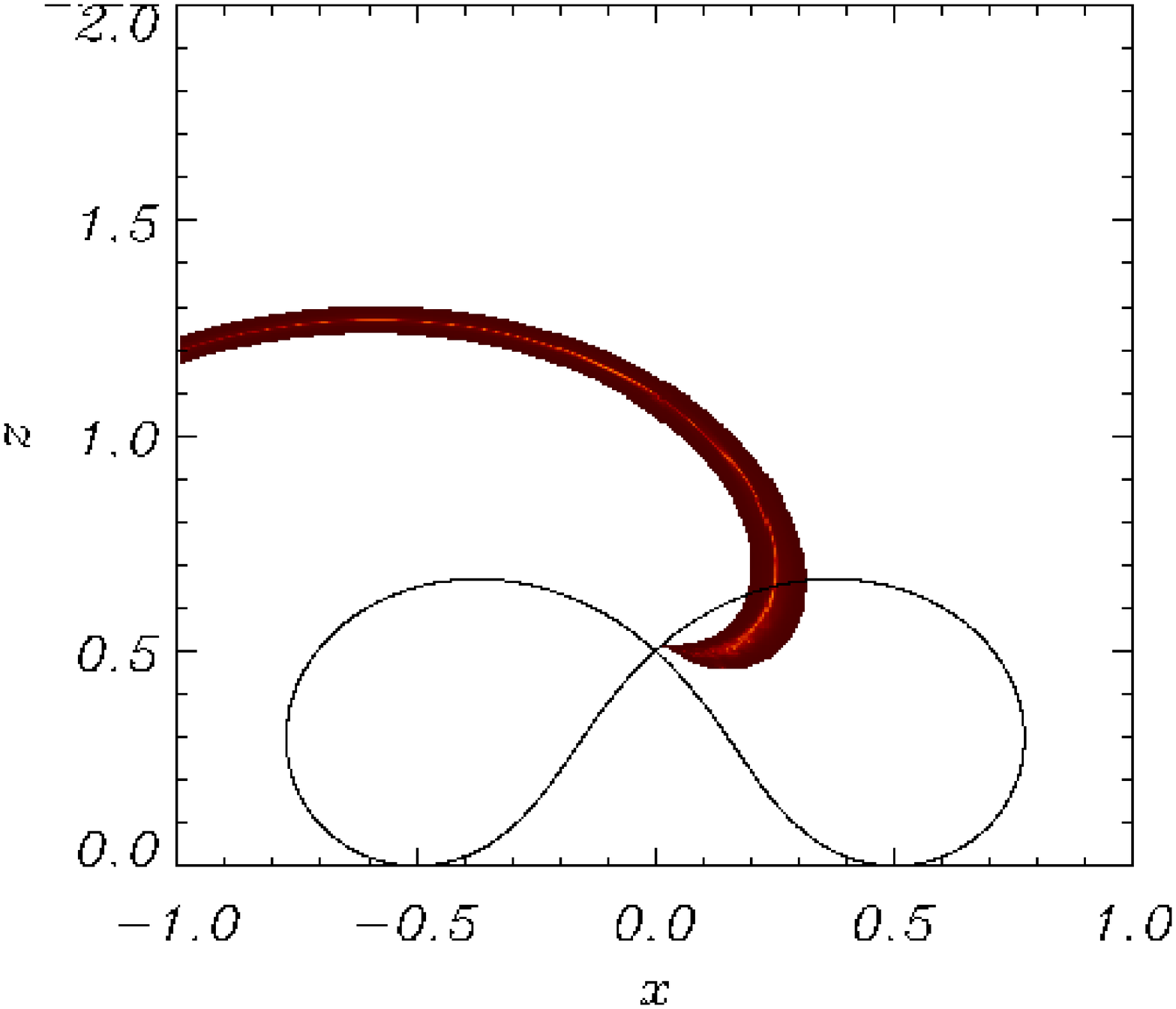}\\
\hspace{0in}
\vspace{0.1in}
%\hspace{0.2in}
\includegraphics[width=1.2in]{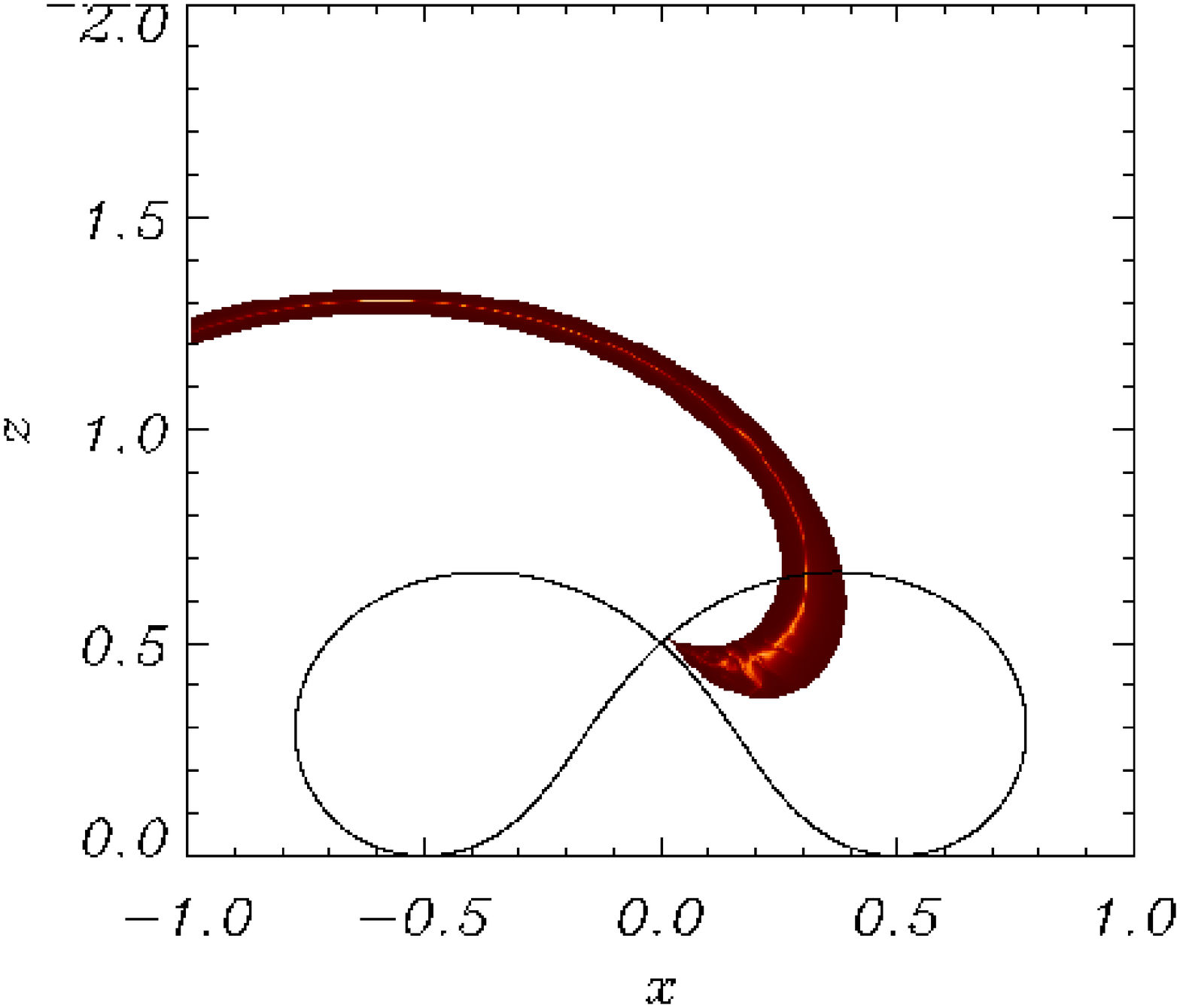}
\hspace{0.0in}
\includegraphics[width=1.2in]{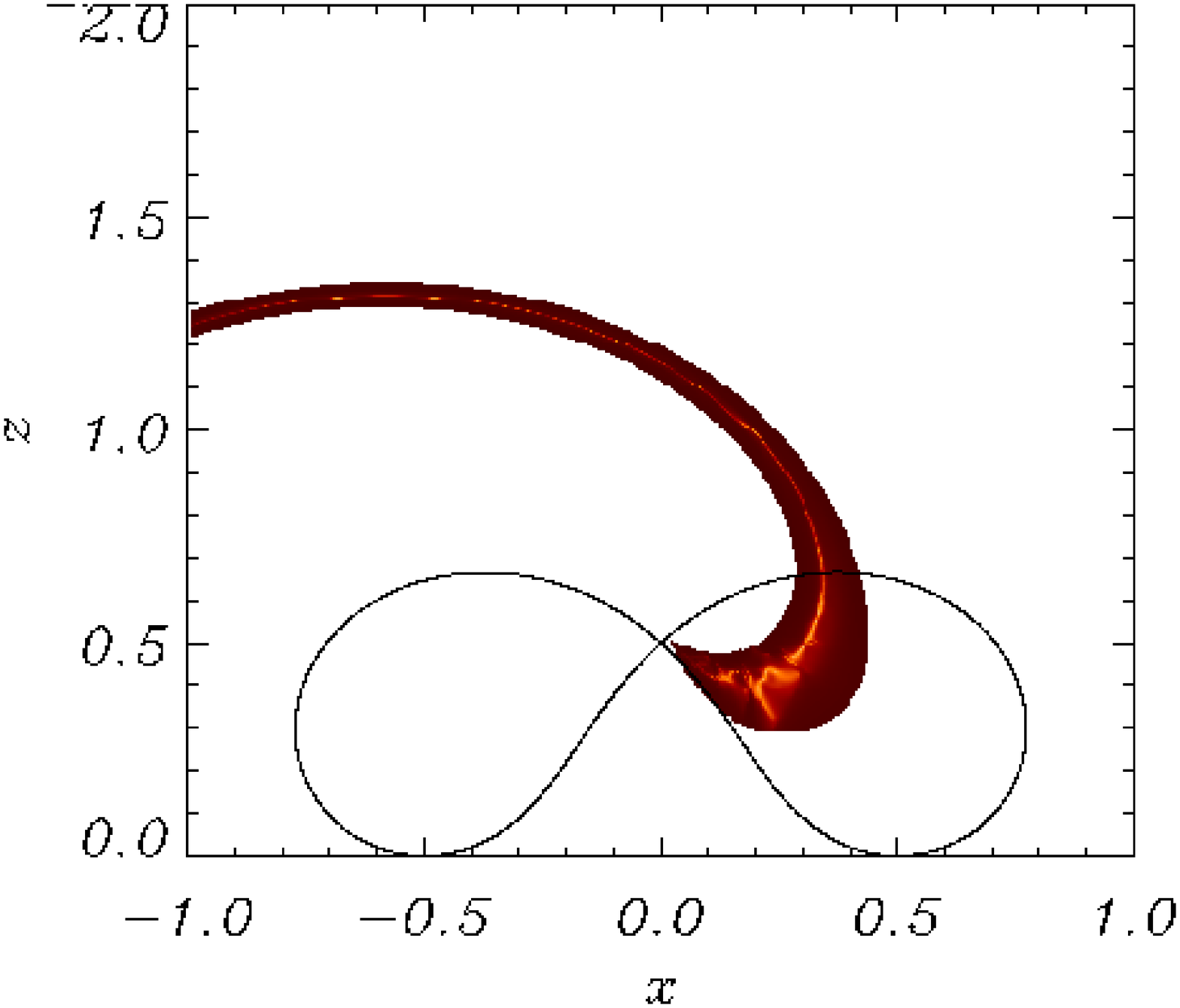}
\hspace{0.0in}
\includegraphics[width=1.2in]{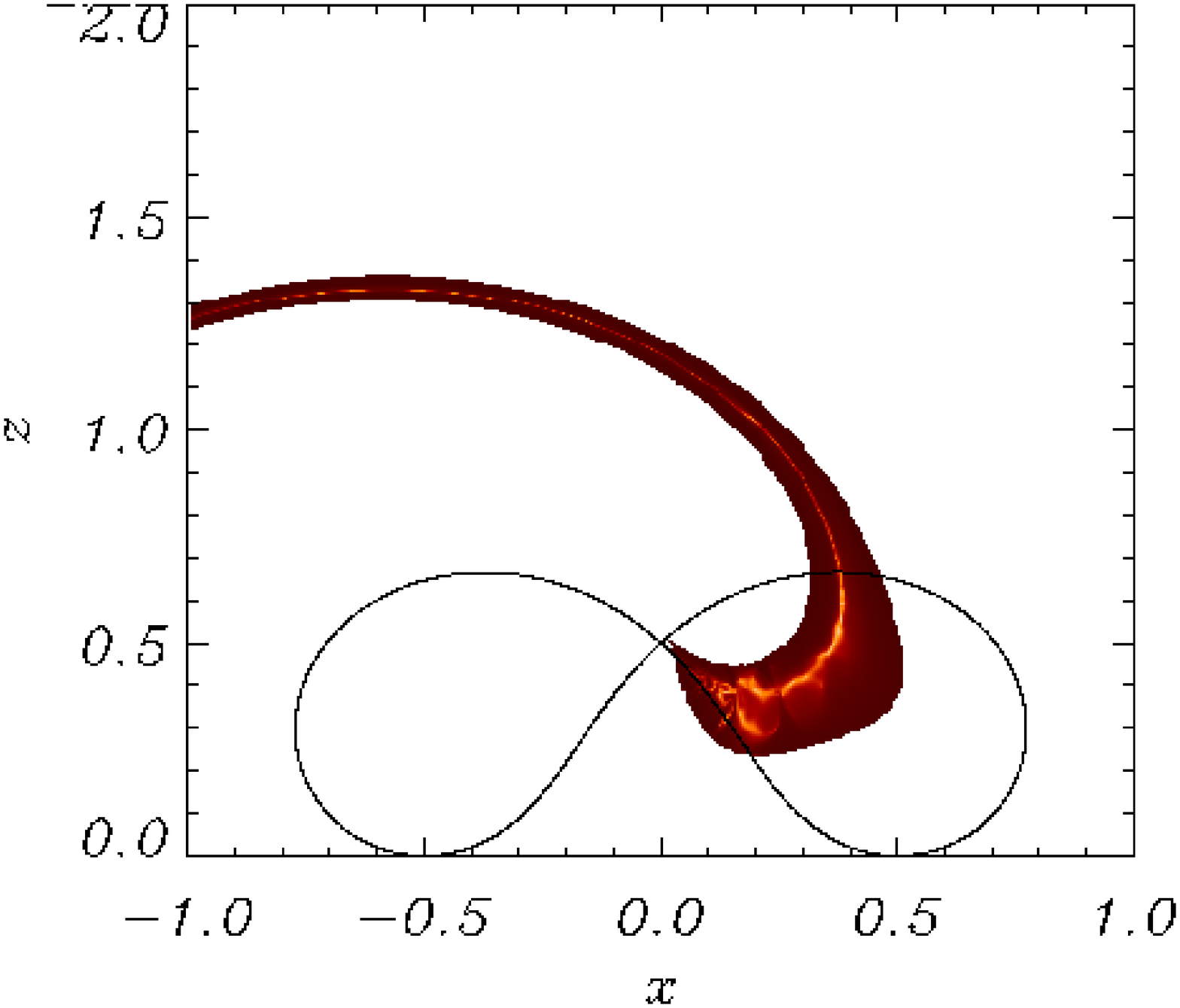}
\hspace{0.0in}
\includegraphics[width=1.2in]{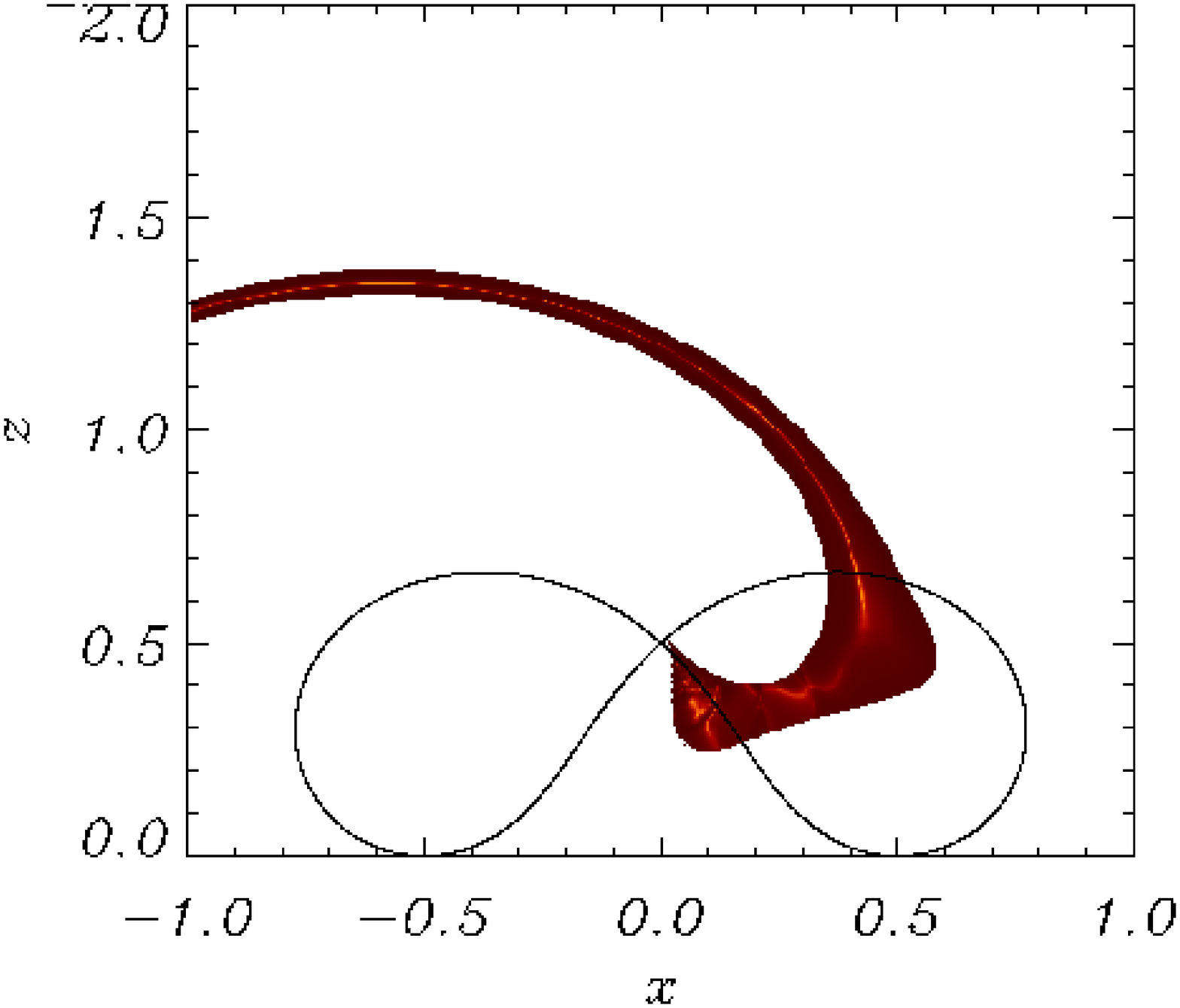}\\
\hspace{0in}
\vspace{0.1in}
%\hspace{0.2in}
\includegraphics[width=1.2in]{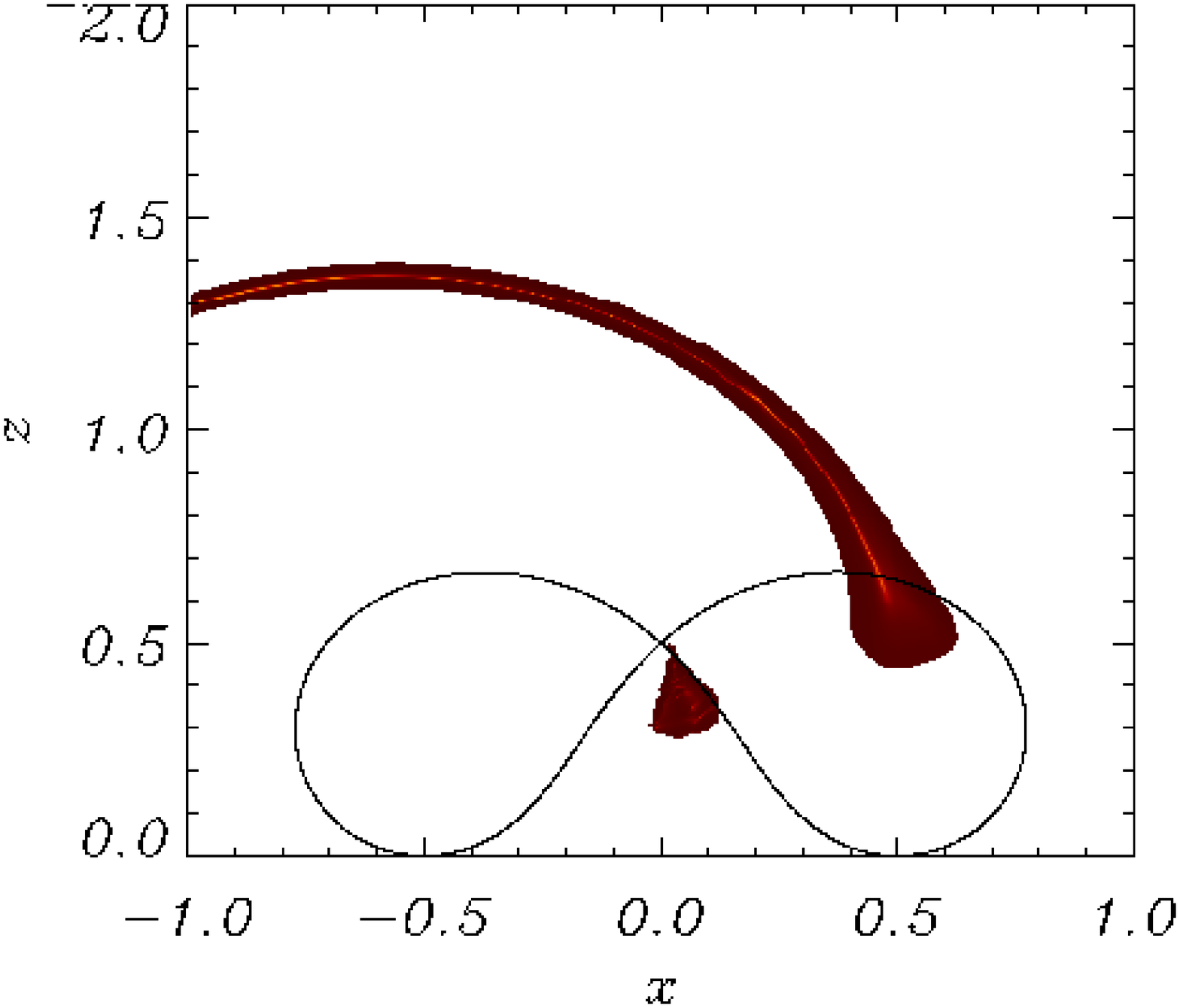}
\hspace{0.0in}
\includegraphics[width=1.2in]{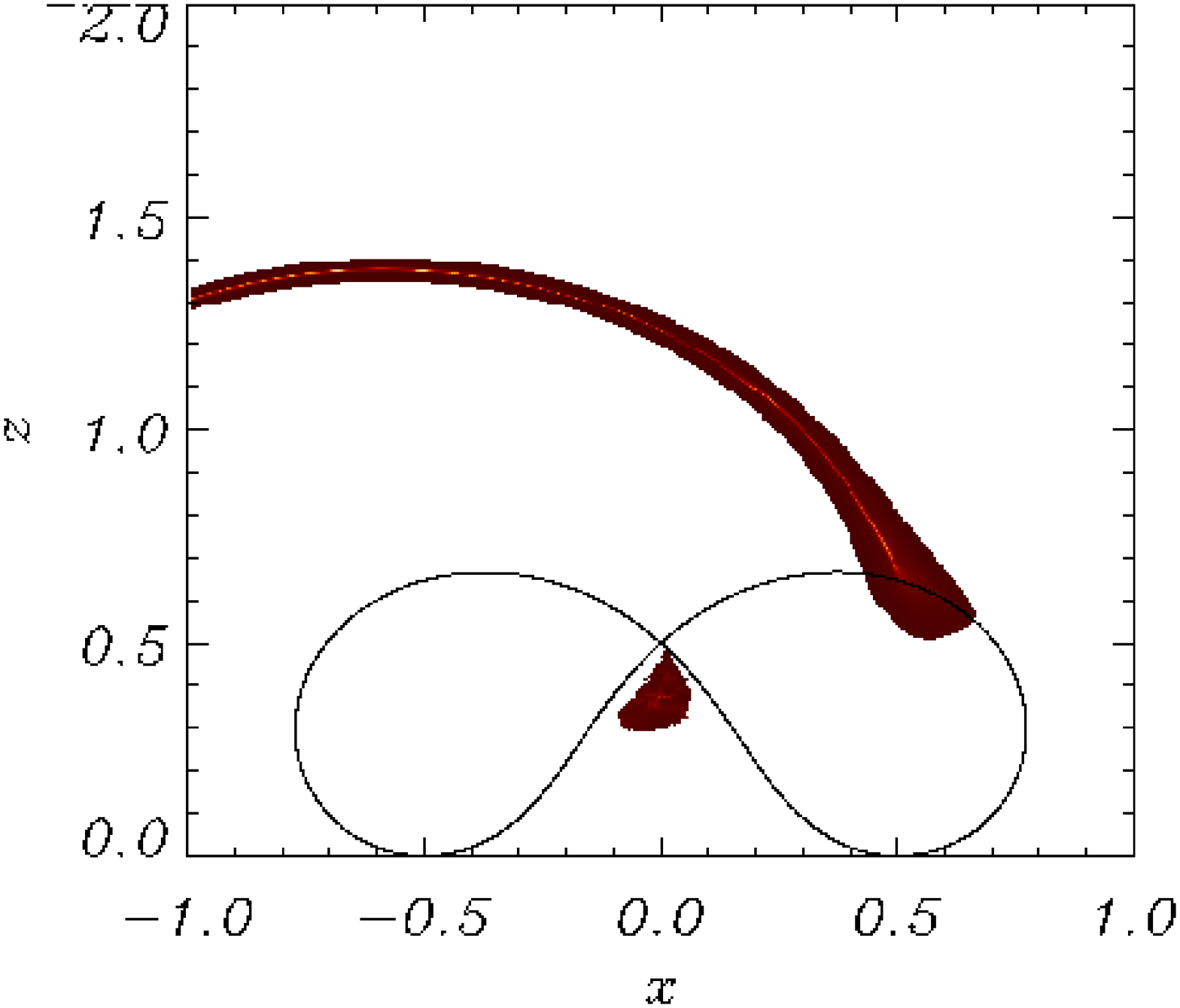}
\hspace{0.0in}
\includegraphics[width=1.2in]{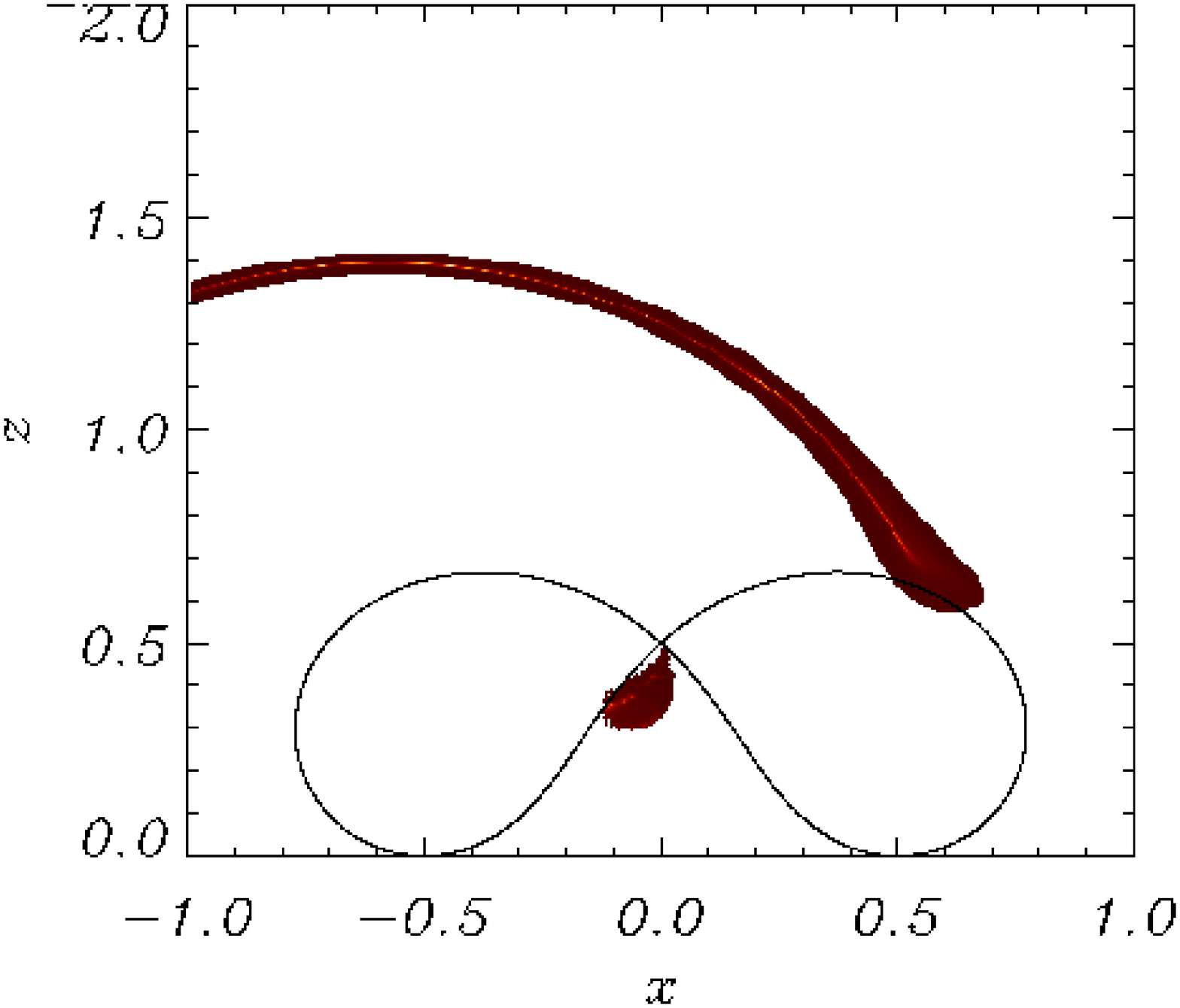}
\hspace{0.0in}
\includegraphics[width=1.2in]{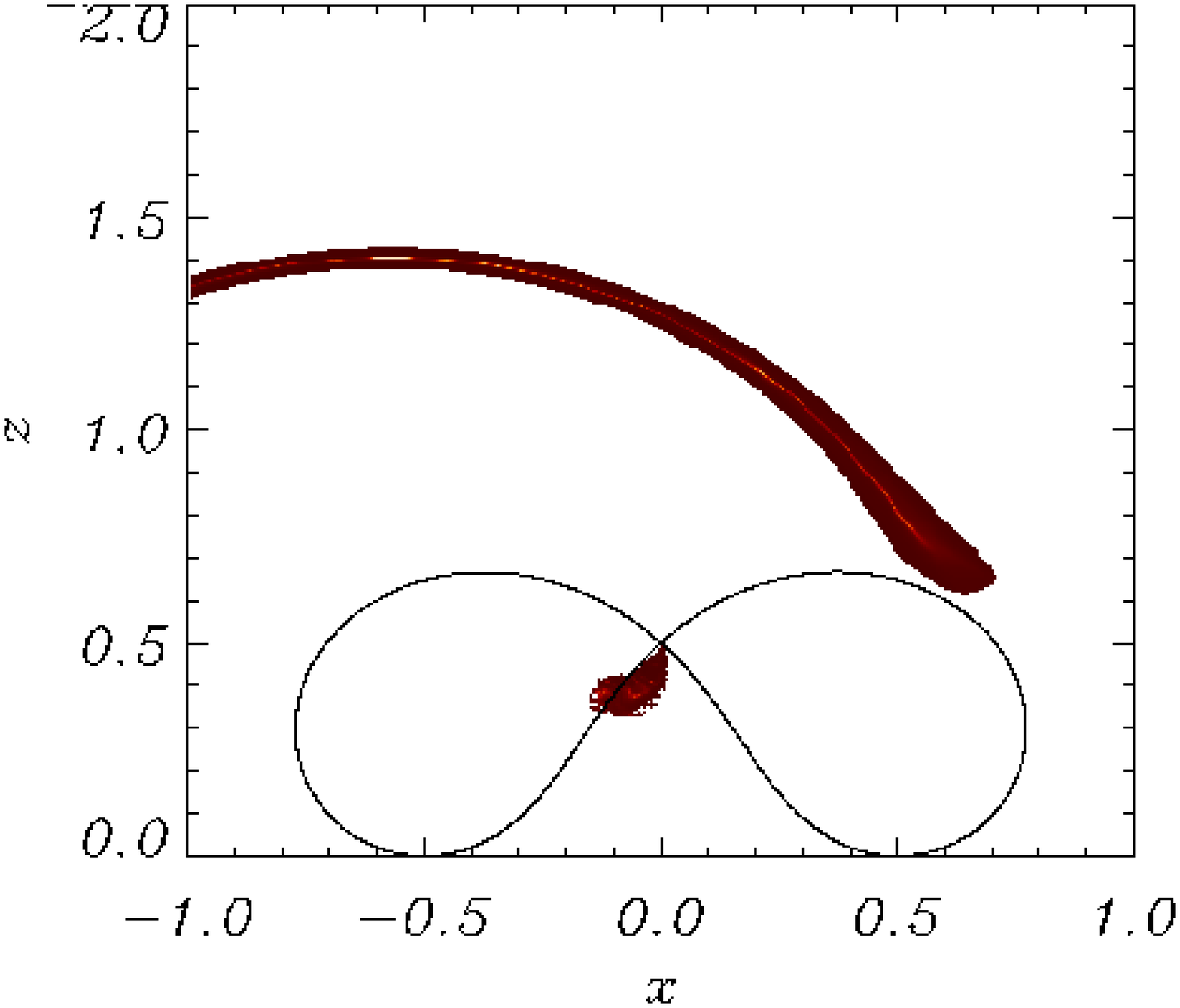}\\
\hspace{0in}
\vspace{0.1in}
%\hspace{0.2125in}
\includegraphics[width=1.2in]{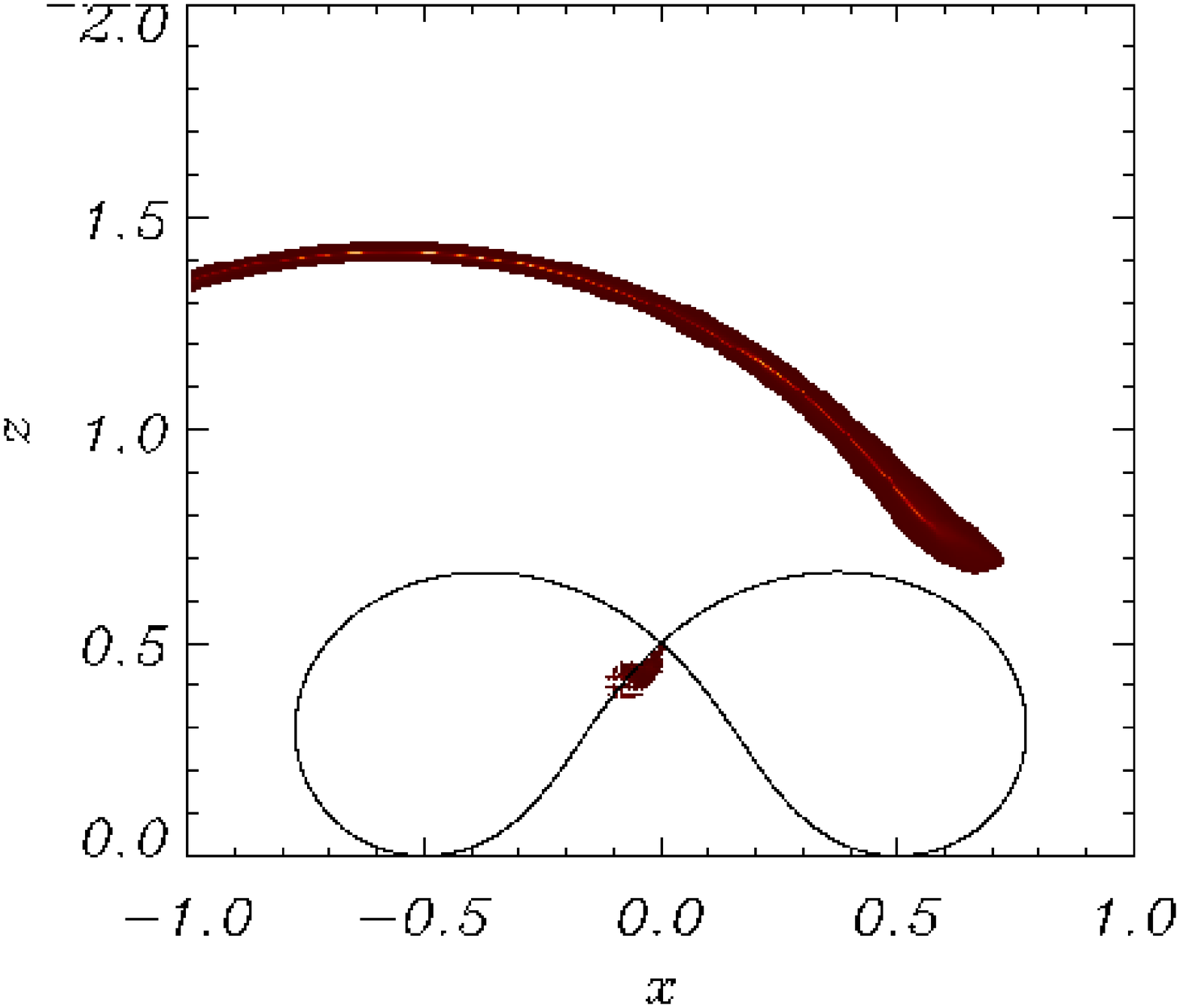}
\hspace{0.0in}
\includegraphics[width=1.2in]{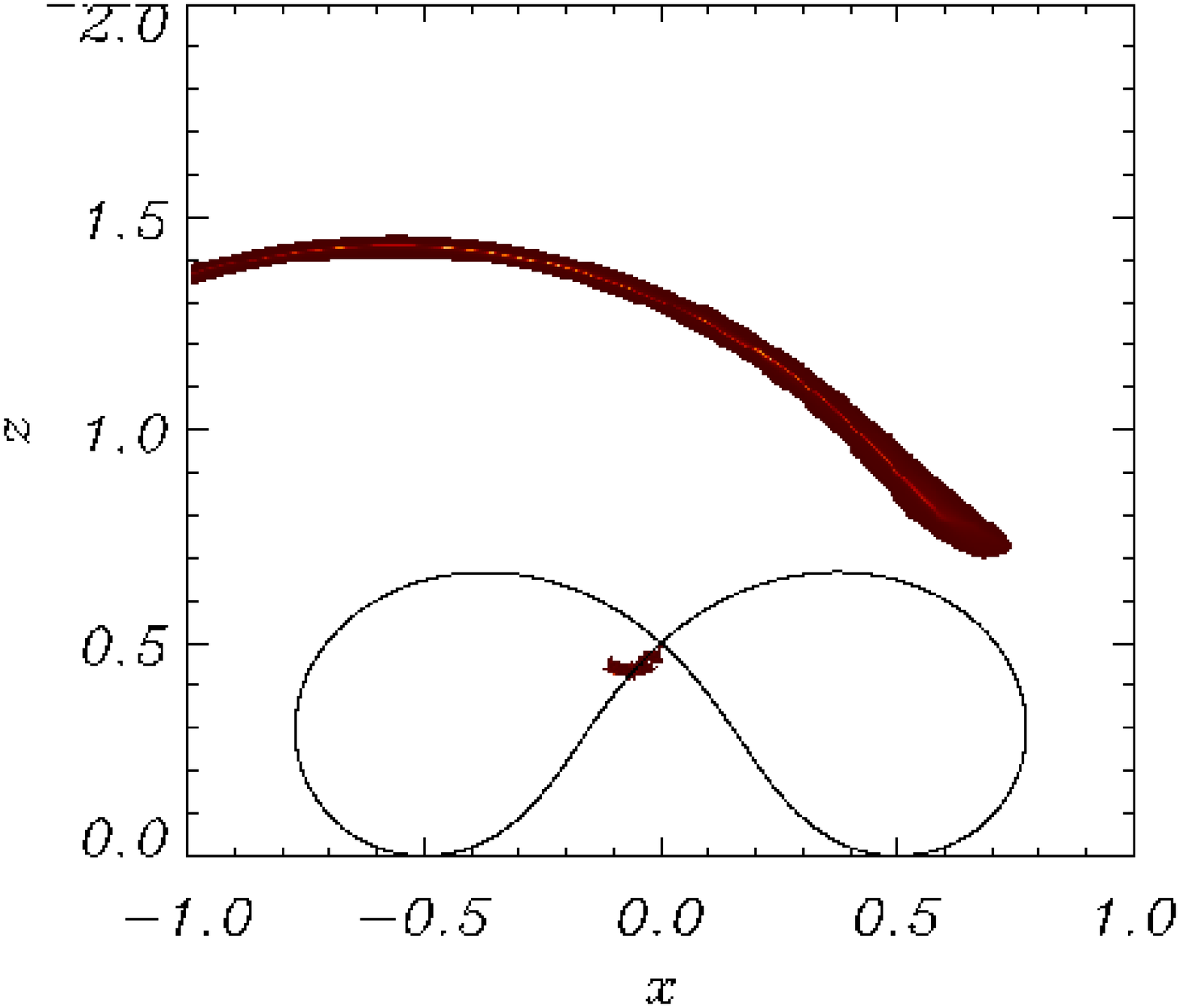}
\hspace{0.0in}
\includegraphics[width=1.2in]{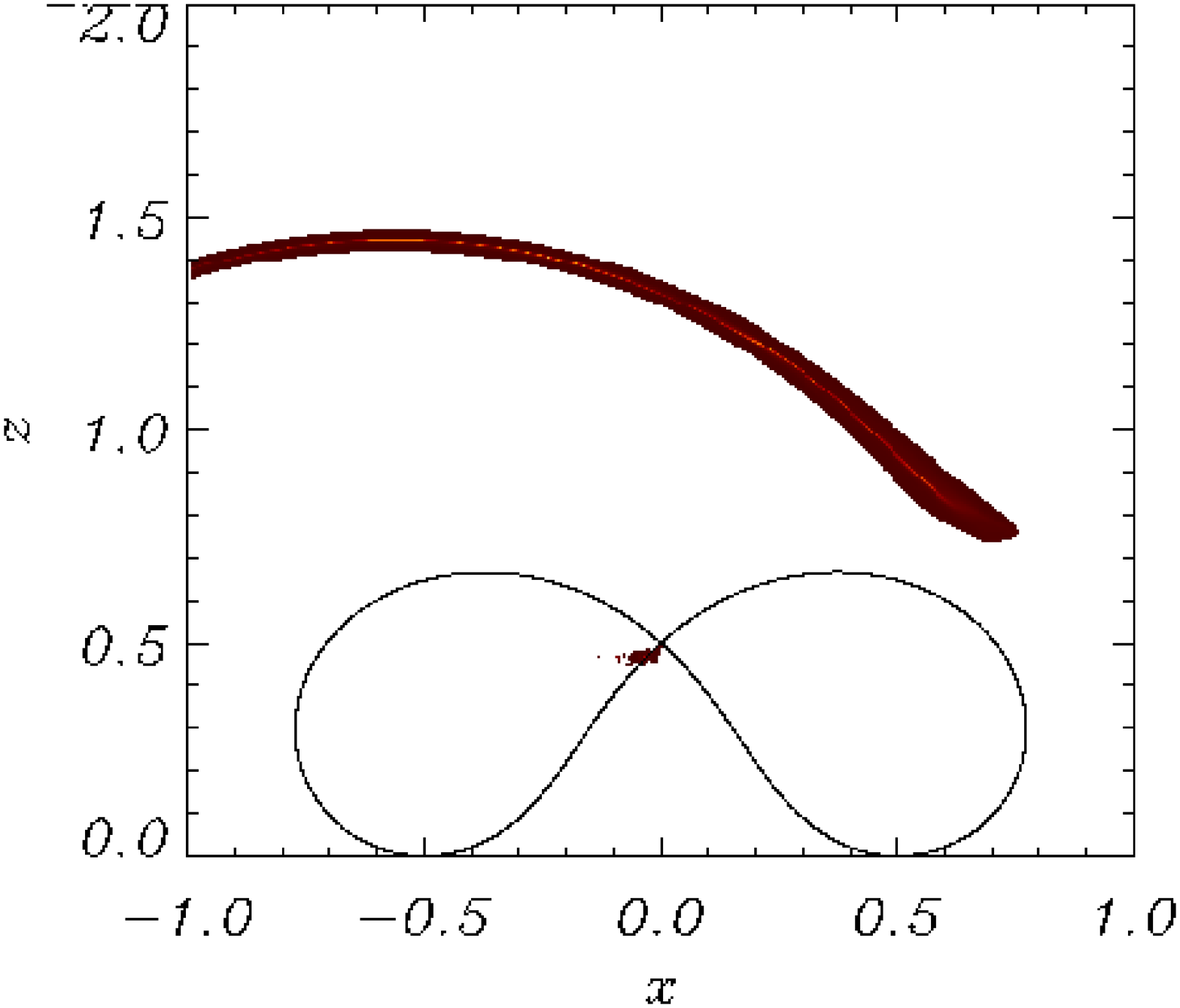}
\hspace{0.0in}
\includegraphics[width=1.55in]{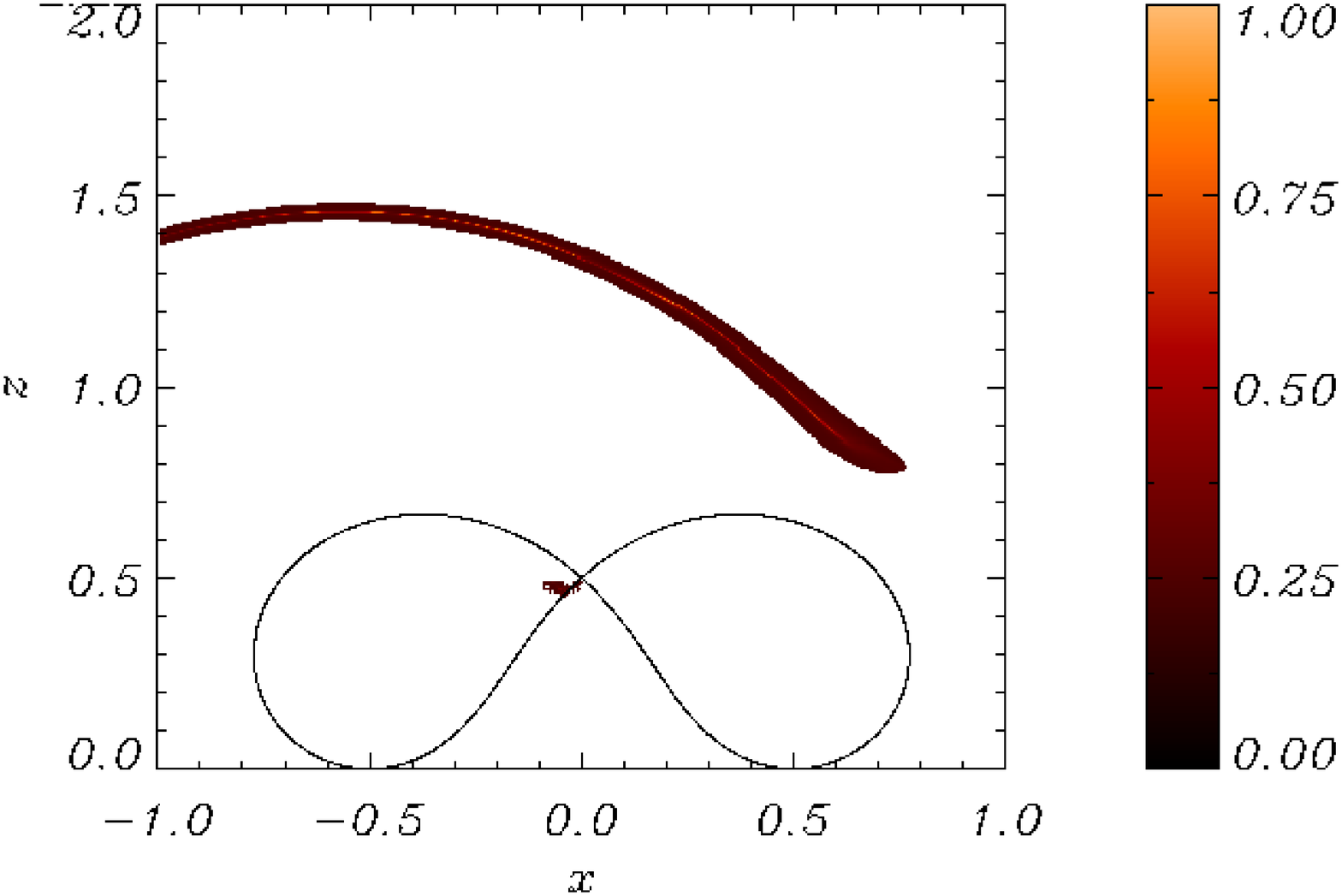}
\caption{Contours of numerical simulation of $V$ for a fast wave sent in from lower boundary for $-1 \leq x \leq 0$, $z=0.1$ and its resultant propagation at times $(a)$ $t$=0.055, $(b)$ $t$=0.11, $(c)$ $t$=0.165, $(d)$ $t=$0.22, $(e)$ $t$=0.275 and $(f)$ $t$=0.33, $(g)$ $t$=0.385, $(h)$ $t$=0.44, $(i)$ $t$=0.495, $(j)$ $t=$0.55, $(k)$ $t$=0.605, $(l)$ $t$=0.66, $(m)$ $t$=0.72, $(n)$ $t$=0.74, $(o)$ $t$=0.76, $(p)$ $t=$0.78, $(q)$ $t$=0.8 and $(r)$ $t$=0.82, $(s)$ $t$=0.84, $(t)$ $t$=0.86, $(u)$ $t$=0.88, $(v)$ $t=$0.9, $(w)$ $t$=0.92 and $(x)$ $t$=0.94, labelling from top left to bottom right.}
\label{fig:4.5.2.3}
\end{figure*}

The results can be seen in Figure \ref{fig:4.5.2.3}.  We find that the localised pulse starting between $-1 \leq x \leq0$ rises, but not all parts rise at the same speed. The central part of the pulse rises much faster, with the maximum occuring over $x=-0.5$. This is due to the high Alfv\'en speed localised in that area (as seen in Section \ref{jeangrey}). This deforms the wave pulse from its original planar form. This area of high speed is a new effect not previously seen in Papers I or II.

\begin{figure}
\begin{center}
\includegraphics[width=2.0in]{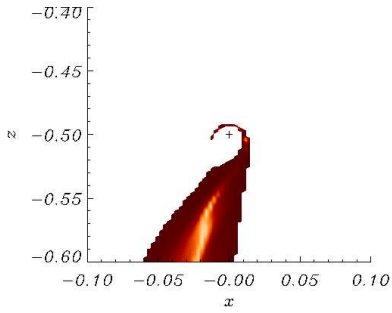}
\caption{Contour of the  numerical simulation of $V$ for a fast wave sent in from lower boundary  for  $-1 \leq x \leq 0 $ after time $t=0.44$. Figure shows blow-up of region around the null point. The image has been reflected in the line $z=-x$.}           
\label{fig:stillwrapsaround}
\end{center}
\end{figure}

Part of the wave pulse also approaches the null (extreme right of the generated wave), and this part appears to get caught around the X-point. As in Simulation One, the wave cannot cross the null and refracts around it. This also occurs here; part of the wave gets caught and wraps around the null (again and again). However, since the wave is being stretched out and thinned as it encircles the null,  our resolution runs out at after some point. Thus, we cannot clearly see the wave wrapping around the null in  Figures \ref{fig:4.5.2.3}. However, in  Figure \ref{fig:stillwrapsaround}  we can see a blow-up of the region around the X-point and thus can see that the effect proceeds in a similar way to the asymmetric pulse described in Paper I. We have plotted $(x,z)\rightarrow (-x,-z)$ to help the comparison (i.e. reflected the image in the line $z=-x$). Thus, this part of the fast wave pulse accumulates at the X-point. The resolution in Figure \ref{fig:stillwrapsaround} is coarse since it is a blow-up from the resolution of the simulation.

Meanwhile, the rest of the wave continues to propagate upwards  and spread out.  This part proceeds at a slower speed than before, again due to the change in Alfv\'en speed. Once above the magnetic skeleton, the wave appears to simply continue to rise and spread out. The wave is no longer heavily influenced by the magnetic null point; it has, in effect, escaped the refraction effect. This is again a new feature not  seen  previously in Papers I or II.

 However, since part of the wave is wrapping around the null and a second part is rising away from the magnetic region, there must come a point where the wave splits. This can be seen in  the lower set  of subfigures in Figure \ref{fig:4.5.2.3}. Turning our attention to the behaviour on the right of the null, we see that the wave continues to spread out; part of it appears to be travelling to the top right corner and part of it is hooked around the null. We see the wave is stretched between these two goals. The wave then splits, with part of it ultimately wrapping around the null (and continuing much in the same way as the asymmetric wave pulse in Paper I) and the other part of the wave then continues to spread out and travel away from the magnetic region. This can be seen in the bottom two rows of subfigures. The actual wave splitting itself occurs when the wave enters the right hand region of high Alfv\'en speed (large $v_A^2$ around $(\pm0.5,0)$). This thins the wave massively and forces the two parts to split.

Thus, a wave pulse starting between $-1 \leq x \leq 0$ travels into the magnetic region (unevenly due to the change in speed profile). Part of the wave experiences a slight refraction effect due to the Alfv\'en speed profile, but ultimately spreads out and propagates away. However, part of the wave is caught by the null and refracted into the X-point. This part of the wave accumulates at the X-point.

%**********************************************************************

%\vspace{0.3in}

%**********************************************************************

%\vspace{1cm}

\subsubsection{WKB approximation}\label{psyy}

It is clear from Simulation Two that quite a complicated splitting effect is in operation. We can, however, use a WKB solution to gain extra insight into the numerical simulation. Substituting $V = e^{i \phi (x,z) } \cdot e^{-i \omega t}$ into equation (\ref{fastbeta}) and assuming $\omega \gg 1$ (WKB approximation) leads to a first order PDE of the form $ \mathcal{F} \left( x,z,\phi,p,q \right)=0$. Applying the method of characteristics, we generate the equations:
\begin{eqnarray}
  \frac {d \phi }{ds} &=&  \omega ^2  \nonumber \\
 \frac {dp}{ds} &=& - \left( B_x \; \frac{\partial}{\partial x} B_x + B_z \; \frac{\partial}{\partial x} B_z \right)\;\left( p^2+q^2 \right)\nonumber\\
\frac {dq}{ds} &=& -\left( B_x \; \frac{\partial}{\partial z} B_x + B_z \; \frac{\partial}{\partial z} B_z \right)\;\left( p^2+q^2 \right)\nonumber\\
 \frac {dx}{ds} &=& \left( B_x^2+B_z^2 \right) p  \nonumber \\
 \frac {dz}{ds}  &=& \left( B_x^2+B_z^2 \right) q \label{fast_dipole_characteristics}
\end{eqnarray}
where  $p=\frac {\partial \phi} {\partial x}$, $q=\frac {\partial \phi} {\partial z}$, $B_x$ and $B_z$ are the components of our equilibrium field, $\omega$ is the frequency of our wave and $s$ is some parameter along the characteristic. These five ODEs were solved numerically using a fourth-order Runge-Kutta method.

%Contours of constant $\phi$ can be thought of as defining the position of the edge of the wave pulse, so with correct choices of $s$ the solution can be directly compared to our numerical solution. 

%Figures {\ref{figure23}}, {\ref{figure17}}, {\ref{figure30}} and {\ref{figure37}} all show comparisons of the numerical simulation and corresponding WKB solutions. In each case, the agreement between the analytic model and numerical results is very good.

\begin{figure*}[t]
\hspace{0in}
\vspace{0.1in}
\hspace{0.2in}
\includegraphics[width=1.2in]{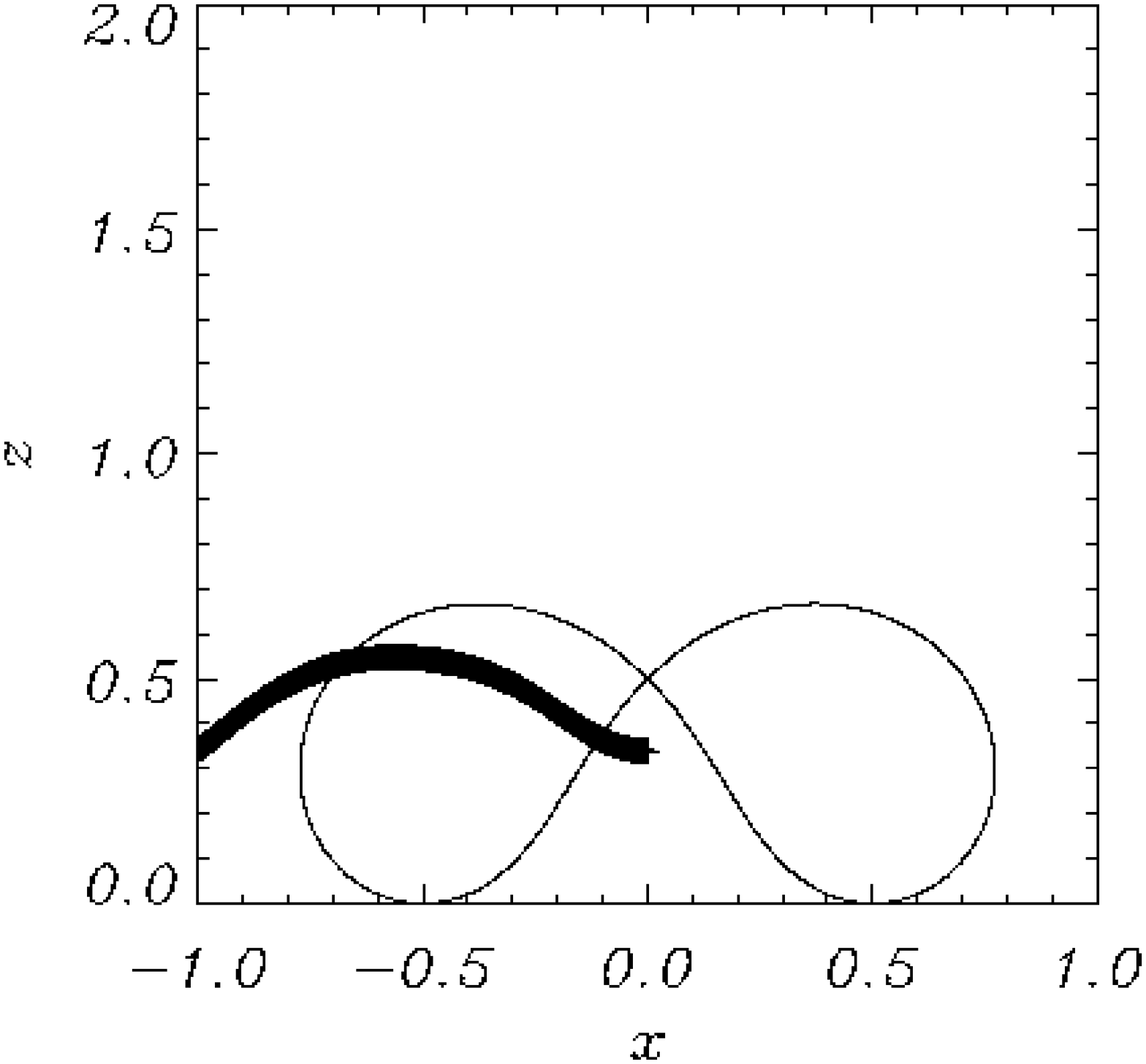}
\hspace{0.0in}
\includegraphics[width=1.2in]{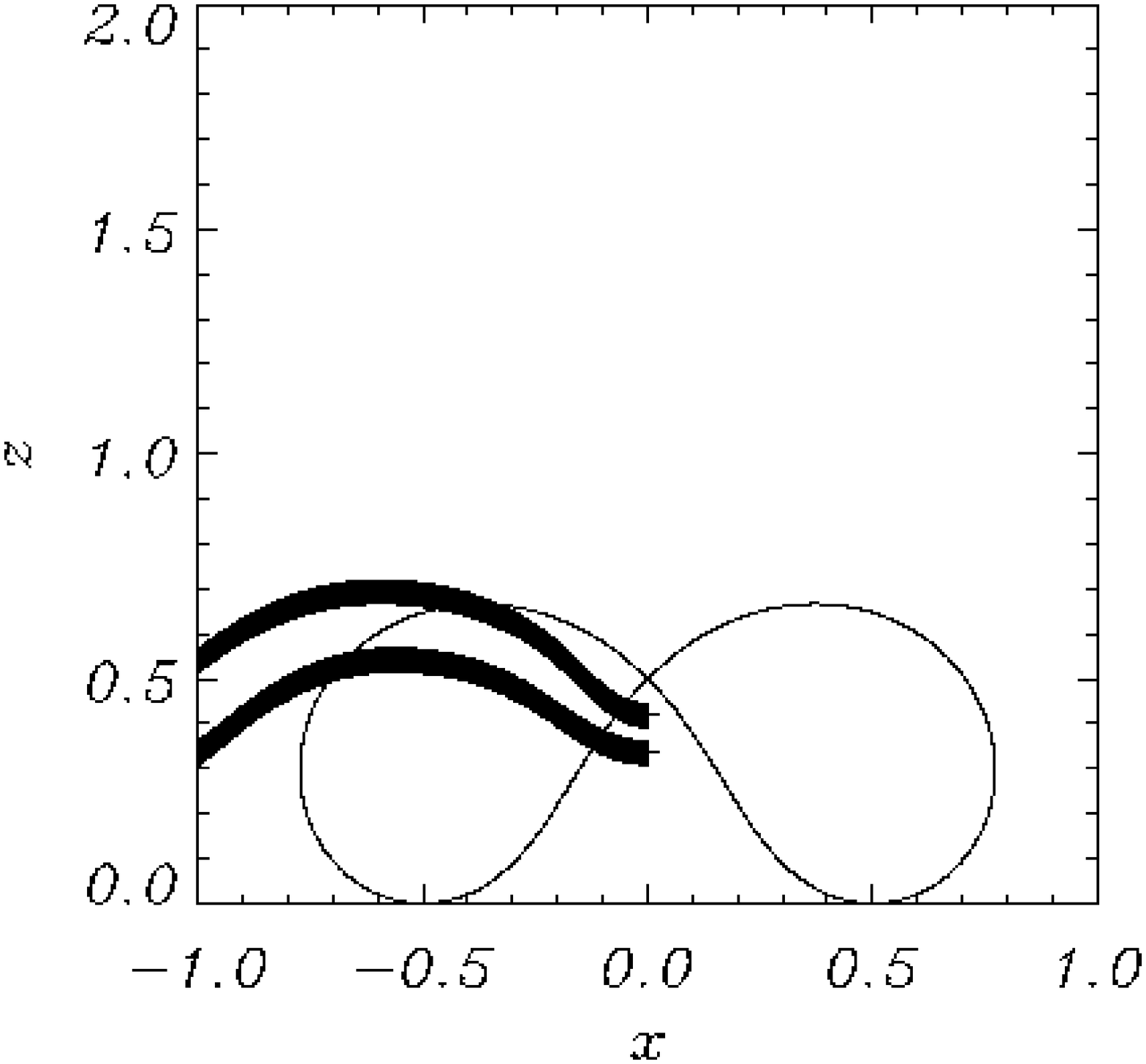}
\hspace{0.0in}
\includegraphics[width=1.2in]{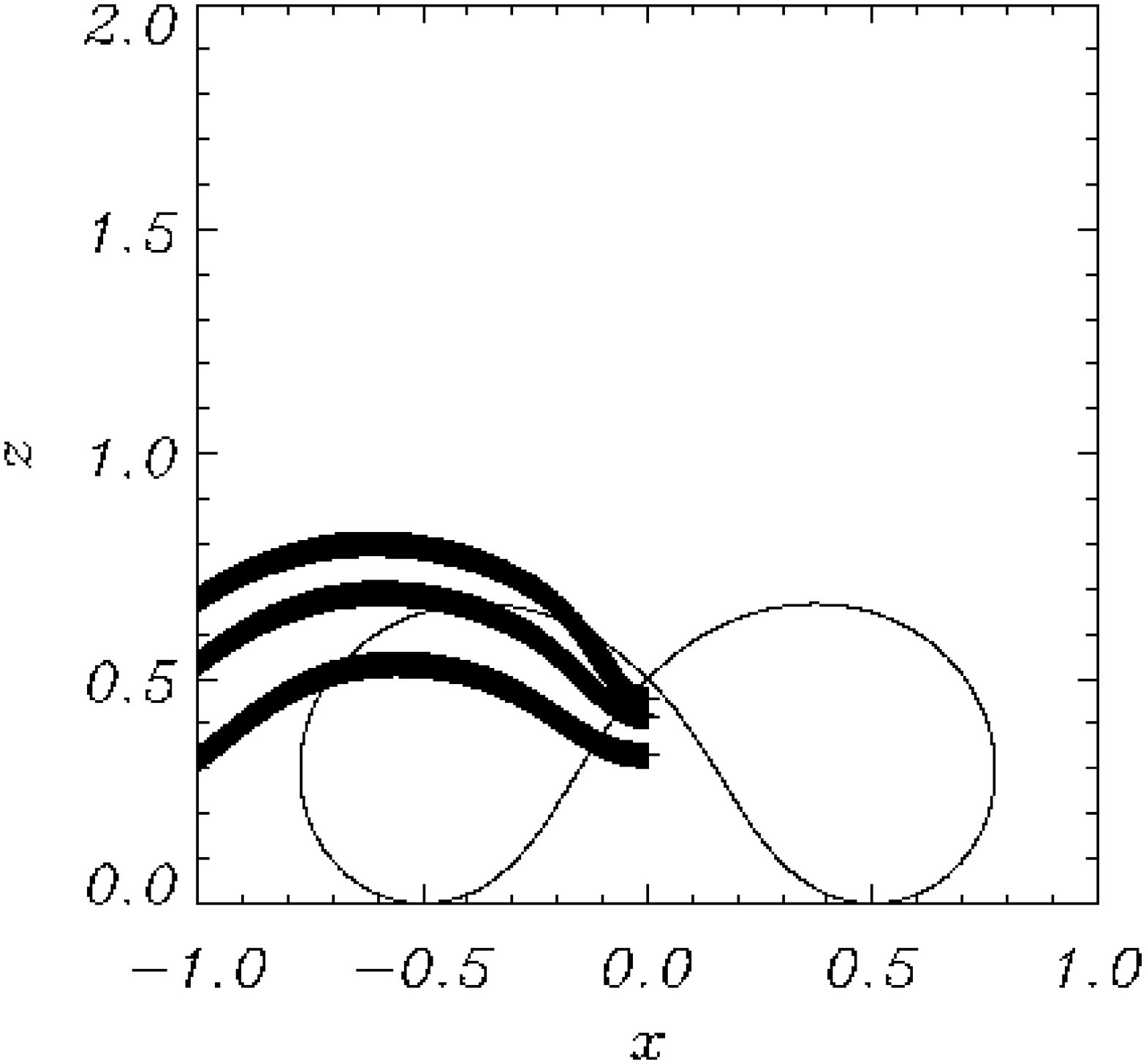}
\hspace{0.0in}
\includegraphics[width=1.2in]{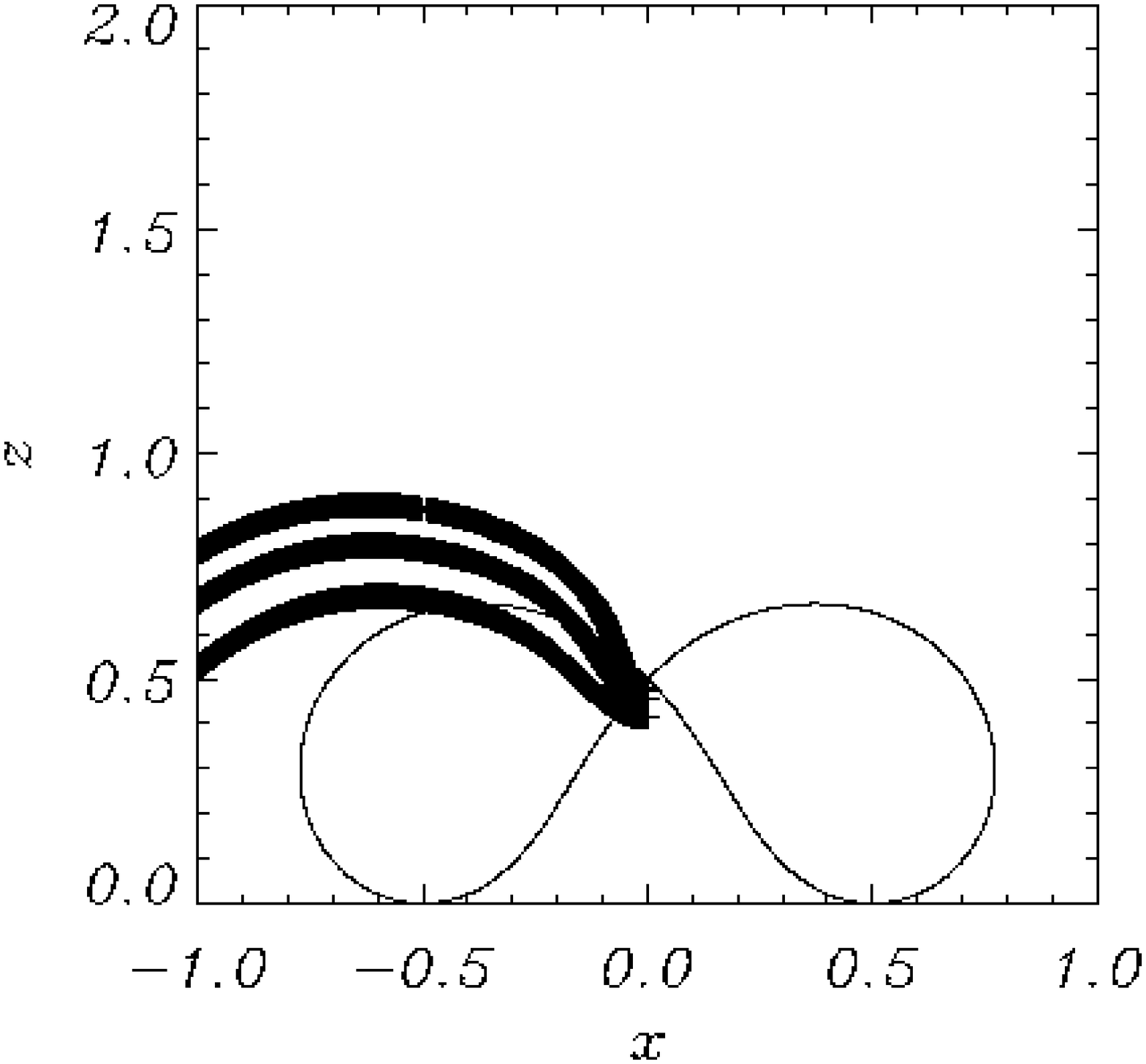}\\
\hspace{0in}
\vspace{0.1in}
\hspace{0.2in}
\includegraphics[width=1.2in]{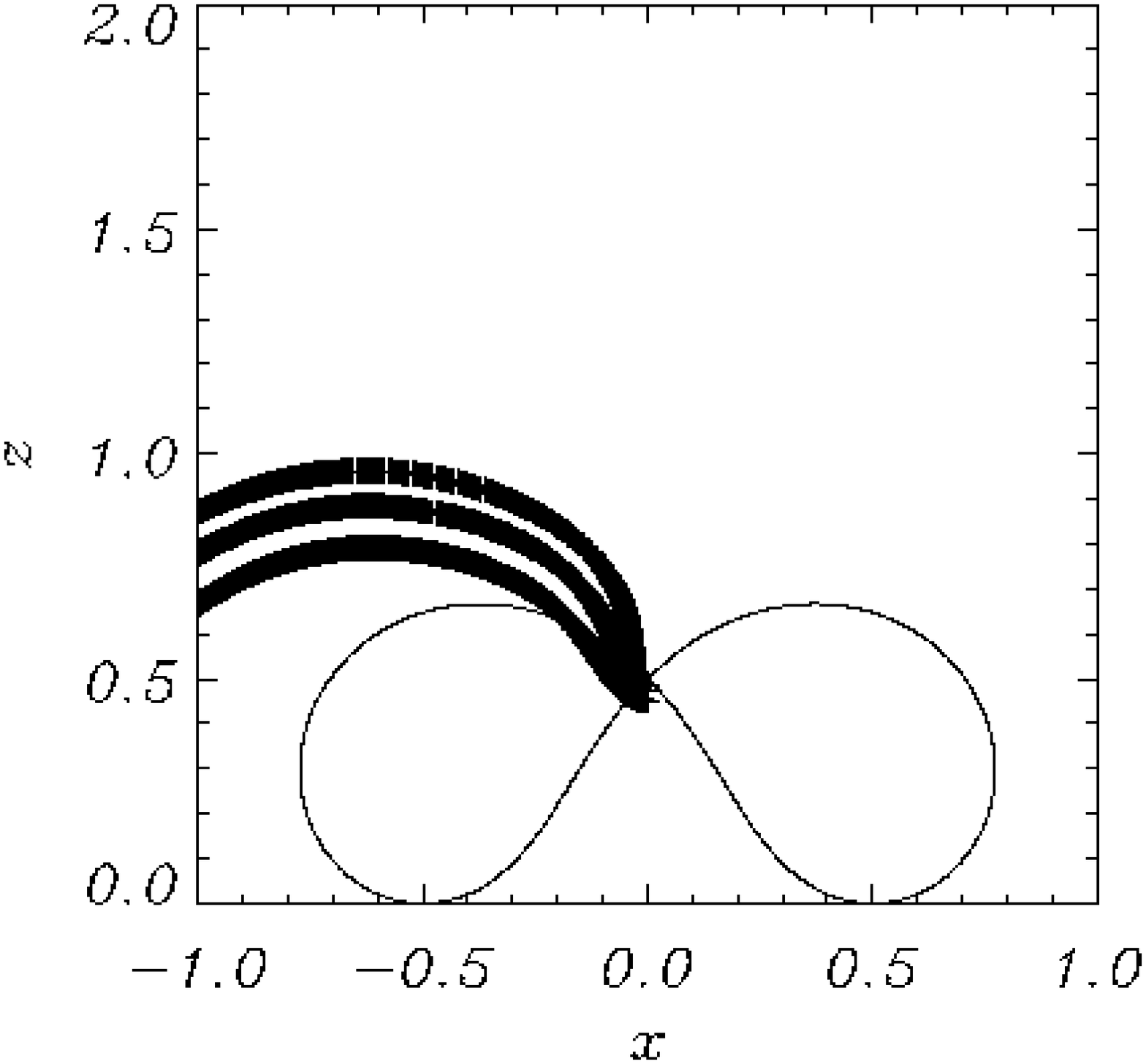}
\hspace{0.0in}
\includegraphics[width=1.2in]{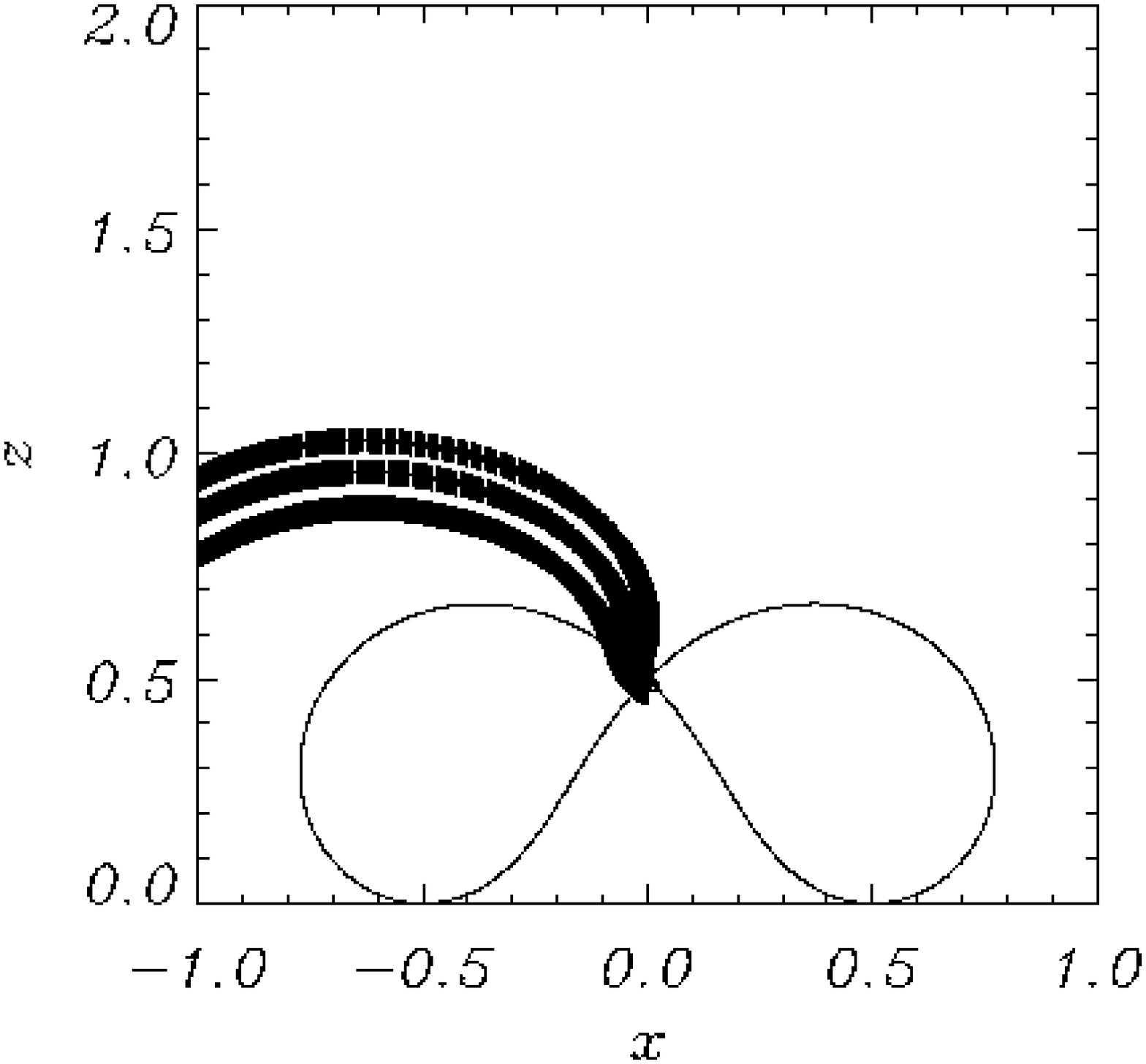}
\hspace{0.0in}
\includegraphics[width=1.2in]{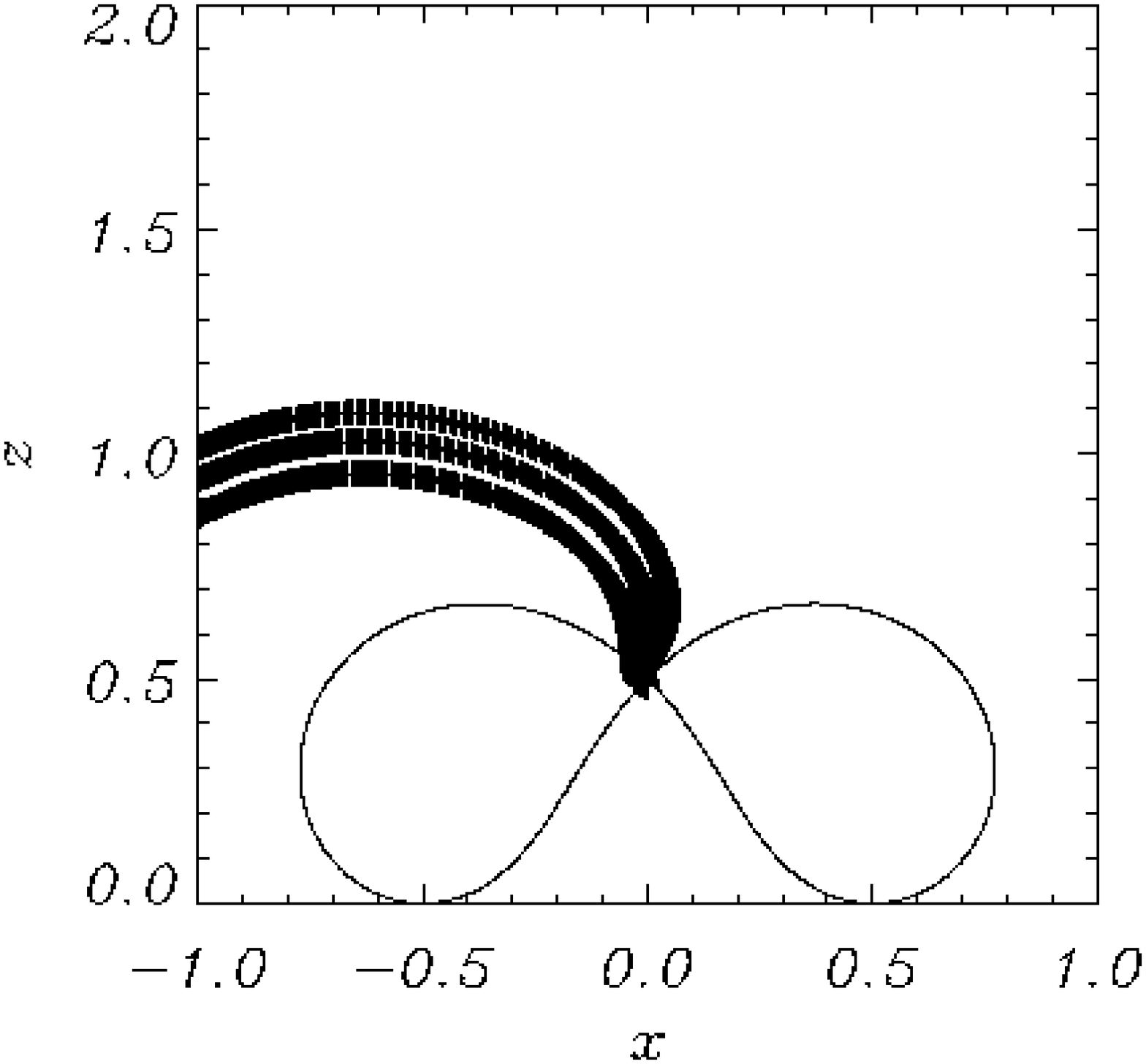}
\hspace{0.0in}
\includegraphics[width=1.2in]{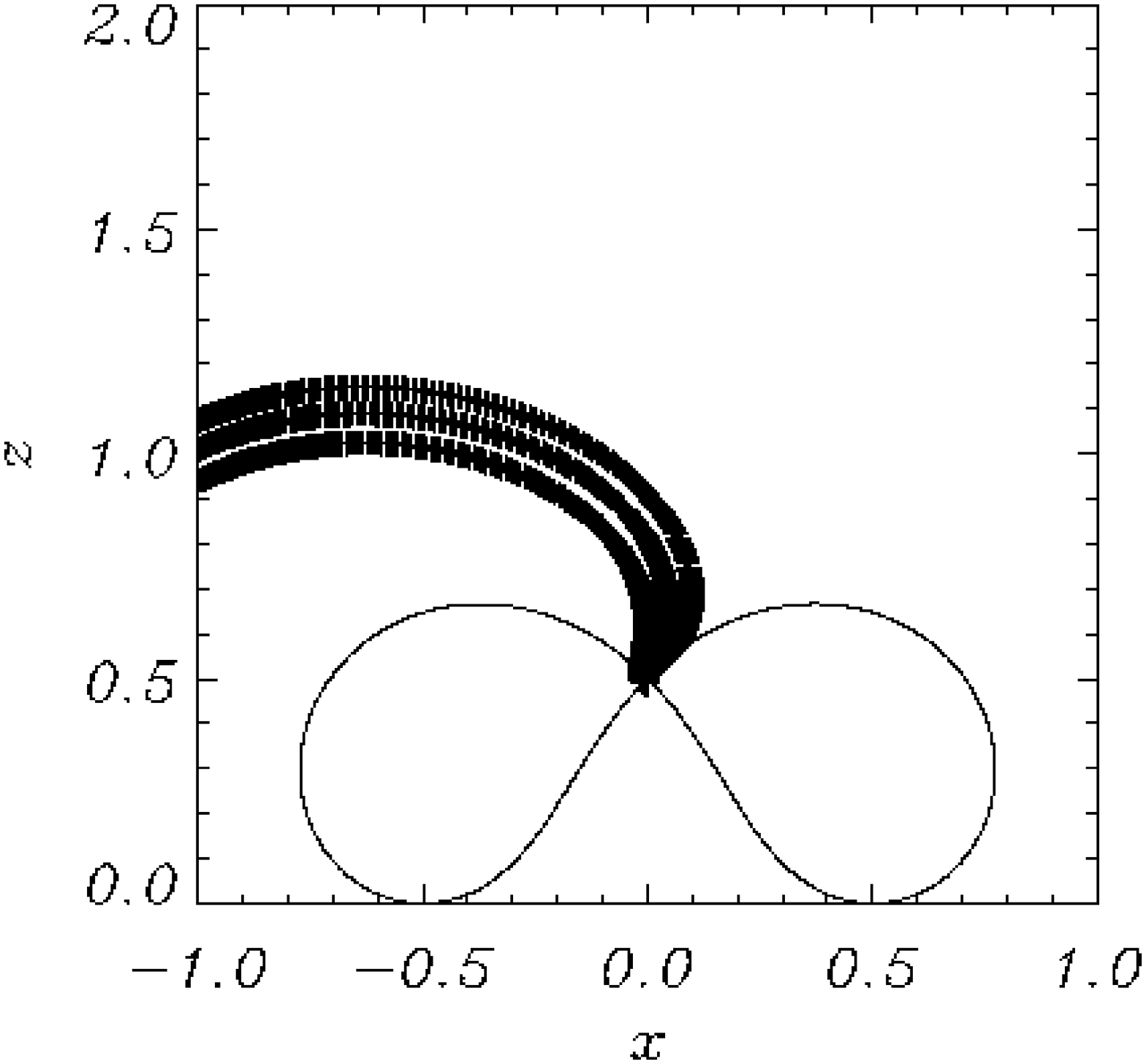}\\
\hspace{0in}
\vspace{0.1in}
\hspace{0.2in}
\includegraphics[width=1.2in]{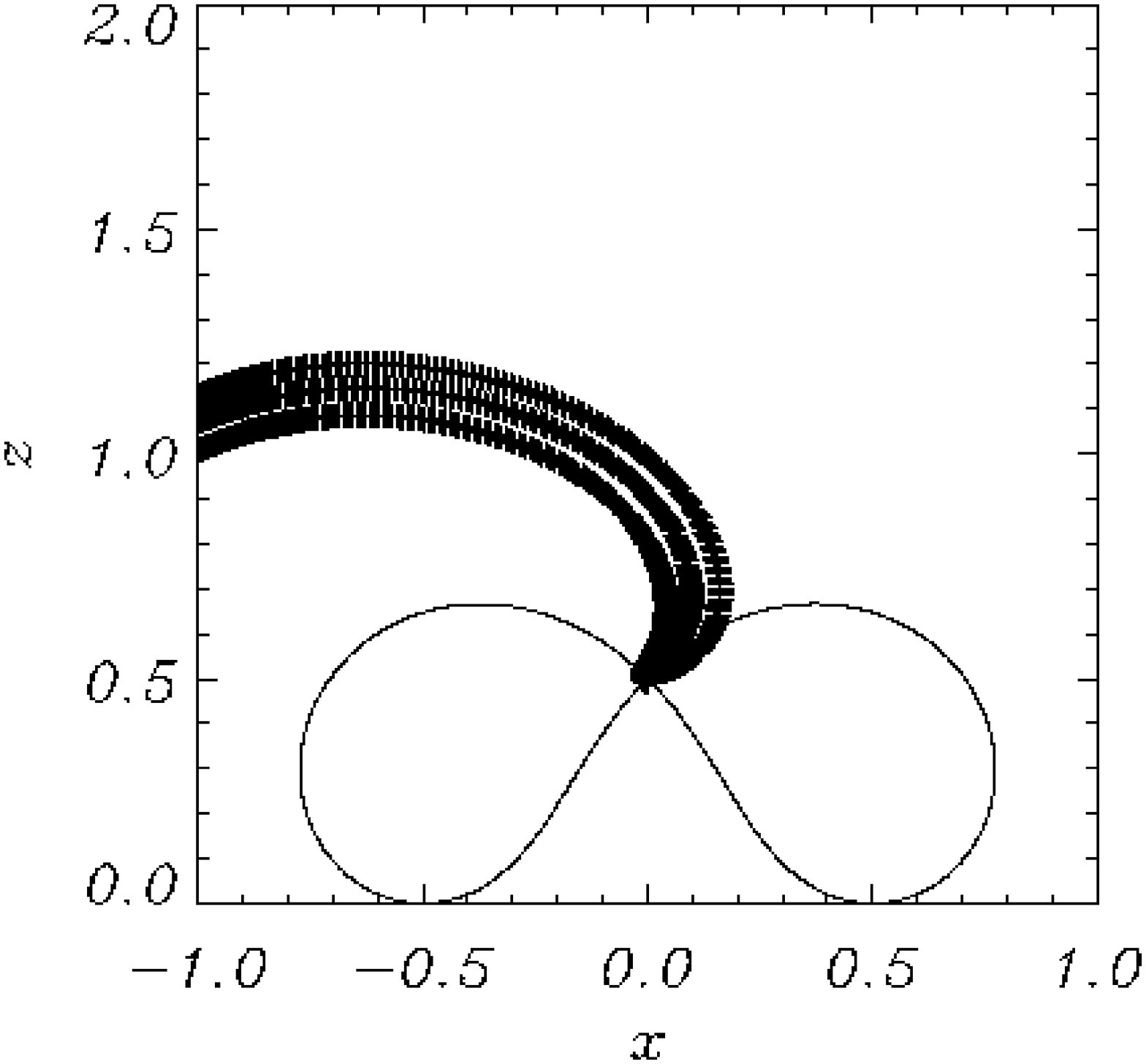}
\hspace{0.0in}
\includegraphics[width=1.2in]{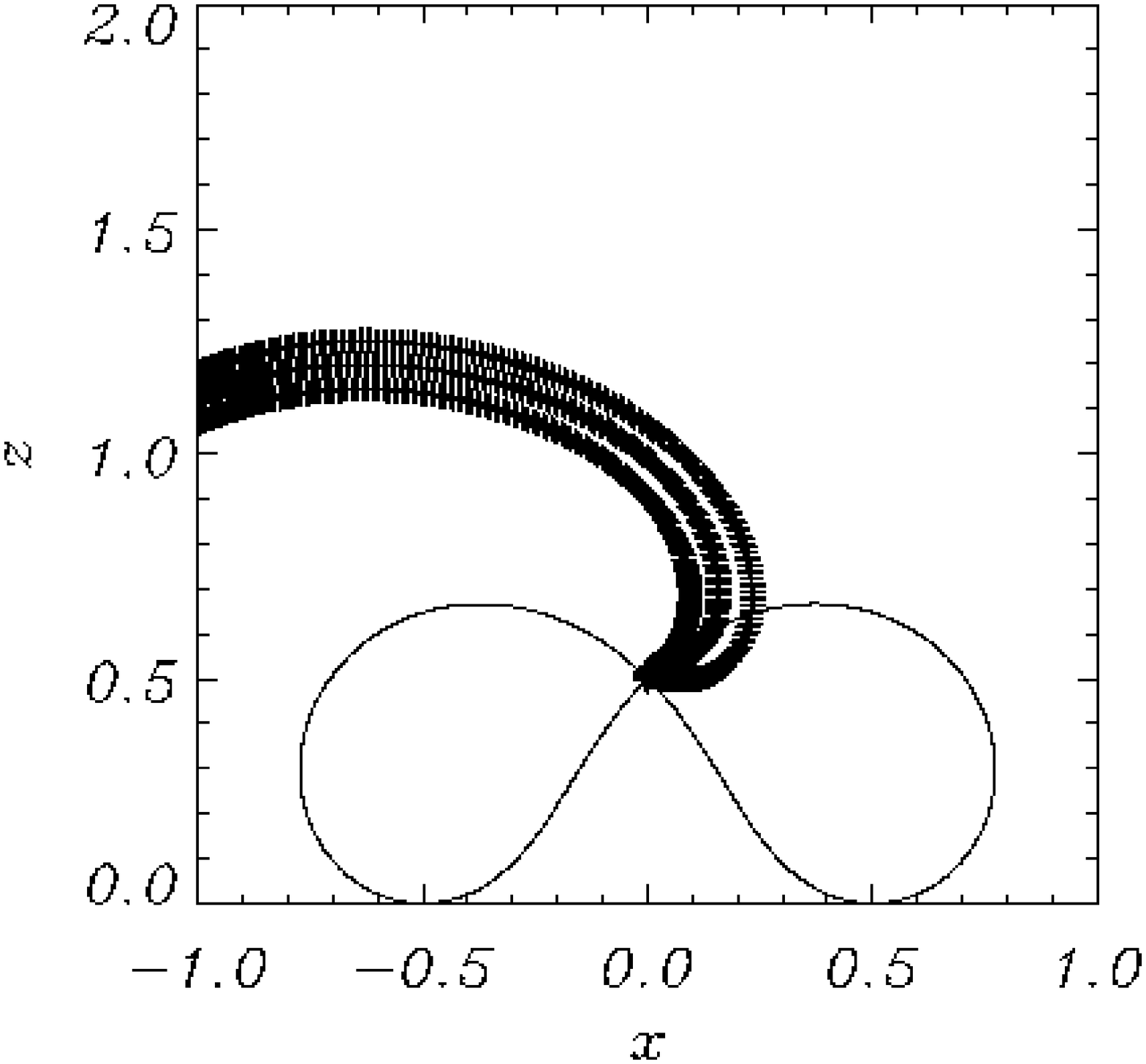}
\hspace{0.0in}
\includegraphics[width=1.2in]{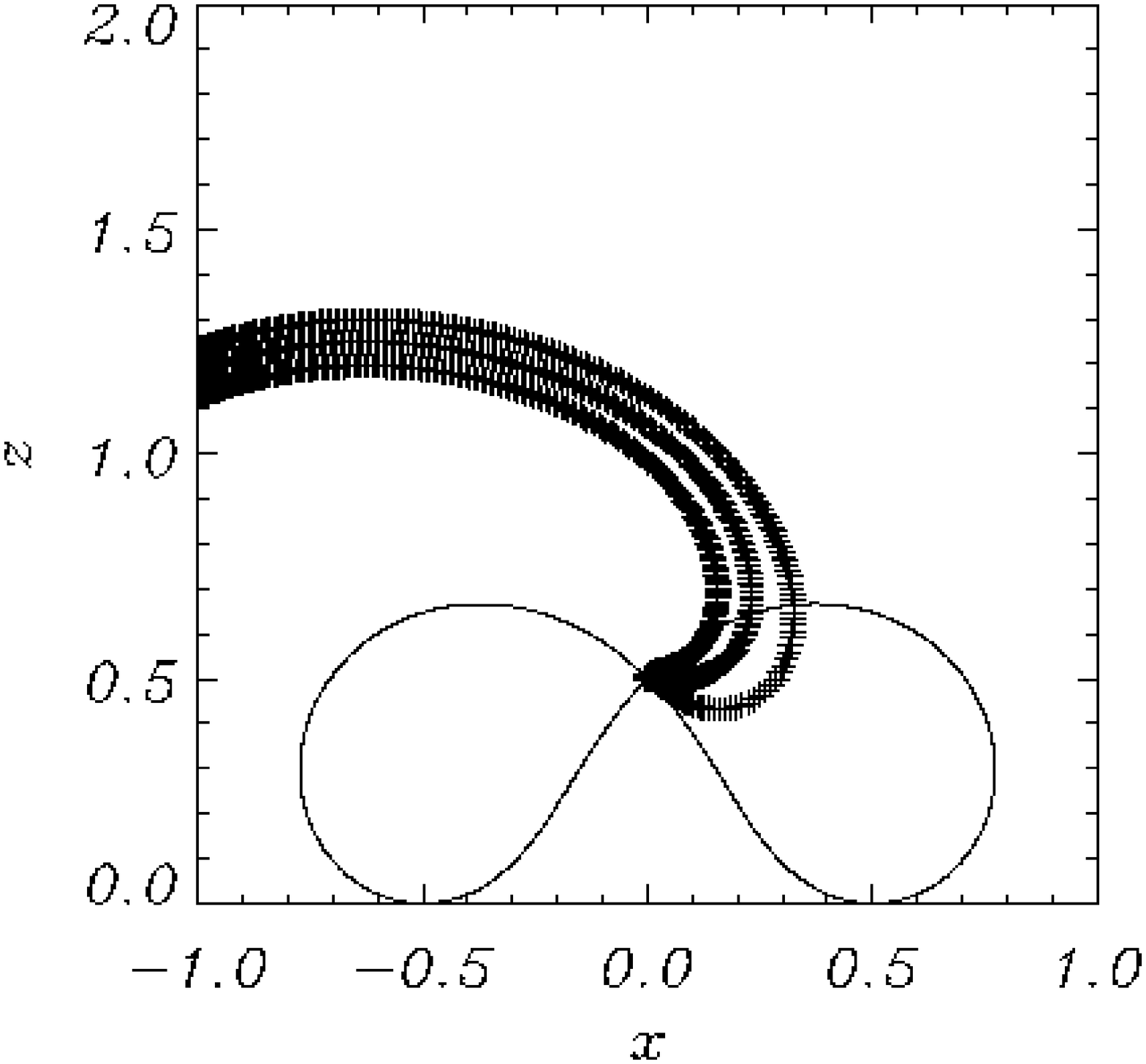}
\hspace{0.0in}
\includegraphics[width=1.2in]{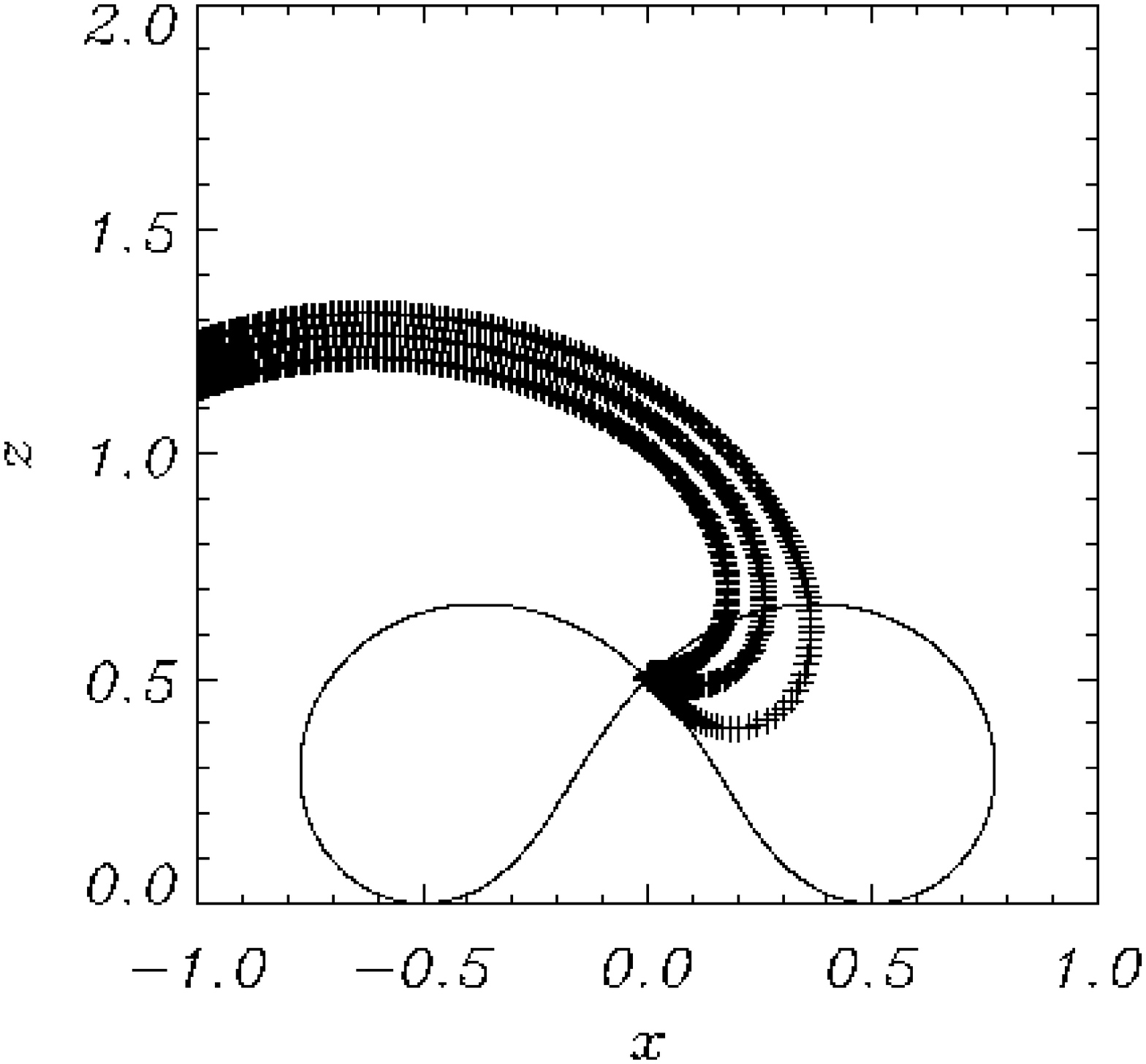}\\
\hspace{0in}
\vspace{0.1in}
\hspace{0.2in}
\includegraphics[width=1.2in]{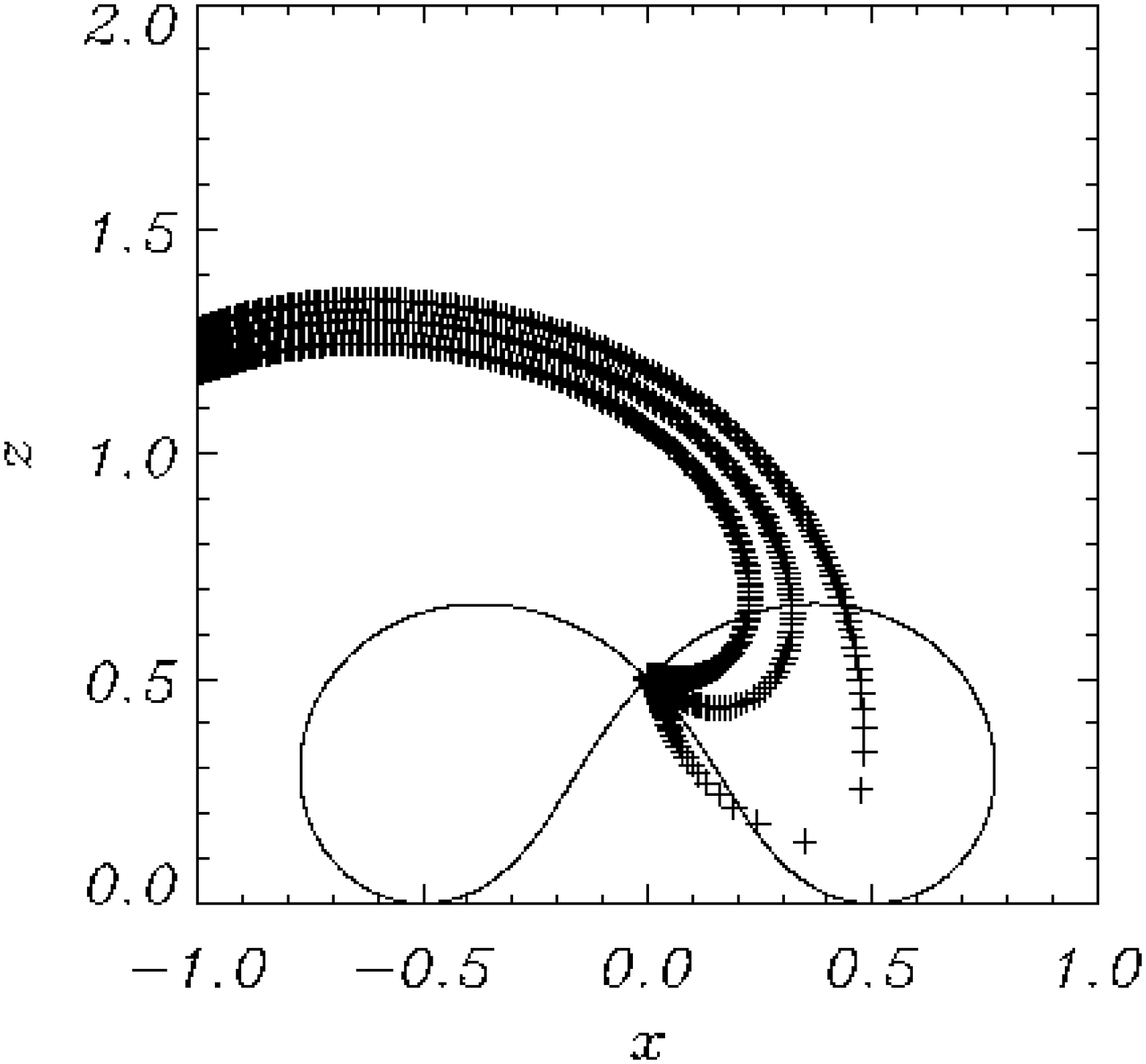}
\hspace{0.0in}
\includegraphics[width=1.2in]{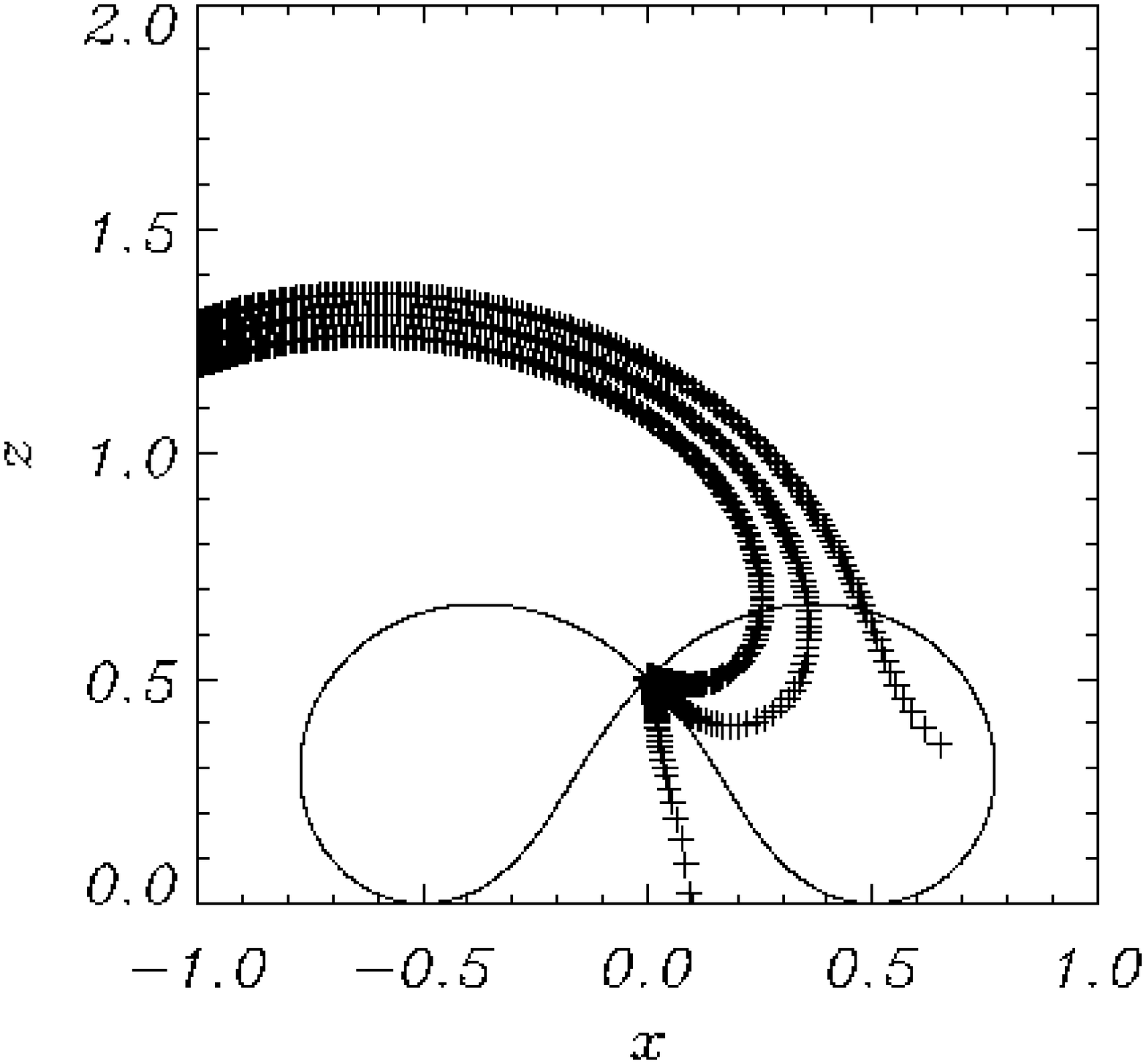}
\hspace{0.0in}
\includegraphics[width=1.2in]{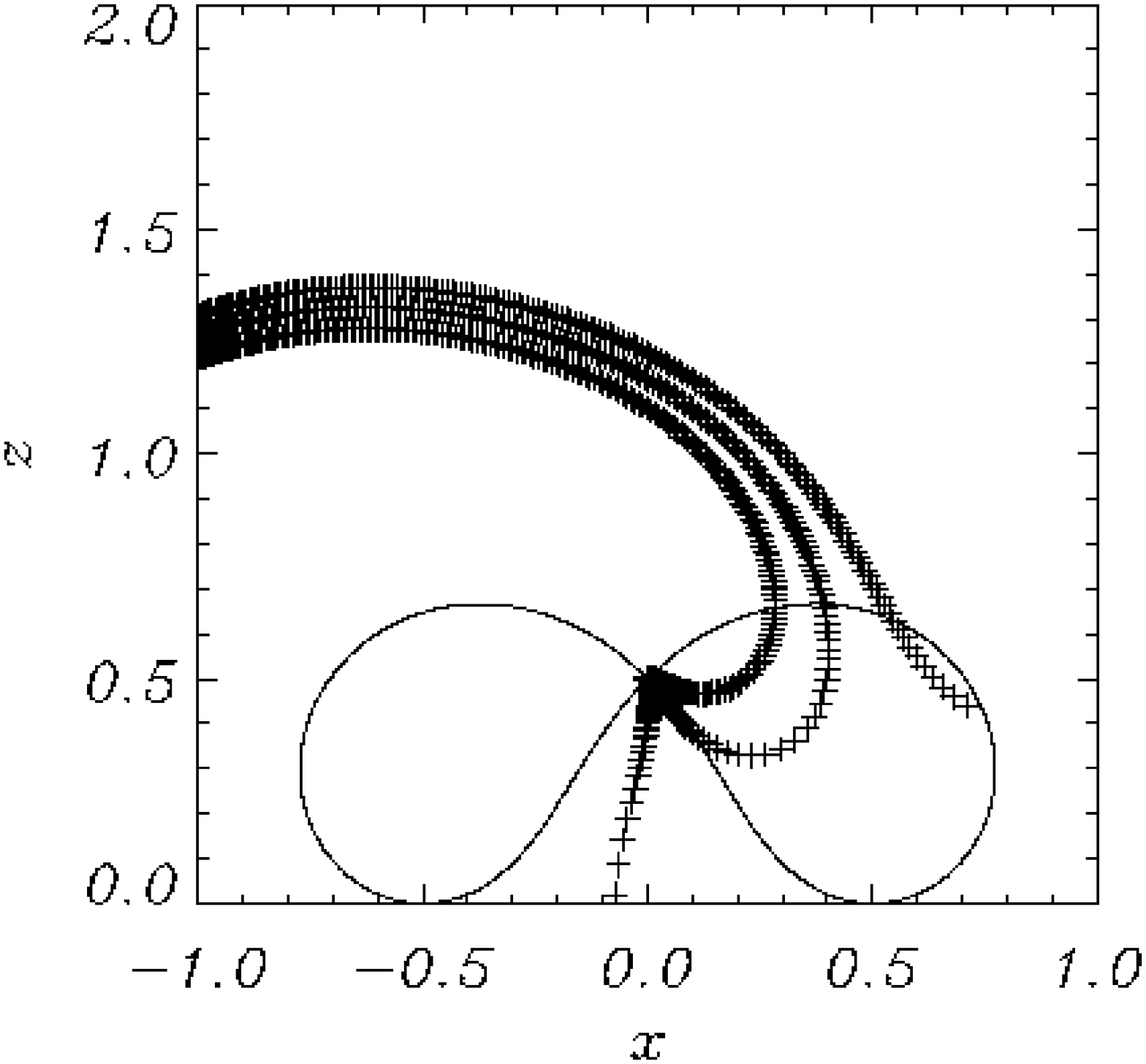}
\hspace{0.0in}
\includegraphics[width=1.2in]{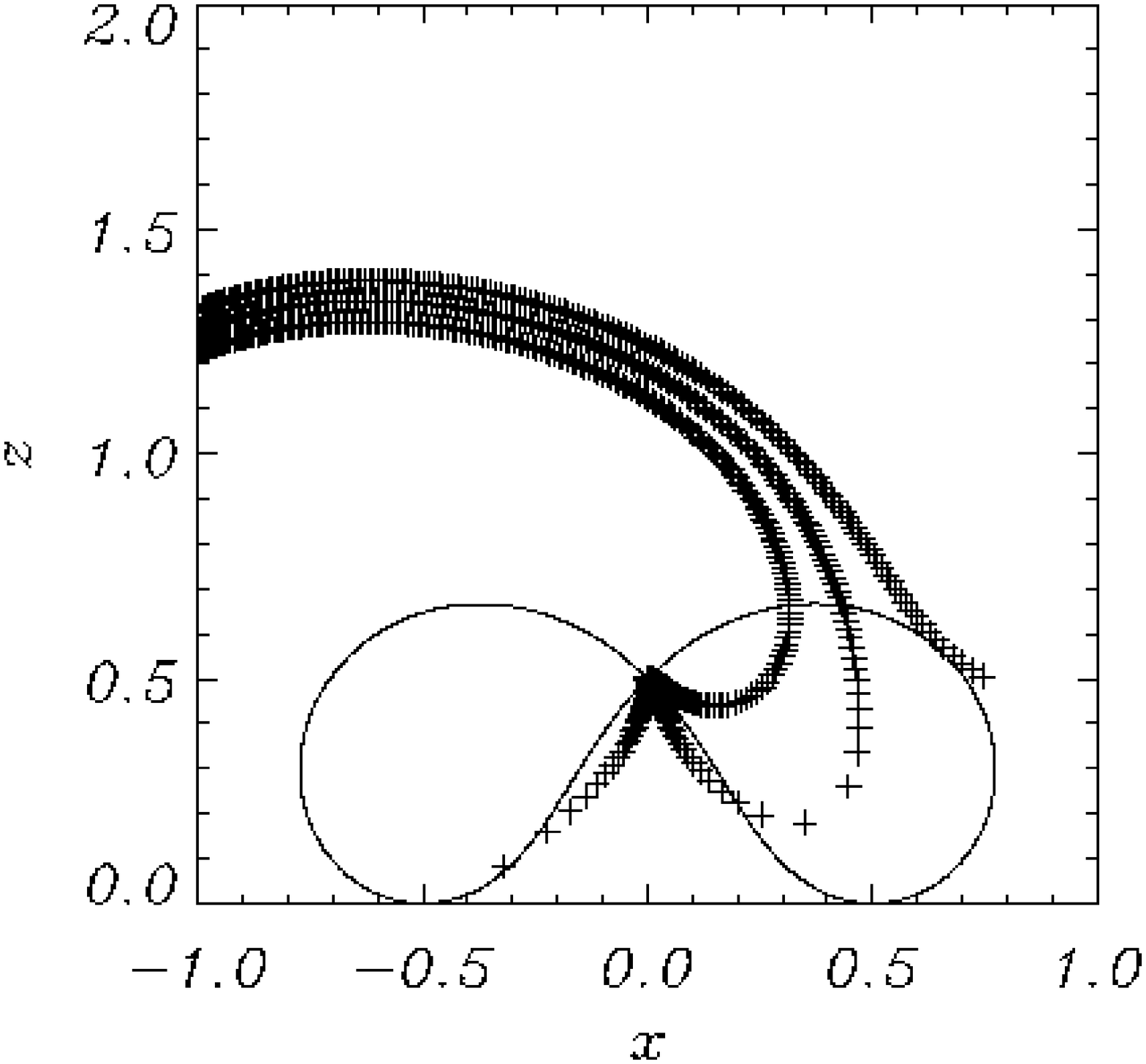}\\
\hspace{0in}
\vspace{0.1in}
\hspace{0.2in}
\includegraphics[width=1.2in]{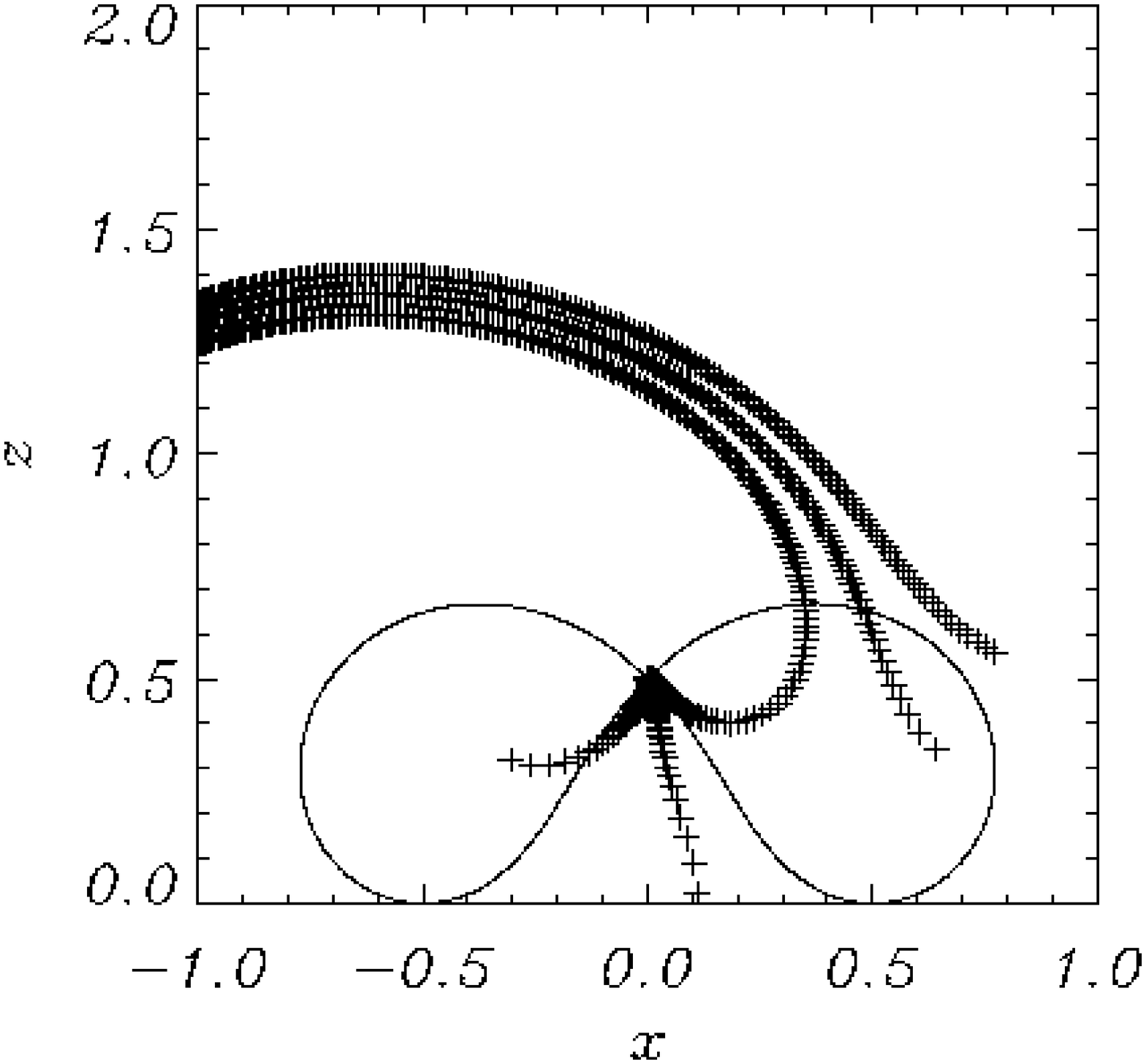}
\hspace{0.0in}
\includegraphics[width=1.2in]{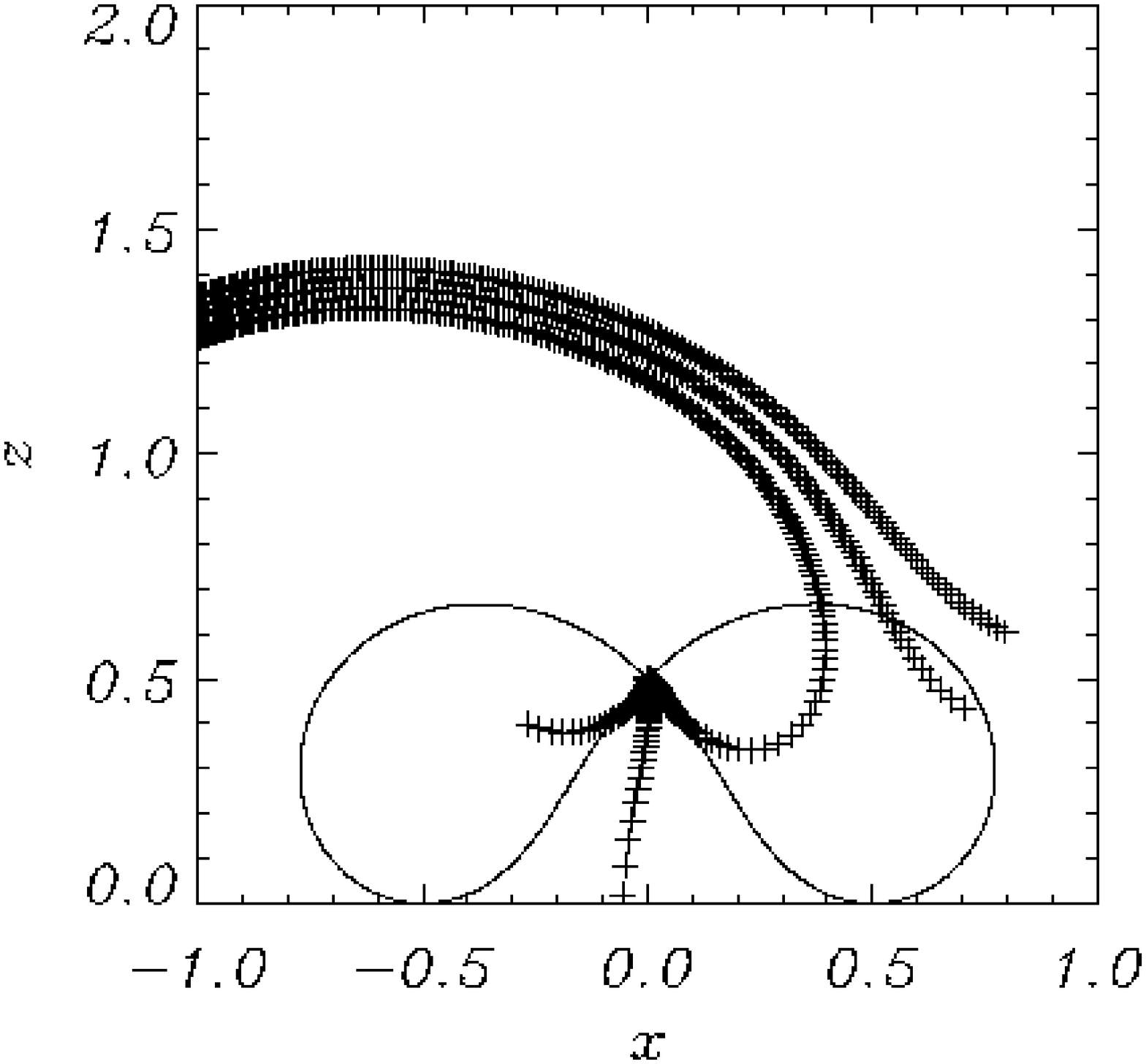}
\hspace{0.0in}
\includegraphics[width=1.2in]{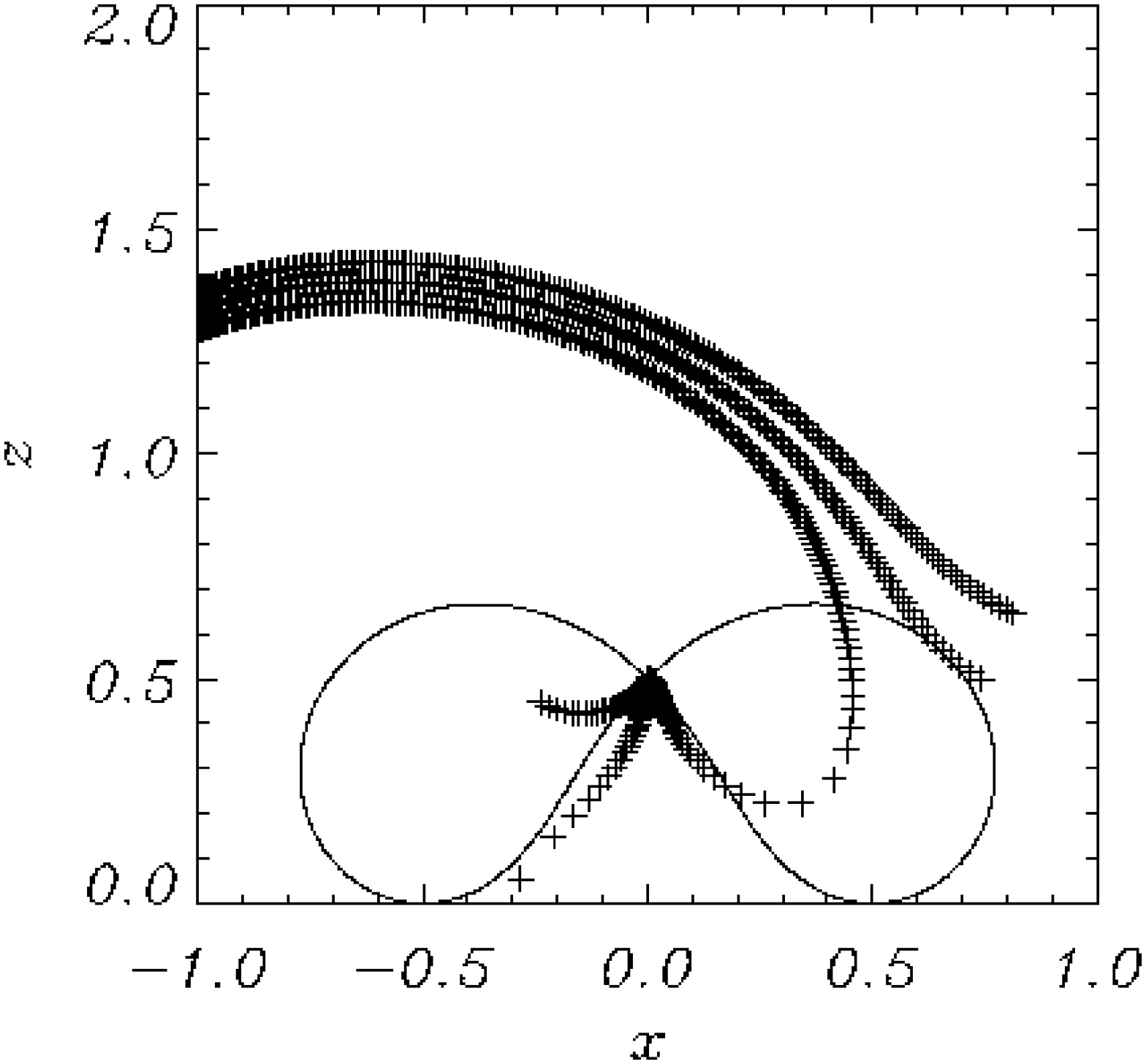}
\hspace{0.0in}
\includegraphics[width=1.2in]{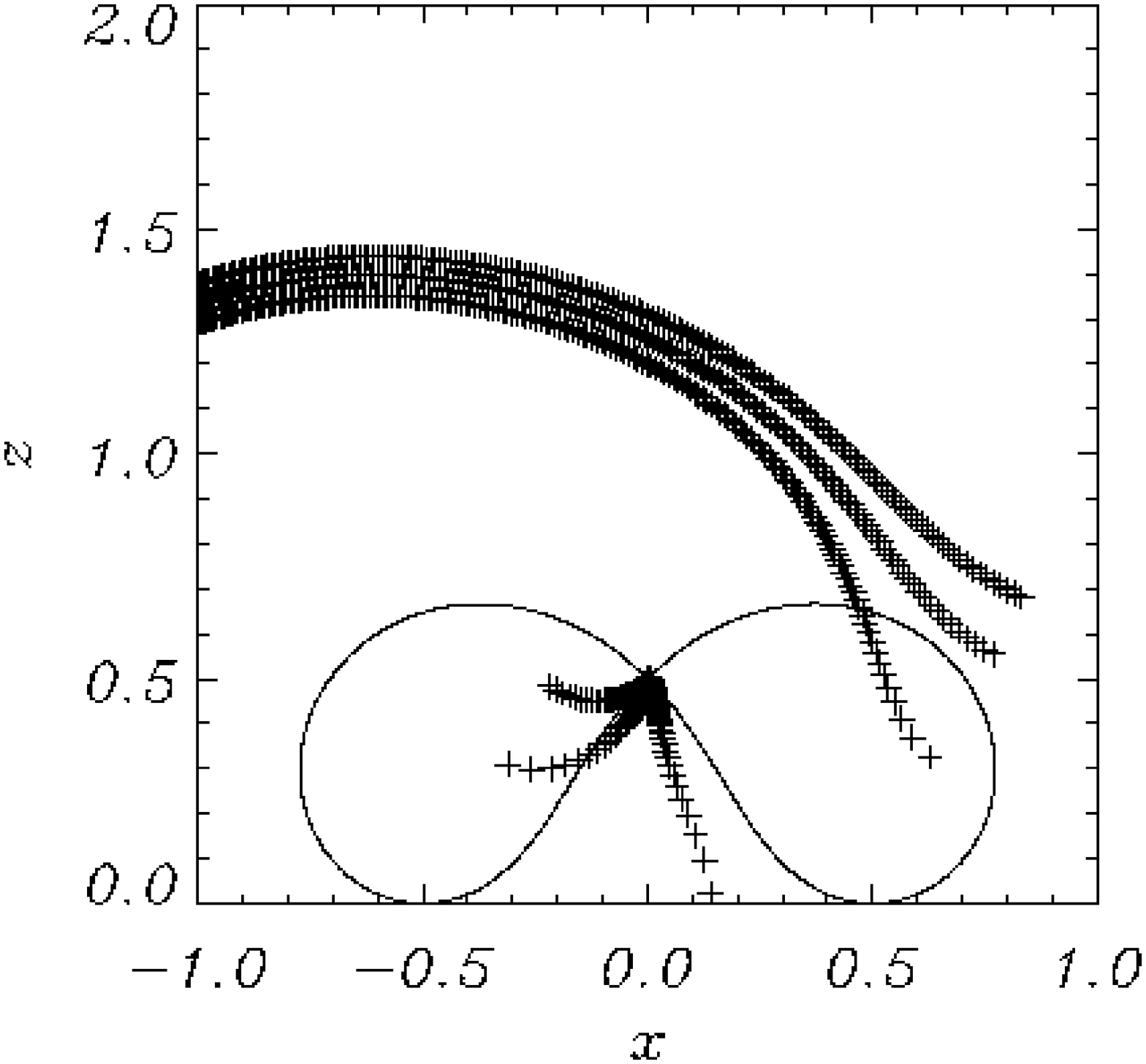}\\
\hspace{0in}
\vspace{0.1in}
\hspace{0.2in}
\includegraphics[width=1.2in]{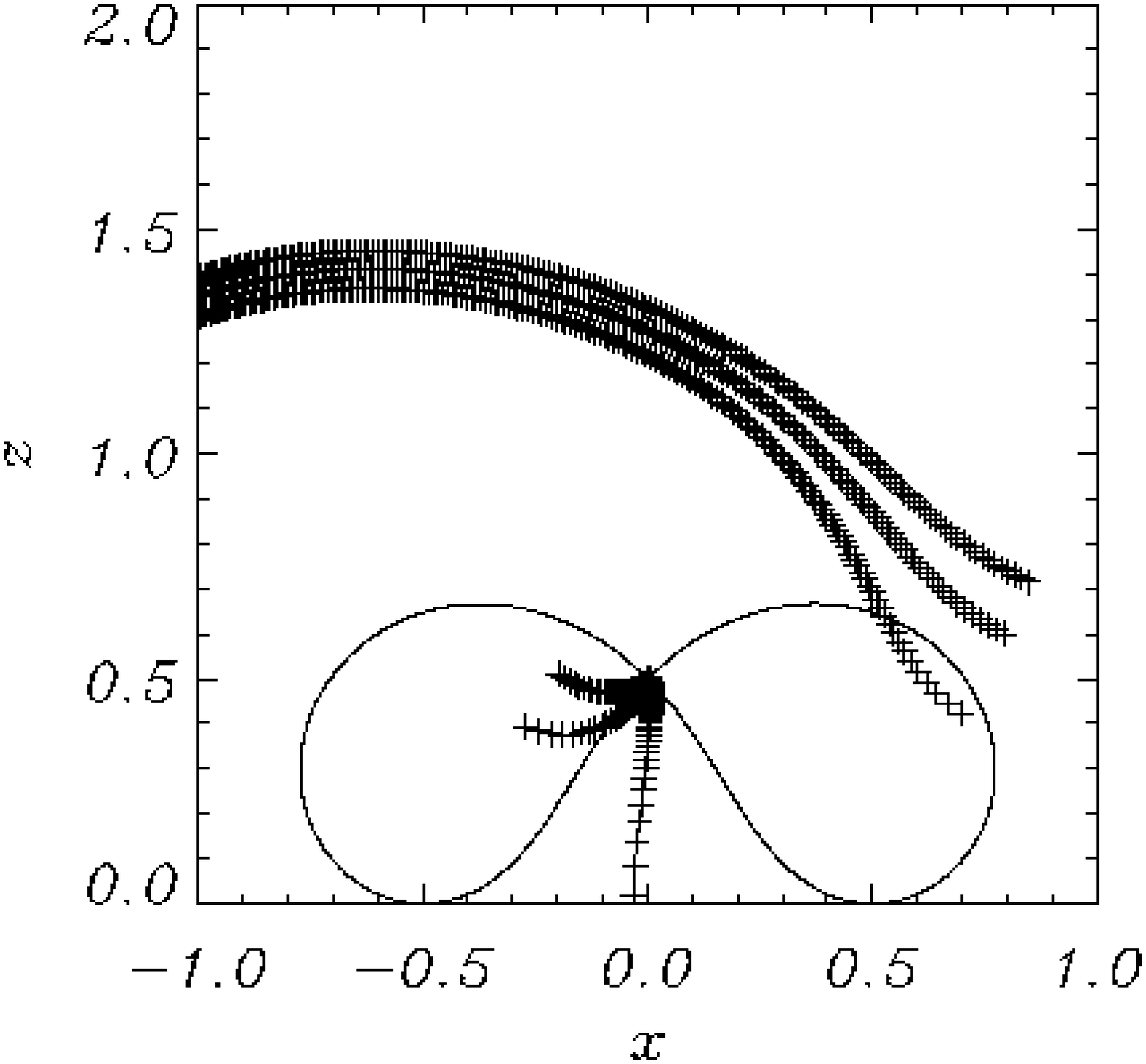}
\hspace{0.0in}
\includegraphics[width=1.2in]{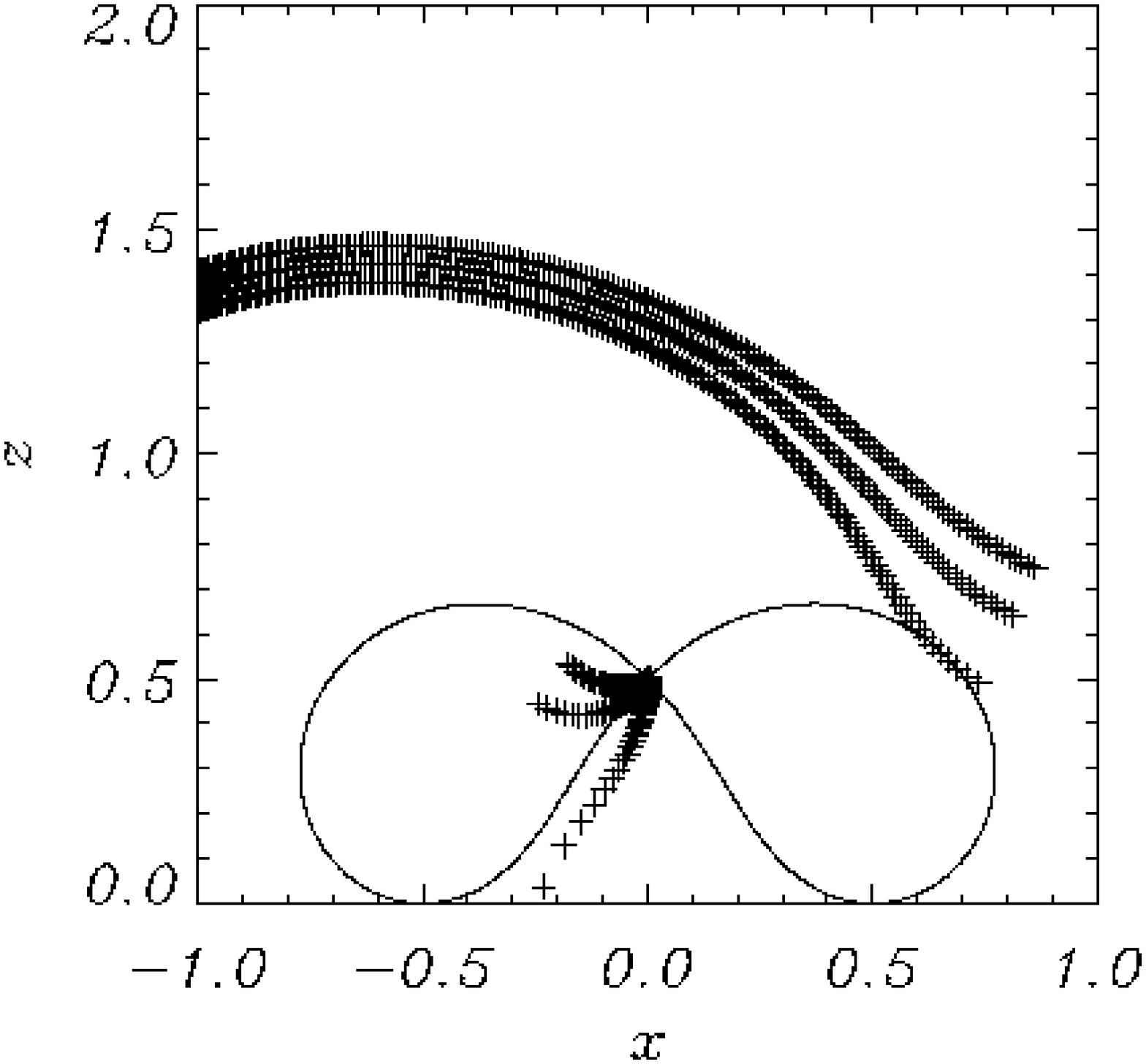}
\hspace{0.0in}
\includegraphics[width=1.2in]{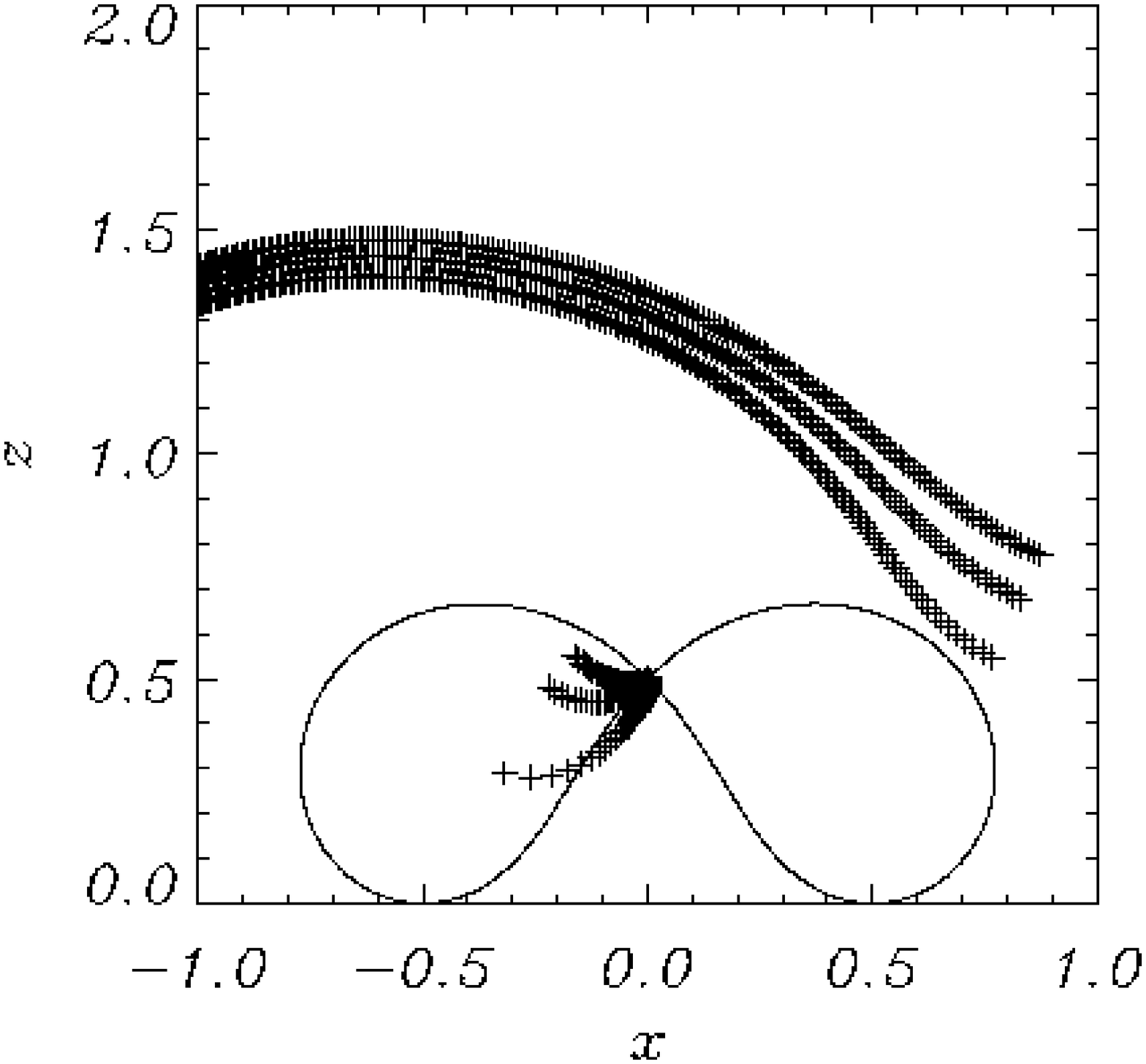}
\hspace{0.0in}
\includegraphics[width=1.2in]{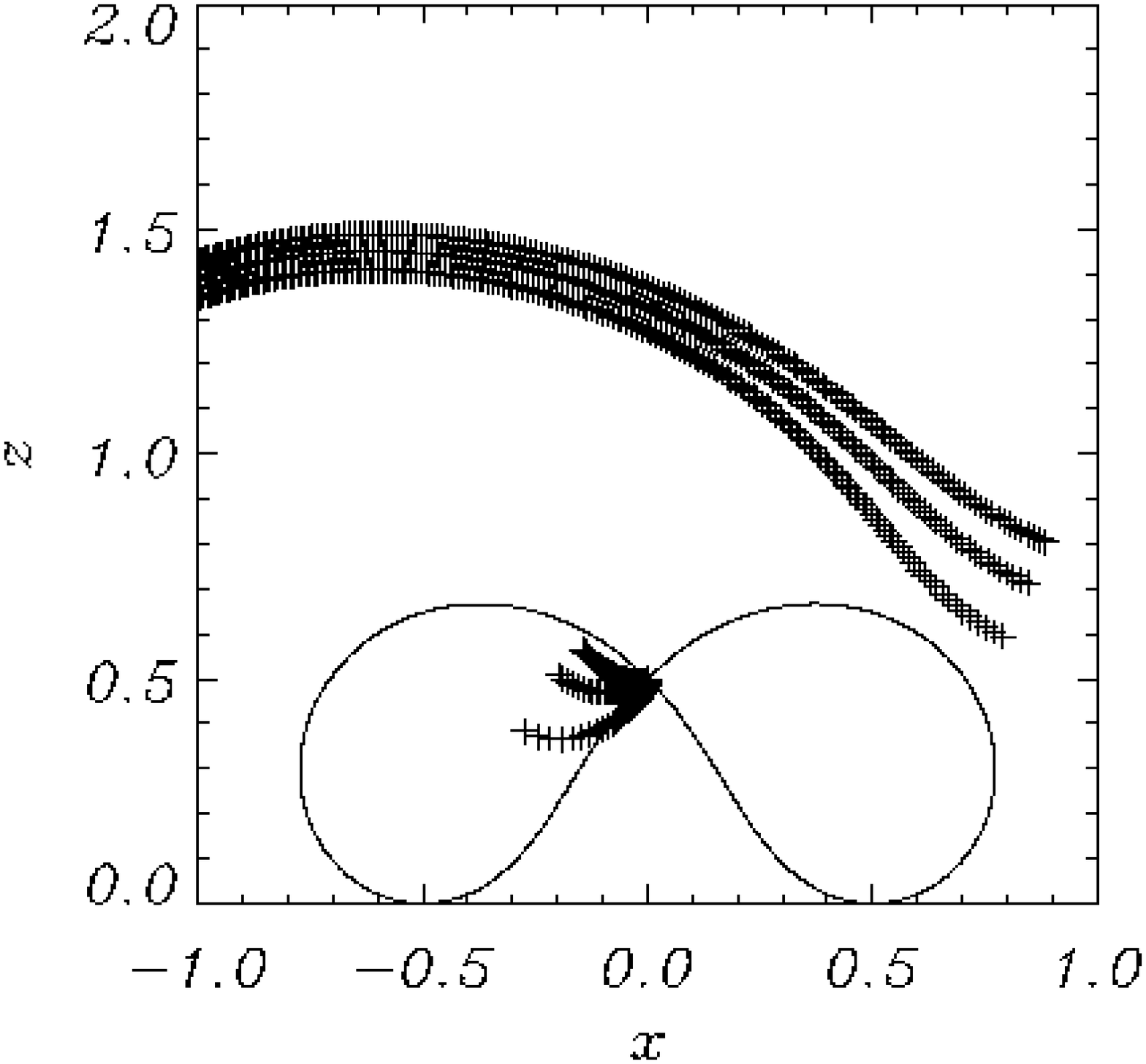}
\caption{Analytical solution of $V$ for WKB approximation of a fast wave sent in from lower boundary for $-1 \leq x \leq 0 $, $z=0.1$ and its resultant propagation at  times  $(a)$ $t$=0.055, $(b)$ $t$=0.11, $(c)$ $t$=0.165, $(d)$ $t=$0.22, $(e)$ $t$=0.275 and $(f)$ $t$=0.33, $(g)$ $t$=0.385, $(h)$ $t$=0.44, $(i)$ $t$=0.495, $(j)$ $t=$0.55, $(k)$ $t$=0.605, $(l)$ $t$=0.66,  $(m)$ $t$=0.72, $(n)$ $t$=0.74, $(o)$ $t$=0.76, $(p)$ $t=$0.78, $(q)$ $t$=0.8, $(r)$ $t$=0.82, $(s)$ $t$=0.84, $(t)$ $t$=0.86, $(u)$ $t$=0.88, $(v)$ $t=$0.9, $(w)$ $t$=0.92 and $(x)$ $t$=0.94, labelling from top left to bottom right. The  curves consisting of the crosses represent the front, middle and back edges of the WKB wave solution, where the pulse enters from the top of the box. The magnetic skeleton is also shown.}
\label{fig:4.5.2.5.hans}
\end{figure*}

Thus, by only plotting the fieldlines coming from $-1 \leq x_0 \leq 0$, we can construct the analytical equivalent of  Simulation Two. This can be seen in Figure \ref{fig:4.5.2.5.hans}.  Also, with correct choices of $s$ the solution can be directly compared to our numerical solution. The agreement between the numerical work and WKB approximation is very good (the contours essentially lie on top of each other). Following the splitting of the fast wave in Figure   \ref{fig:4.5.2.5.hans} is particularly interesting. In the figure, the position of each wavefront is indicated by many tiny crosses. This allows the reader to more clearly follow see the wave stretching and splitting.

%**********************************************************************

%\vspace{0.3in}

%**********************************************************************

\begin{figure*}[t]
\begin{center}
\includegraphics[width=2.0in]{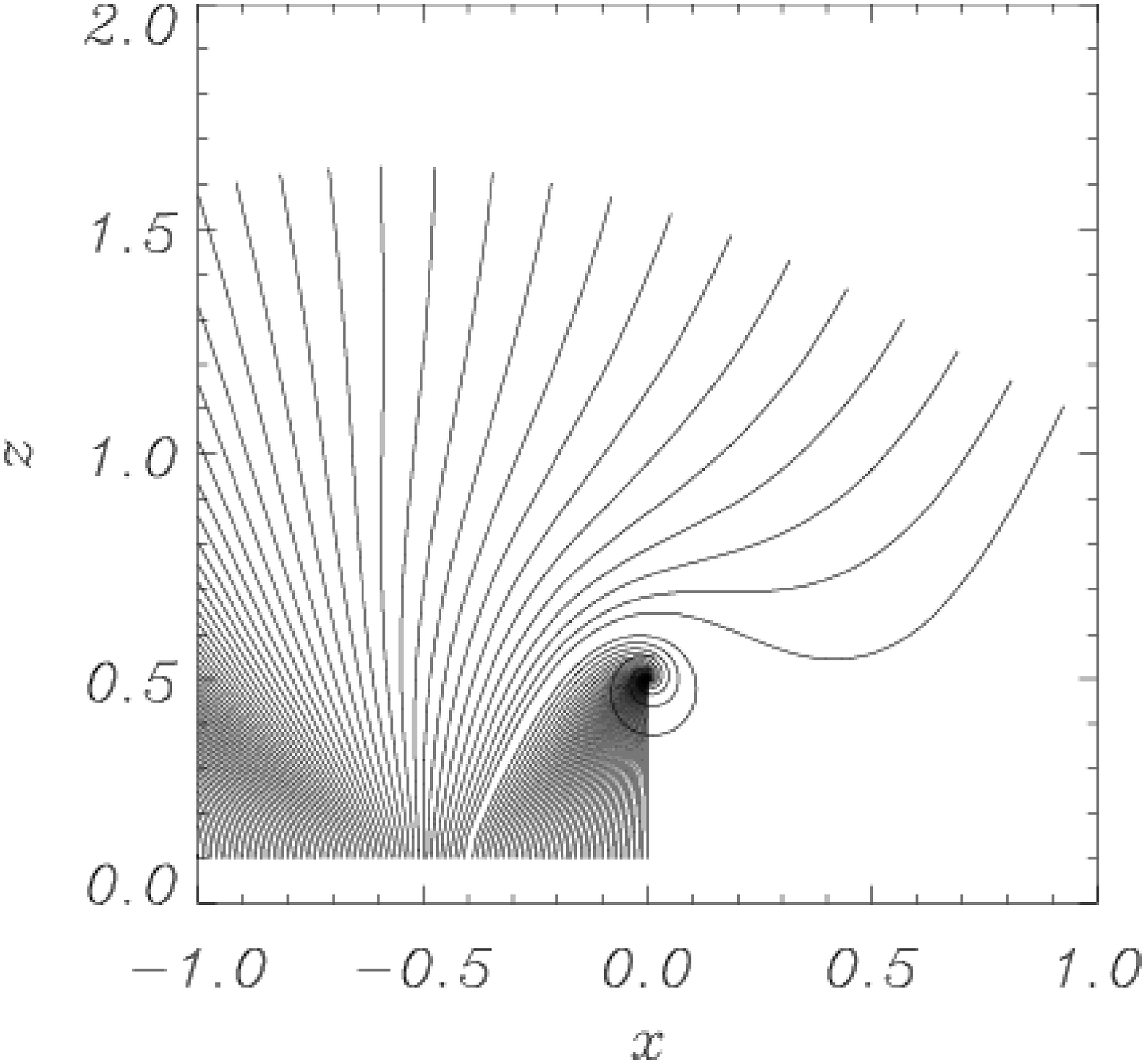}
\hspace{0.15in}
\includegraphics[width=2.0in]{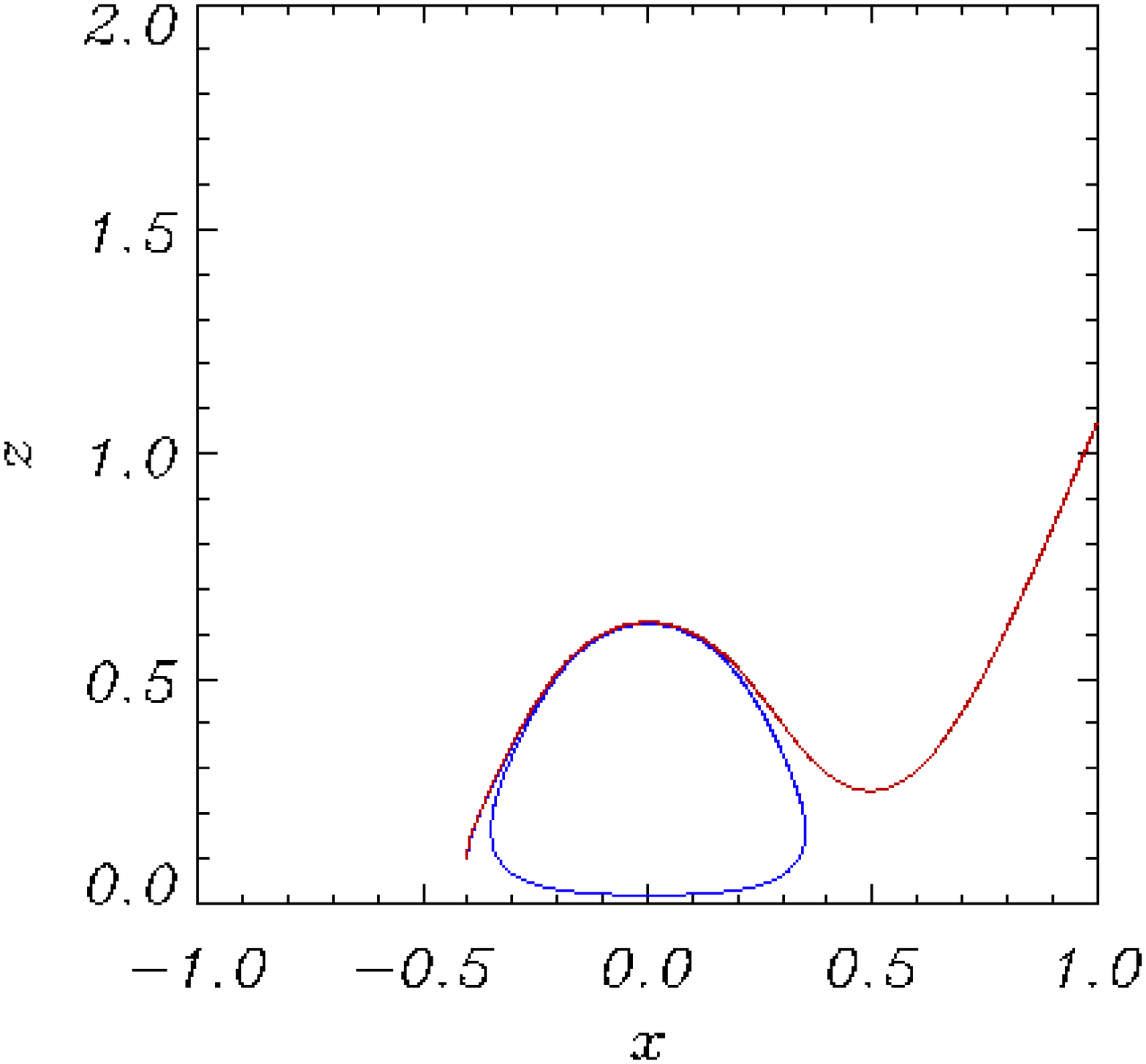}
\hspace{0.15in}
\includegraphics[width=2.0in]{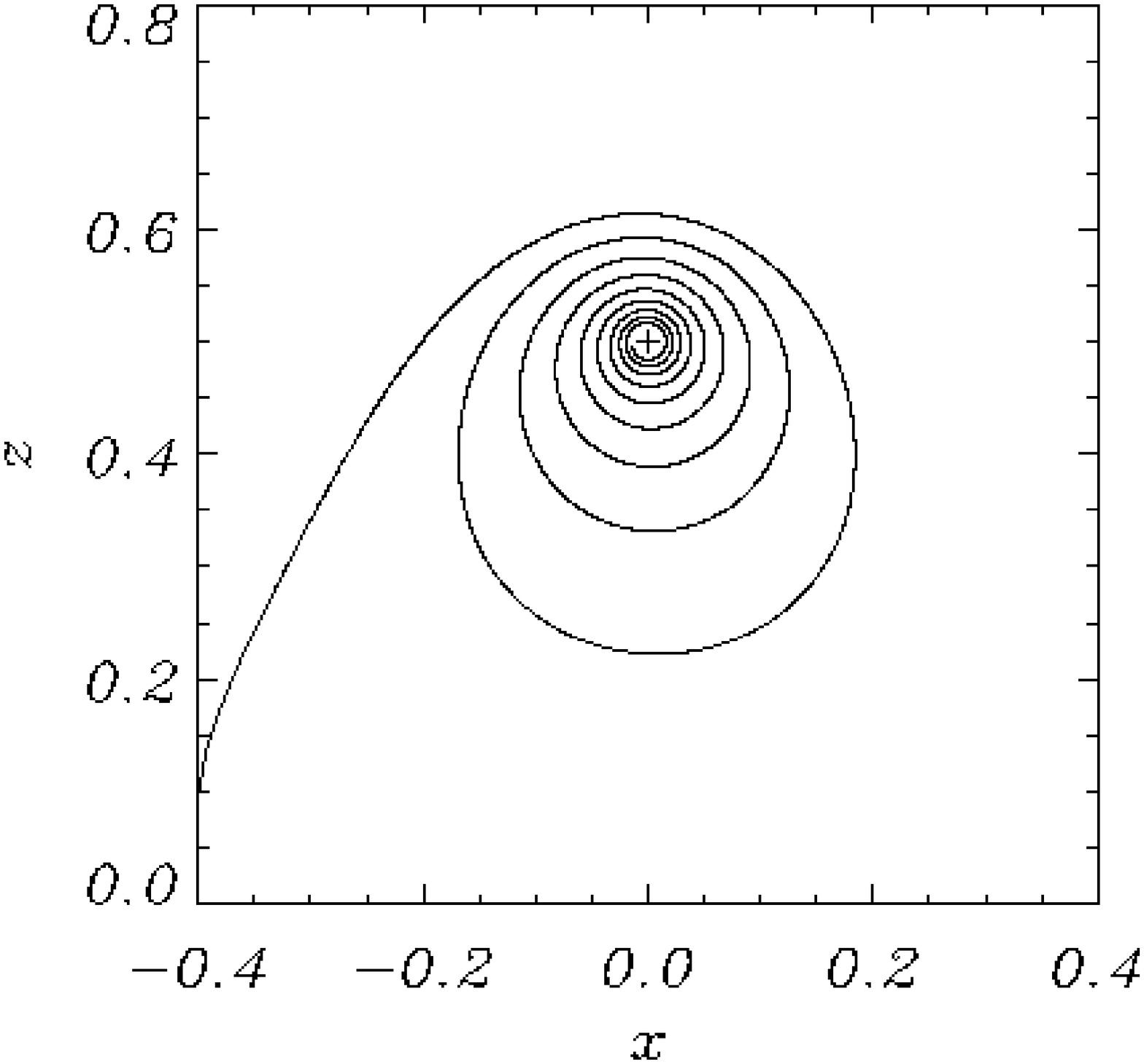}
\caption{\emph{Left} particle paths for starting points of $-1 \leq x_0 \leq 0$ except $x_0=-0.4$ set at intervals of $0.01$. \emph{Centre} particle paths for $x_0=-0.403$ (red) and  $x_0=-0.402$ (blue). \emph{Right} particle paths for starting point of $x_0=-0.4$ (note change in axes).}
\label{WKBtwodipolesparticlepaths}
\end{center}
\end{figure*}

We can also use our WKB solution to plot the particle paths of individual elements from the initial wave. This can be seen in Figure \ref{WKBtwodipolesparticlepaths}.

In the left subfigure, we see the particle paths for starting points of $-1 \leq x_0 \leq 0$ set at intervals of $0.01$. The line $x_0=-0.4$ has been removed (this starting point leads to a spiralling particle path that would confuse the plot). We see that  the lines for  $x_0 \leq -0.5$ do not appear to be influenced by the null and simply propagate at varying angles. The particle paths for starting points of $x_0 \ge -0.5$ and (approximately) $x_0 \leq -0.4$ are influenced by the null, but only in so much as to deflect the ray. However, for starting points greater than (approximately) $x_0 = -0.4$, the particle paths spiral in towards the X-point and are trapped. Thus, {\emph{there is a critical starting point that divides these two types of behaviour}}. The central figure shows the particle paths for $x_0=-0.403$ (red) and  $x_0=-0.402$ (blue). We can see that the critical starting point is thus somewhere in the middle of these two. Increasing the resolution shows the critical starting point to be $x_0=-0.40236$ to five decimal places. Finally, the right hand figure shows the particle path for a starting point of $x_0=-0.4$ (note the change in axes). This clearly demonstrates the particle spiralling into the null.

\subsection{Simulation Three}

Our final simulation again solves equations (\ref{fastalpha}) using our two-step Lax-Wendroff scheme, but this time for a wave pulse coming in from the entire lower boundary.  Using the magnetic field in Figure \ref{fig:dipolemagneticfield}, the boundary conditions were set such that:
\begin{eqnarray*}
V(x, z_0,t) = \left\{  \begin{array}{cl} 
 \sin { \omega t } \;\sin{\left[\frac{\pi}{2}\left( {x +1}\right)\right]}  & {\mathrm{for}} \;  \left\{  \begin{array}{c}
{-1 \leq x \leq 1}\\
  {0 \leq t \leq \frac {\pi}{\omega}}\end{array} \right.\\
0 & { \mathrm{otherwise} }\end{array} \right. \\
\left.\frac {\partial V} {\partial x }  \right| _{x=-1.4} = 0 \; , \quad \left.\frac {\partial V} {\partial x }    \right| _{x=1.4} = 0 \; , \quad \left.\frac {\partial  V } {\partial z }  \right| _{z=1.7}  = 0 \; .
\end{eqnarray*}

\begin{figure*}[t]
\hspace{0in}
\vspace{0.1in}
%\hspace{0.2in}
\includegraphics[width=1.2in]{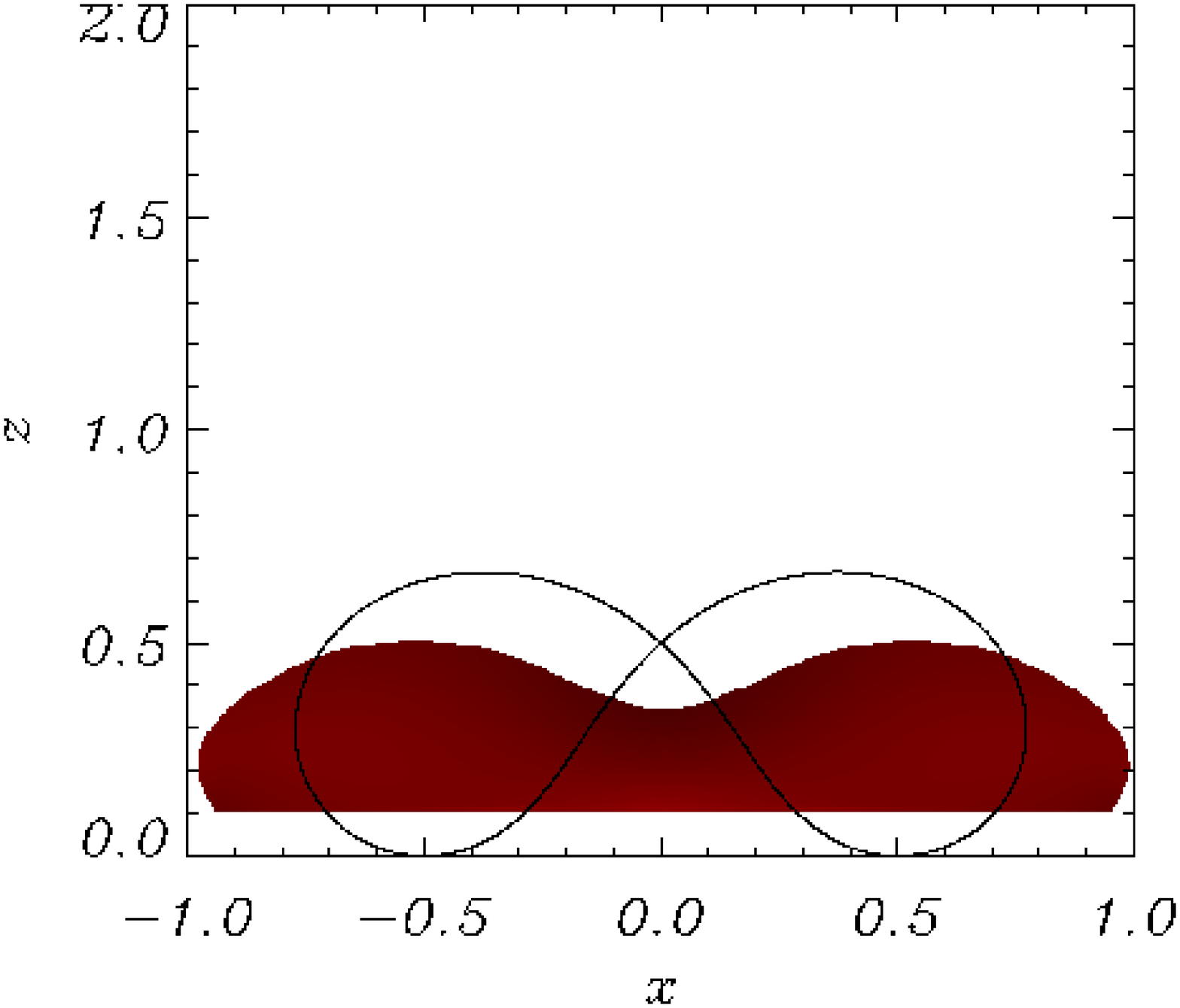}
\hspace{0.0in}
\includegraphics[width=1.2in]{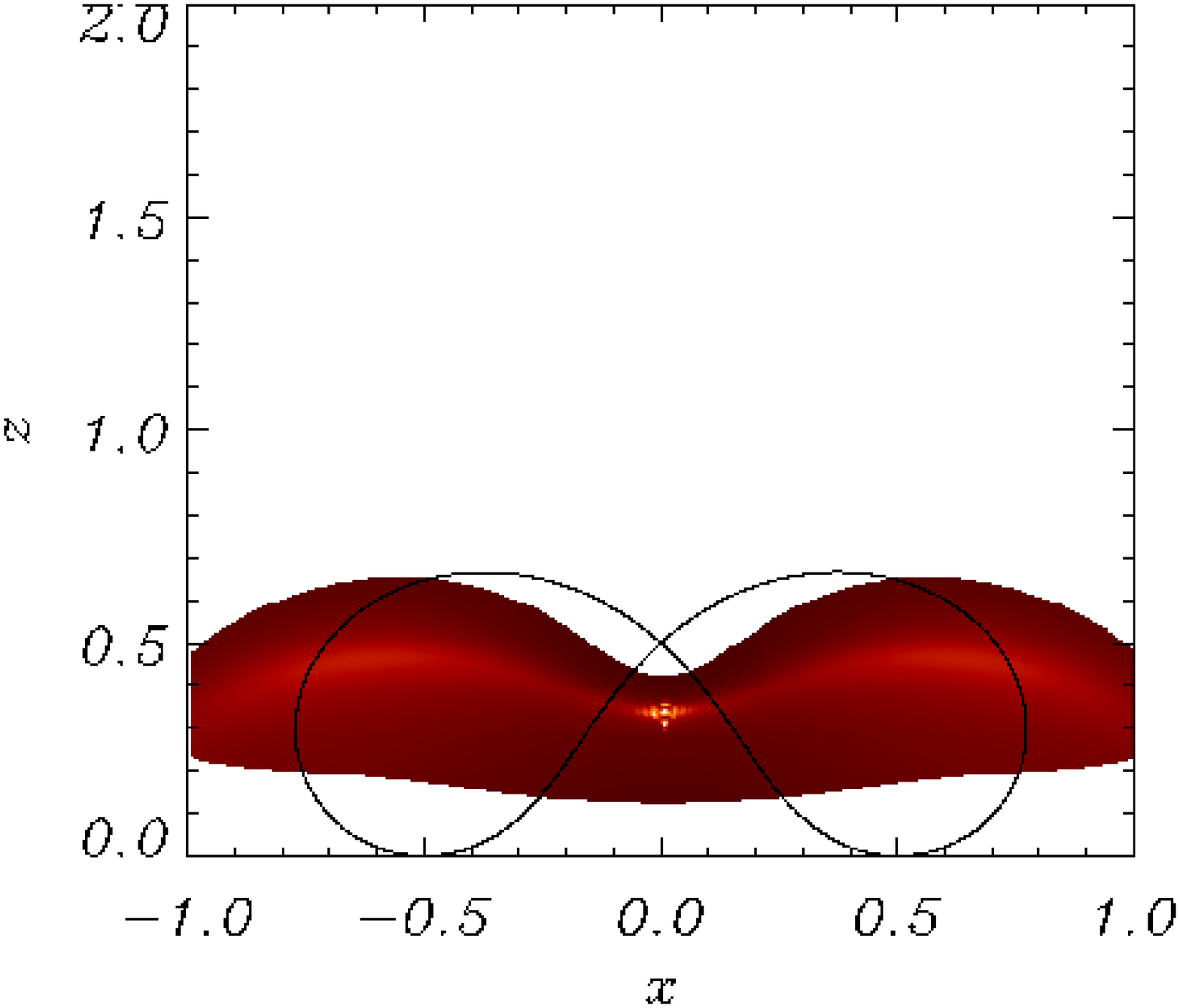}
\hspace{0.0in}
\includegraphics[width=1.2in]{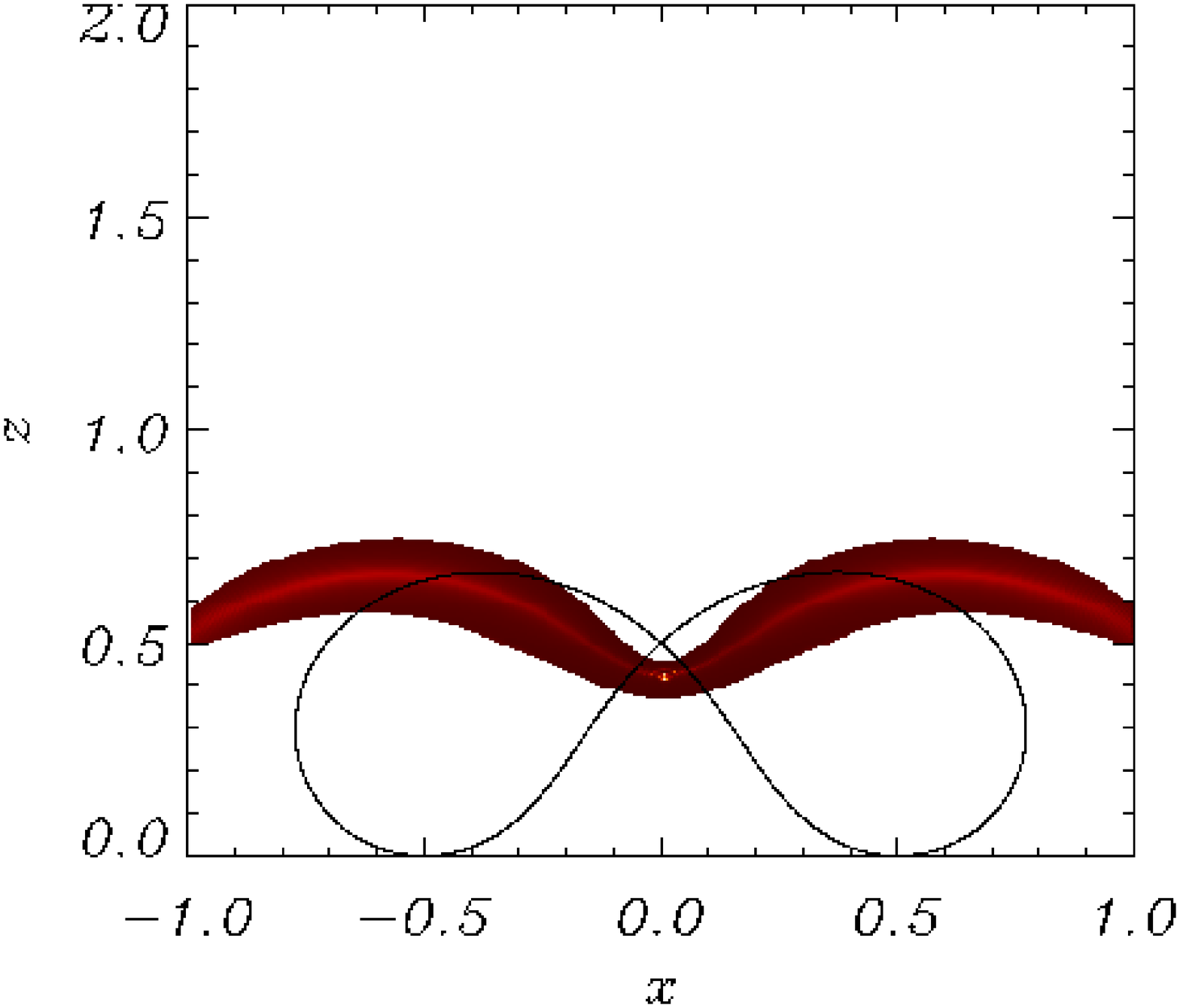}
\hspace{0.0in}
\includegraphics[width=1.2in]{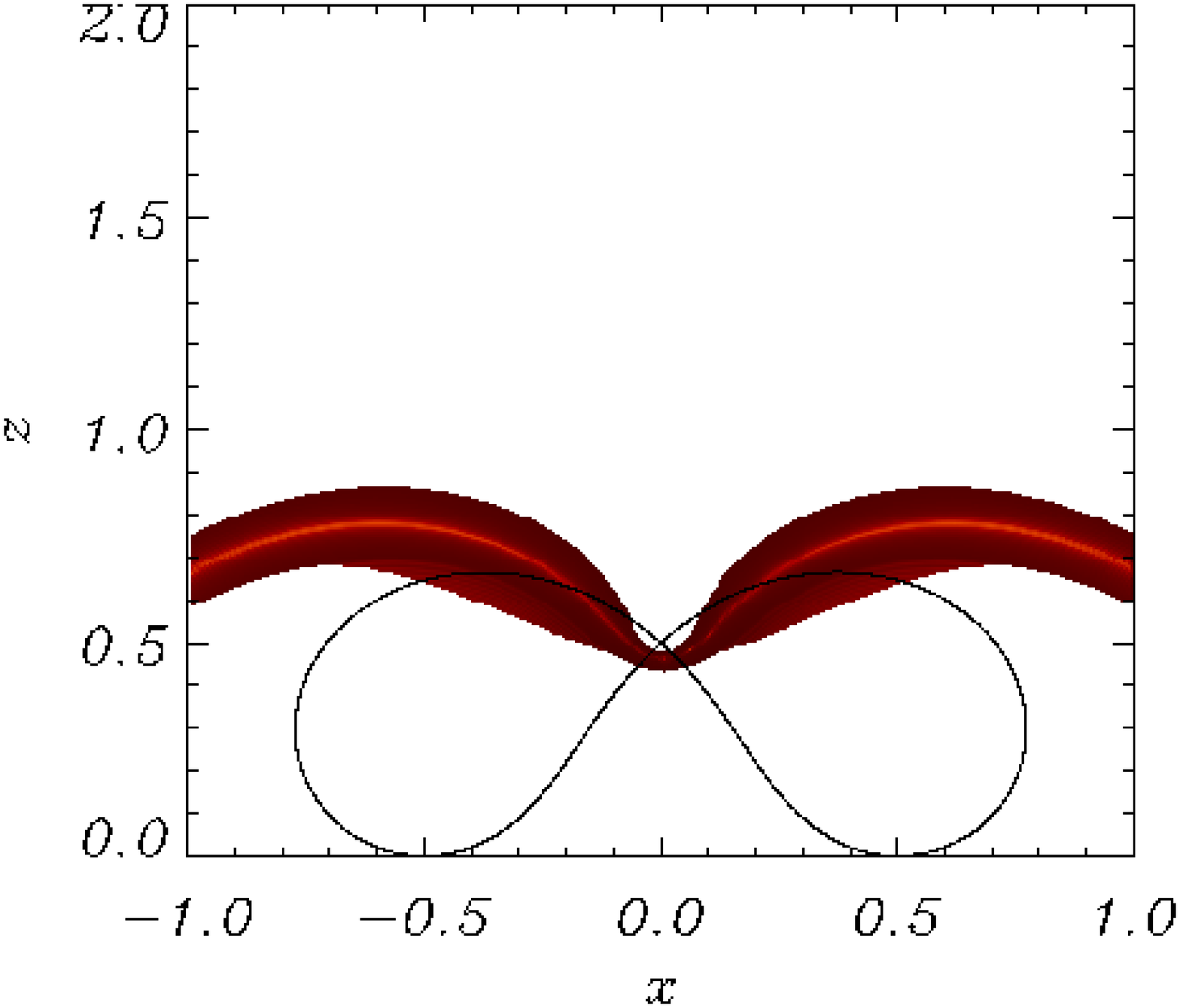}\\
\hspace{0in}
\vspace{0.1in}
%\hspace{0.2in}
\includegraphics[width=1.2in]{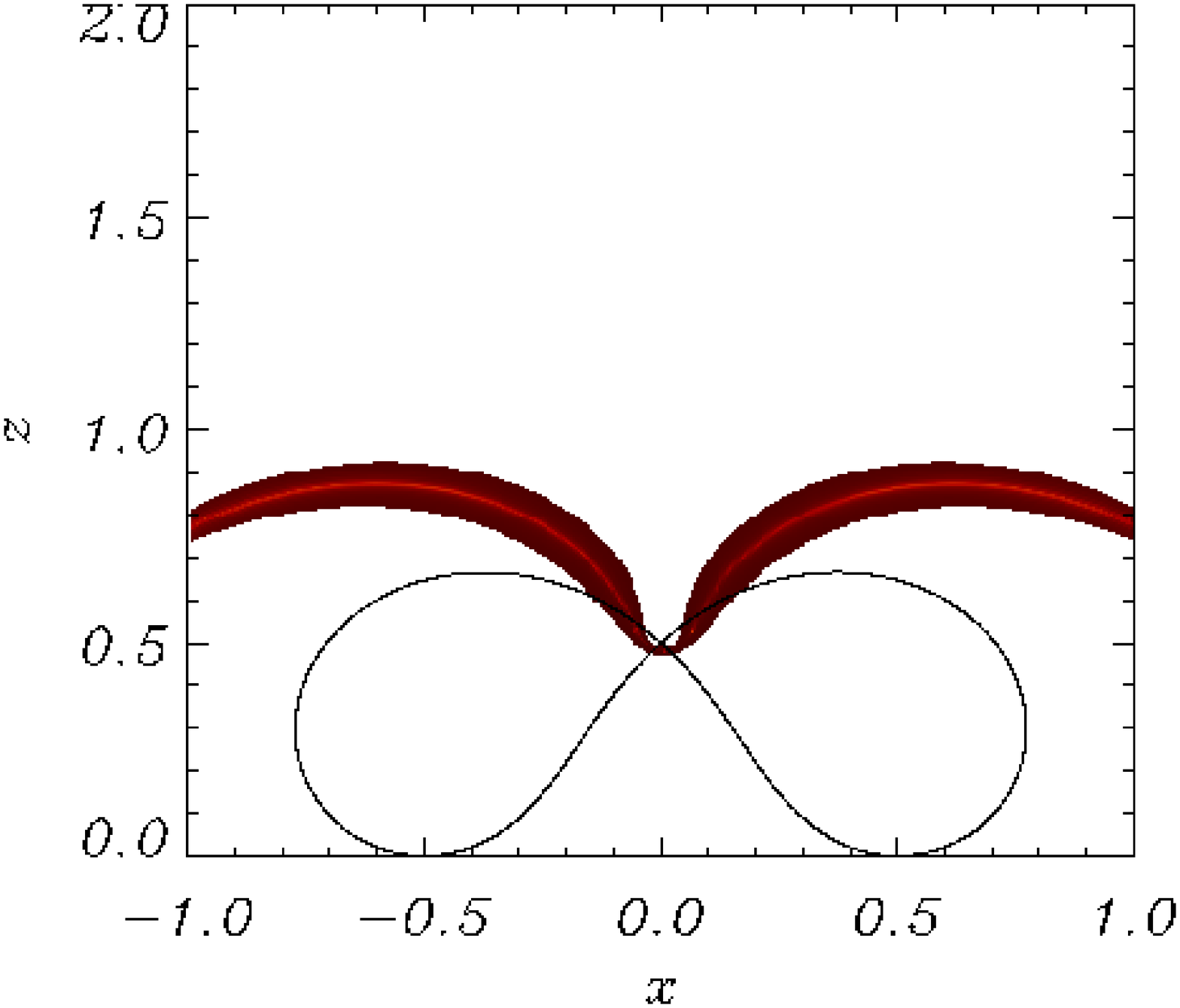}
\hspace{0.0in}
\includegraphics[width=1.2in]{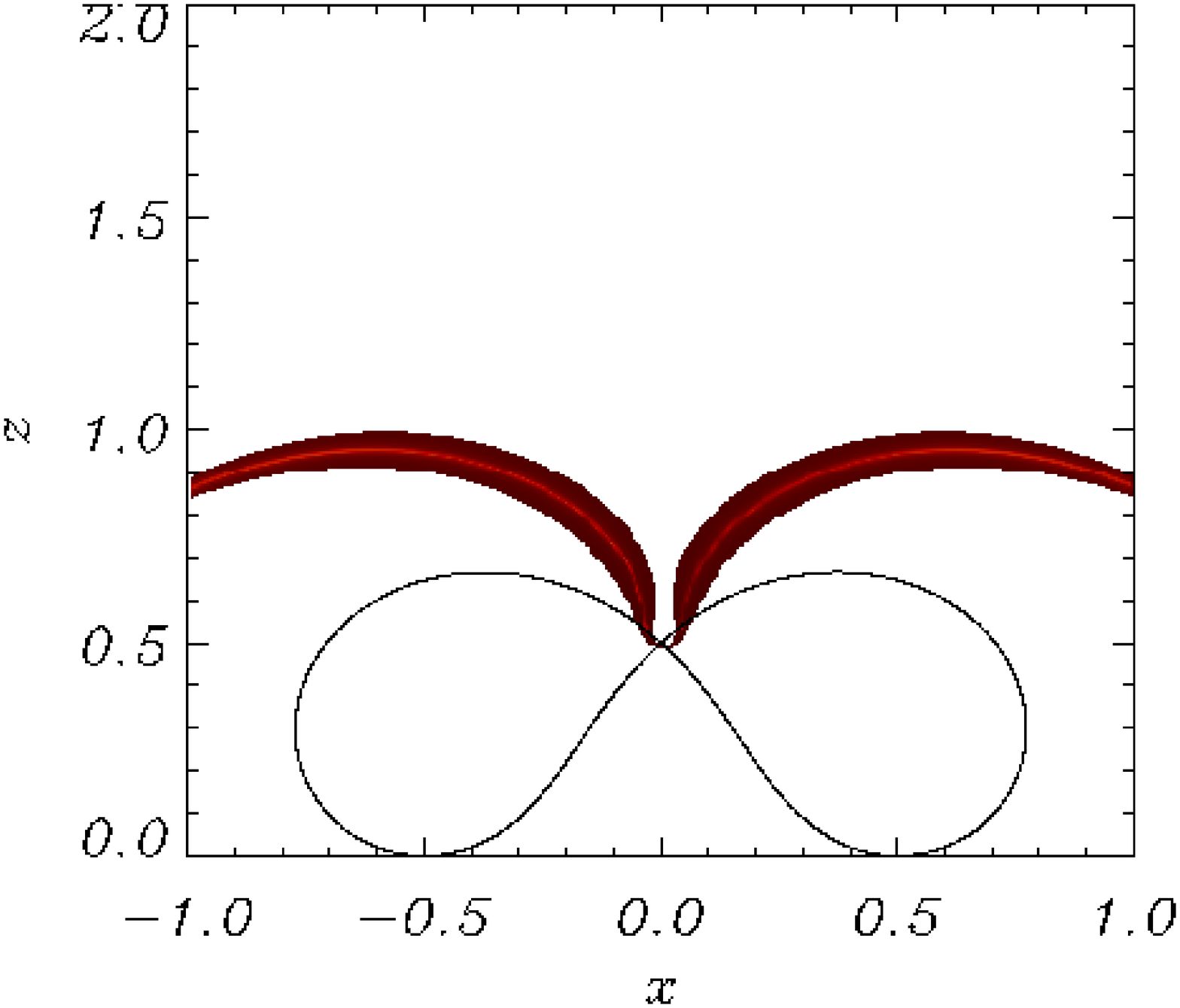}
\hspace{0.0in}
\includegraphics[width=1.2in]{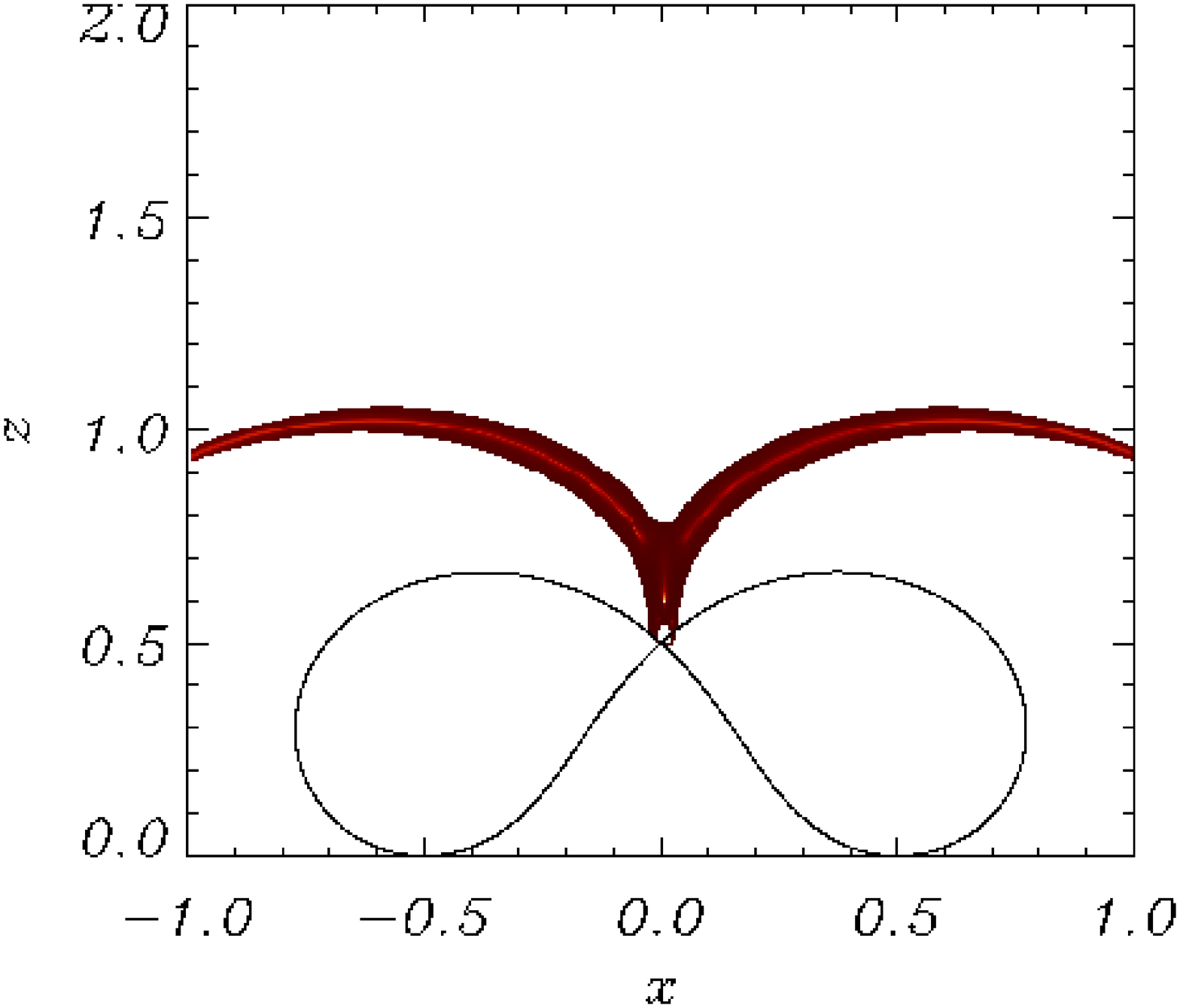}
\hspace{0.0in}
\includegraphics[width=1.2in]{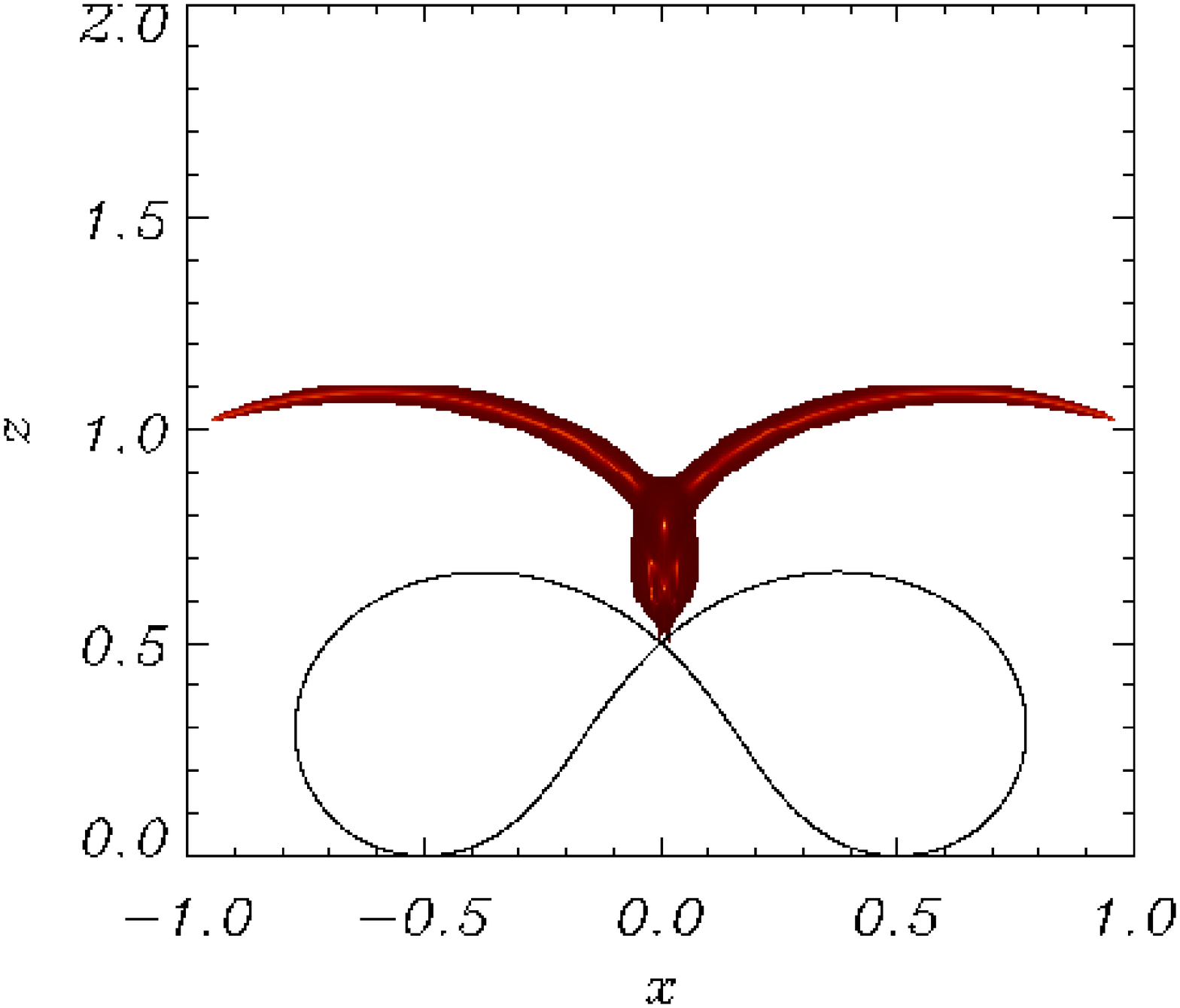}\\
\hspace{0in}
\vspace{0.1in}
%\hspace{0.2125in}
\includegraphics[width=1.2in]{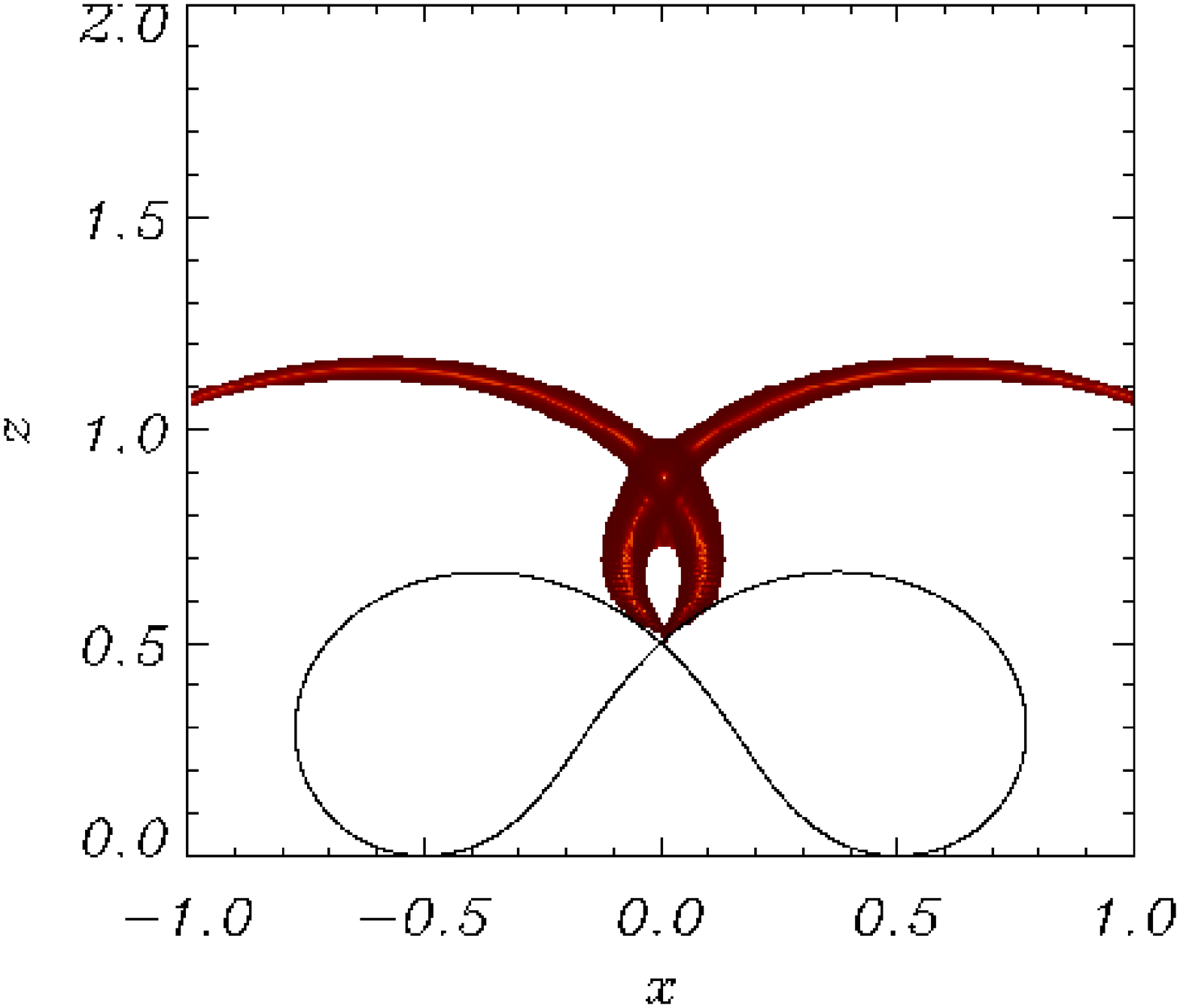}
\hspace{0.0in}
\includegraphics[width=1.2in]{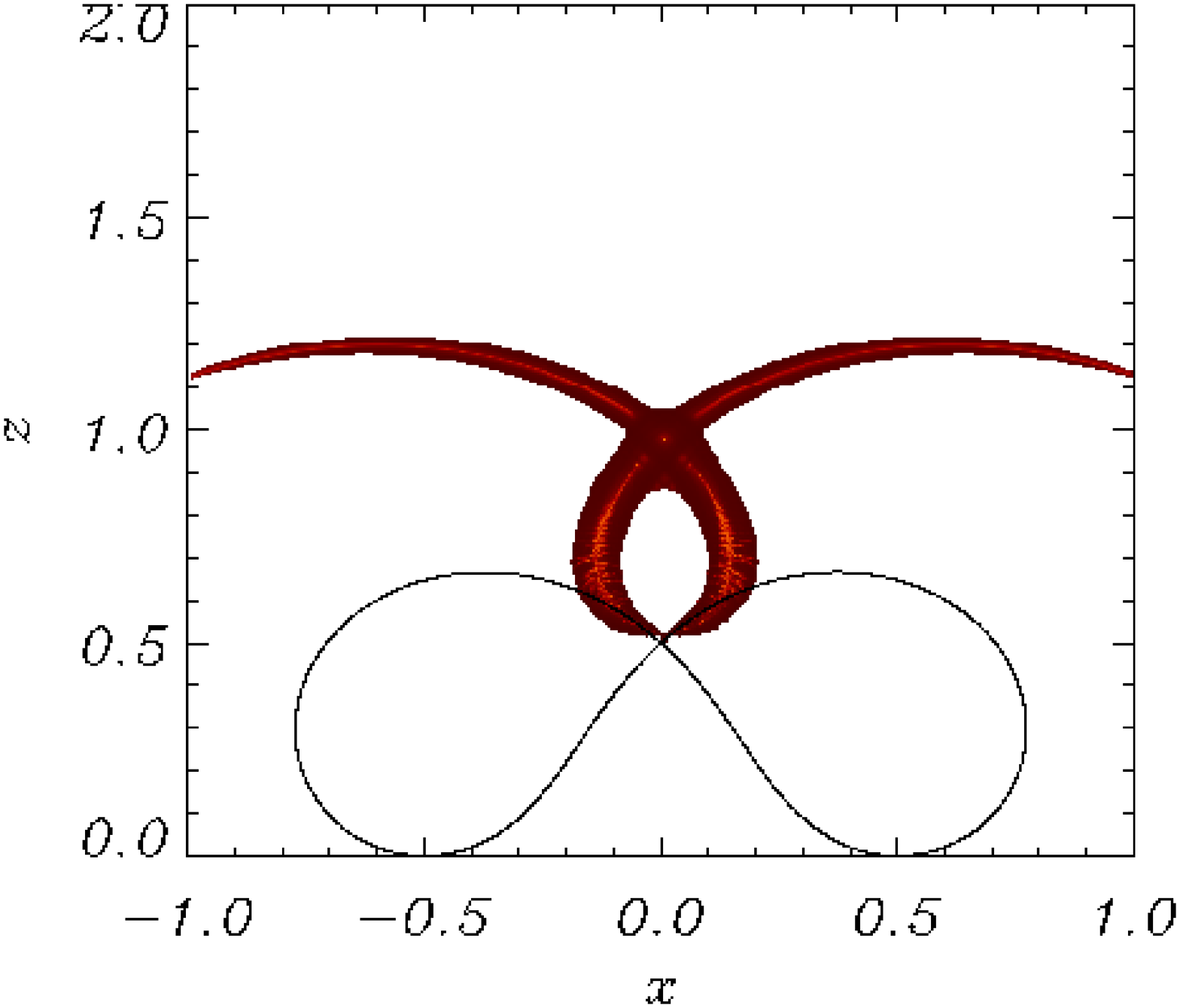}
\hspace{0.0in}
\includegraphics[width=1.2in]{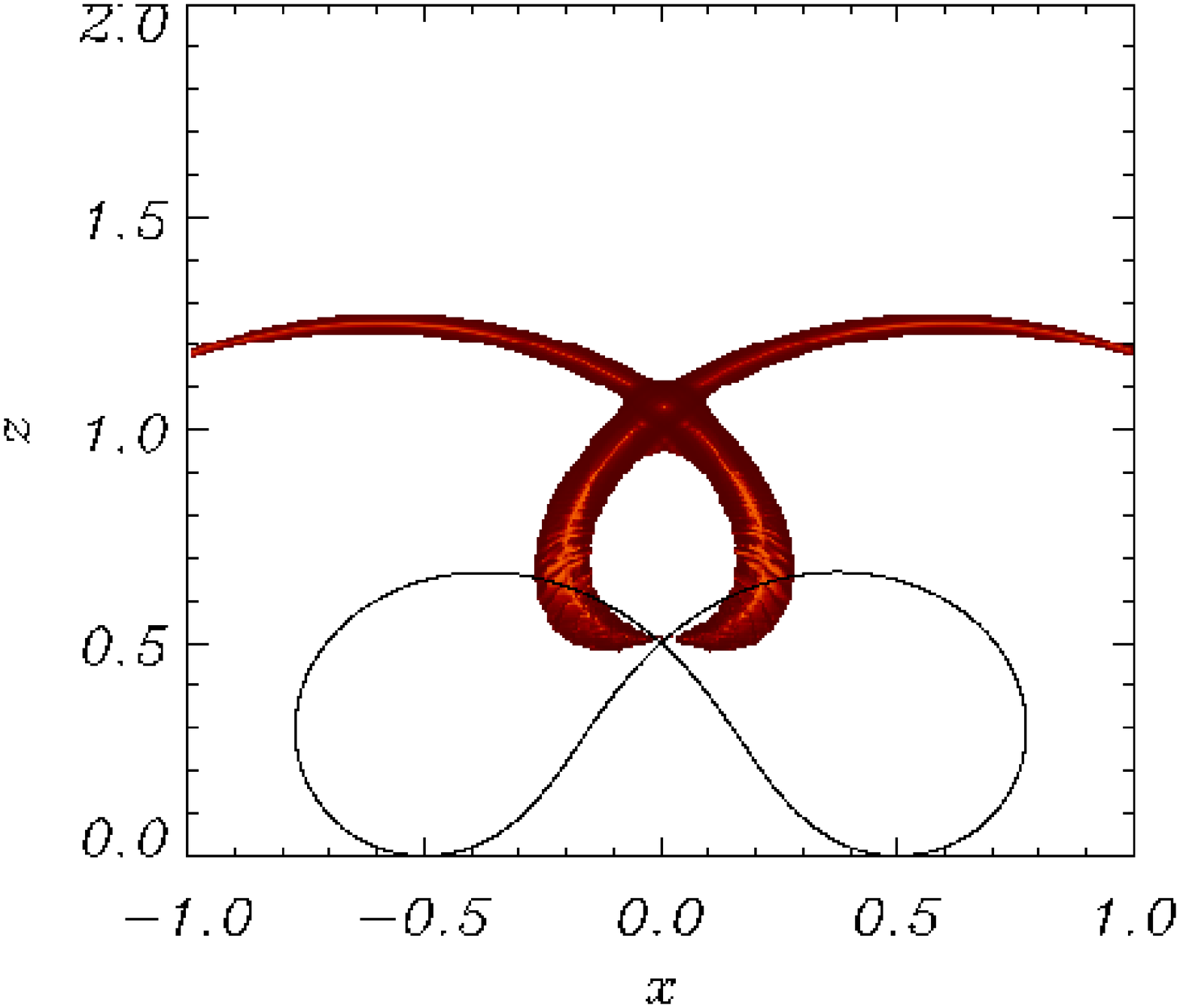}
\hspace{0.0in}
\includegraphics[width=1.2in]{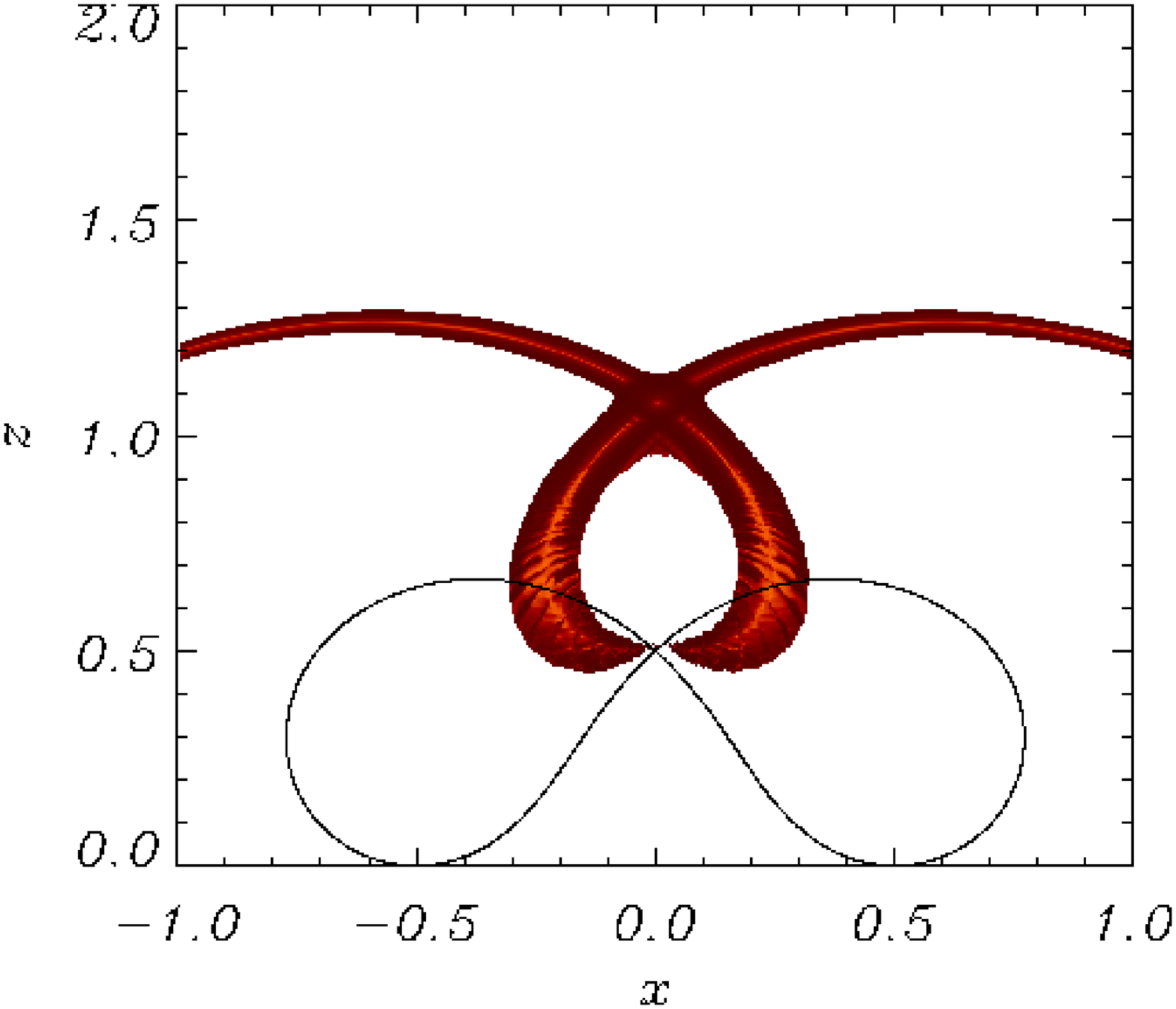}\\
\hspace{0in}
\vspace{0.1in}
%\hspace{0.2in}
\includegraphics[width=1.2in]{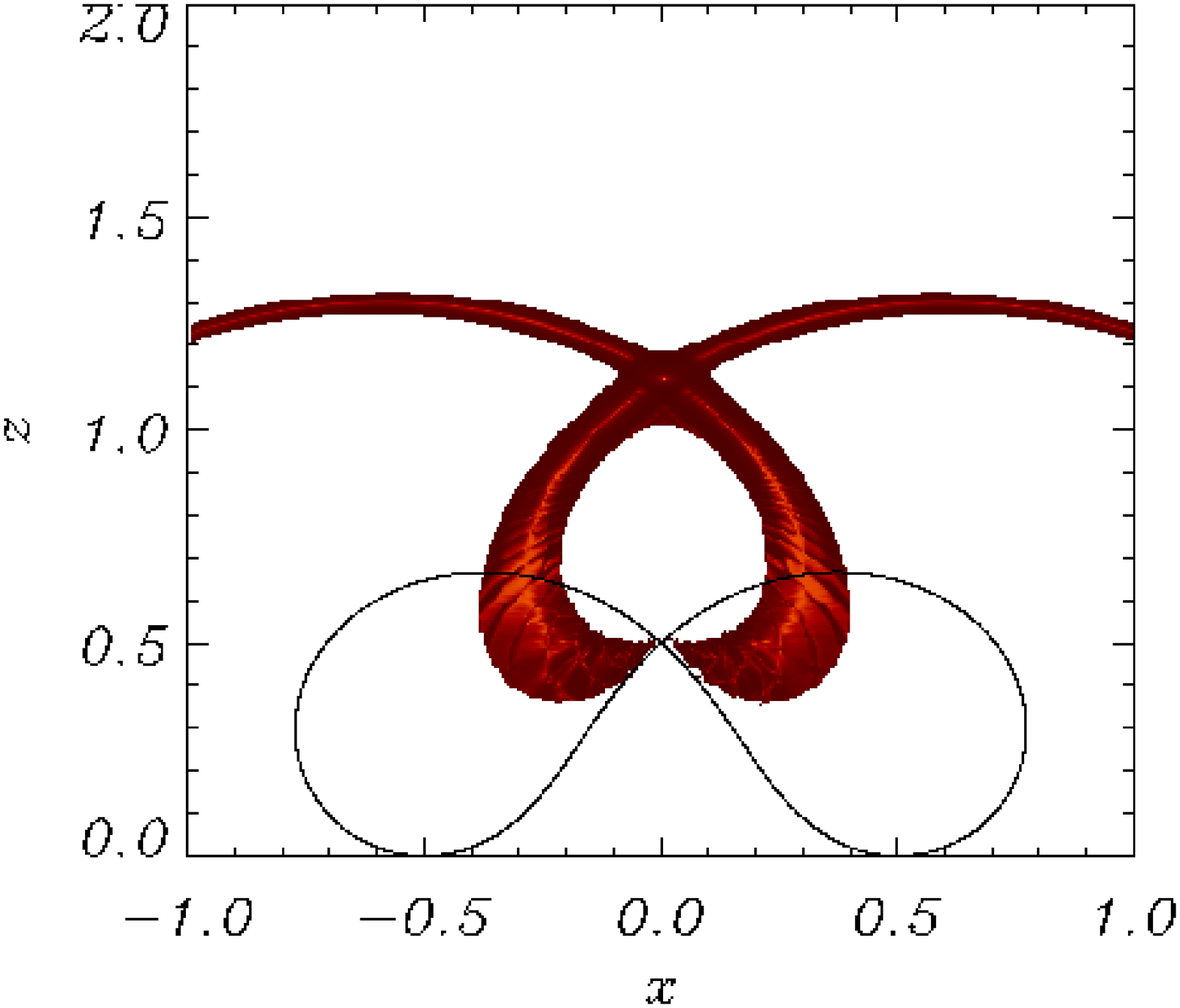}
\hspace{0.0in}
\includegraphics[width=1.2in]{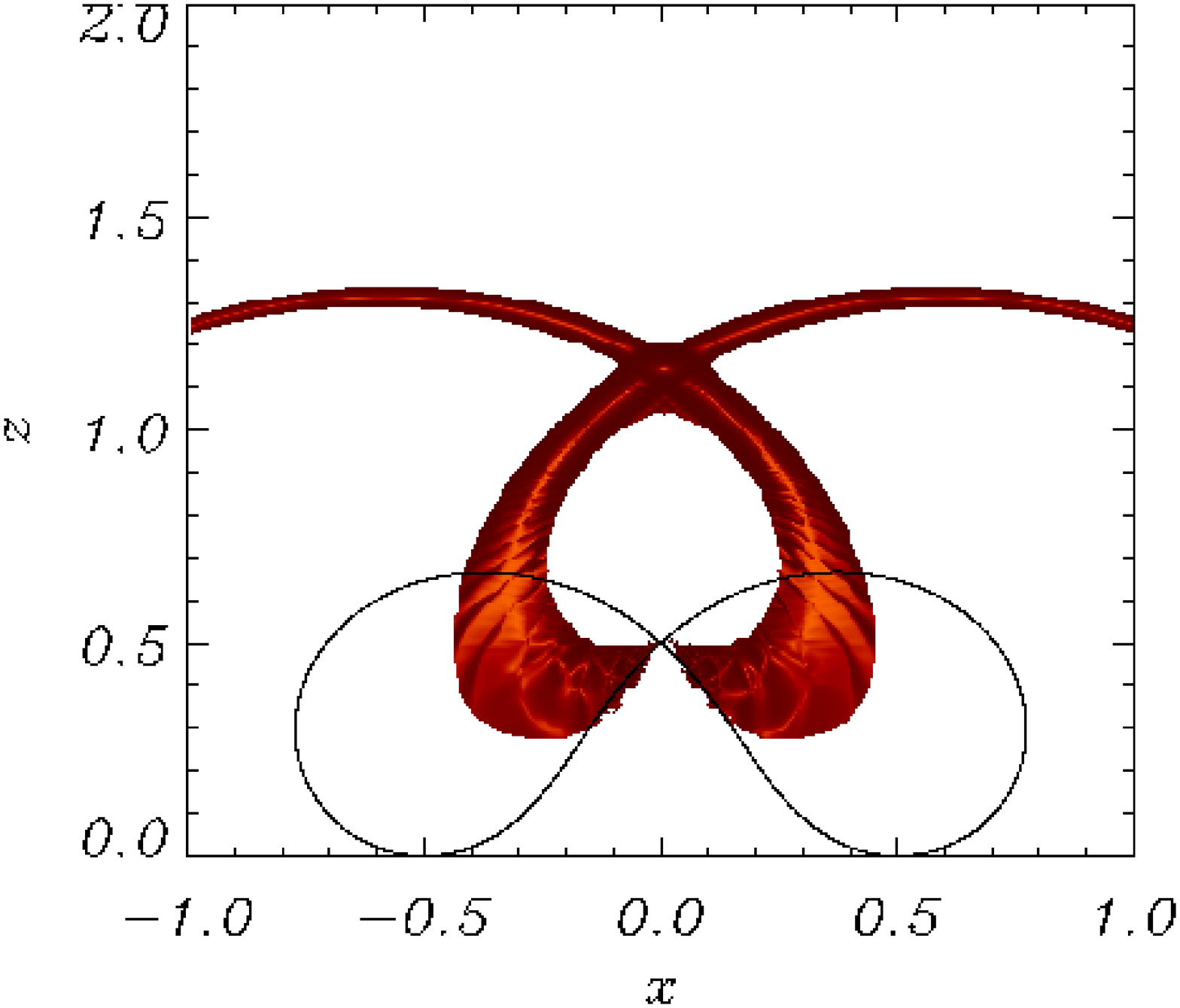}
\hspace{0.0in}
\includegraphics[width=1.2in]{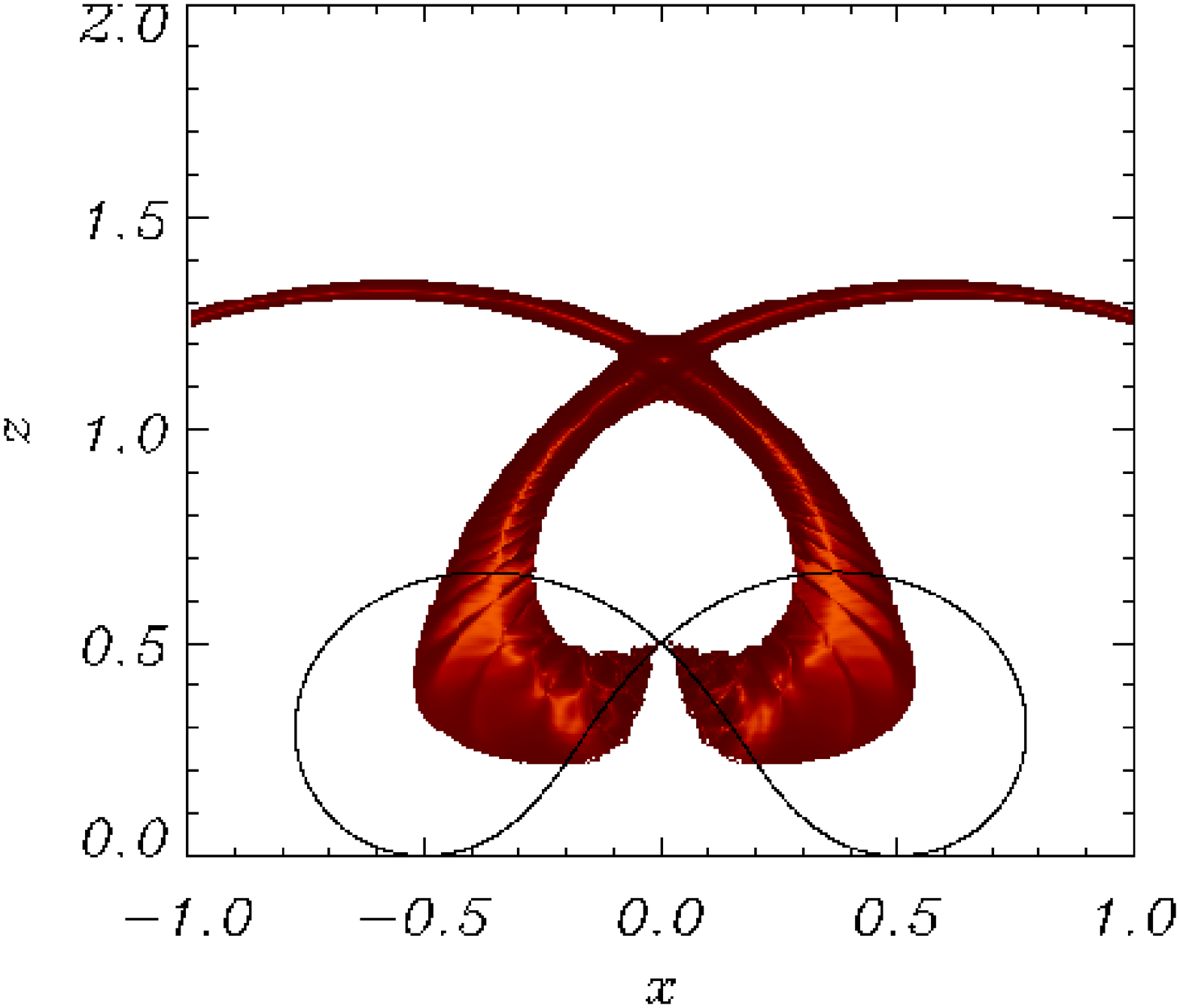}
\hspace{0.0in}
\includegraphics[width=1.2in]{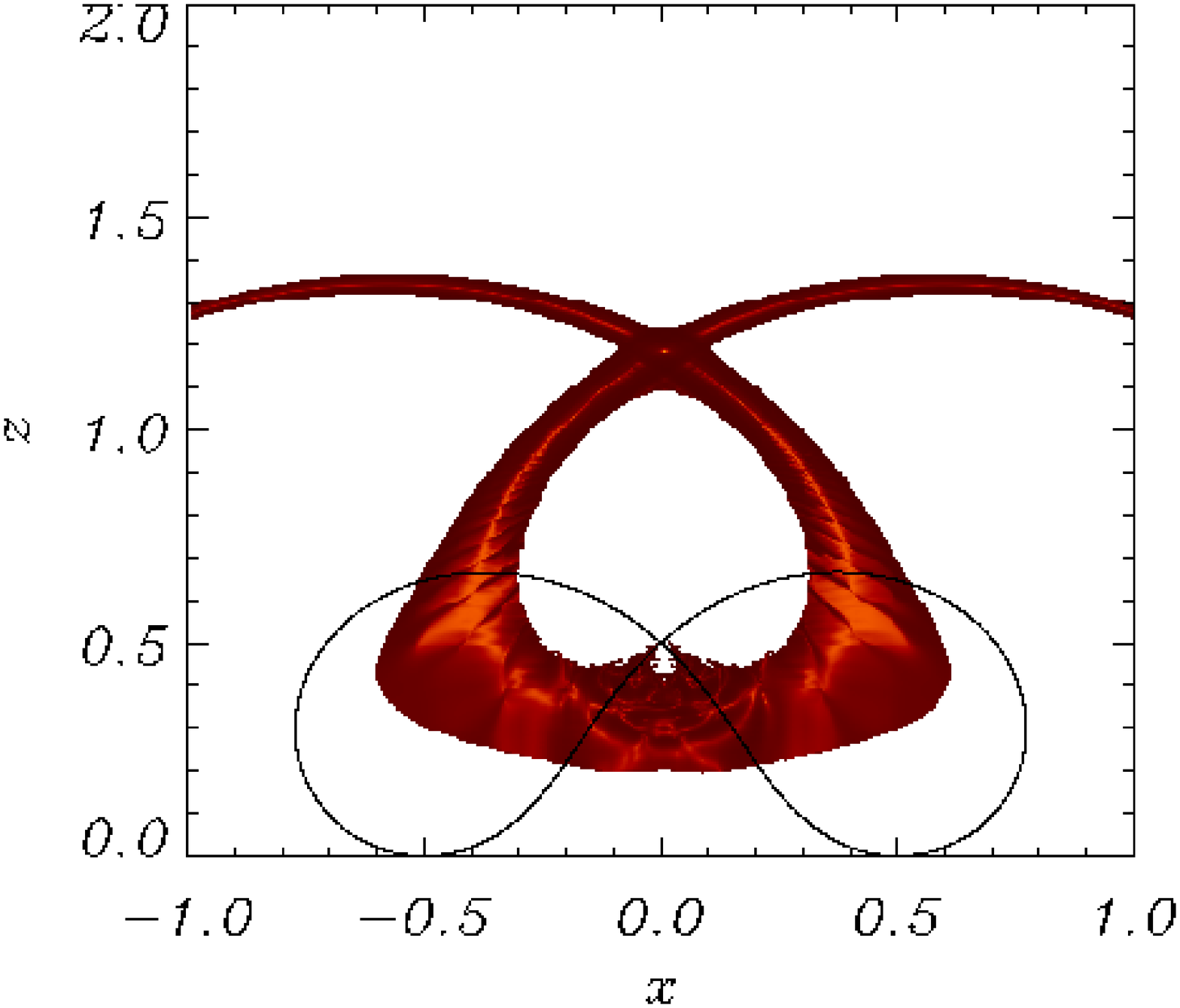}\\
\hspace{0in}
\vspace{0.1in}
%\hspace{0.2in}
\includegraphics[width=1.2in]{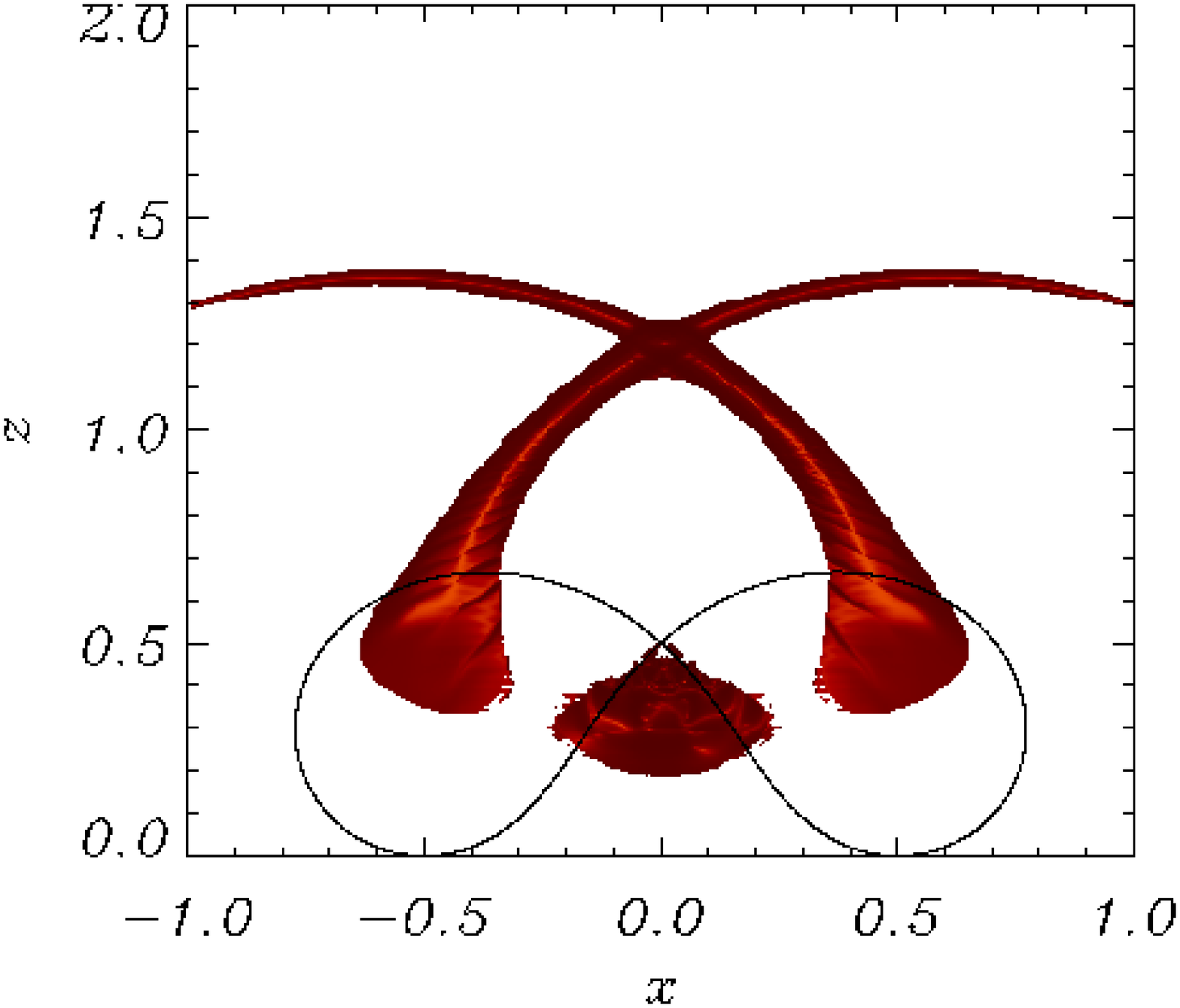}
\hspace{0.0in}
\includegraphics[width=1.2in]{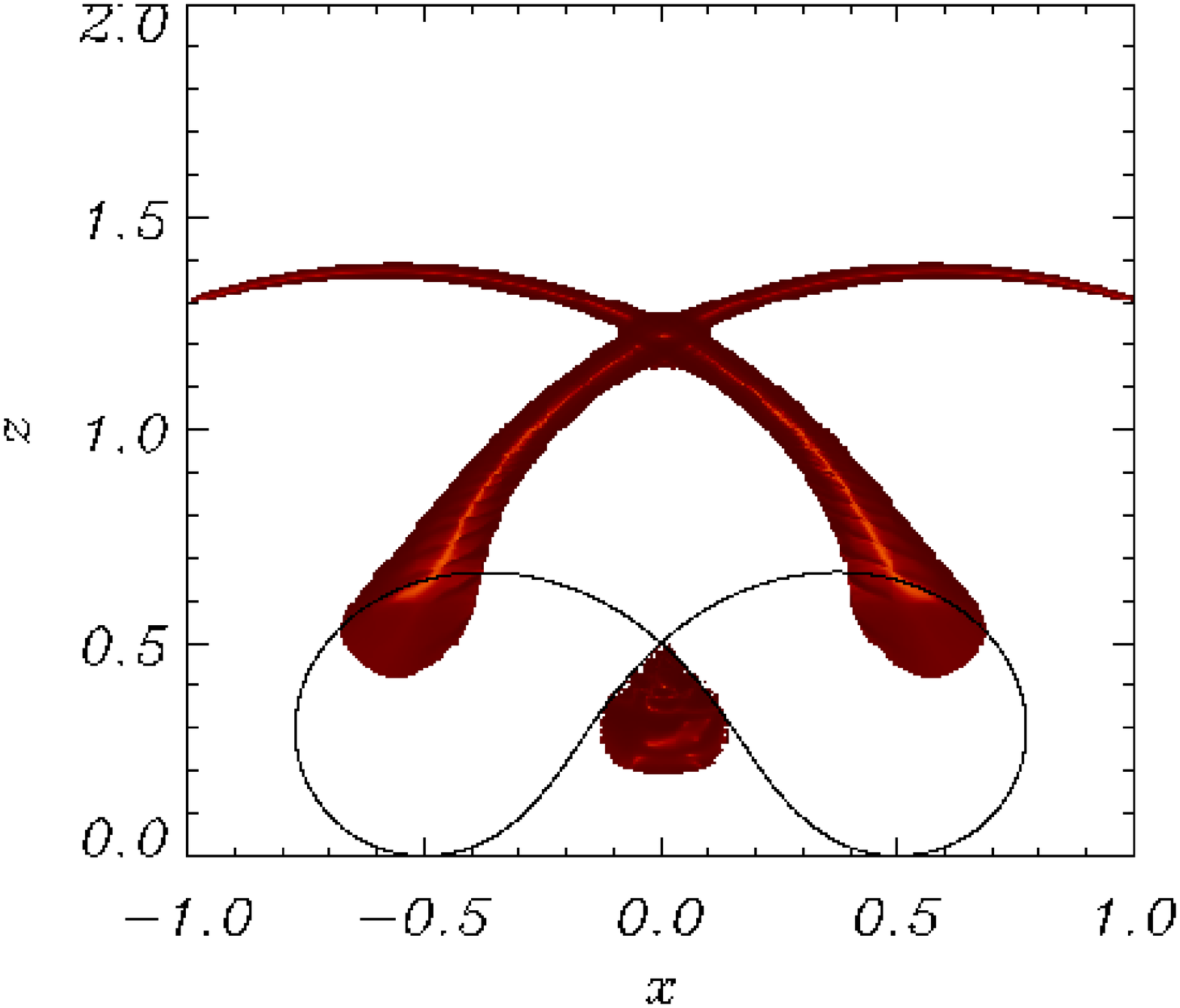}
\hspace{0.0in}
\includegraphics[width=1.2in]{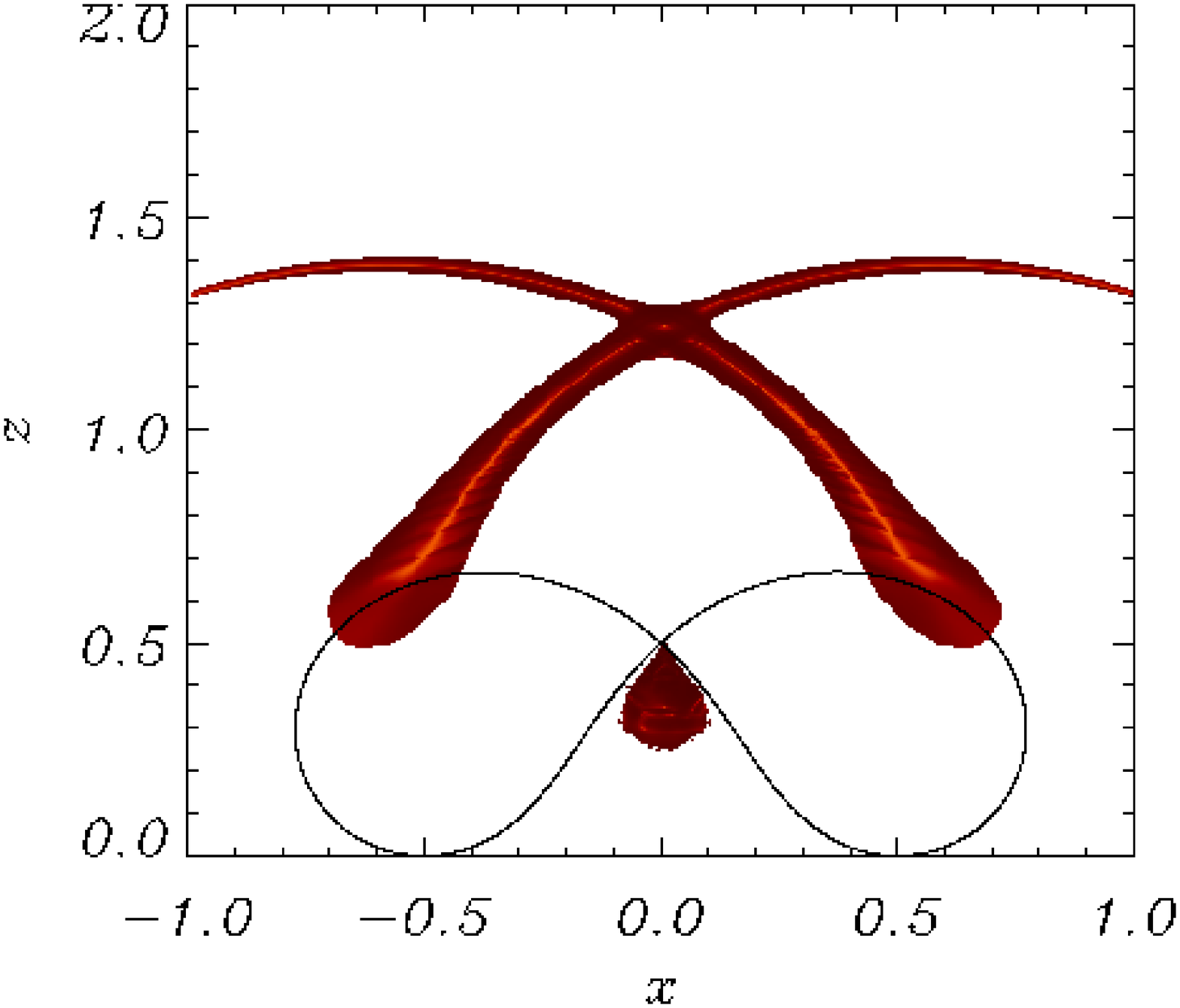}
\hspace{0.0in}
\includegraphics[width=1.2in]{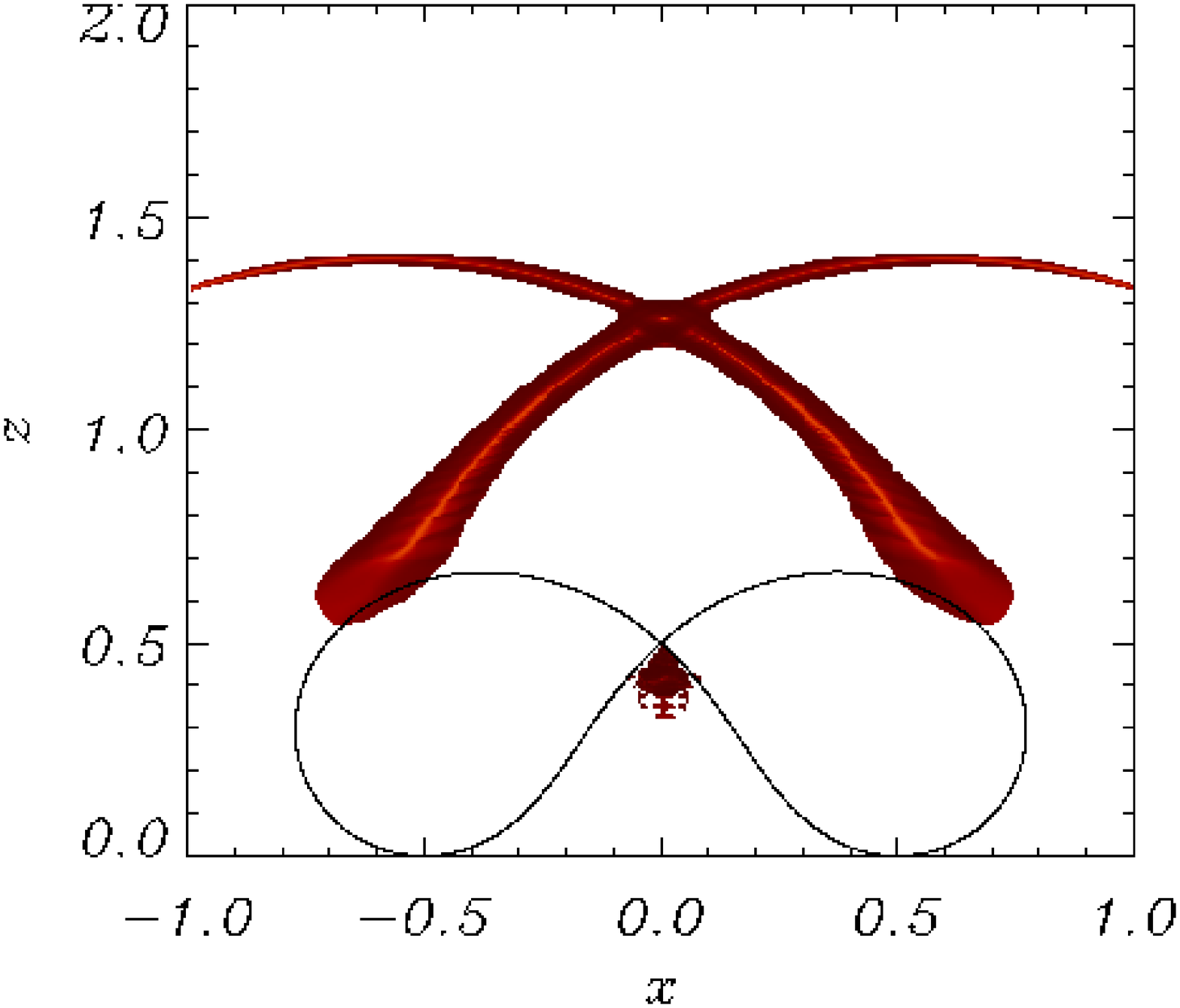}\\
\hspace{0in}
\vspace{0.1in}
%\hspace{0.2125in}
\includegraphics[width=1.2in]{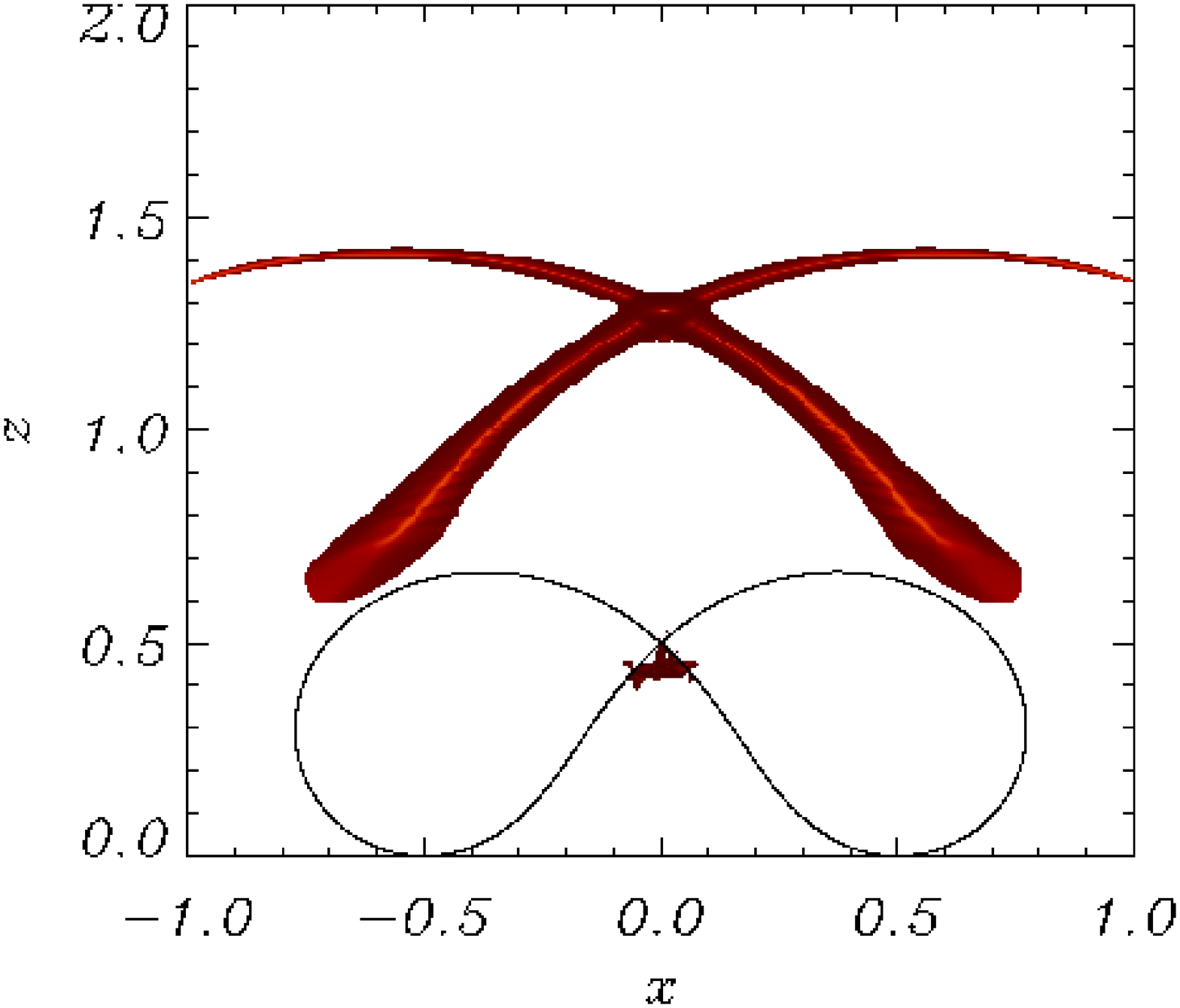}
\hspace{0.0in}
\includegraphics[width=1.2in]{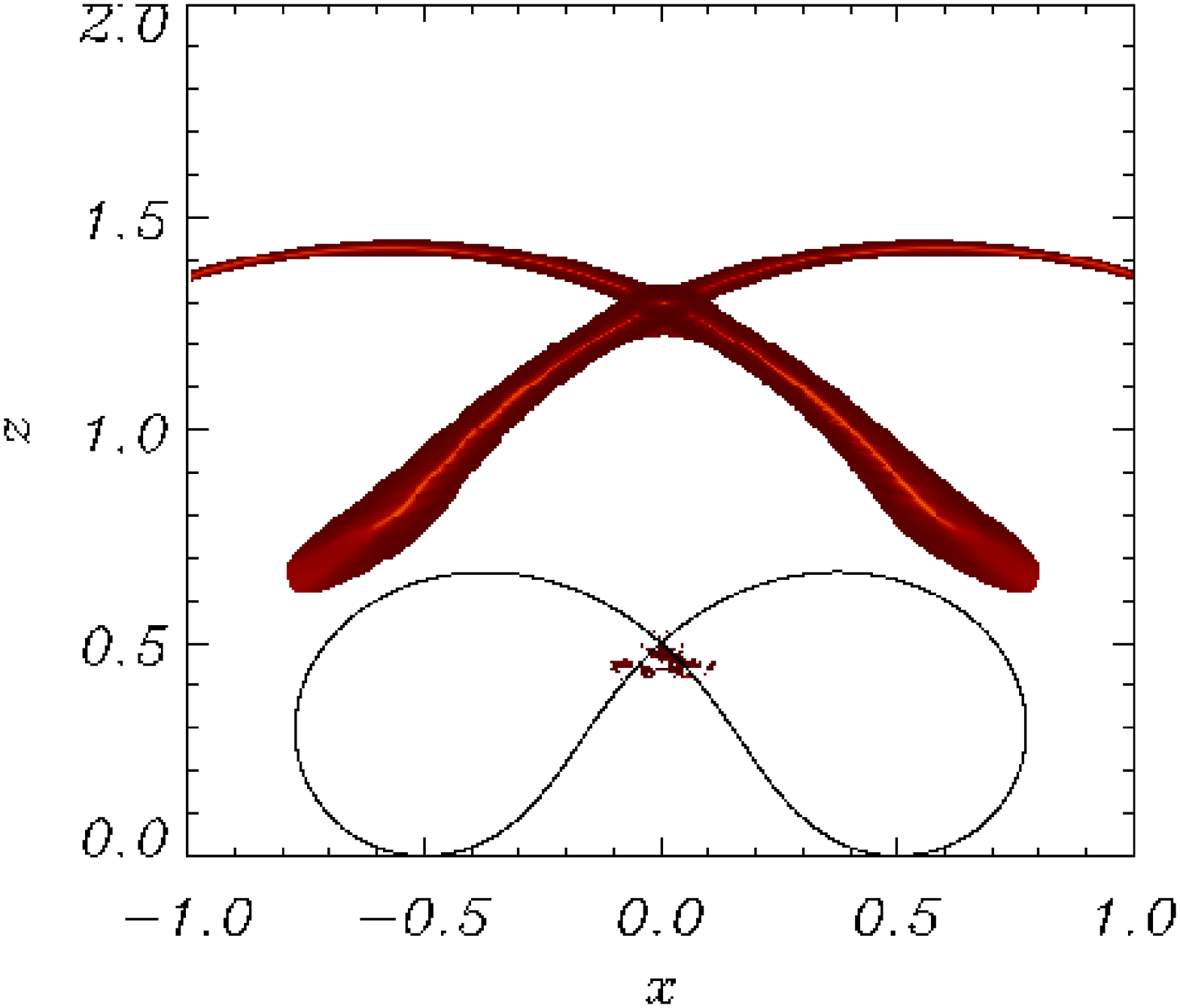}
\hspace{0.0in}
\includegraphics[width=1.2in]{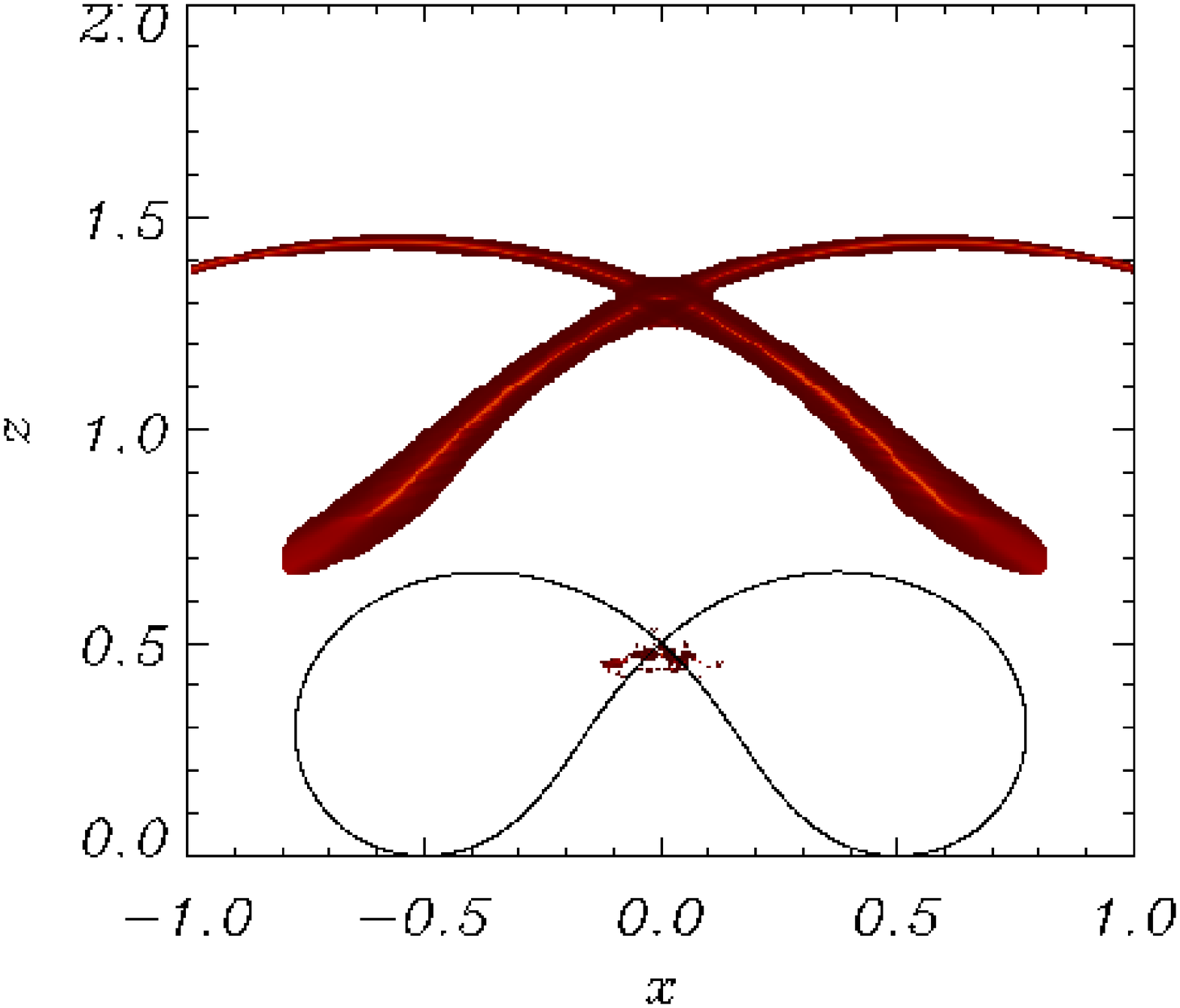}
\hspace{0.0in}
\includegraphics[width=1.55in]{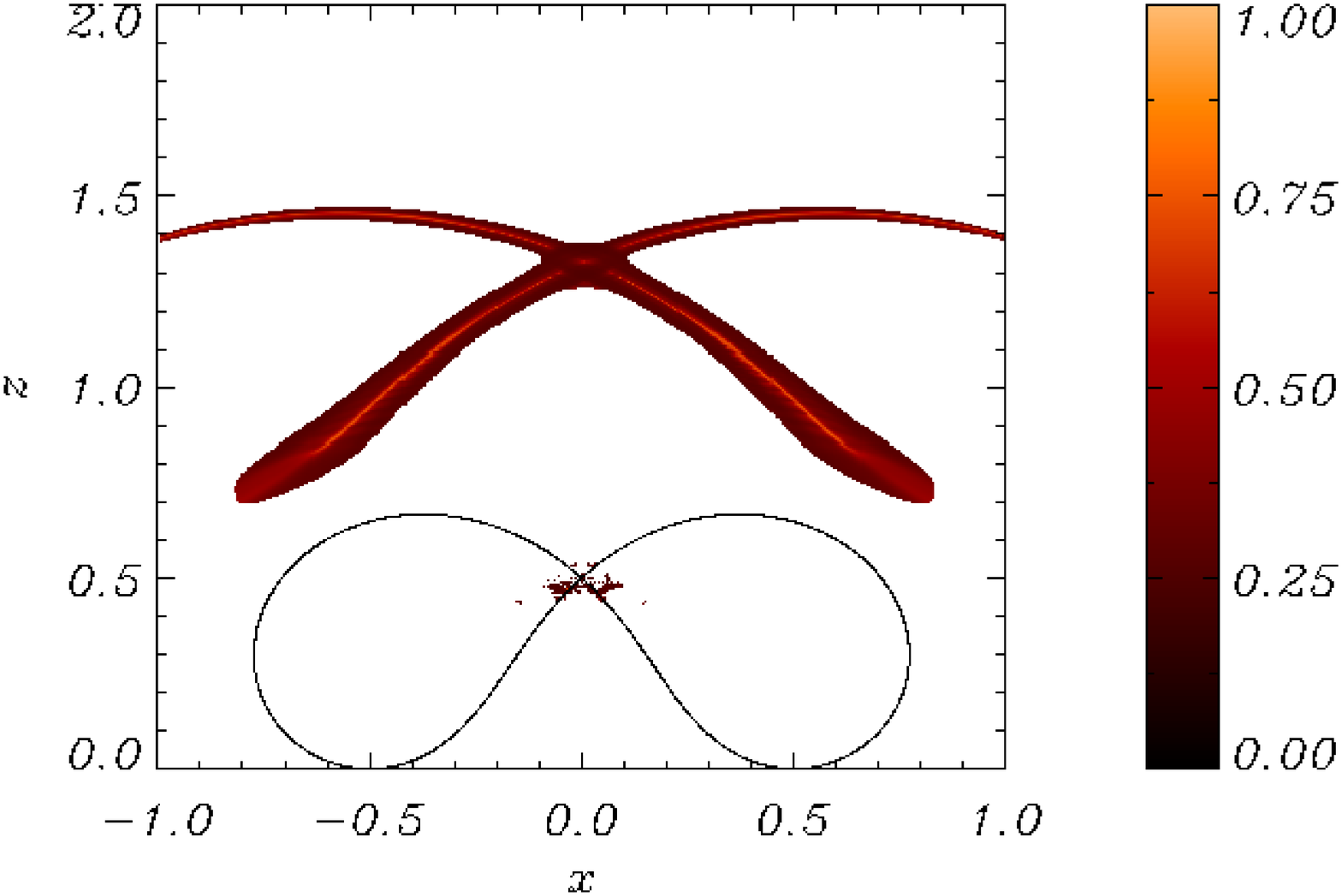}
\caption{Contours of numerical simulation of $V$ for a fast wave sent in from lower boundary for $-1 \leq x \leq 1$, $z=0.1$ and its resultant propagation at times $(a)$ $t$=0.055, $(b)$ $t$=0.11, $(c)$ $t$=0.165, $(d)$ $t=$0.22, $(e)$ $t$=0.275 and $(f)$ $t$=0.33, $(g)$ $t$=0.385, $(h)$ $t$=0.44, $(i)$ $t$=0.495, $(j)$ $t=$0.55, $(k)$ $t$=0.605, $(l)$ $t$=0.66, $(m)$ $t$=0.72, $(n)$ $t$=0.74, $(o)$ $t$=0.76, $(p)$ $t=$0.78, $(q)$ $t$=0.8 and $(r)$ $t$=0.82, $(s)$ $t$=0.84, $(t)$ $t$=0.86, $(u)$ $t$=0.88, $(v)$ $t=$0.9, $(w)$ $t$=0.92 and $(x)$ $t$=0.94, labelling from top left to bottom right.}
\label{fig:4.5.2.4}
\end{figure*}

We find that the linear, fast MHD wave travels up from the lower boundary and has its propagation influenced by the magnetic configuration. The results can be seen in Figure \ref{fig:4.5.2.4}. The resultant propagation is identical to the behaviour described in Simulation Two, since the magnetic region is symmetrical about $x=0$. Initially, we see the wave distorts due to the two regions of enhanced Alfv\'en speed. Then, the central part of the wave slows down and refracts around the null point (as seen in Simulation One). The rest of the wave (the \emph{wings}) continue to propagate upwards and because the wave is spreading out (as seen most clearly in the particle paths of Figure \ref{WKBtwodipolesparticlepaths}), the wave gives the impression of pivoting about the null. As this pivoting continues, the two wings eventually cross each other. However, due to the linear nature of the model, both wings pass each other without interacting.

% and do not affect each other.

\begin{figure}
\begin{center}
\includegraphics[width=2.0in]{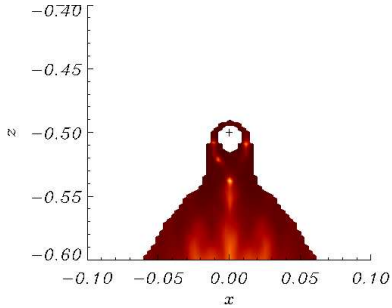}
\caption{Contour of the  numerical simulation of $V$ for a fast wave sent in from lower boundary for  $-1 \leq x \leq 1 $ after time $t=0.44$. Figure shows blow-up of region around the null point. The image has been reflected in the line $z=-x$.}  
\label{fig:stillwrapsaround_2}
\end{center}
\end{figure}

The two wings continue to pivot, and near to the null part of the wave wraps tightly around it. This can be seen clearly in Figure \ref{fig:stillwrapsaround_2}. This shows a blow-up of the wave accumulation around the null. Even though the resolution is coarse in this area, it is clear that this is the same behaviour seen in Paper I  for the fast wave wrapping around the simple X-point.  In Figure \ref{fig:stillwrapsaround_2} we have let $(x,z)\rightarrow (-x,-z)$ to help the comparison (i.e. again reflected  the image in the  line $z=-x$).

%In Figure \ref{fig:4.5.2.5}, the propagation continues. 
In the lower half of subfigures of Figure \ref{fig:4.5.2.4},  the wave is eventually stretched so much in the region $-0.6 \leq x \leq 0.6$, $0.2 \leq z \leq 0.5$ that it splits; part of the wave remains close to the X-point (and ultimately accumulates there) and part of the wave is repelled by the regions of high Alfv\'en speed and is ejected away from the magnetic skeleton. These parts then continue to propagate upwards and outwards, forming a distinctive cross shape.

%**********************************************************************

%\vspace{0.3in}

%**********************************************************************

\subsubsection{WKB approximation}

\begin{figure*}[t]
\hspace{0in}
\vspace{0.1in}
\hspace{0.2in}
\includegraphics[width=1.2in]{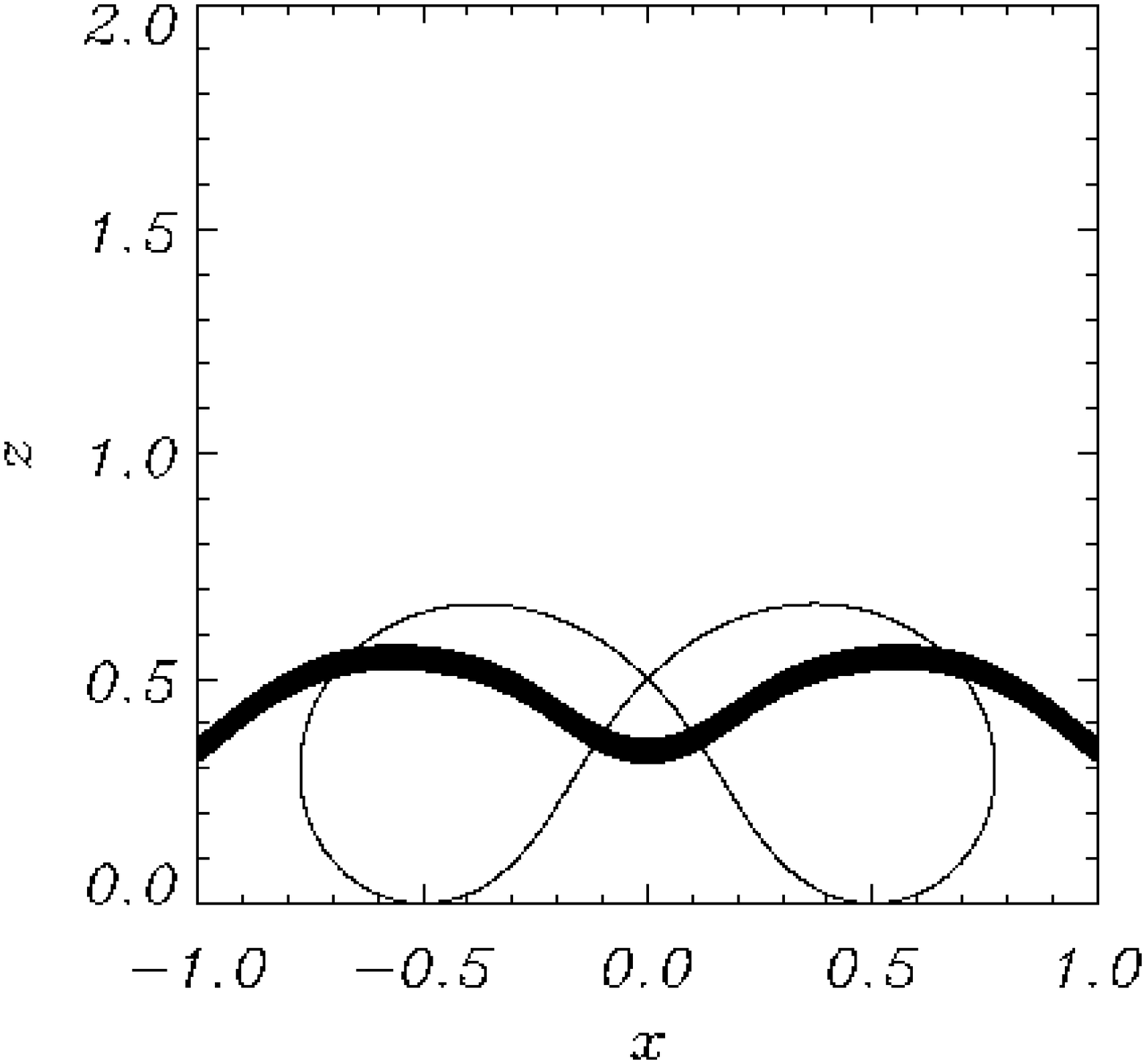}
\hspace{0.0in}
\includegraphics[width=1.2in]{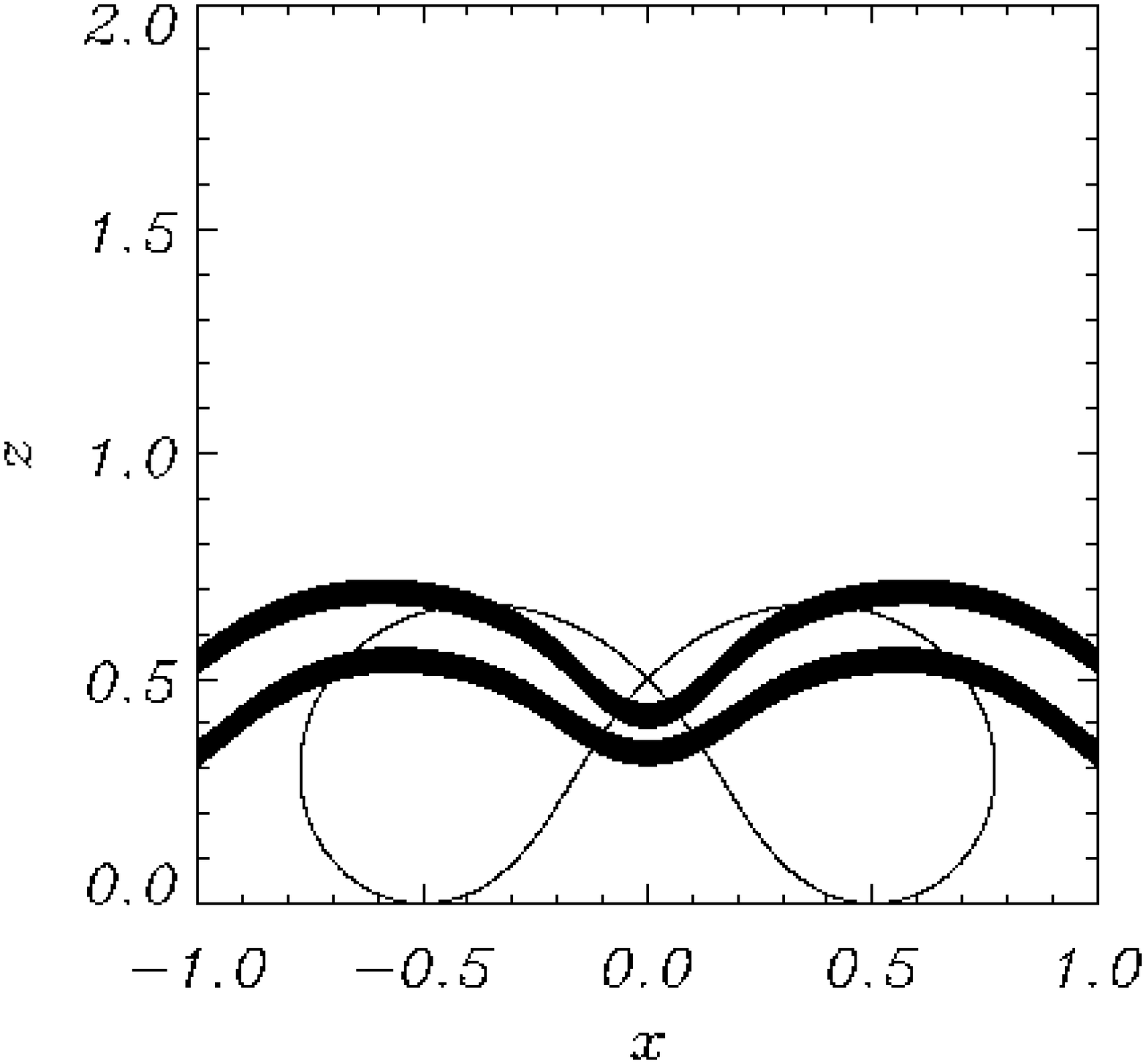}
\hspace{0.0in}
\includegraphics[width=1.2in]{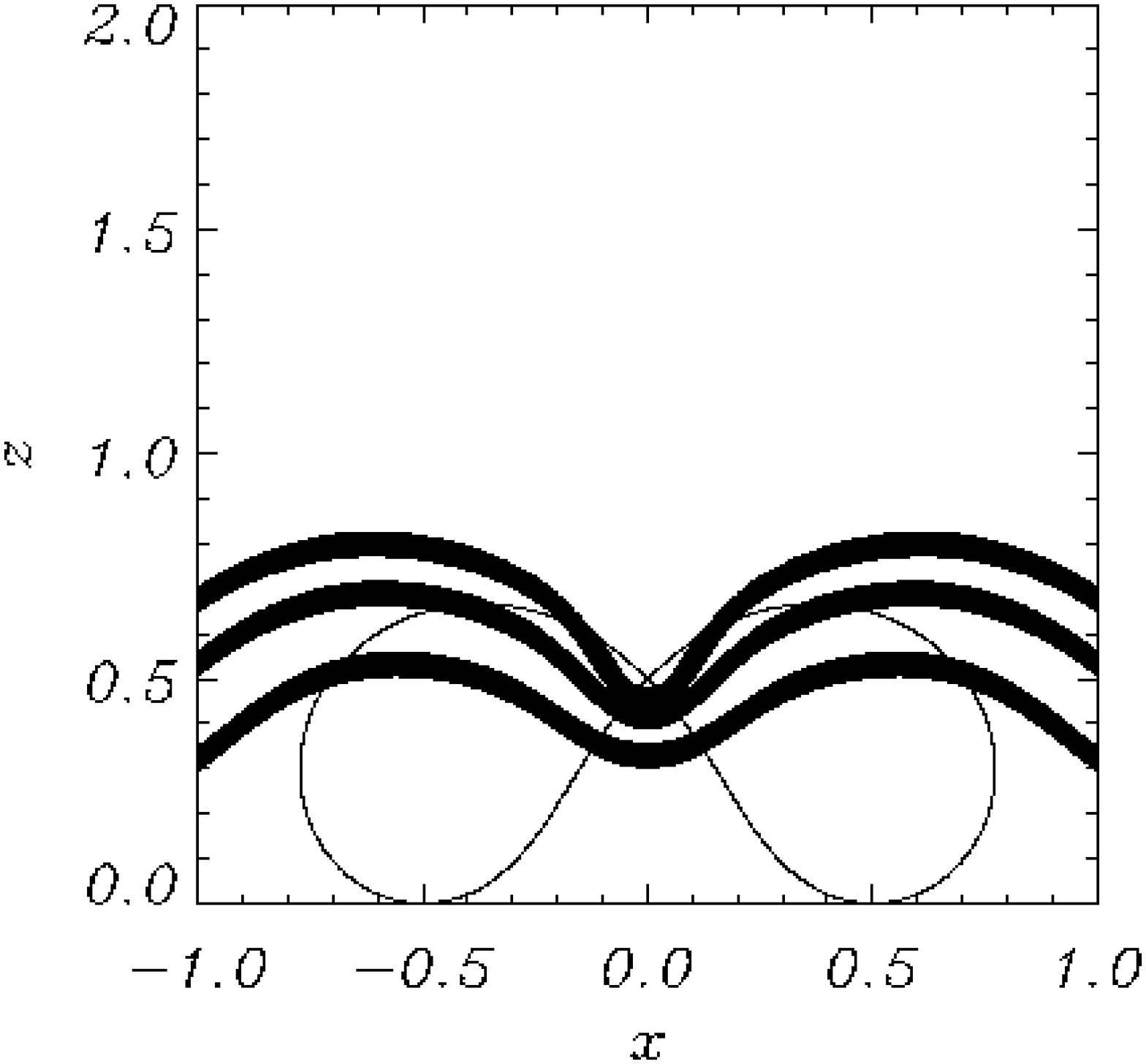}
\hspace{0.0in}
\includegraphics[width=1.2in]{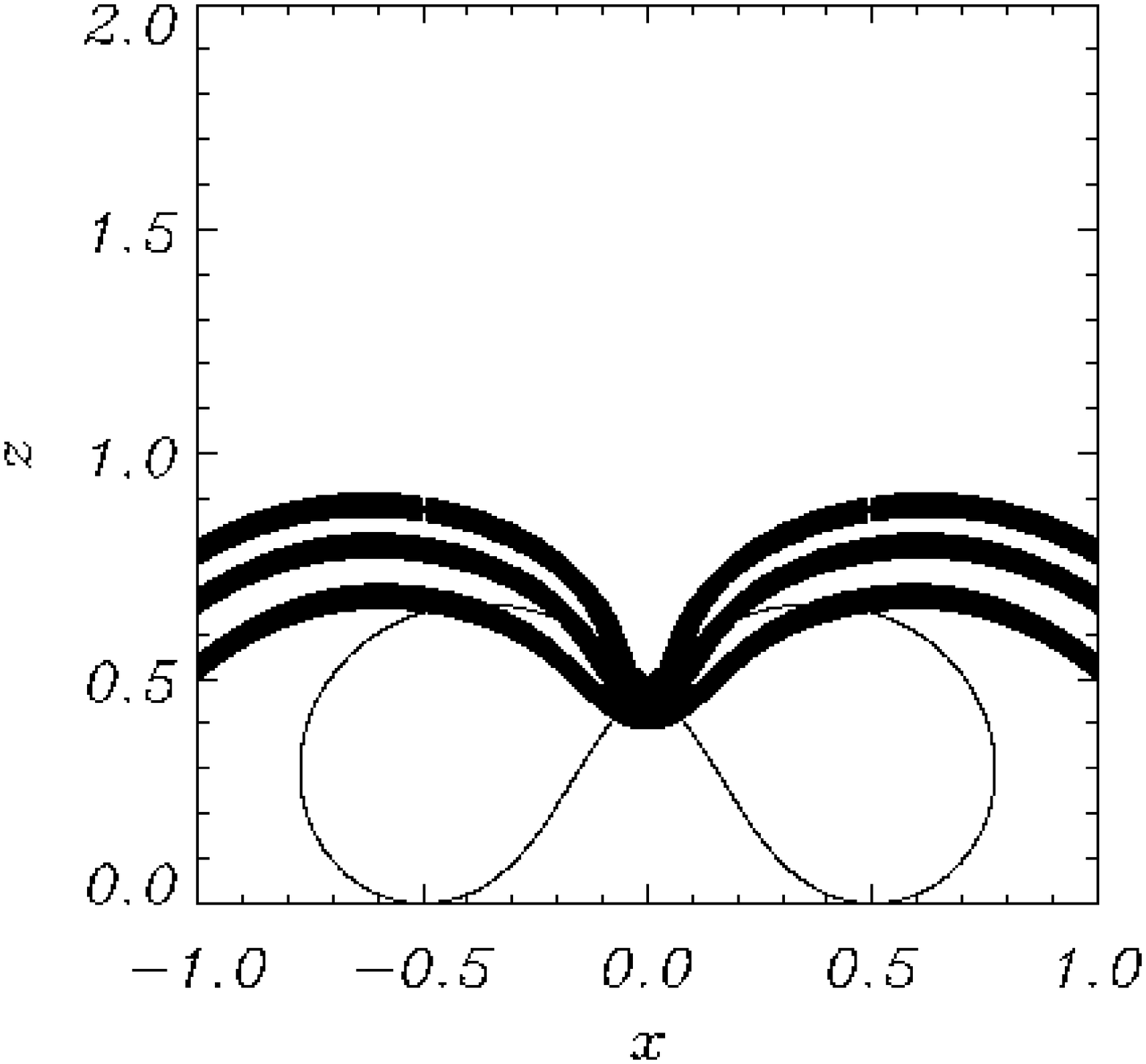}\\
\hspace{0in}
\vspace{0.1in}
\hspace{0.2in}
\includegraphics[width=1.2in]{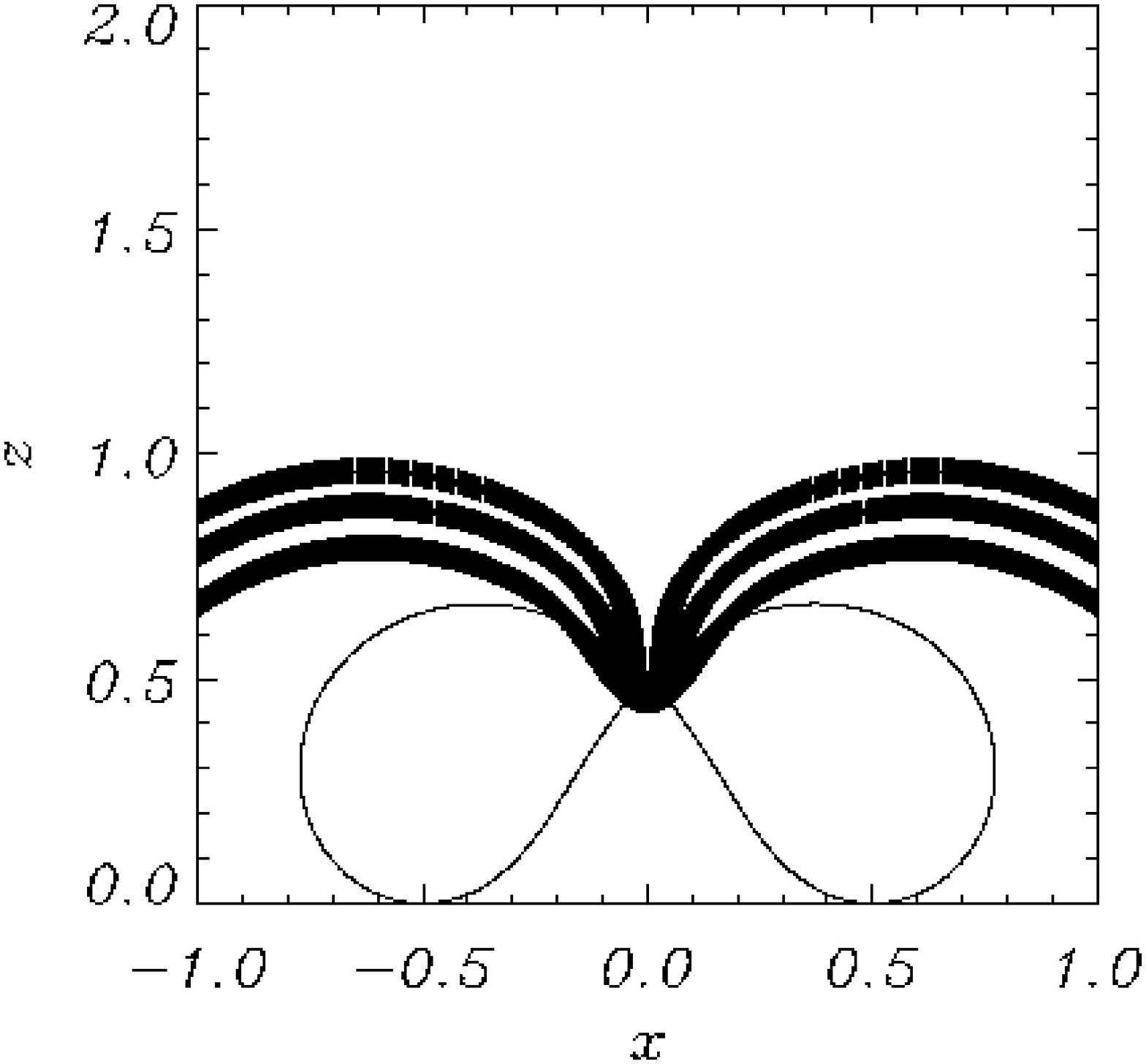}
\hspace{0.0in}
\includegraphics[width=1.2in]{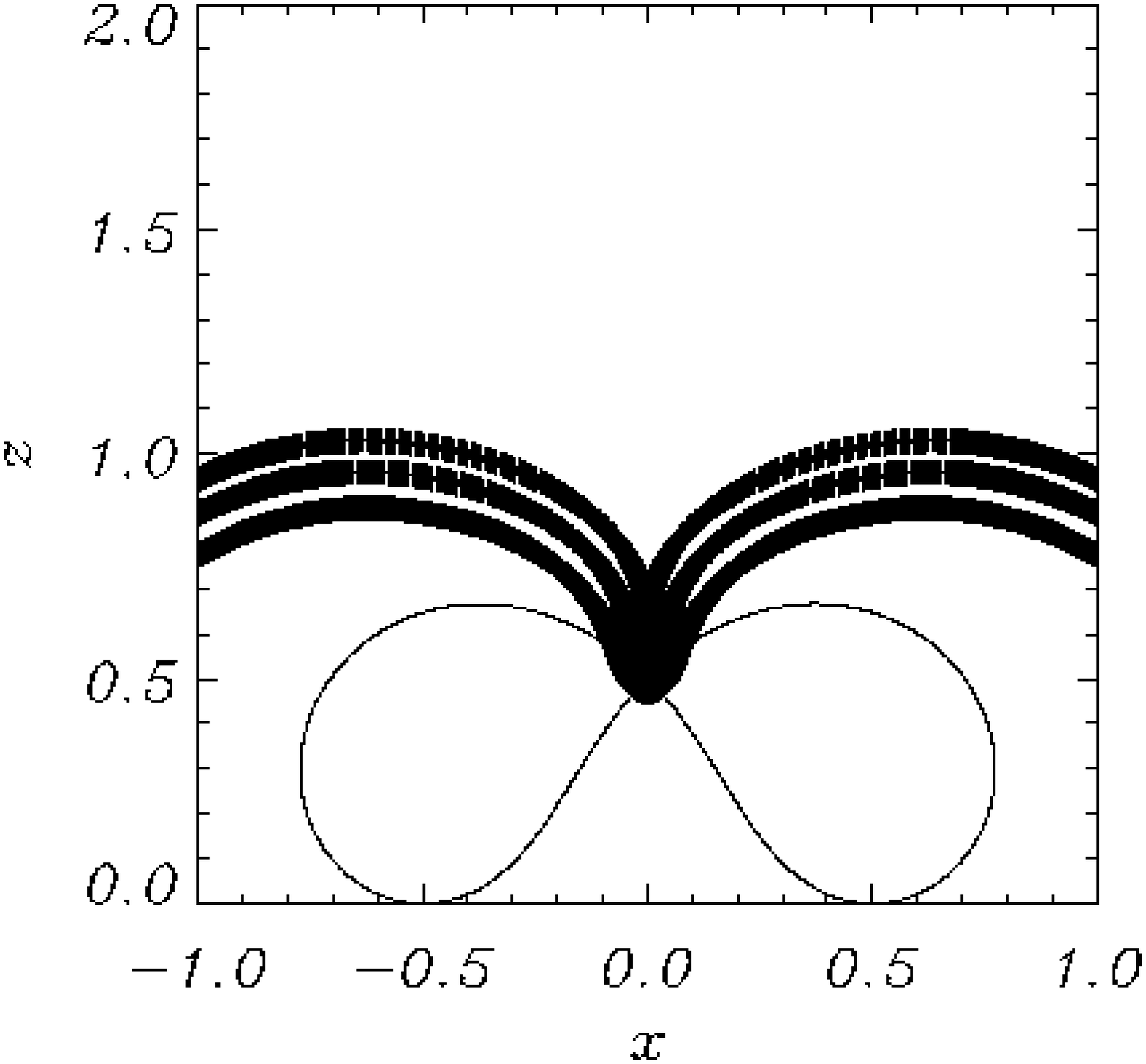}
\hspace{0.0in}
\includegraphics[width=1.2in]{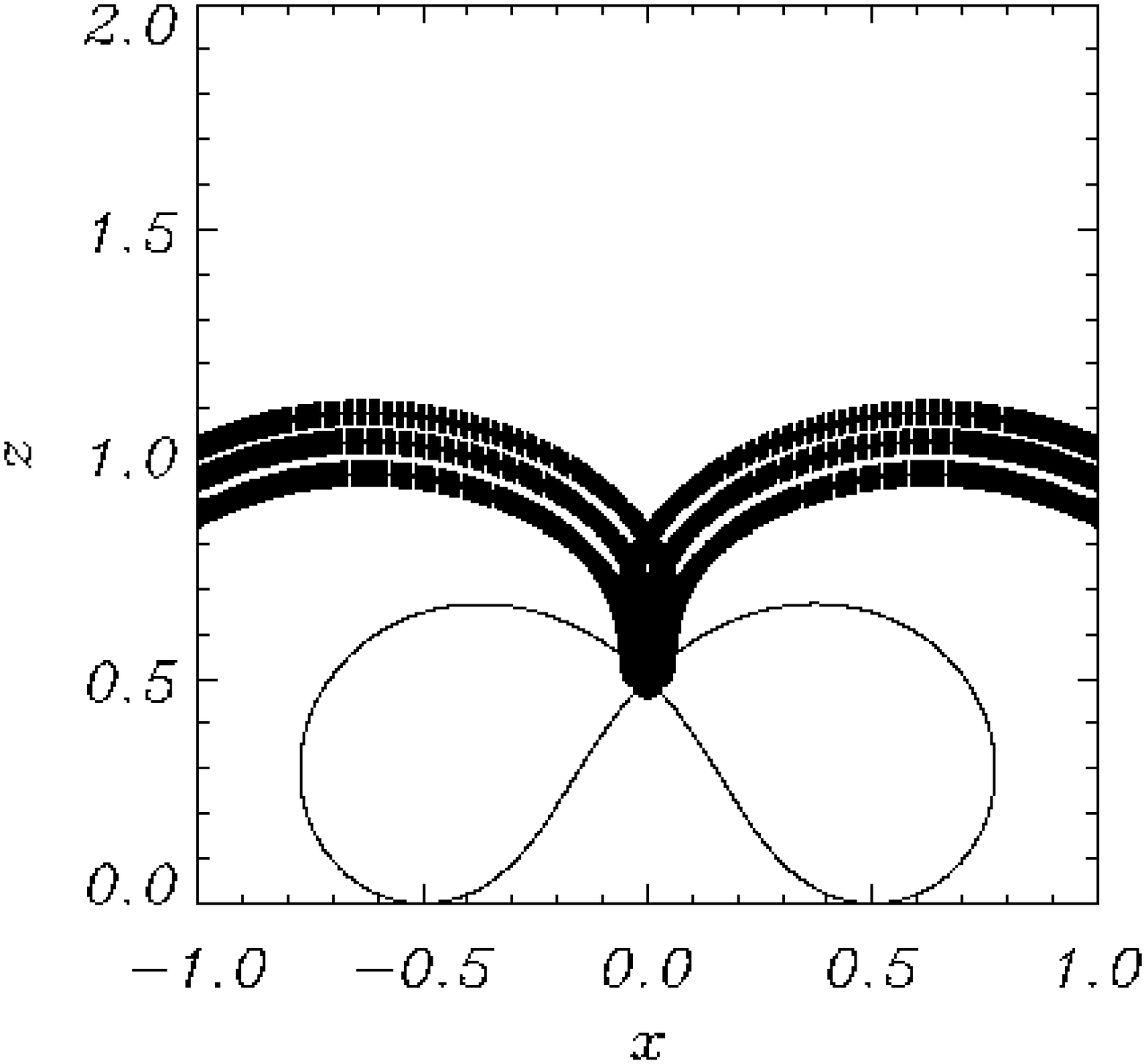}
\hspace{0.0in}
\includegraphics[width=1.2in]{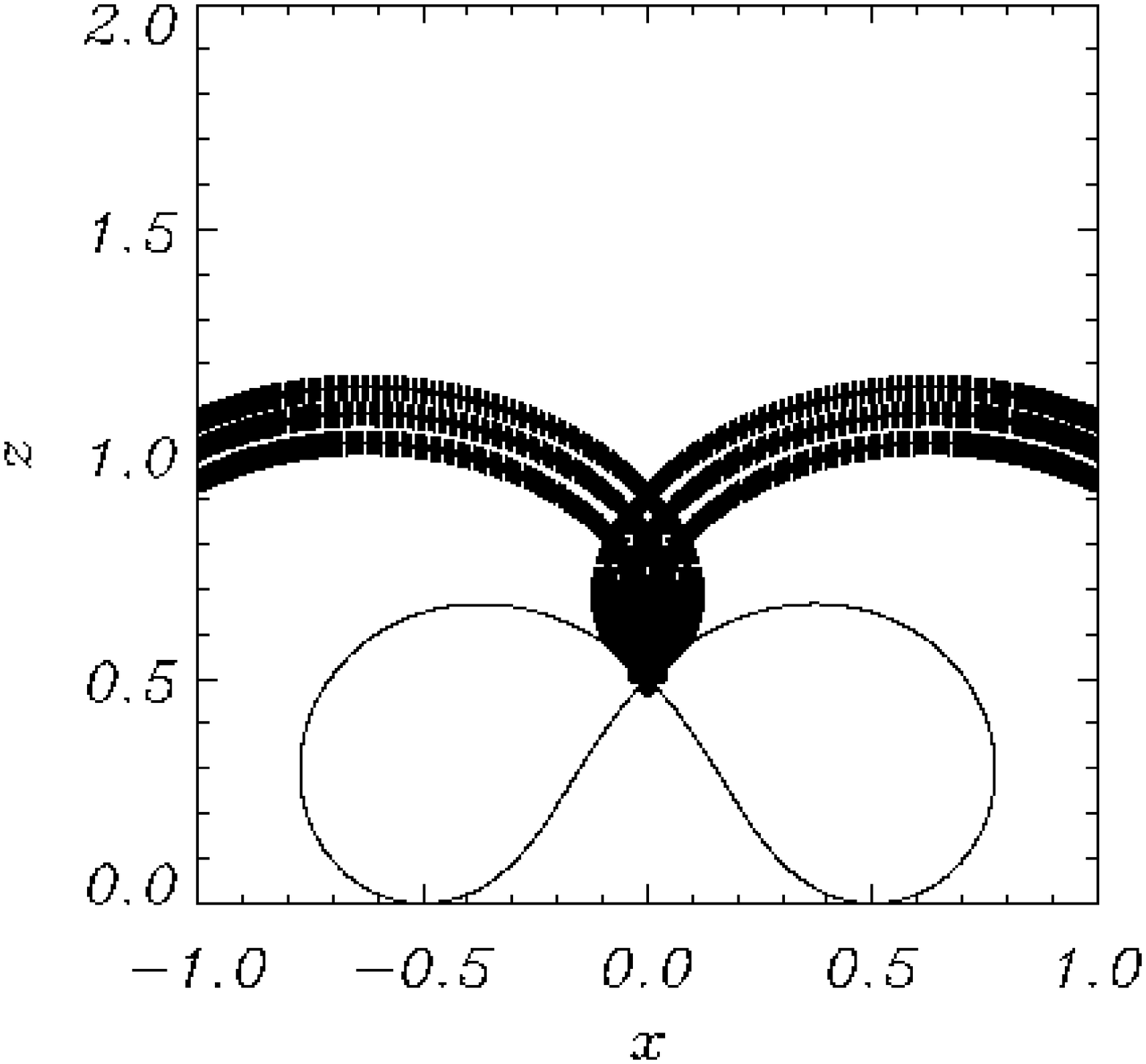}\\
\hspace{0in}
\vspace{0.1in}
\hspace{0.2in}
\includegraphics[width=1.2in]{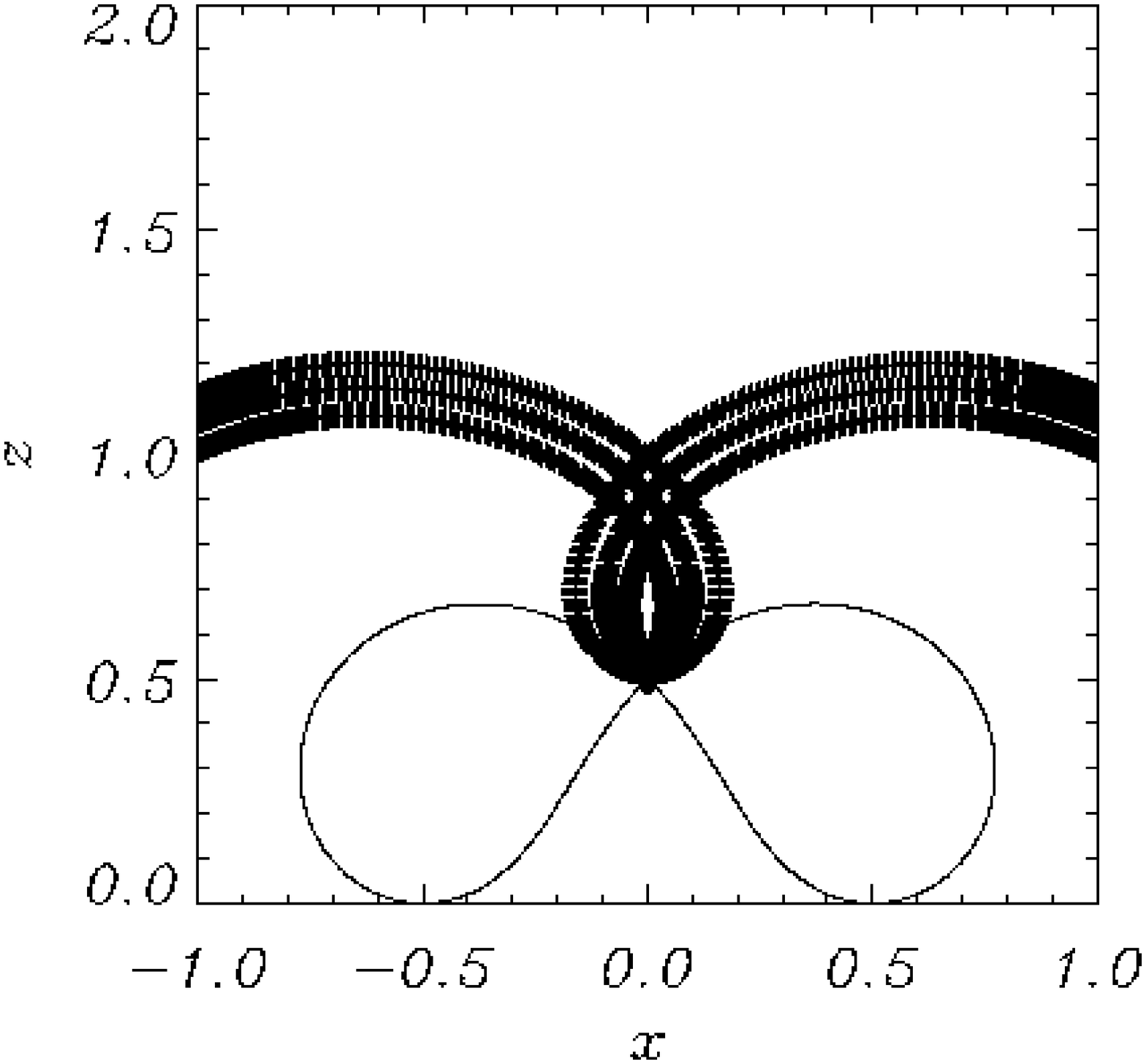}
\hspace{0.0in}
\includegraphics[width=1.2in]{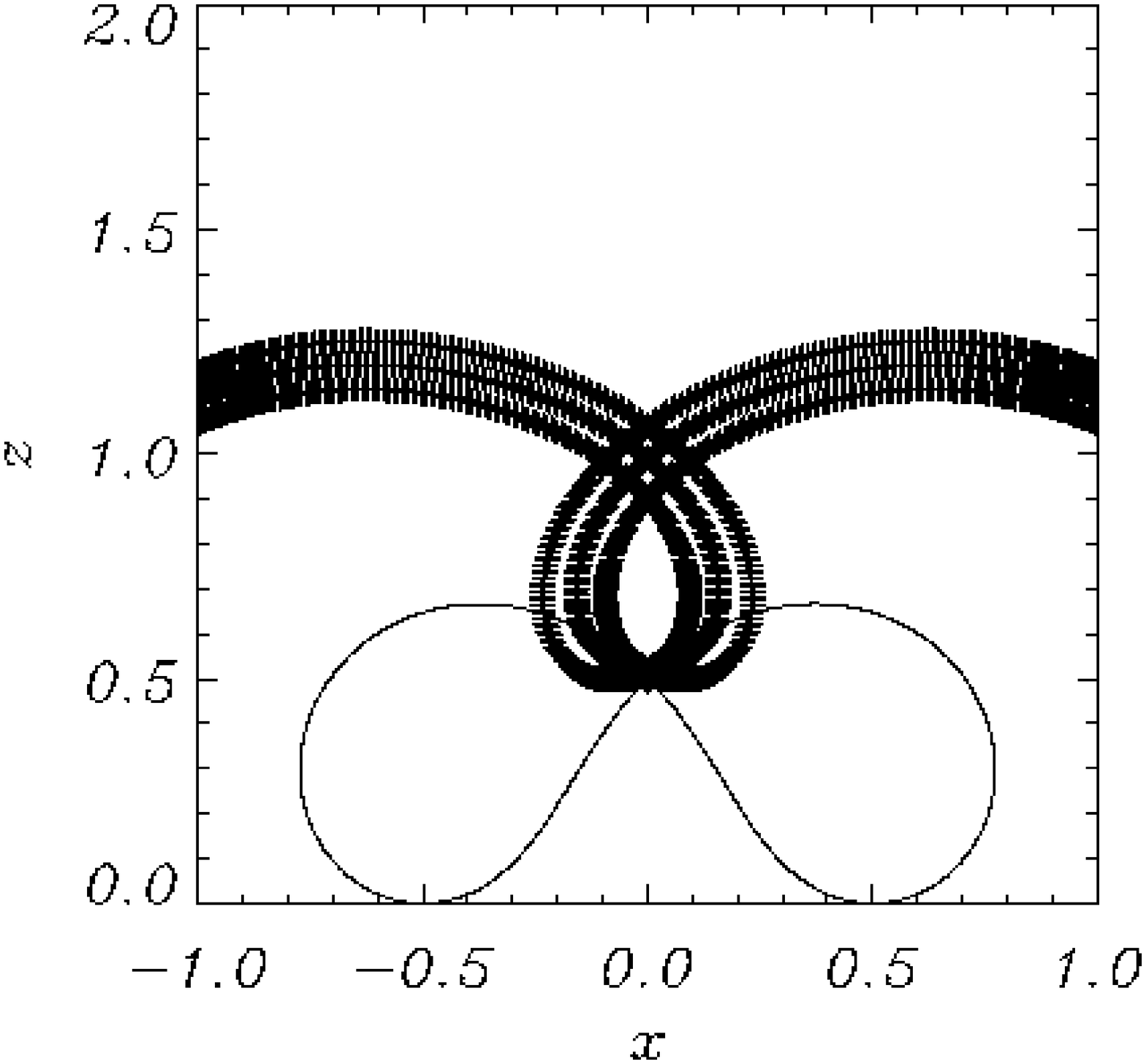}
\hspace{0.0in}
\includegraphics[width=1.2in]{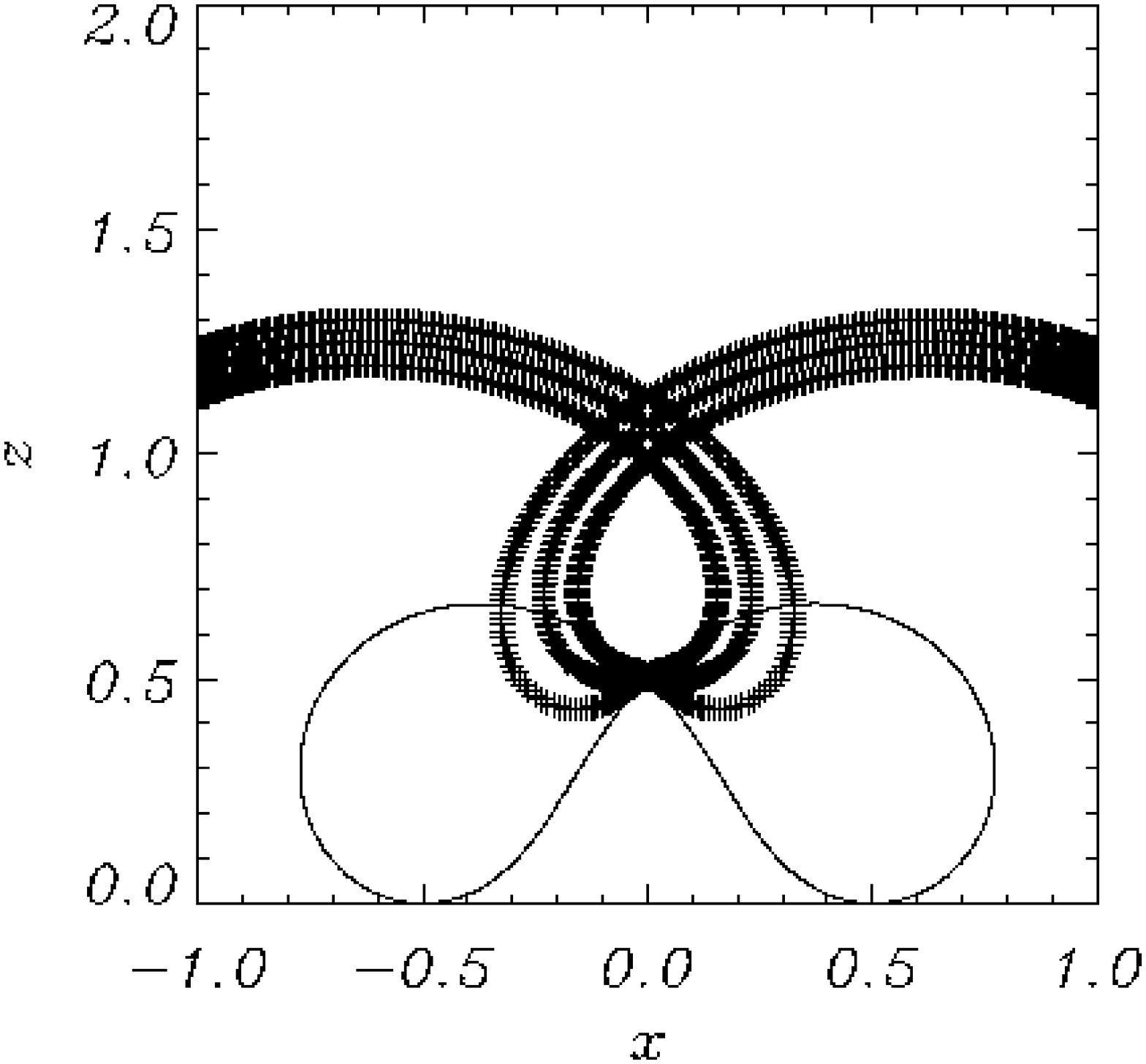}
\hspace{0.0in}
\includegraphics[width=1.2in]{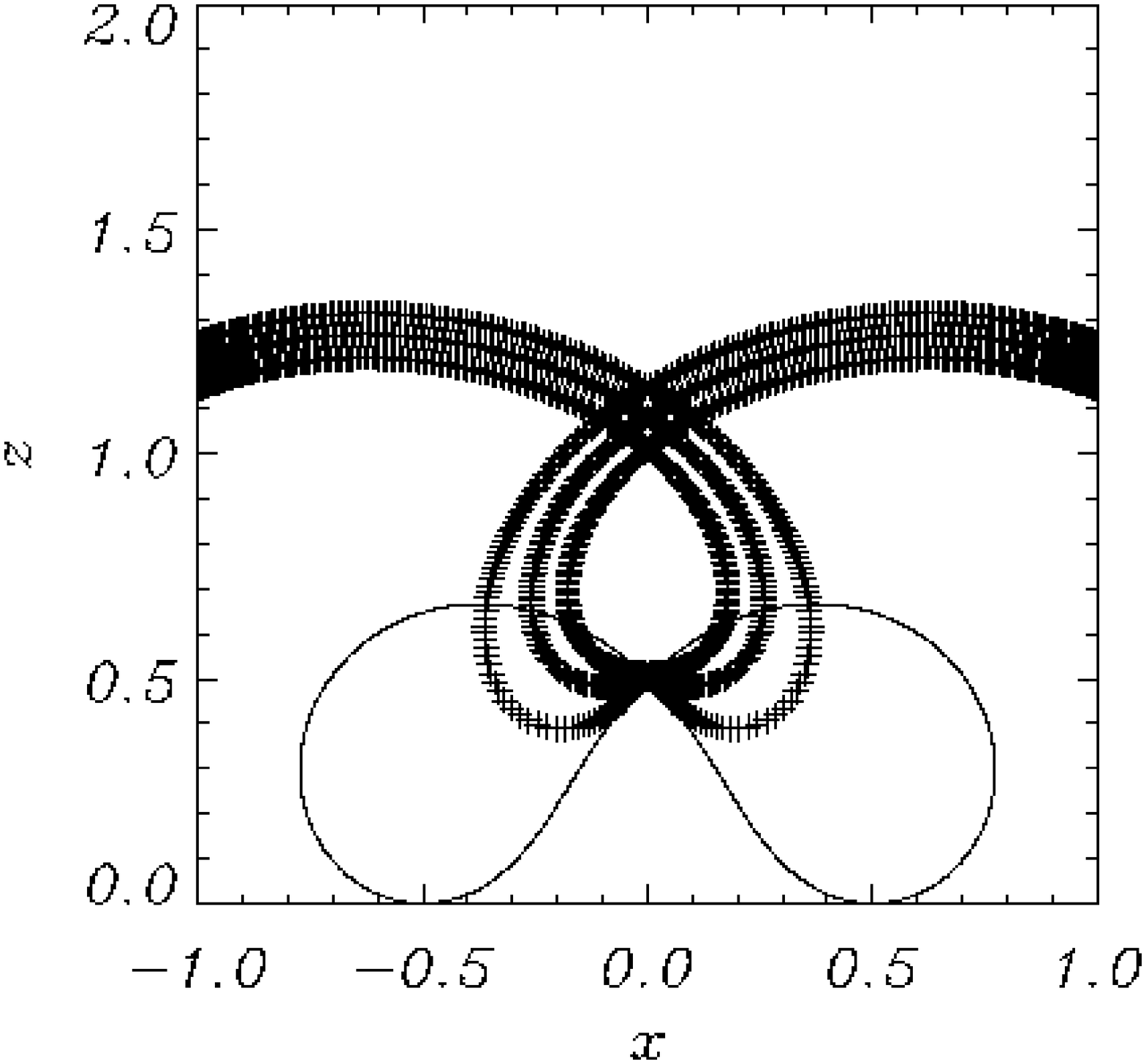}\\
\hspace{0in}
\vspace{0.1in}
\hspace{0.2in}
\includegraphics[width=1.2in]{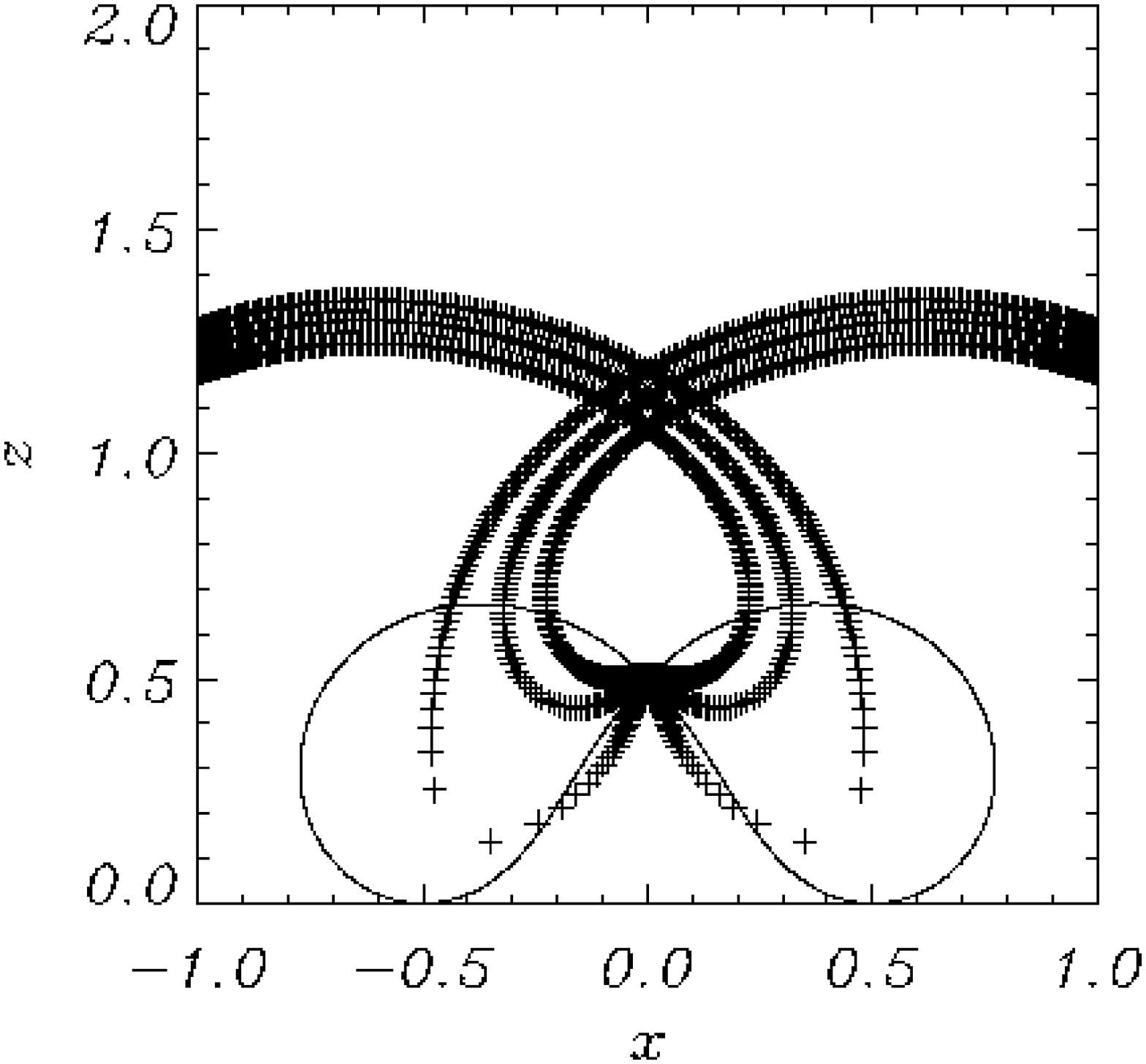}
\hspace{0.0in}
\includegraphics[width=1.2in]{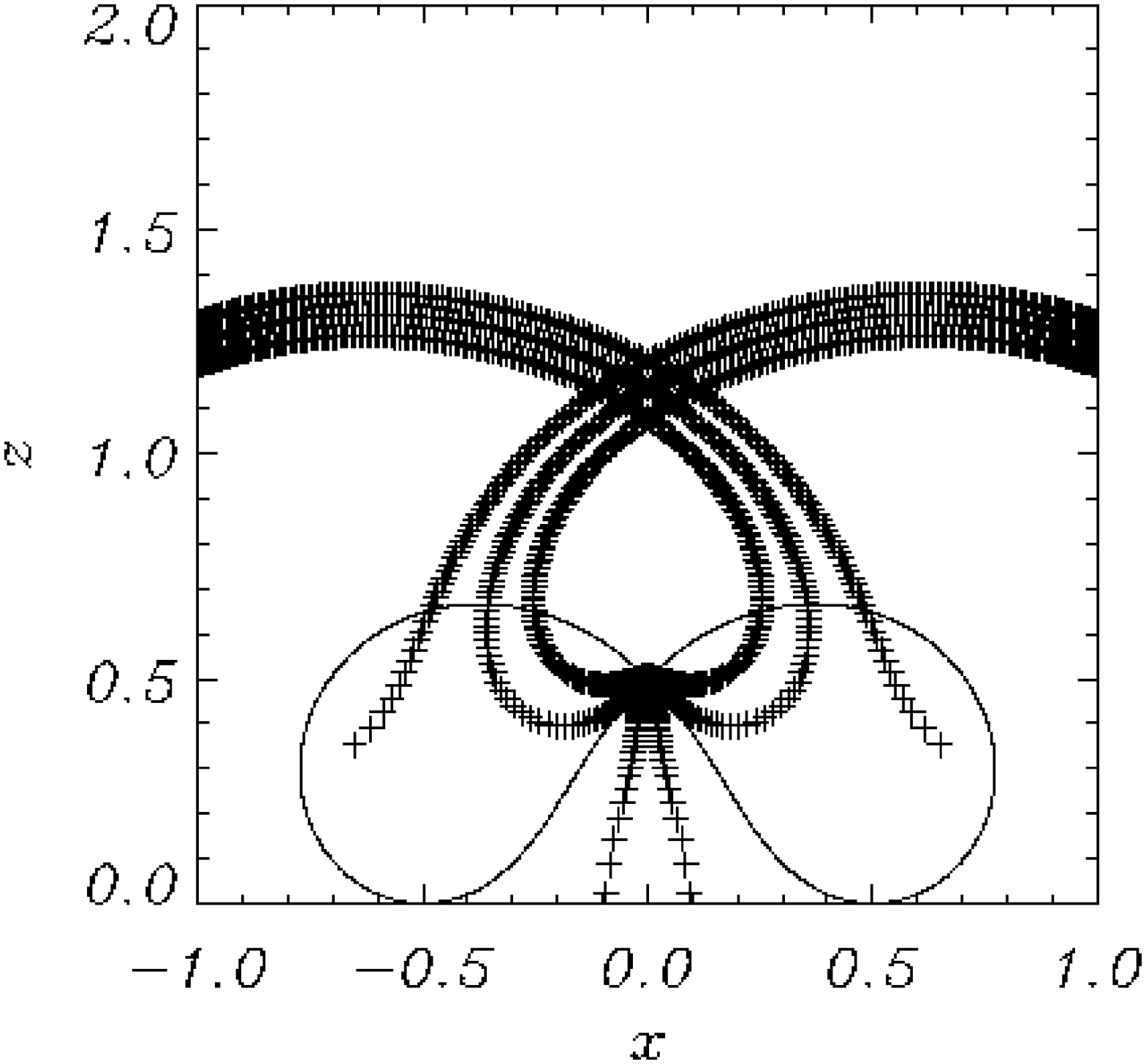}
\hspace{0.0in}
\includegraphics[width=1.2in]{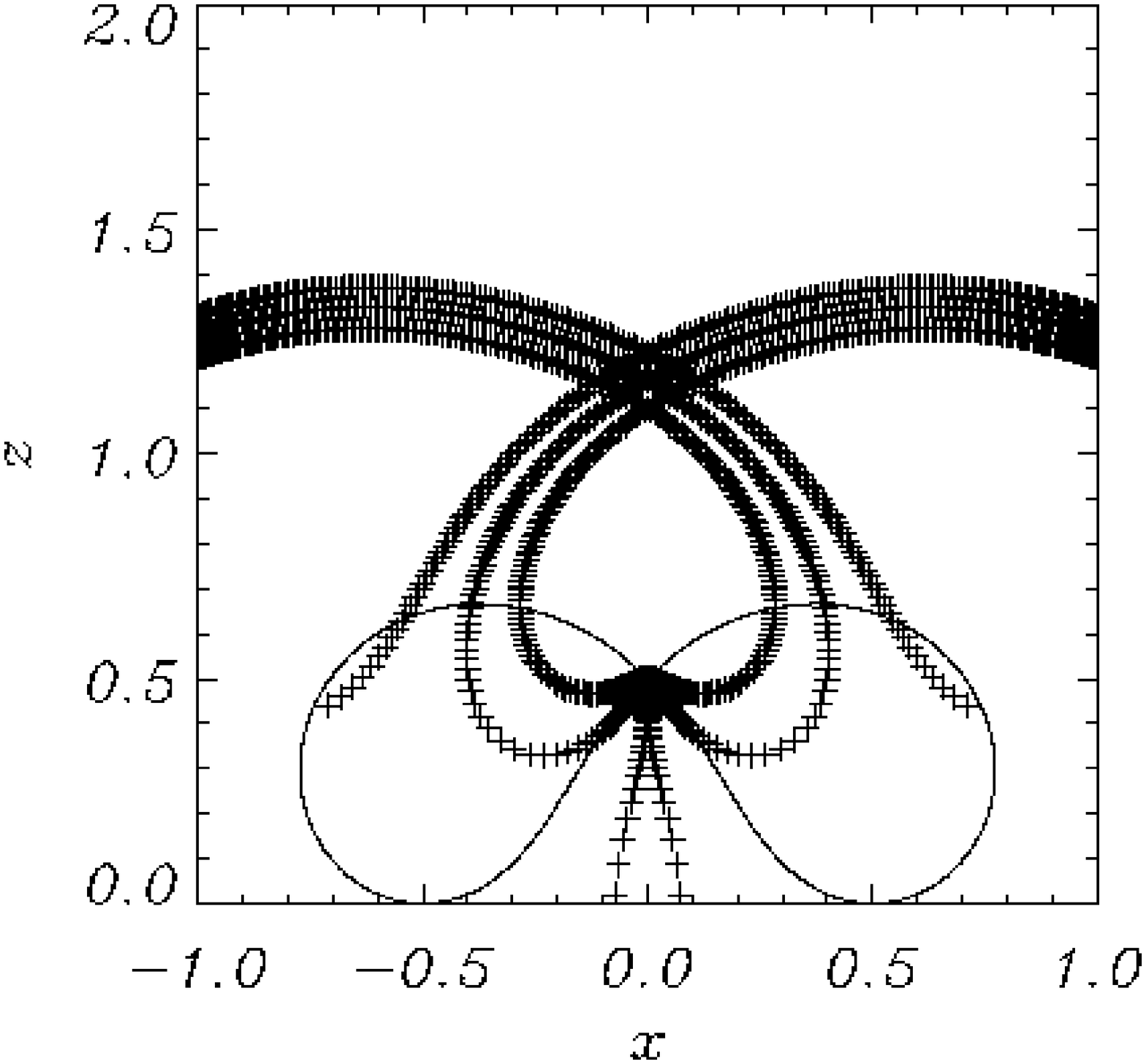}
\hspace{0.0in}
\includegraphics[width=1.2in]{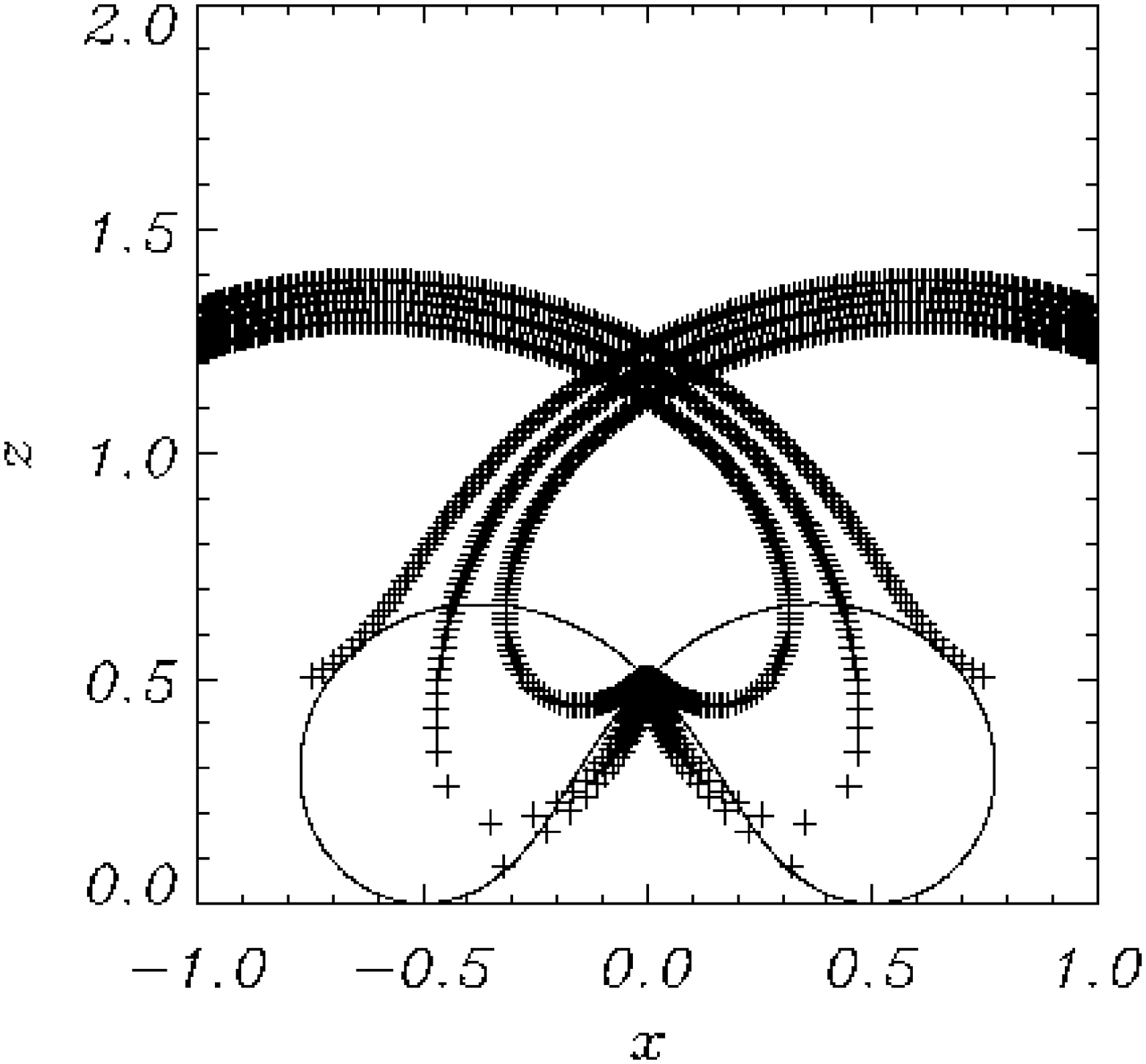}\\
\hspace{0in}
\vspace{0.1in}
\hspace{0.2in}
\includegraphics[width=1.2in]{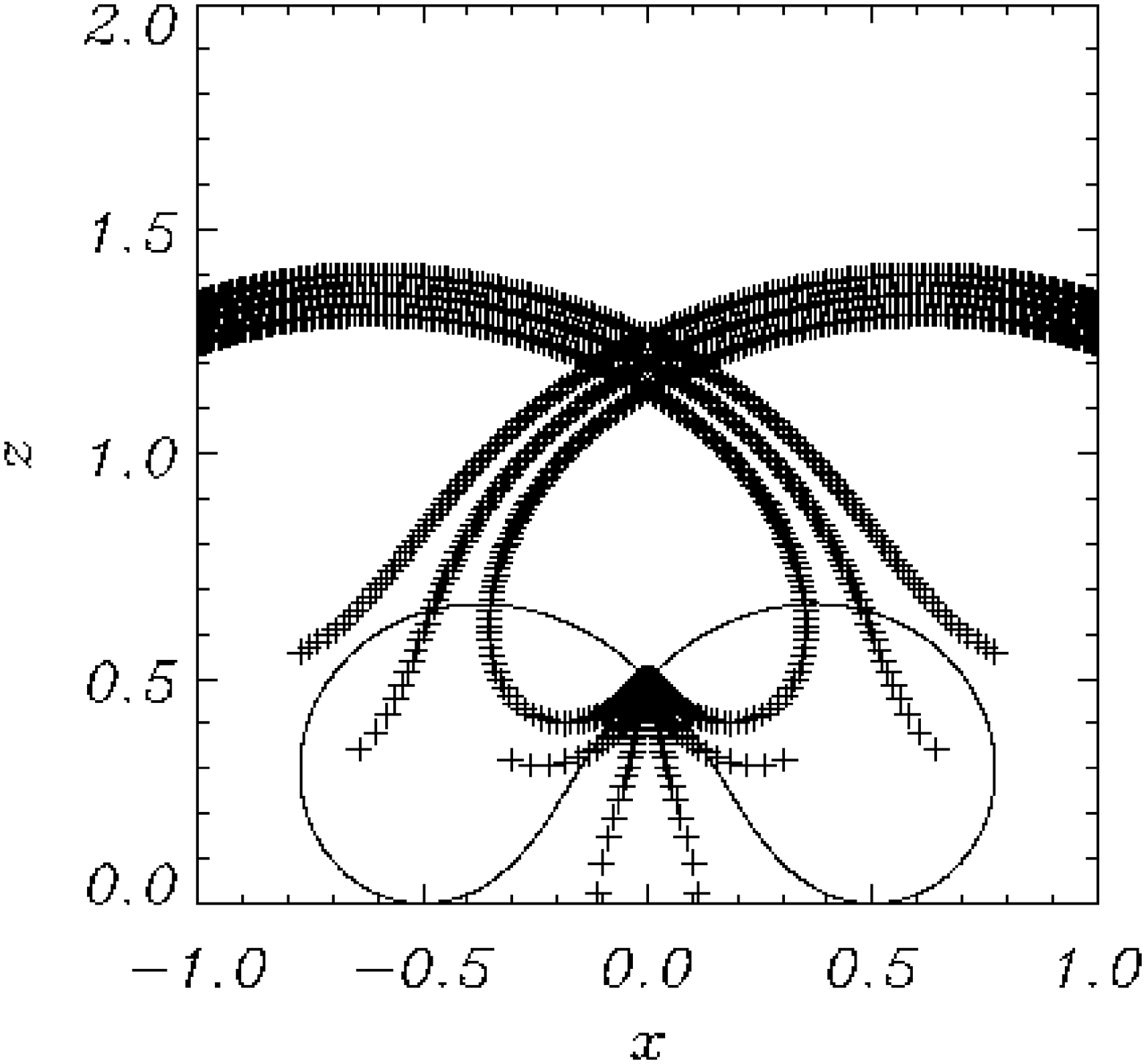}
\hspace{0.0in}
\includegraphics[width=1.2in]{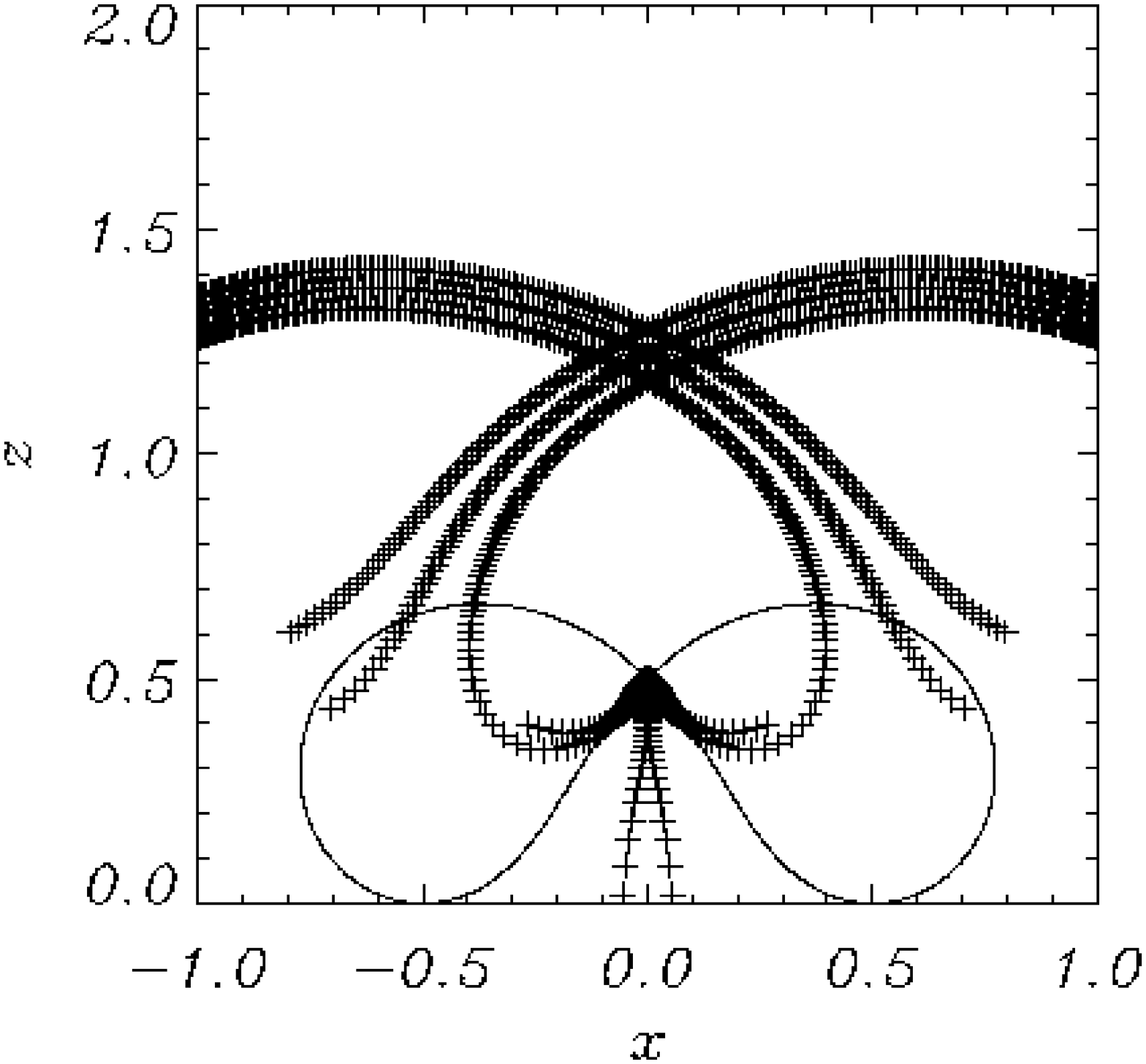}
\hspace{0.0in}
\includegraphics[width=1.2in]{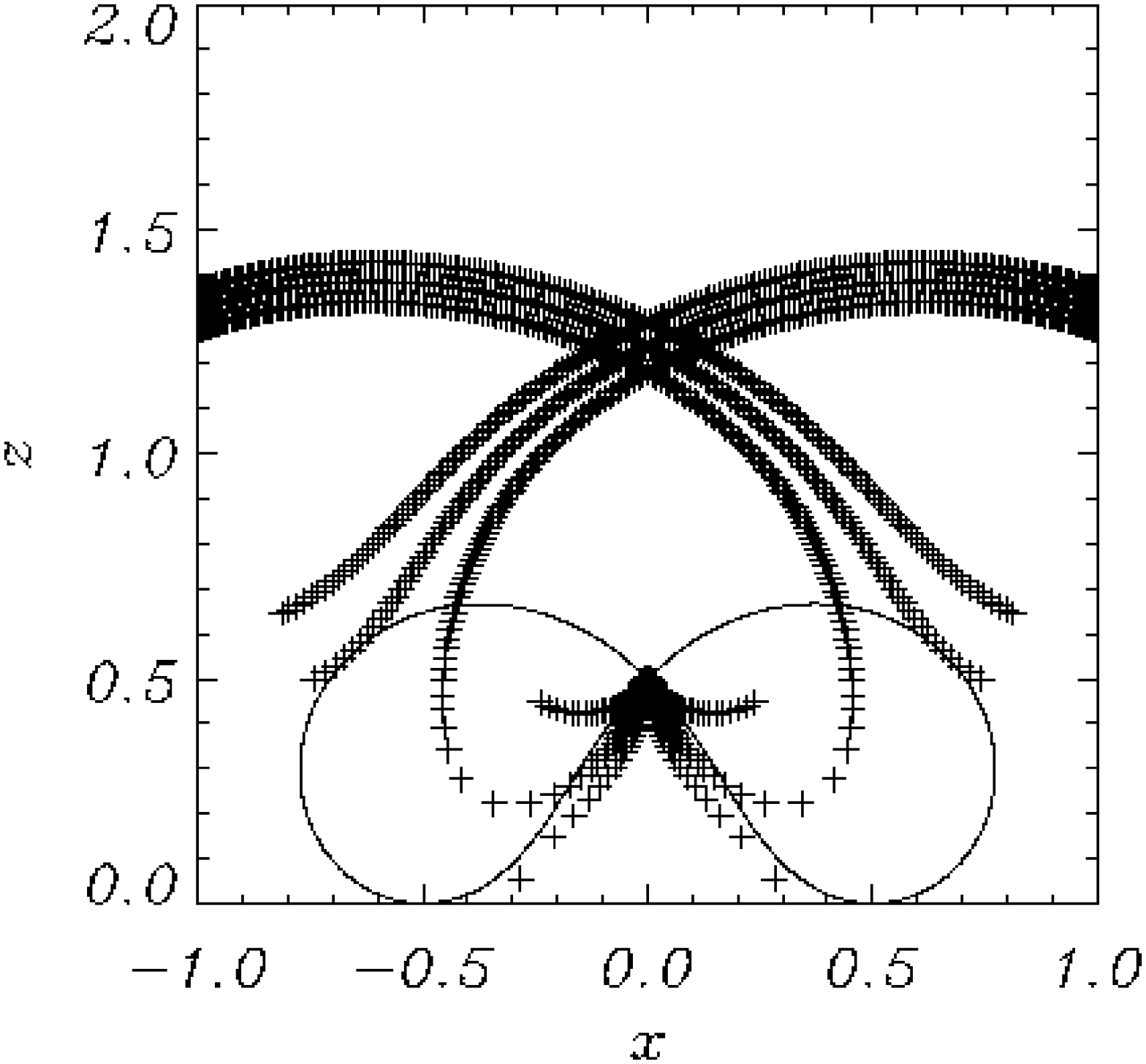}
\hspace{0.0in}
\includegraphics[width=1.2in]{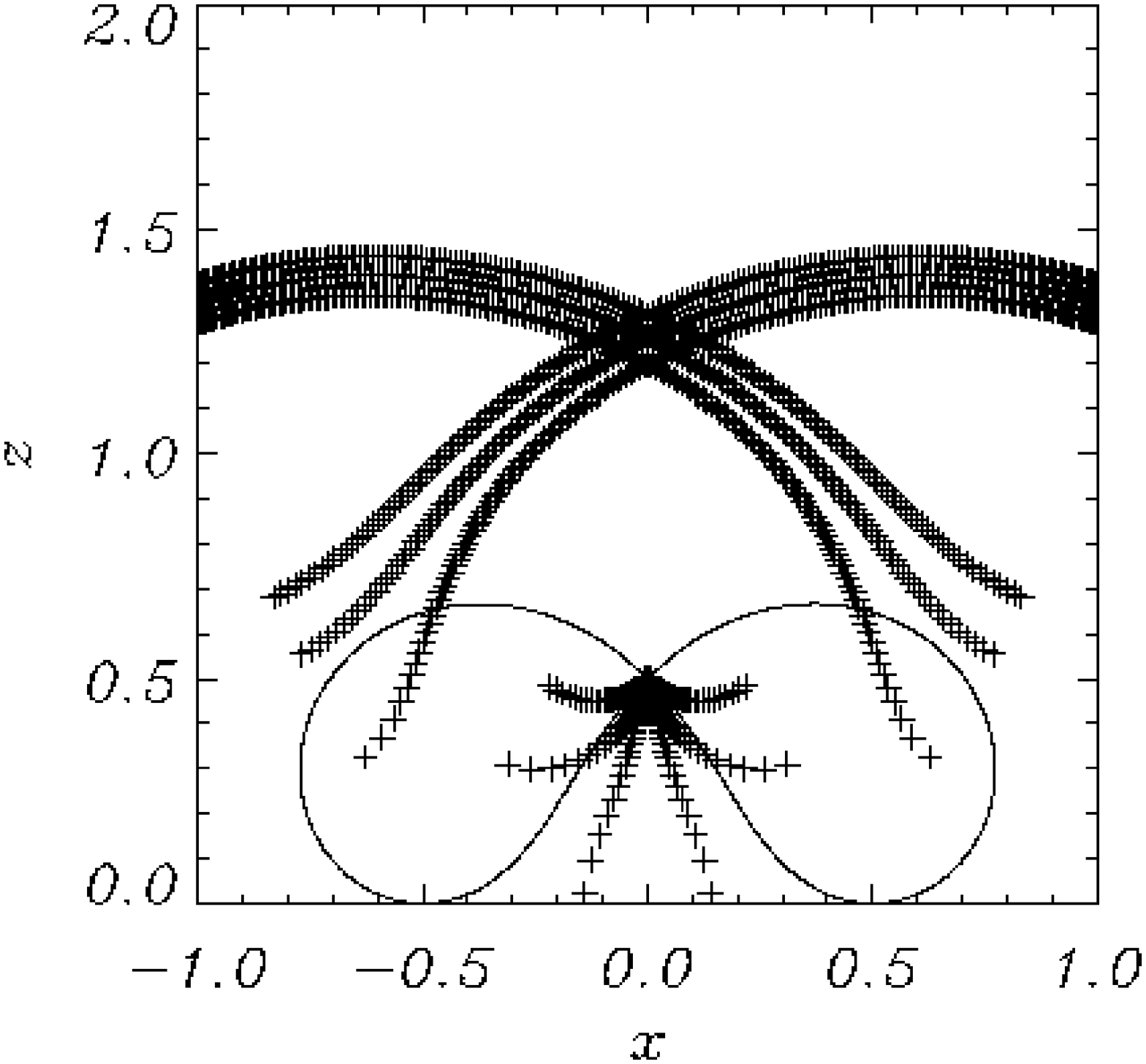}\\
\hspace{0in}
\vspace{0.1in}
\hspace{0.2in}
\includegraphics[width=1.2in]{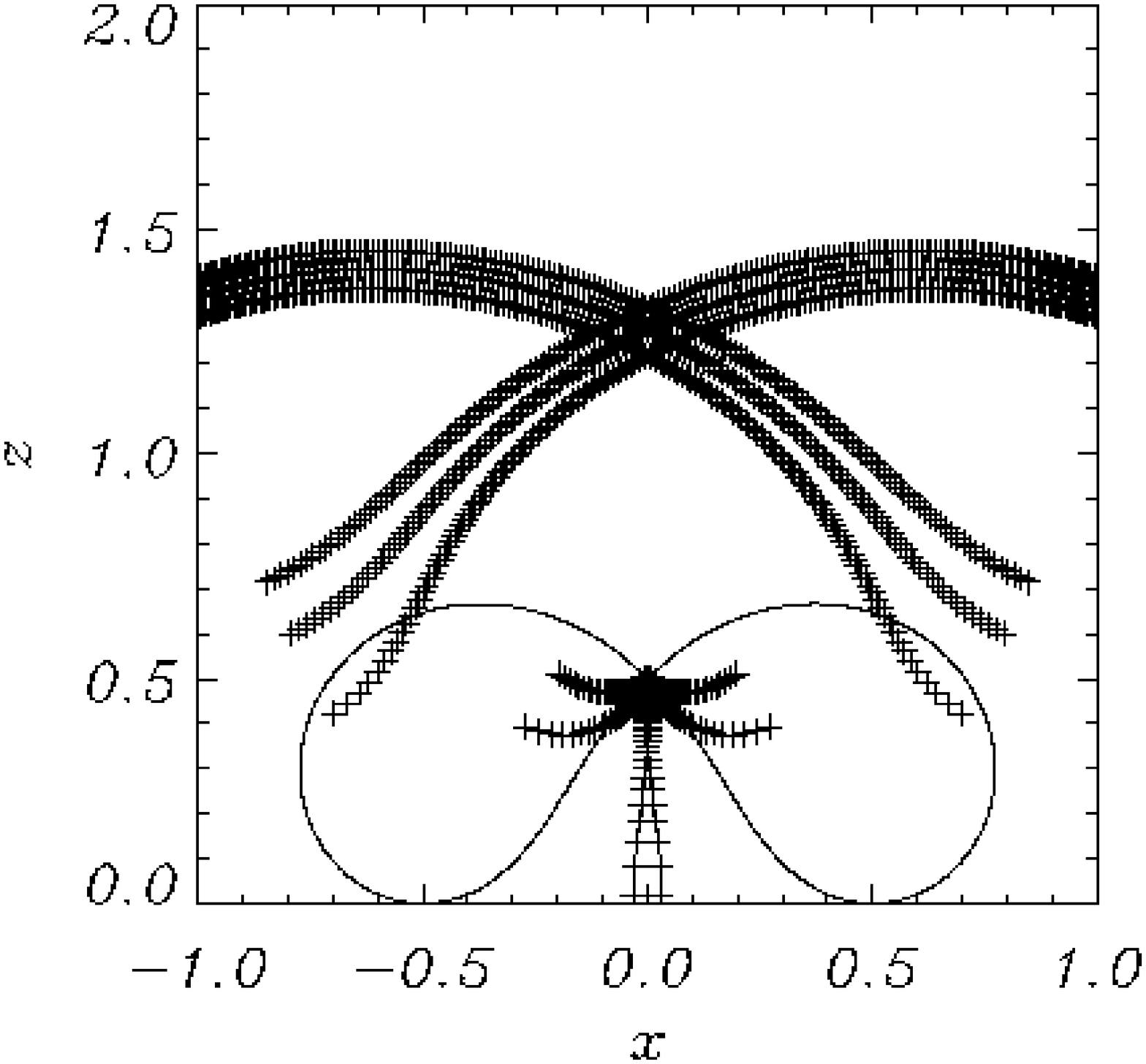}
\hspace{0.0in}
\includegraphics[width=1.2in]{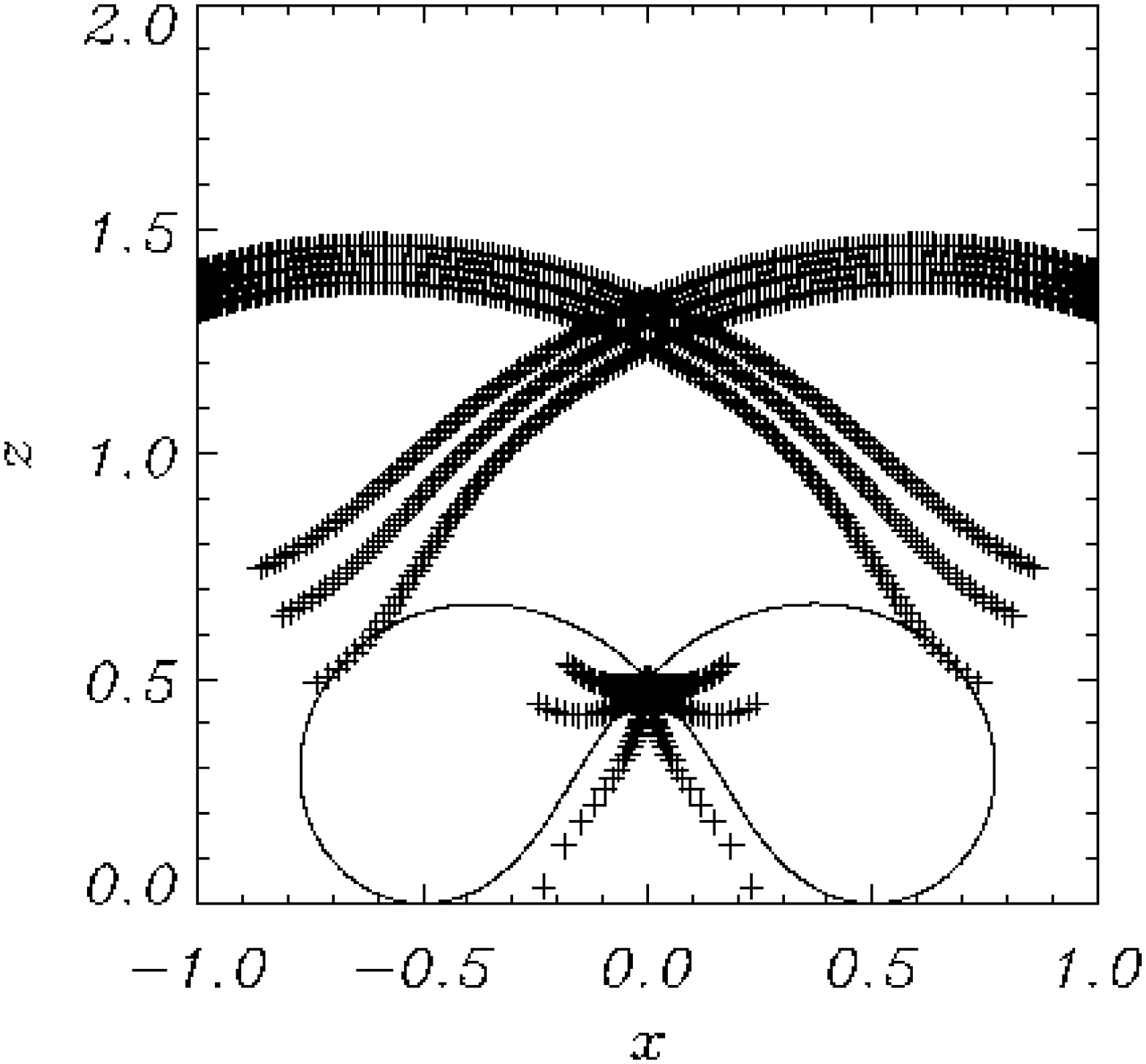}
\hspace{0.0in}
\includegraphics[width=1.2in]{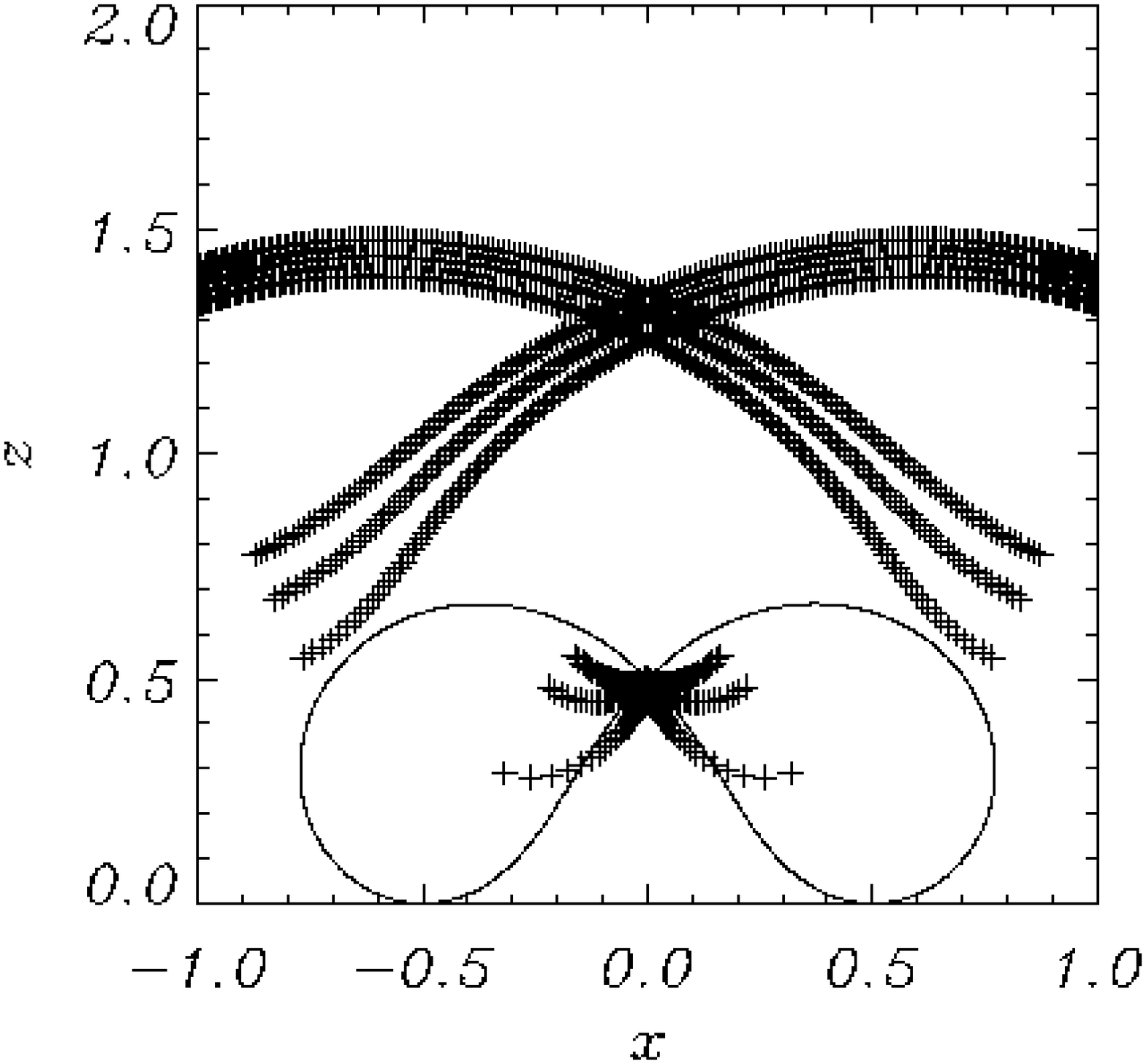}
\hspace{0.0in}
\includegraphics[width=1.2in]{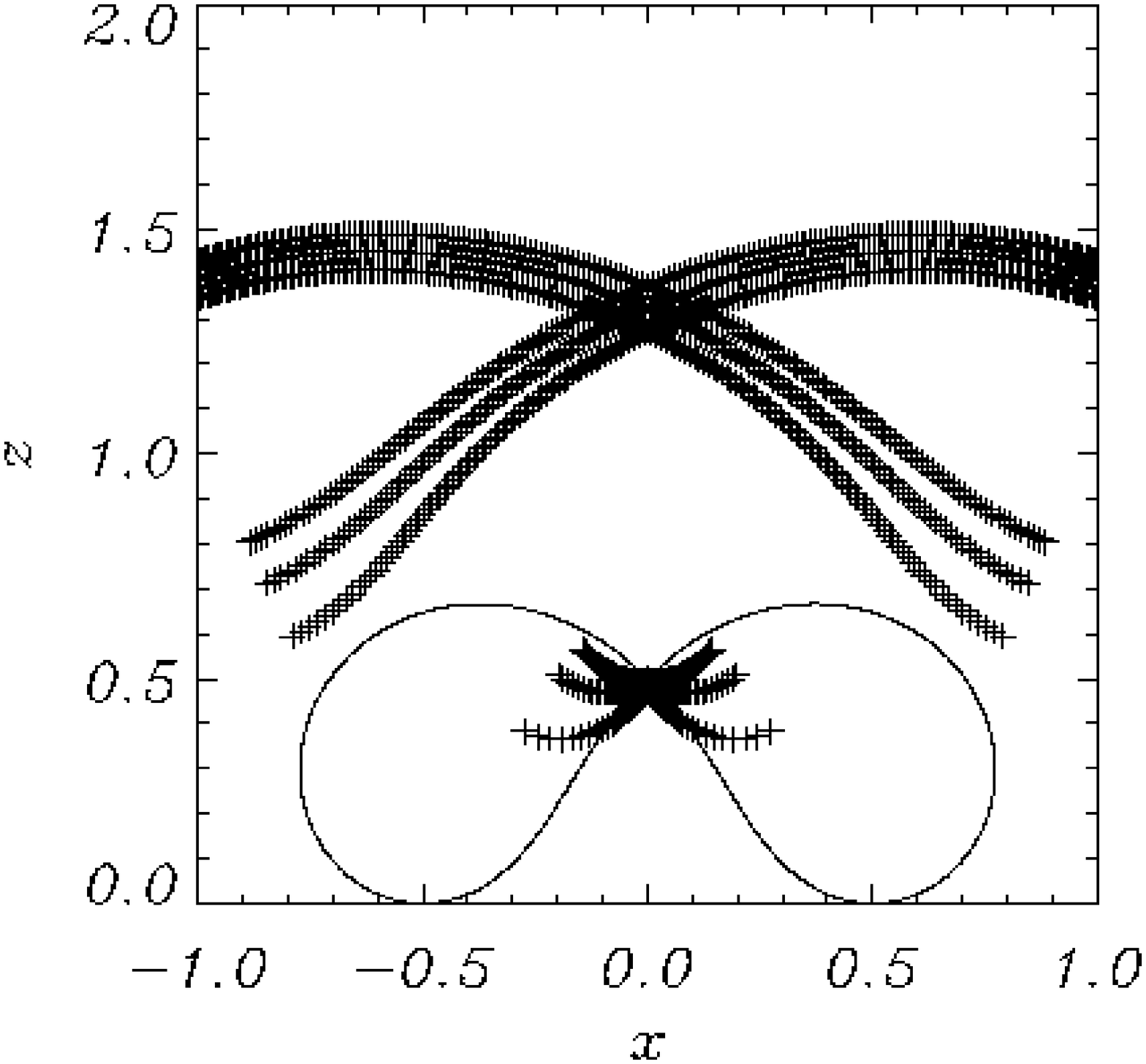}
\caption{Analytical solution of $V$ for WKB approximation of a fast wave sent in from lower boundary for $-1 \leq x \leq 1$, $z=0.1$ and its resultant propagation at  times  $(a)$ $t$=0.055, $(b)$ $t$=0.11, $(c)$ $t$=0.165, $(d)$ $t=$0.22, $(e)$ $t$=0.275 and $(f)$ $t$=0.33, $(g)$ $t$=0.385, $(h)$ $t$=0.44, $(i)$ $t$=0.495, $(j)$ $t=$0.55, $(k)$ $t$=0.605, $(l)$ $t$=0.66,  $(m)$ $t$=0.72, $(n)$ $t$=0.74, $(o)$ $t$=0.76, $(p)$ $t=$0.78, $(q)$ $t$=0.8, $(r)$ $t$=0.82, $(s)$ $t$=0.84, $(t)$ $t$=0.86, $(u)$ $t$=0.88, $(v)$ $t=$0.9, $(w)$ $t$=0.92 and $(x)$ $t$=0.94, labelling from top left to bottom right. The  curves consisting of the crosses represent the front, middle and back edges of the WKB wave solution, where the pulse enters from the top of the box.}
\label{fig:hulk}
\end{figure*}

We can again use the WKB approximation  to find a semi-analytical solution, and thus compare it to Simulation Three. In Section \ref{psyy}, we found a WKB approximation for our fast wave equation (equations \ref{fast_dipole_characteristics}). Thus, by plotting the fieldlines coming from $-1 \leq x_0 \leq 1$, we can make a direct comparison with Simulation Three. This can be seen in Figure \ref{fig:hulk}, where the wavefronts are plotted at the same times as Figure \ref{fig:4.5.2.4}. Again, the agreement is very good. As before, each wavefront consists of many tiny crosses.

Note that in the numerical simulation, the fast wave diffuses slightly and resolution runs out, hence some parts of the wave are unresolved fully. This is not the case in the WKB approximation, where the wavefronts are crisp. Thus, the numerical and WKB results show the same behaviour, but have small differences close to the leading and trailing wavefronts. However, the comparison with the  middle WKB wavefront is excellent.

%\newpage

%\vspace{0.3in}

%**********************************************************************

\begin{figure}
\begin{center}
\includegraphics[width=2.0in]{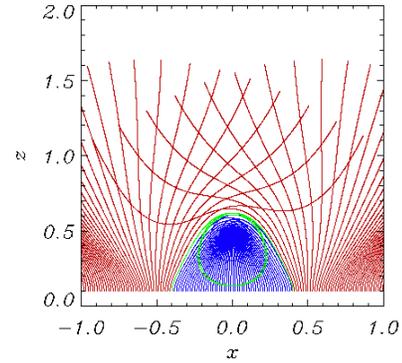}
\caption{Particle paths for starting points of $-1 \leq x_0 \leq 1$ set at intervals of $0.01$. Blue lines shows starting points $-0.4 < x < 0.4$. Green lines show starting points of $x_0=\pm0.4$. All other starting points ($x_0 <0.4$, $x_0 >0.4$) are in red.} 
\label{WKBtwodipolesparticlepaths_2}
\end{center}
\end{figure}

We can also use our WKB solution to map out the particle paths of the wavefront. This can be seen in Figure \ref{WKBtwodipolesparticlepaths_2}. Here, we can see particle paths for starting points of $-1 \leq x_0 \leq 1$ set at intervals of $0.01$. The blue lines shows starting points $-0.4 < x < 0.4$, green lines show starting points of $x_0=\pm0.4$, and all other starting points ($x_0 < -0.4$, $x_0 >0.4$) are in red. We can clearly see that the critical points (where the wave splits to head to the null or away from the magnetic skeleton) are around $x_0=\pm0.4$.

Hence, we see that for fast magnetoacoustic waves in the neighbourhood of two dipoles, a fraction of the wave (and thus wave energy) is trapped by the null with the rest  escaping. The fraction captured by the null will be of the form $\frac{1}{L}{x_{\small{\rm{\textit{crit}}}}\left(a,z_0\right)}$, where $L$ is the length of our lower boundary $\left(  -L\leq x \leq L\right)$ and $x_{\small{\rm{\textit{crit}}}}$ is the critical starting point that divides the particle paths into those that spiral into the null and those that do not (e.g. the starting point connected to the green lines in Figure \ref{WKBtwodipolesparticlepaths_2}). For the linear fast wave investigated in this paper; $L=1$, $z=z_0=0.1$, $a=0.5$ and  $x_{\small{\rm{\textit{crit}}}}=0.4$, which means  $40\%$ of the  wave is trapped by the null.  Hence, $40\%$ of the wave energy accumulates at the null and ohmic dissipation will extract the energy in the wave at this point.

%**********************************************************************

\section{Alfv\'en waves}\label{sec:3}

We now turn our attention to studying the behaviour of the Alfv\'en wave in the neighbourhood of two dipoles. Again, we will be launching a wave into the system from the lower boundary ($z=z_0$). However, we already have a good idea of how the wave will behave due to the work carried out in Paper II, i.e. that the linear Alfv\'en wave is always  confined to propagate along the field line it starts on.

Hence, we will not carry out three similar simulations (as we did above for the fast wave). Instead, we will look a one case for the Alfv\'en wave and find that the behaviour is indead of the same nature as found in Paper II. Thus, this section is included for completeness, but no further cases are shown here (since any other experiments would just show the same type of predictable behaviour).

\subsection{Numerical simulation}

The equations describing the behaviour of the Alfv\'en wave (equations \ref{alfvenalpha}) were solved numerically using a two-step Lax-Wendroff scheme. We initially consider a box ($-1 \leq x \leq 1$, $z_0 \leq z \leq 2$)  using the magnetic field shown in Figure \ref{fig:dipolemagneticfield}. We consider a single wave pulse coming in from the bottom boundary, localised along $-1 \leq x \leq -0.7$, $z=z_0$, where the pulse is chosen so that it crosses a separatrix.  For the single wave pulse, the boundary conditions were set such that:
\begin{eqnarray*}
v_y(x, z_0,t) = \left\{ \begin{array}{cl}
\sin { \omega t } \sin \left[ {\frac{ 10\pi}{3} \left(x+1\right) } \right] & \mathrm{for} \;  \left\{\begin{array}{c}
{-1 \leq x \leq -0.7} \\
{0 \leq t \leq \frac {\pi}{\omega} } \end{array}\right.  \\
{0 } & { \mathrm{otherwise} }\end{array}\right.\\
\left. \frac {\partial } {\partial x } v_y  \right| _{x=-1.4} = 0 \; , \quad \left.\frac {\partial} {\partial x }  v_y \right| _{x=1} = 0 \; , \quad \left.\frac {\partial } {\partial z }v_y  \right| _{z=1.6}  = 0 \; .
\end{eqnarray*}
Also, $z_0=0.2$ was chosen (as opposed to $z_0=0.1$) as it demonstrates the nature more clearly ($z_0=0.1$ gets a bit messy close to $(x,z)=(-0.5,0)$).

\begin{figure*}
\hspace{0in}
\vspace{0.1in}
%\hspace{0.2in}
\includegraphics[width=1.2in]{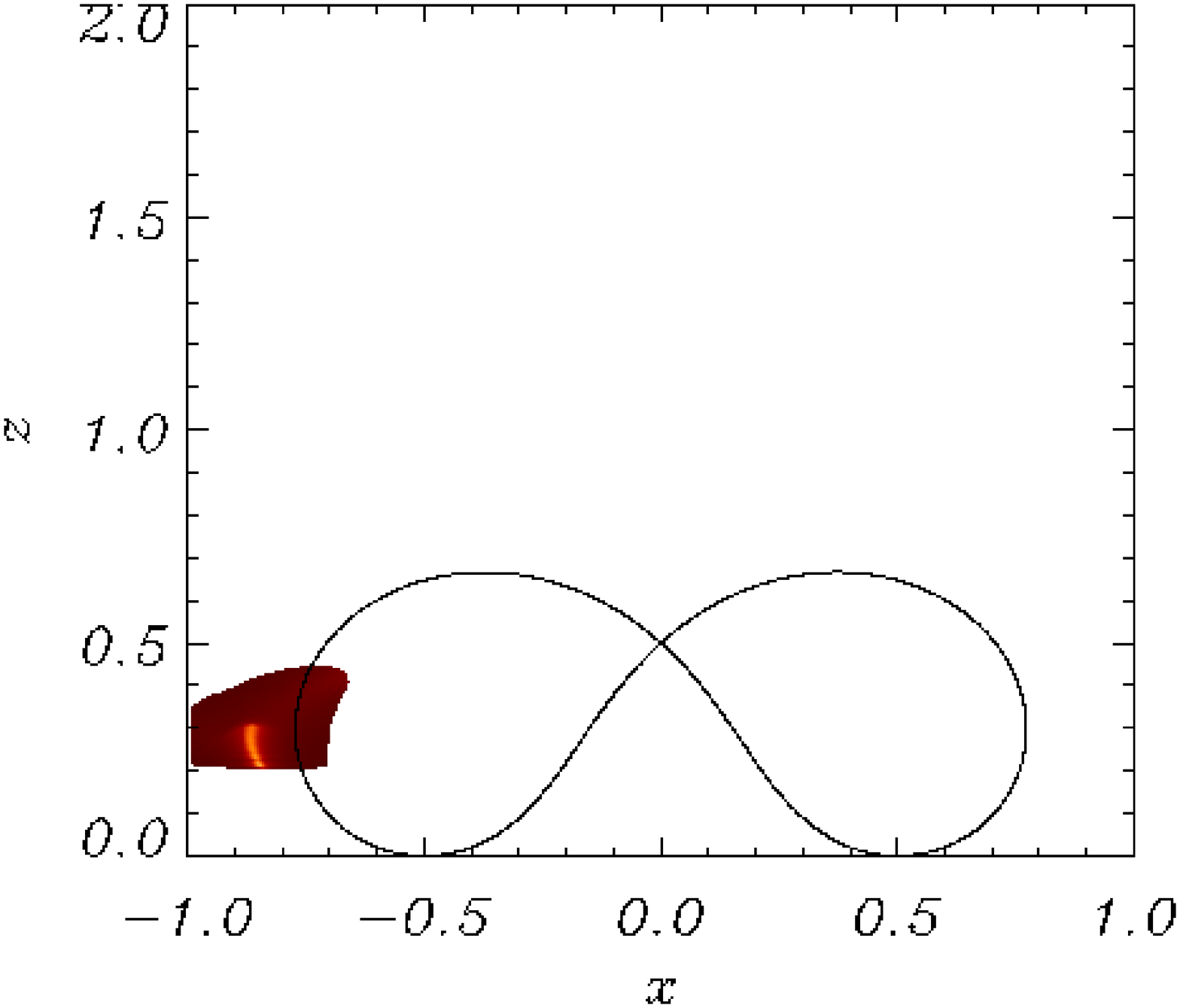}
\hspace{0.0in}
\includegraphics[width=1.2in]{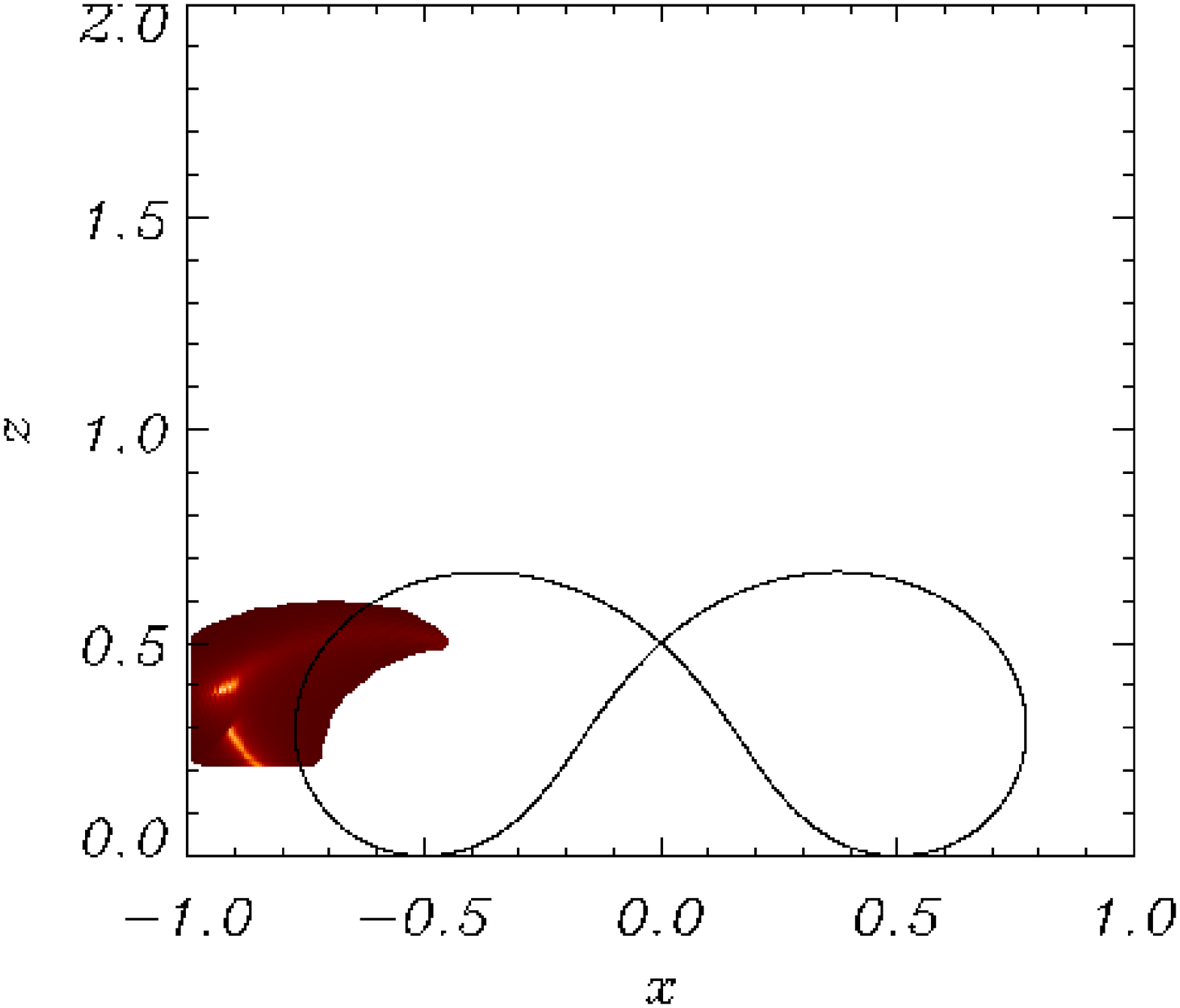}
\hspace{0.0in}
\includegraphics[width=1.2in]{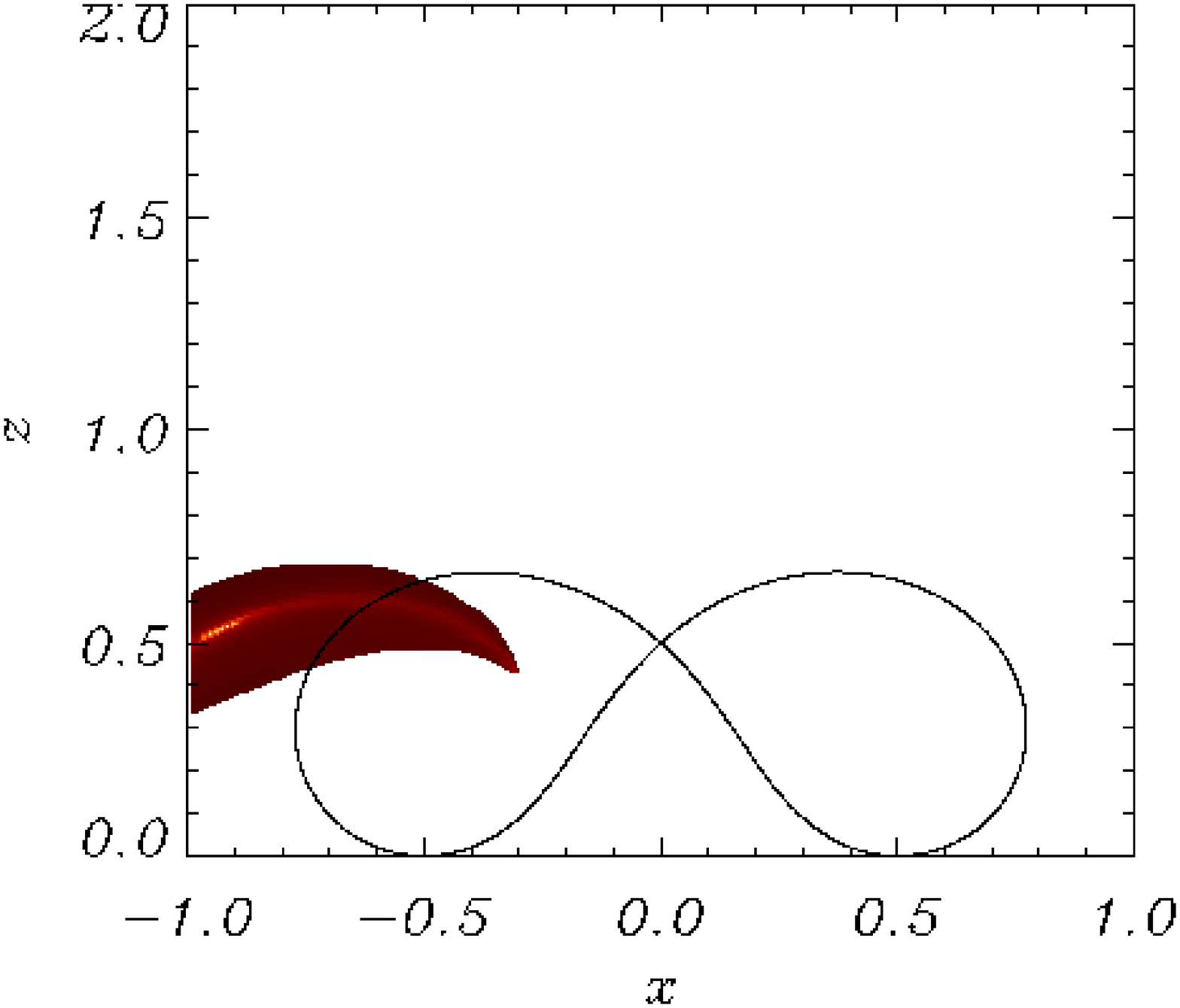}
\hspace{0.0in}
\includegraphics[width=1.2in]{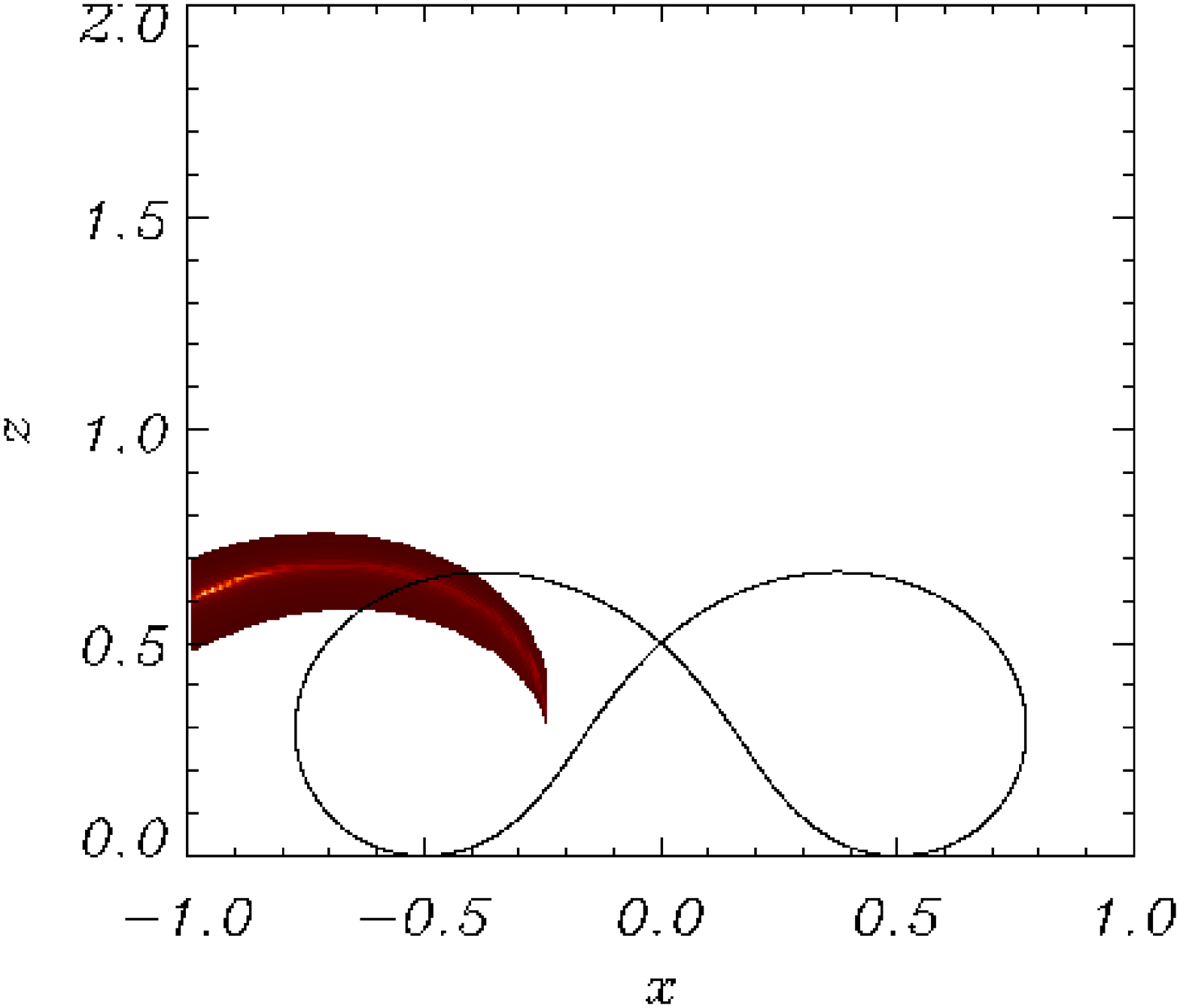}\\
\hspace{0in}
\vspace{0.1in}
%\hspace{0.2in}
\includegraphics[width=1.2in]{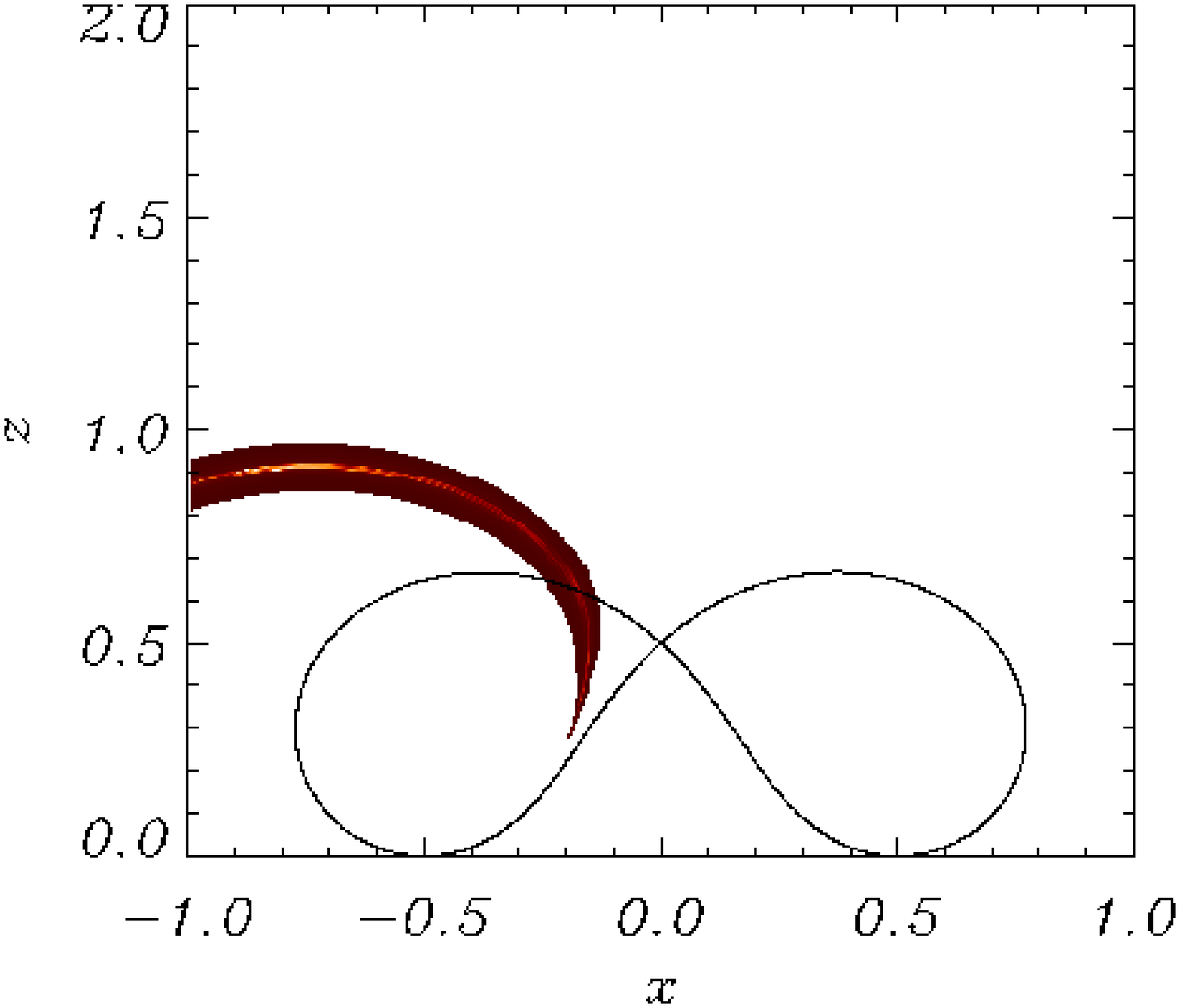}
\hspace{0.0in}
\includegraphics[width=1.2in]{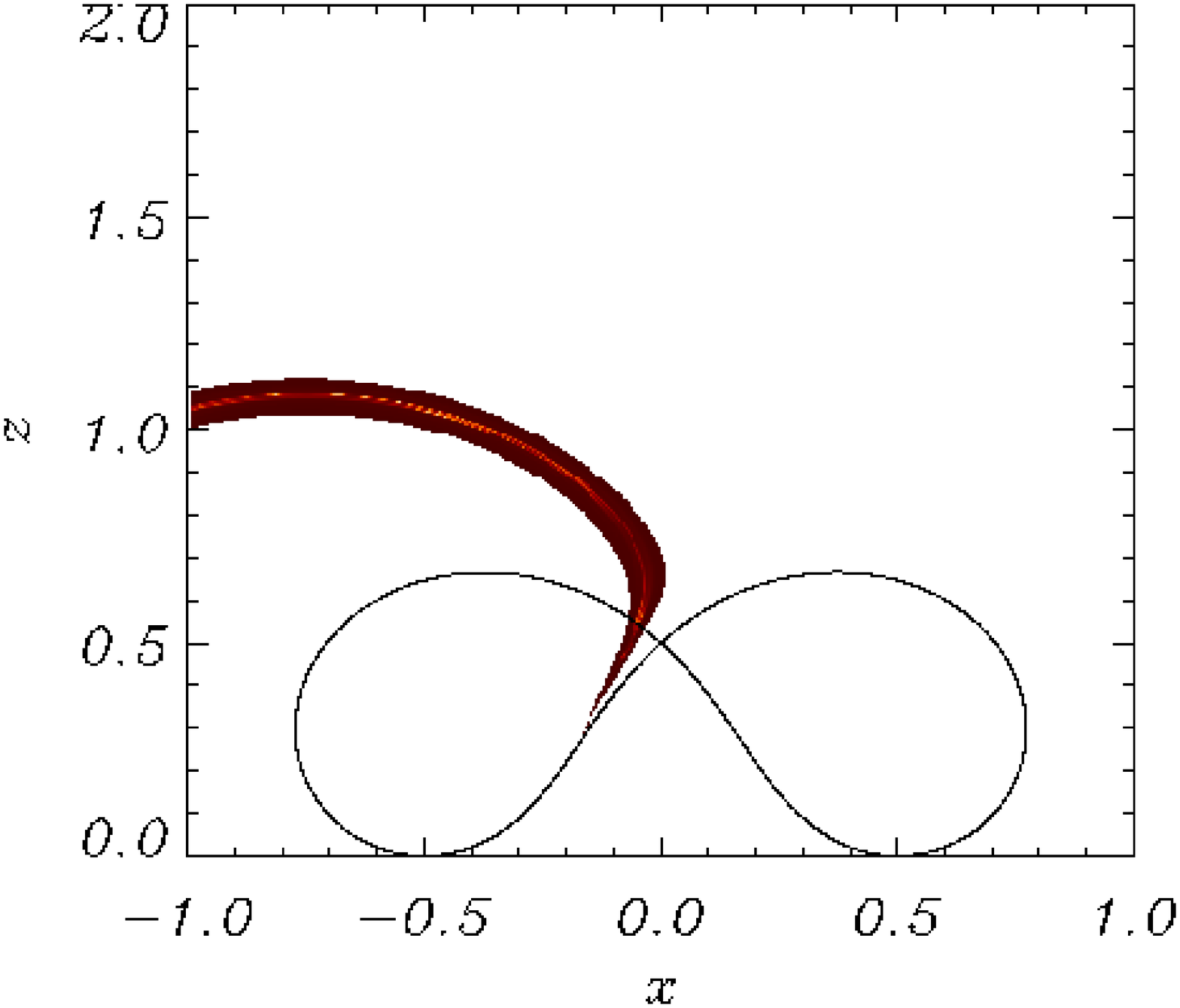}
\hspace{0.0in}
\includegraphics[width=1.2in]{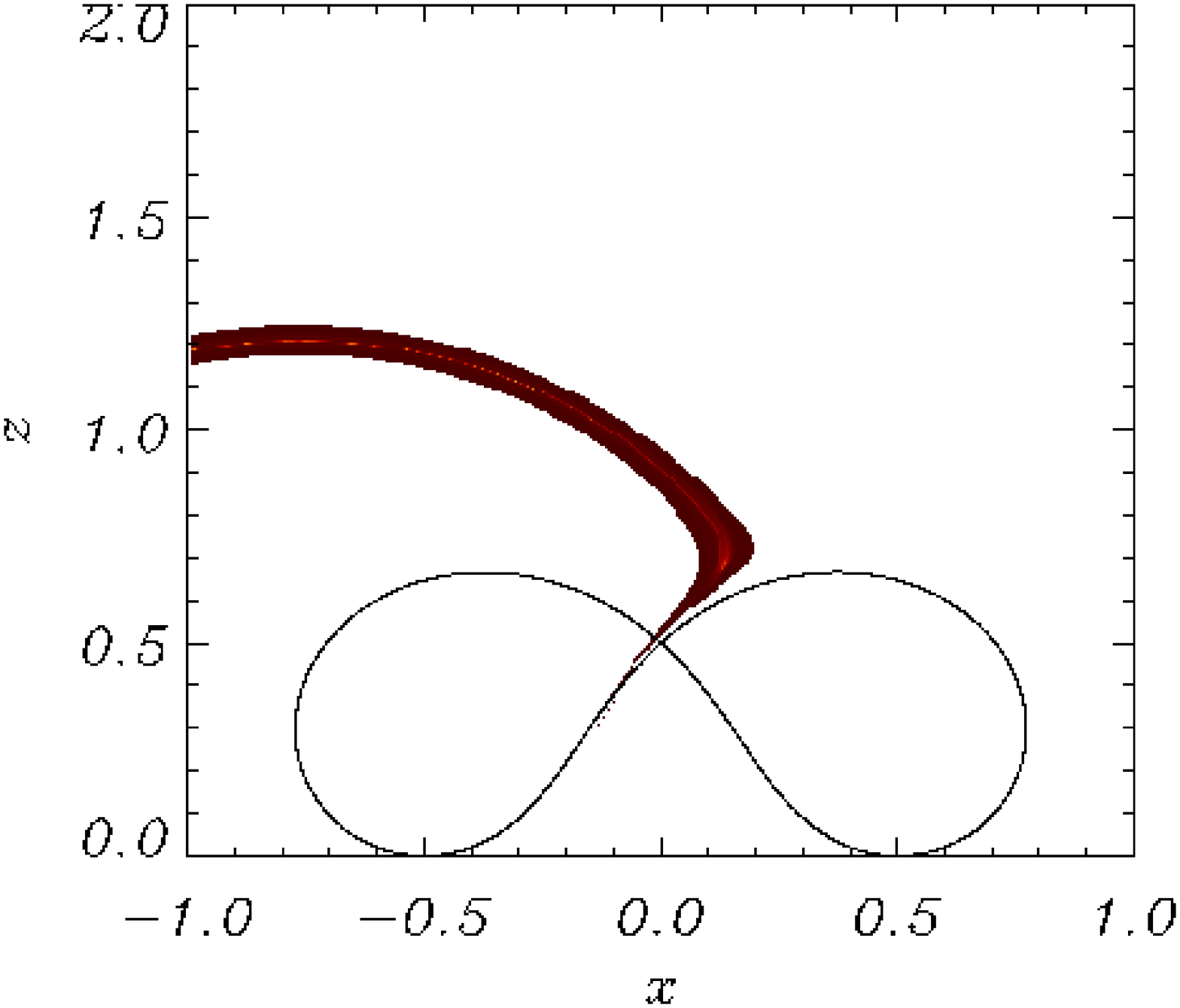}
\hspace{0.0in}
\includegraphics[width=1.2in]{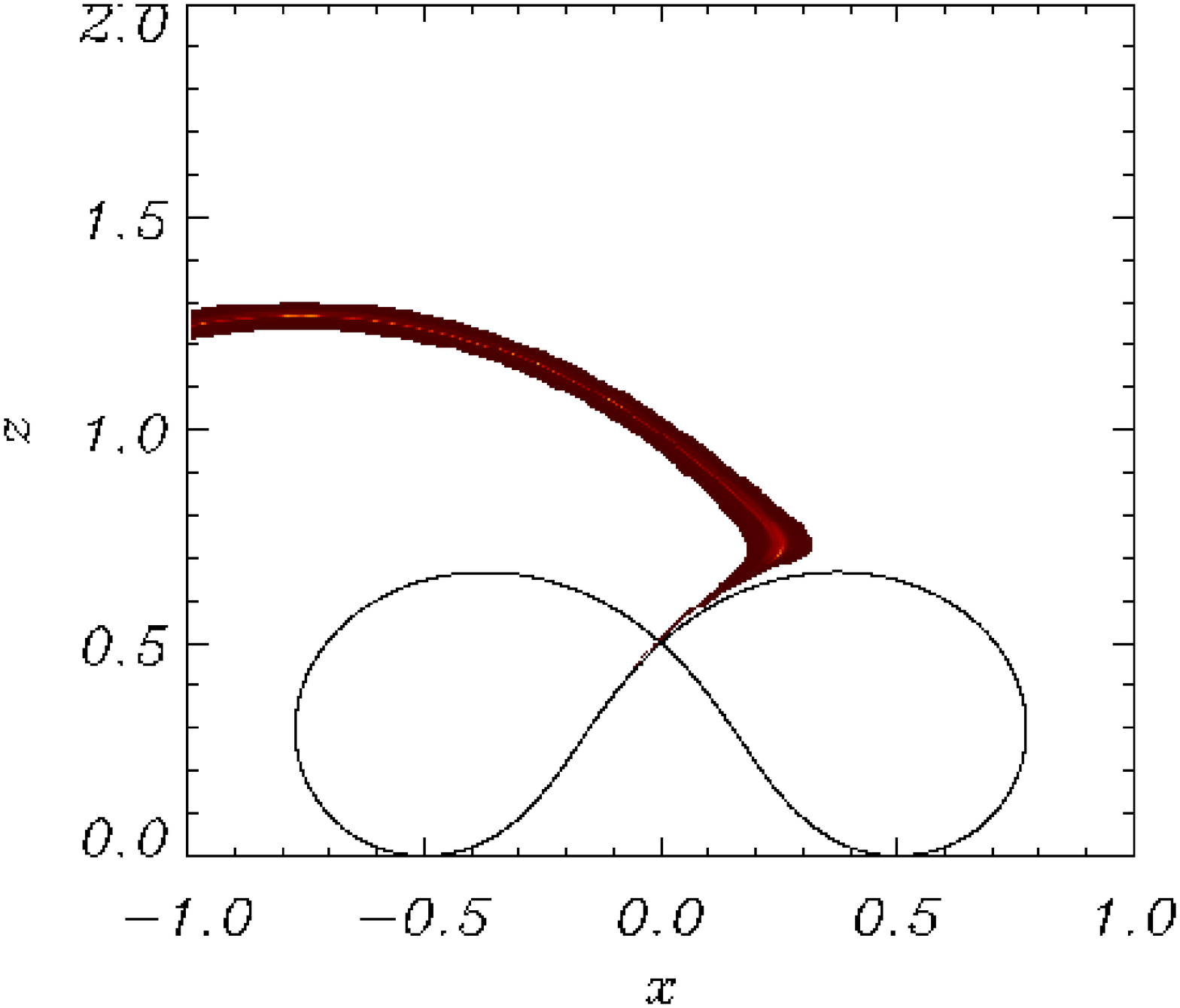}\\
\hspace{0in}
\vspace{0.1in}
%\hspace{0.2in}
\includegraphics[width=1.2in]{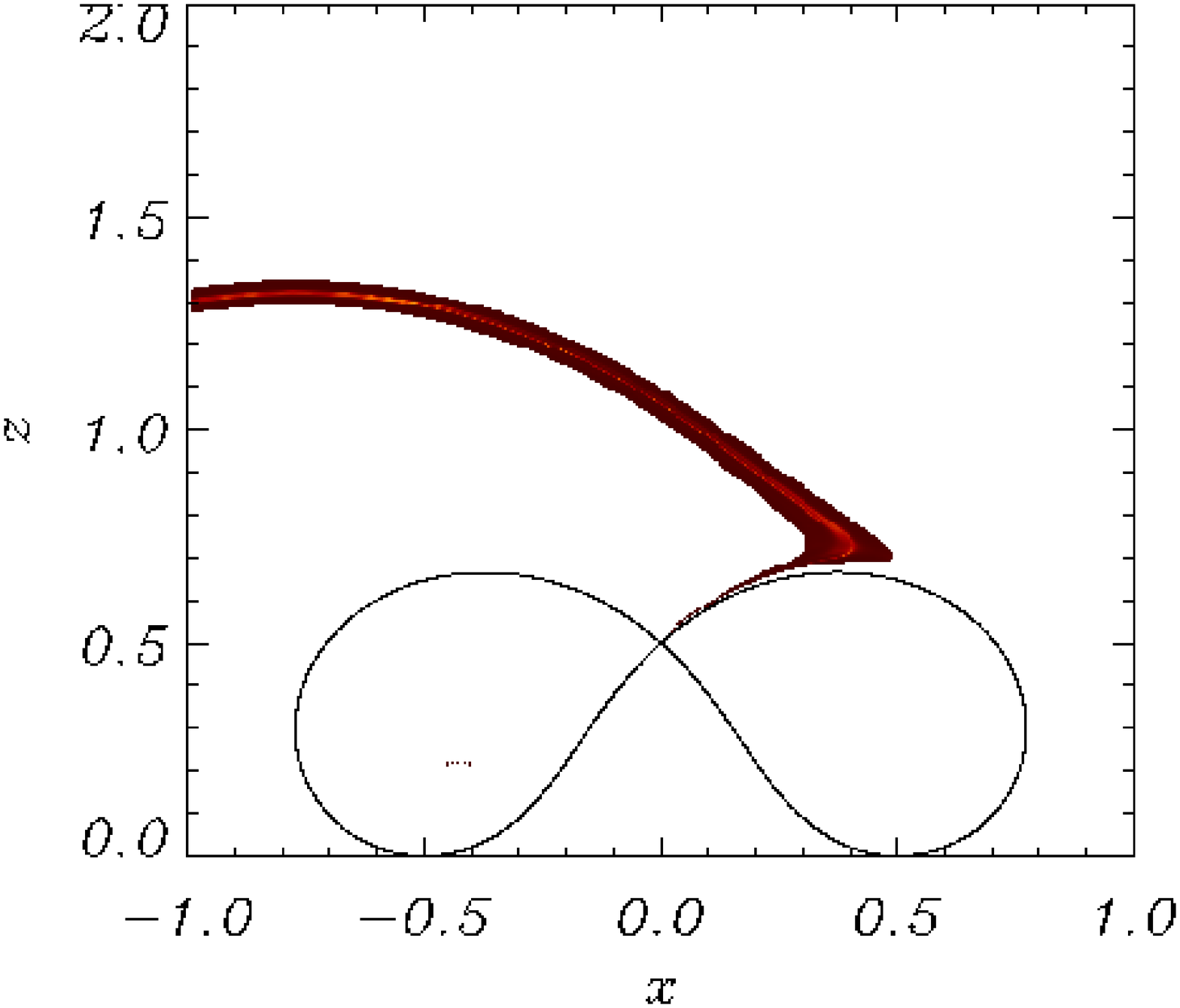}
%\vspace{0.1in}
\hspace{0.0in}
\includegraphics[width=1.2in]{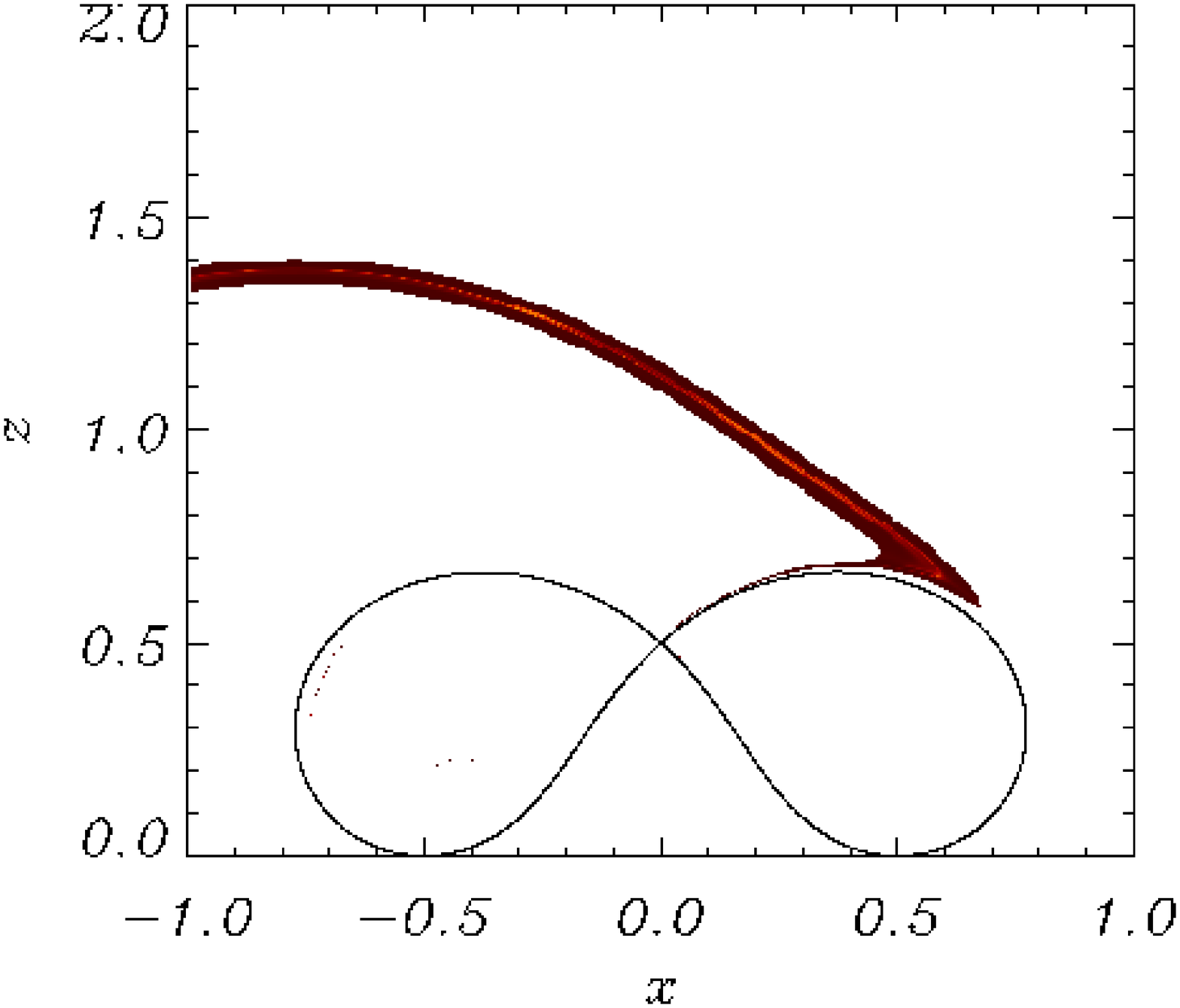}
\hspace{0.0in}
\includegraphics[width=1.2in]{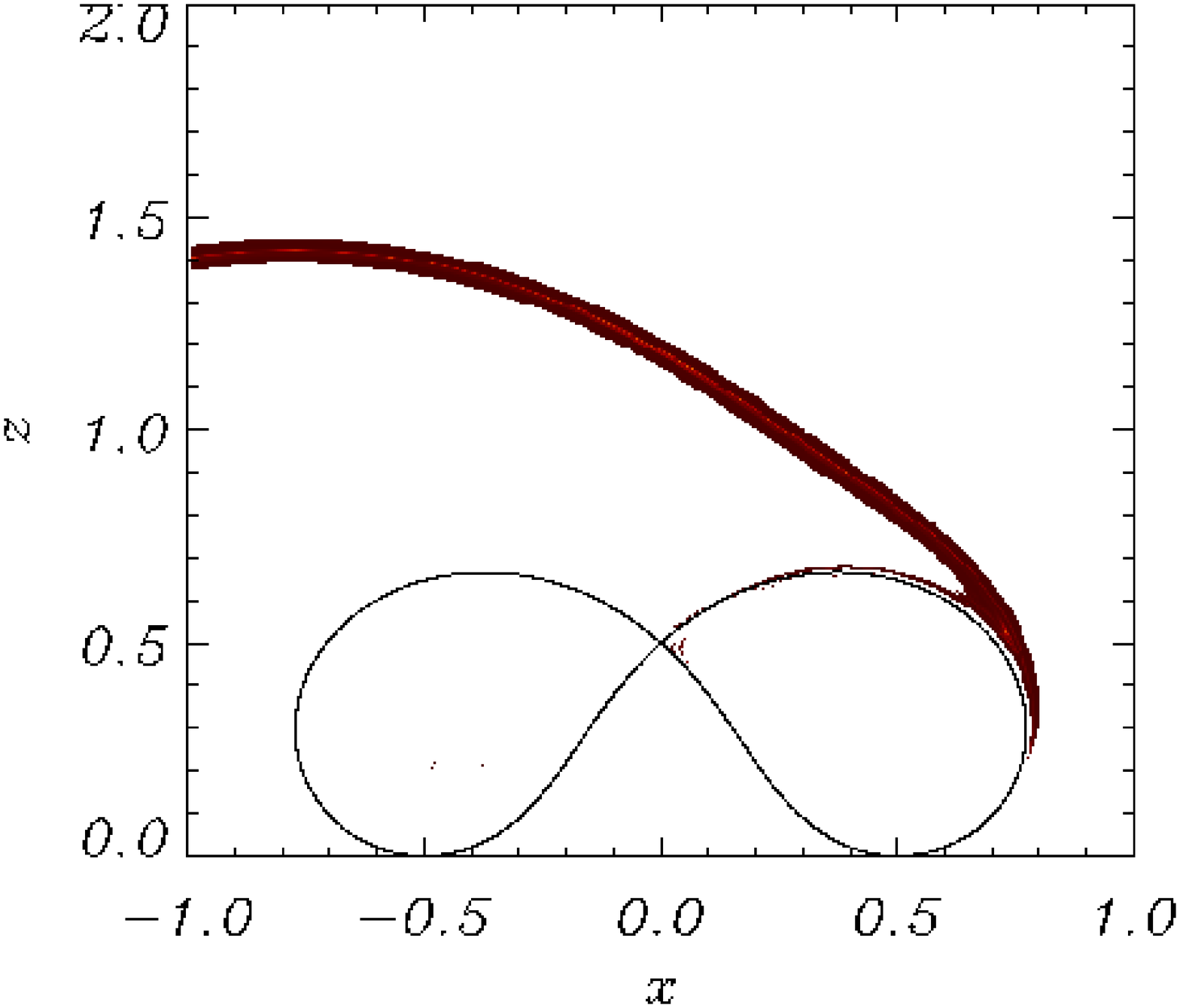}
\hspace{0.0in}
\includegraphics[width=1.55in]{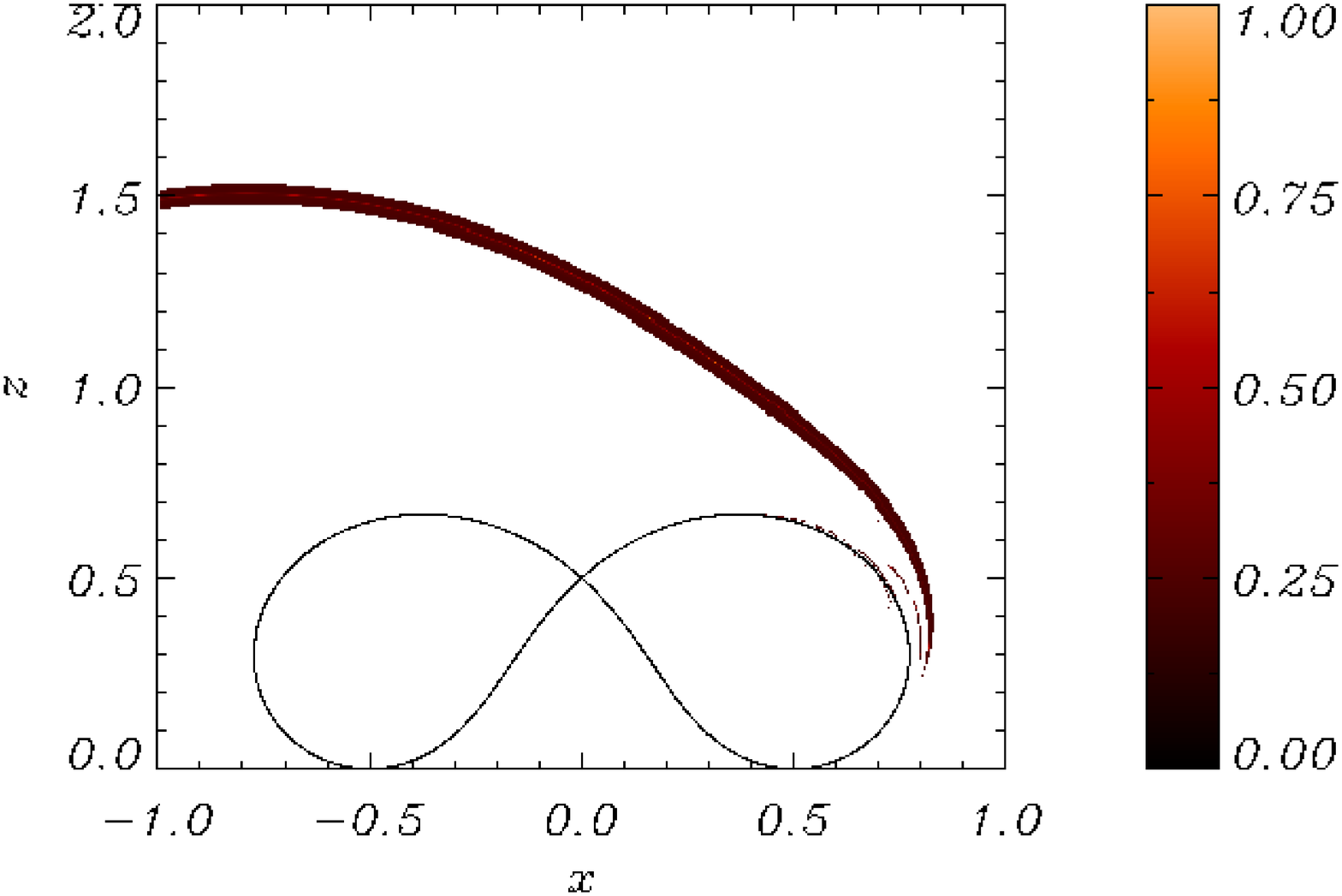}
\caption{Contours of $v_y$ for an Alfv\'en wave sent in from lower boundary for $-1\leq x \leq -0.7$ and its resultant propagation at times $(a)$ $t$=0.05, $(b)$ $t$=0.1, $(c)$ $t$=0.15, $(d)$ $t=0.2$, $(e)$ $t$=0.4, $(f)$ $t$=0.6, $(g)$ $t=0.8$, $(h)$ $t$=0.9, $(i)$ $t$=1.0, $(j)$ $t=1.1$, $(k)$ $t$=1.2, $(l)$ $t$=1.4, labelling from top left to bottom right.}
\label{fig:4.2.buffy}
\end{figure*}

We found that the linear Alfv\'en wave propagates up from the $z=z_0$ boundary and begins to spread out, following the field lines. This can be seen in Figure \ref{fig:4.2.buffy}. As it follows the field lines, part of the wave accumulates along the separatrix and appears to stick to the separatrices (depositing parts of itself as it goes). This behaviour  stretches the wave thinly along the separatrices.

Part of  the wave also appears to be propagating away from the magnetic skeleton  (though it is actually just following the field lines and spreading out). These two goals for different parts of the wave result in a sharp wave front being formed. Note that as the wavefront gets very sharp (as seen in the bottom row of subfigures), the resolution of the simulation is unable to fully resolve the wave behaviour and accumulation. However, it is clear that part of the wave is propagating away from the magnetic skeleton and the other part of the wave has accumulated along the separatrices (thus displaying the same nature as seen in Paper II).

\subsection{WKB approximation}

Again, we can use our WKB approximation to gain insight into the numerical simulation. The Alfv\'en equations we have to solve are:
\begin{eqnarray}
  \frac {\partial ^2 v_y } {\partial t^2 } = \left( B_x\frac {\partial }{\partial x} +B_z \frac {\partial }{\partial z } \right) ^2 v_y \;,\label{buffyTVS}
\end{eqnarray}
Using the magnetic field for the two dipoles (equation \ref{twodipoles}), we can apply the WKB approximation. Substituting $v_y = e^{i \phi (x,z) } \cdot e^{-i \omega t}$ into equation (\ref{buffyTVS}) and assuming that $\omega \gg 1$, leads to a first order PDE of the form $\mathcal{F} \left( x,z,\phi,p,q \right)=0$. Applying the method of characteristics, we generate the equations:
\begin{eqnarray}
  \frac {d \phi }{ds} &=&  \omega ^2  \; ,        \nonumber              \\
  \frac {dp}{ds} &=&  -\left( p \frac{\partial B_x}{\partial x}+q  \frac{\partial B_z}{\partial x} \right) \:\xi \; , \quad  \frac {dx}{ds} = B_x\:\xi\; ,  \nonumber     \\
\frac {dq}{ds} &=&  -\left( p \frac{\partial B_x}{\partial z}+q  \frac{\partial B_z}{\partial z} \right) \:\xi \;, \quad   \frac {dz}{ds} =  B_z\:\xi  \label{alfven_characteristics_twodipoles}
\end{eqnarray}
where $\omega$ is the frequency of our wave, $s$ is some parameter along the characteristic and $\xi = \left(B_xp+B_zq\right)$. These five ODEs were solved numerically using a fourth-order Runge-Kutta method.

\begin{figure*}
\hspace{0in}
\vspace{0.1in}
%\hspace{0.4in}
\includegraphics[width=1.2in]{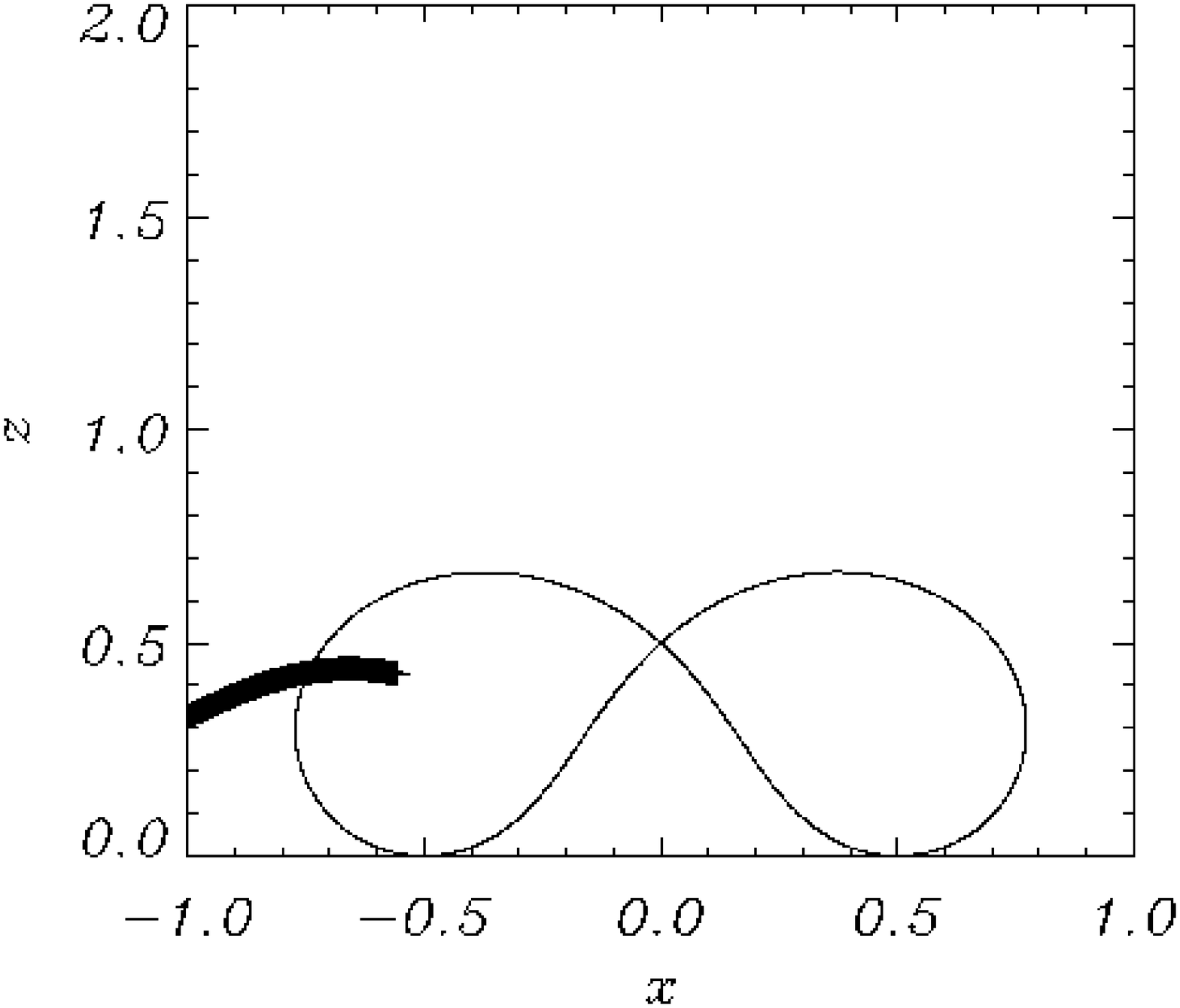}
\hspace{0.0in}
\includegraphics[width=1.2in]{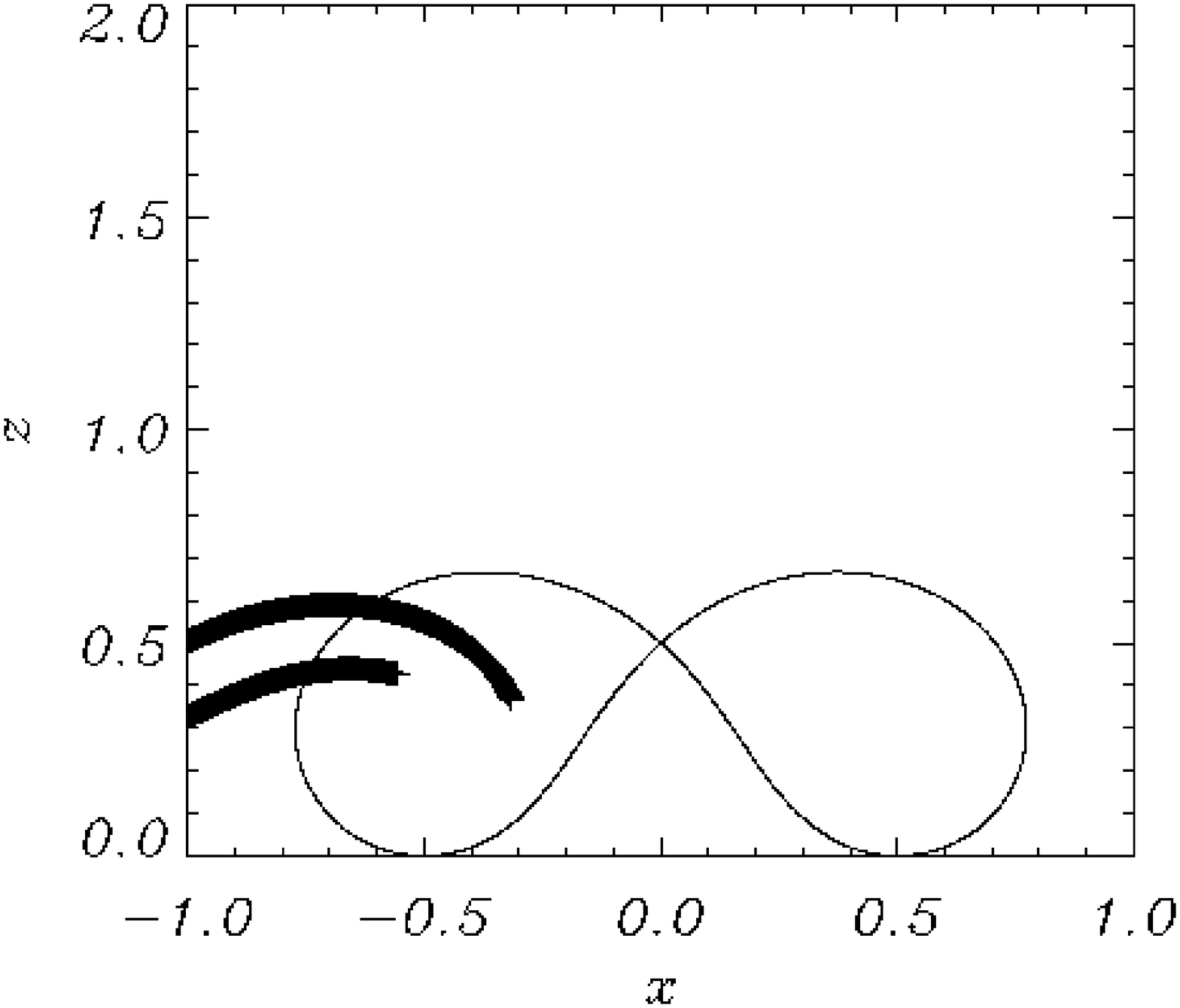}
\hspace{0.0in}
\includegraphics[width=1.2in]{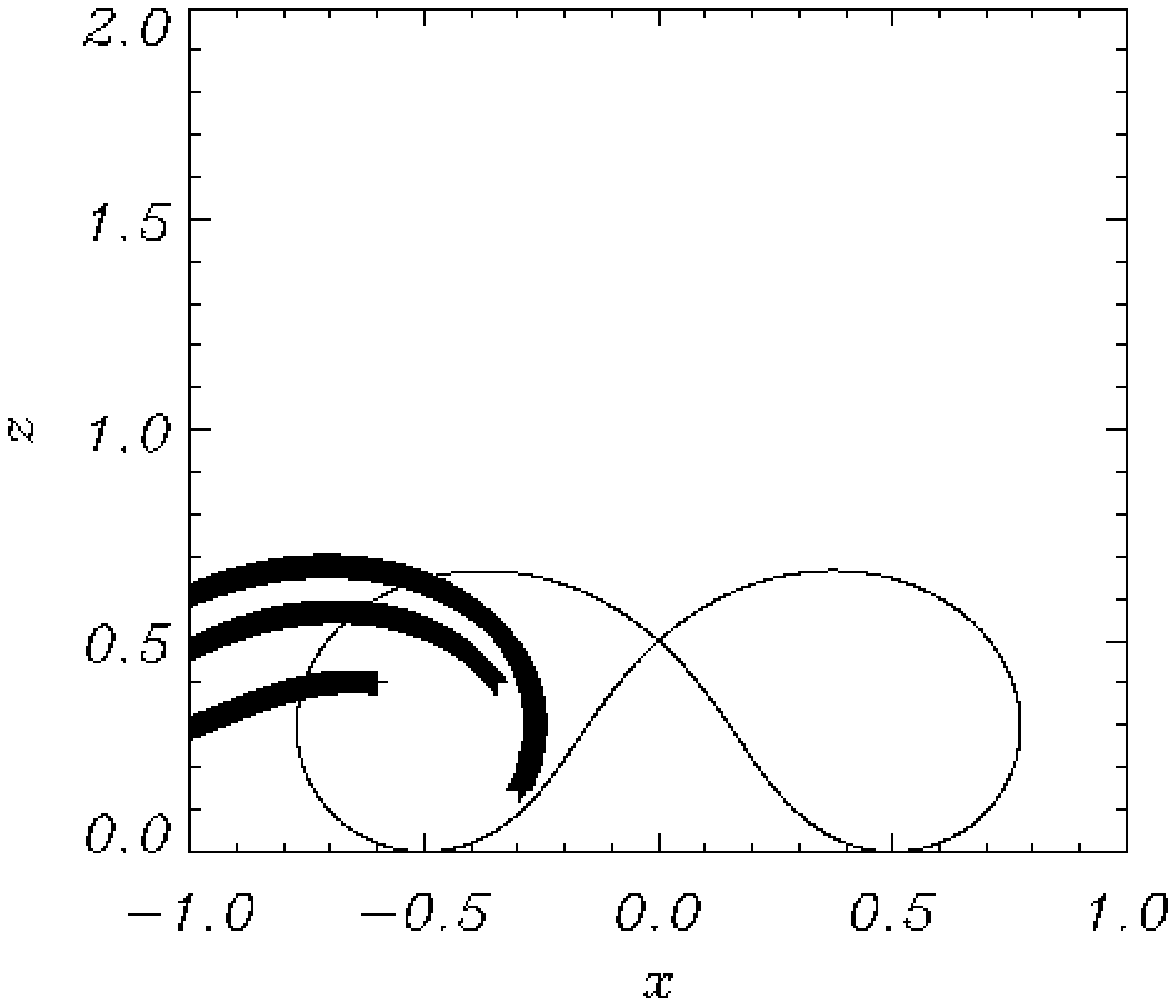}
\hspace{0.0in}
\includegraphics[width=1.2in]{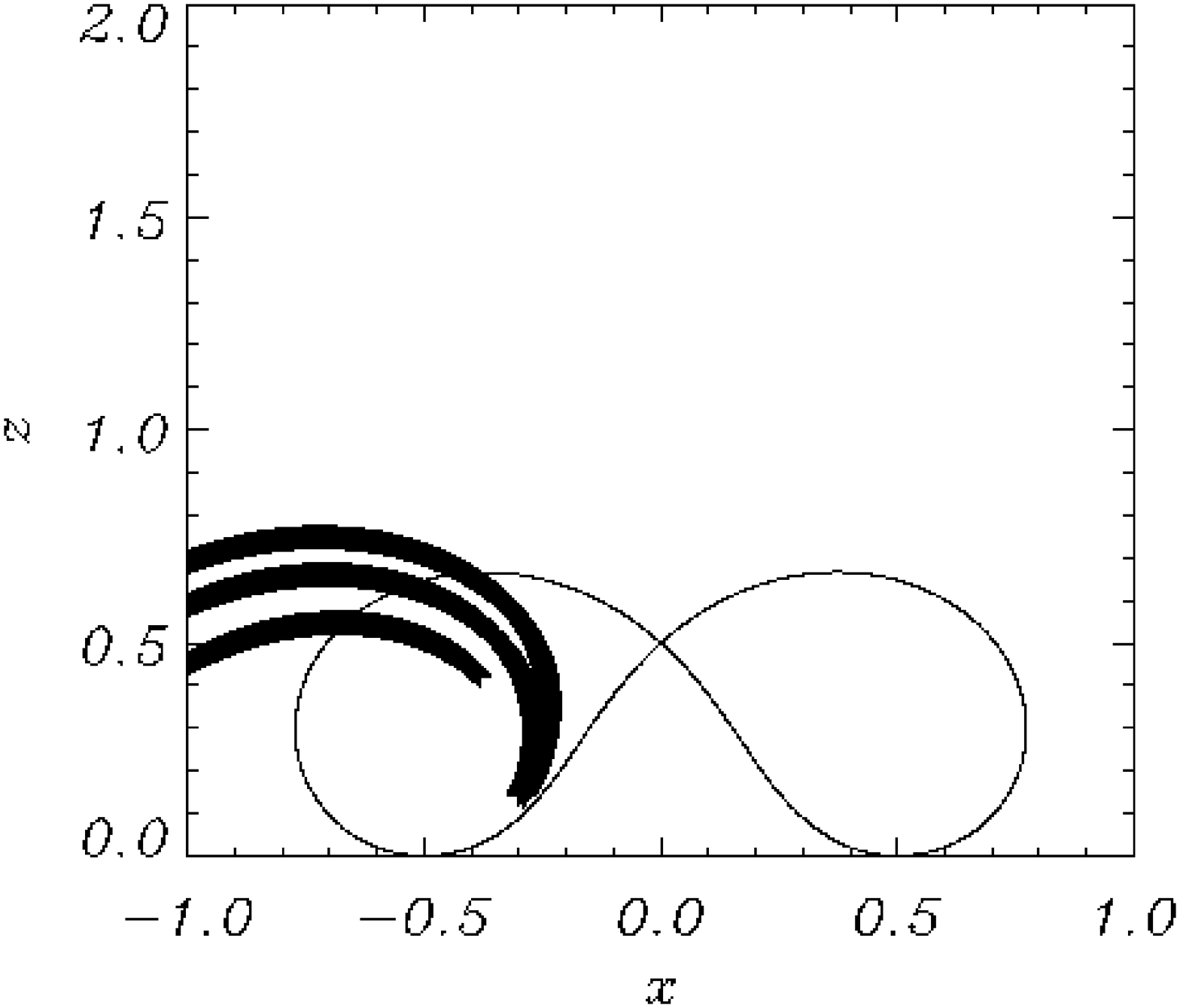}\\
\hspace{0in}
\vspace{0.1in}
%\hspace{0.4in}
\includegraphics[width=1.2in]{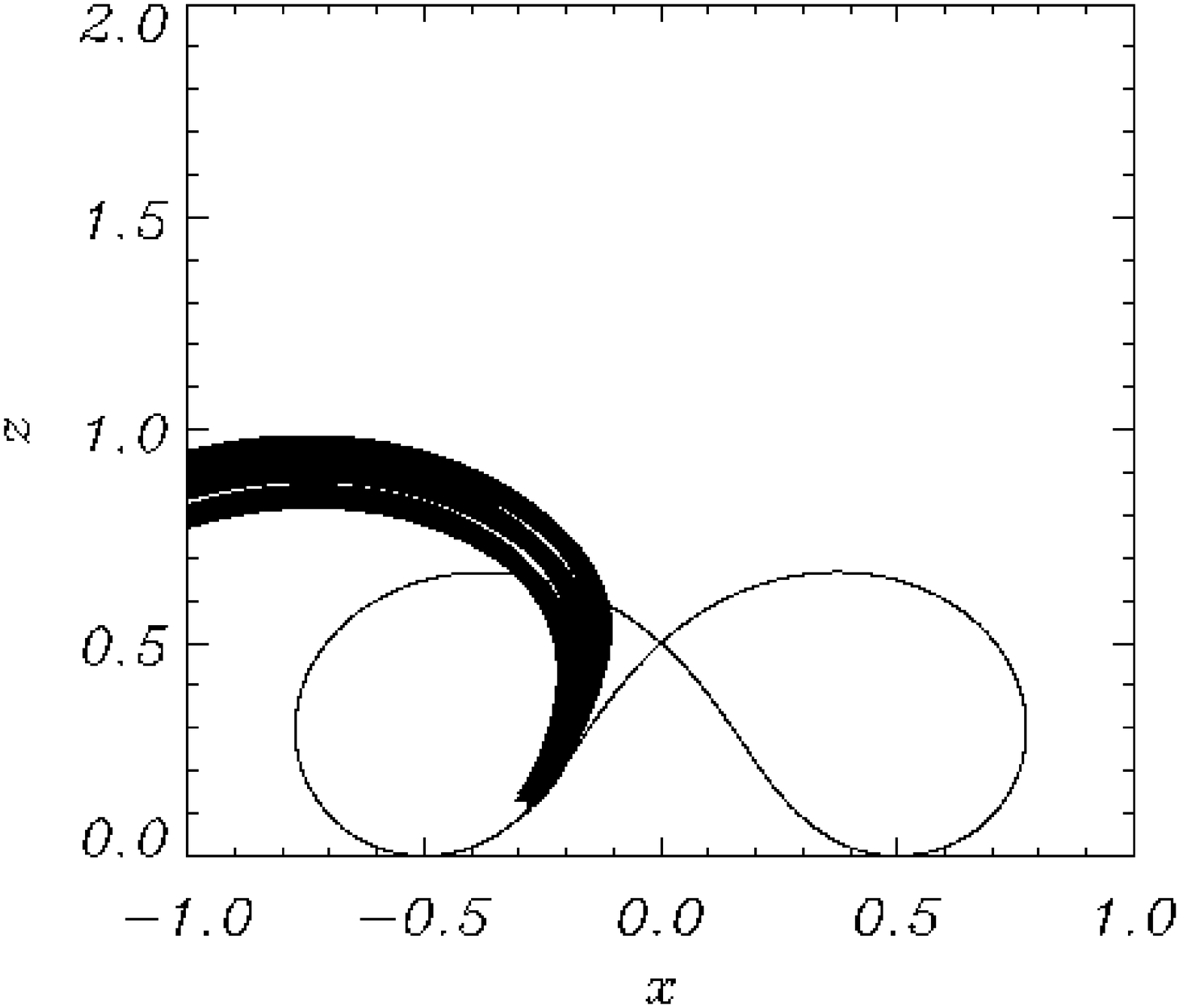}
\hspace{0.0in}
\includegraphics[width=1.2in]{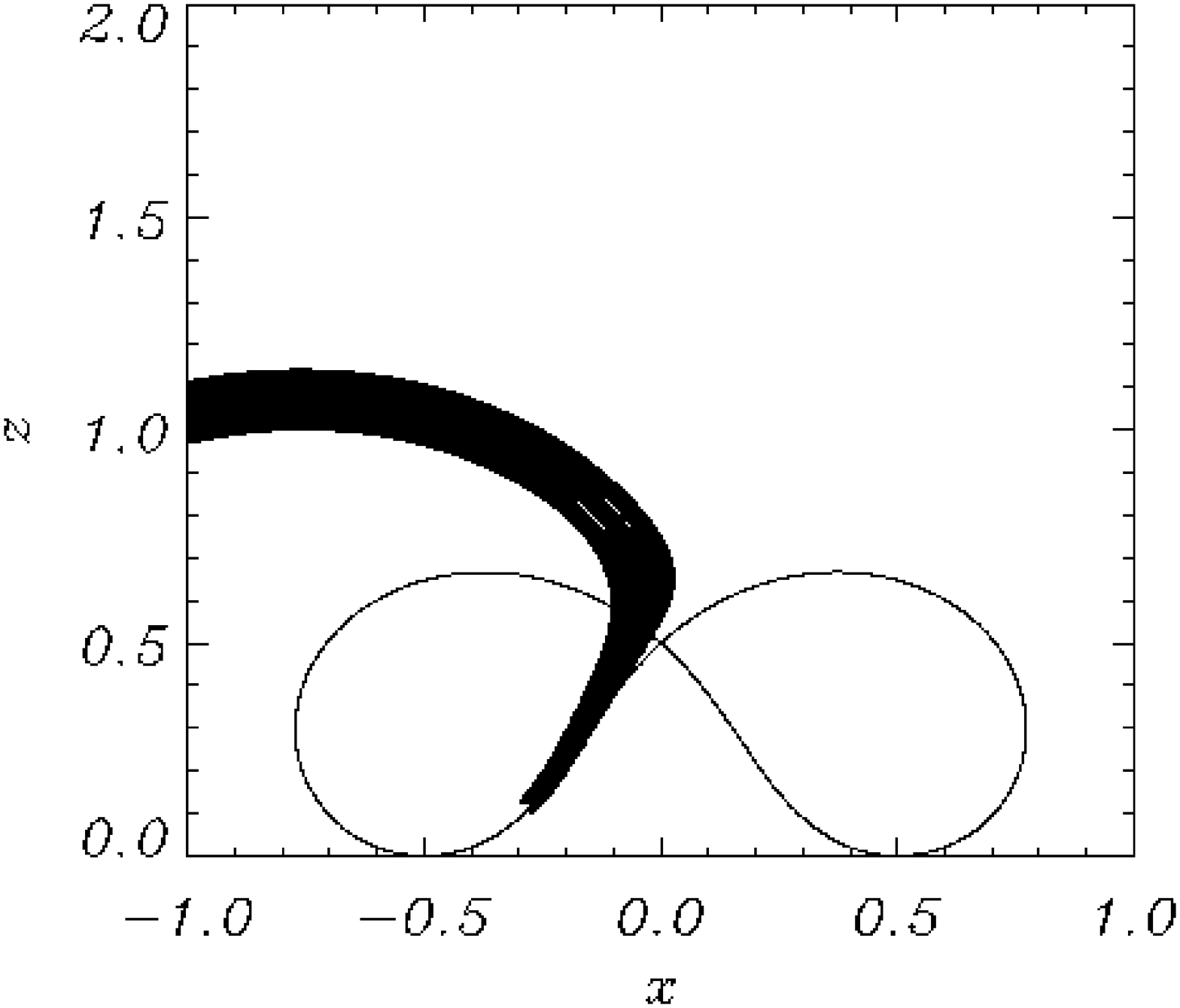}
\hspace{0.0in}
\includegraphics[width=1.2in]{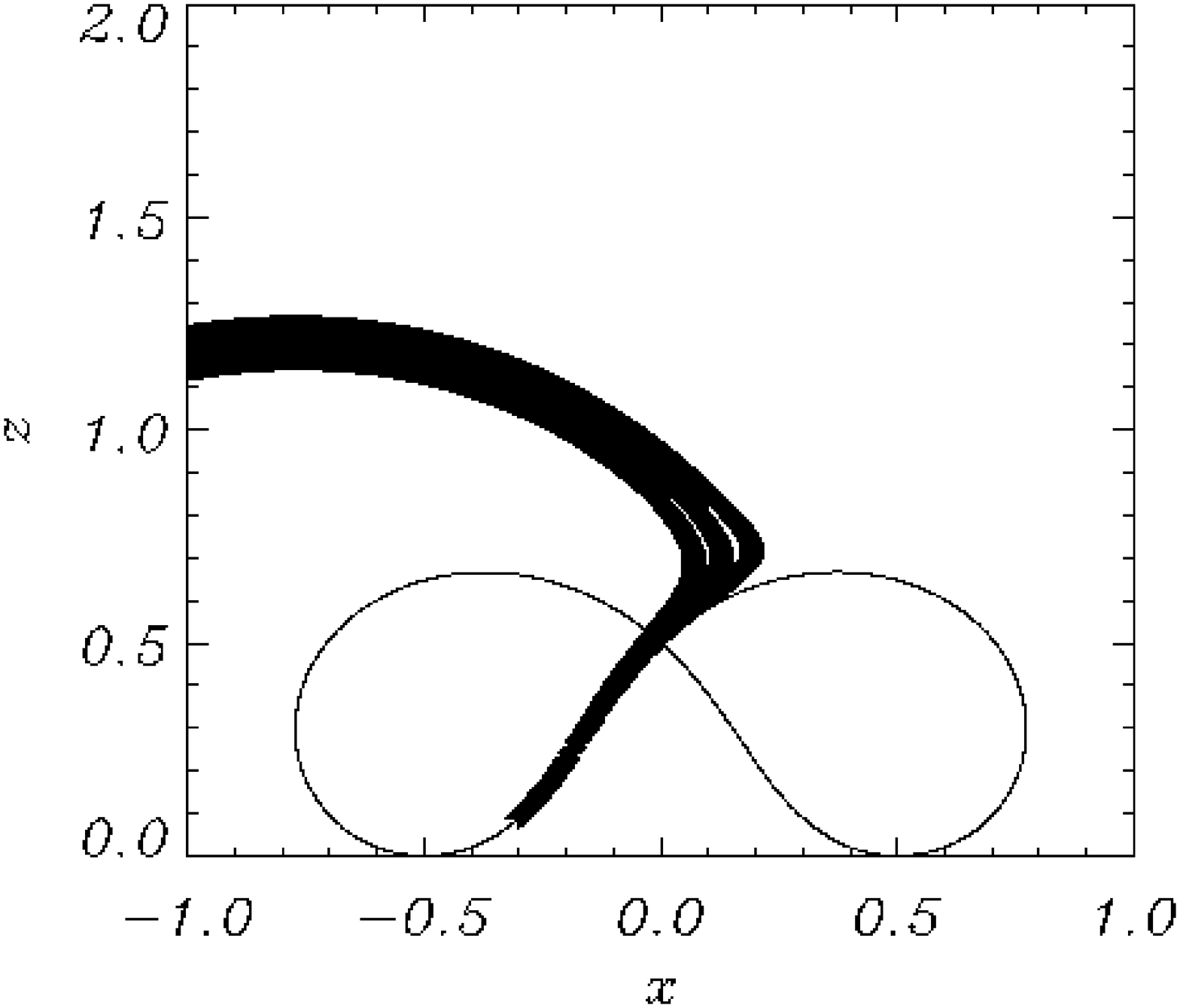}
\hspace{0.0in}
\includegraphics[width=1.2in]{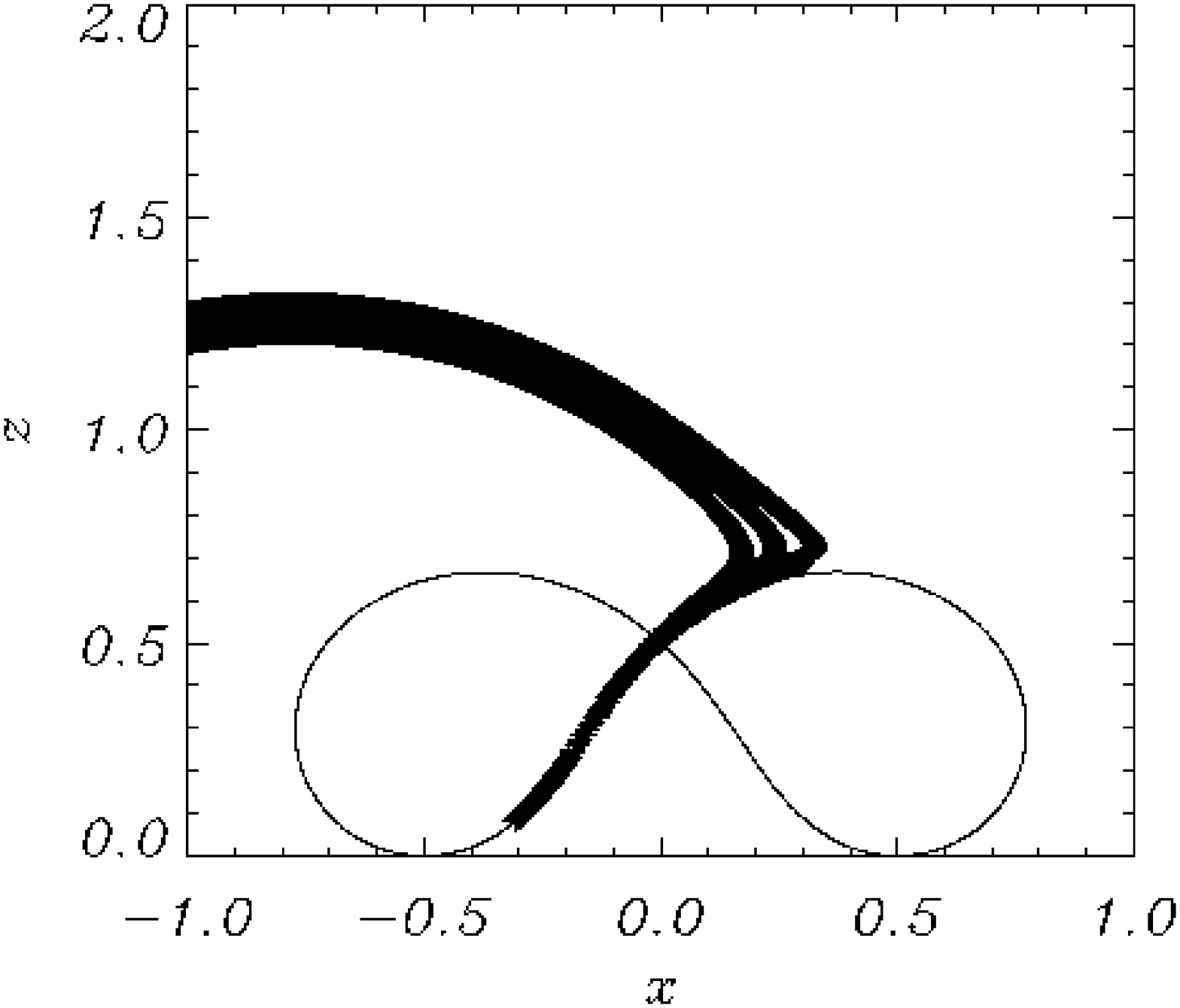}\\
\hspace{0in}
\vspace{0.1in}
%\hspace{0.4in}
\includegraphics[width=1.2in]{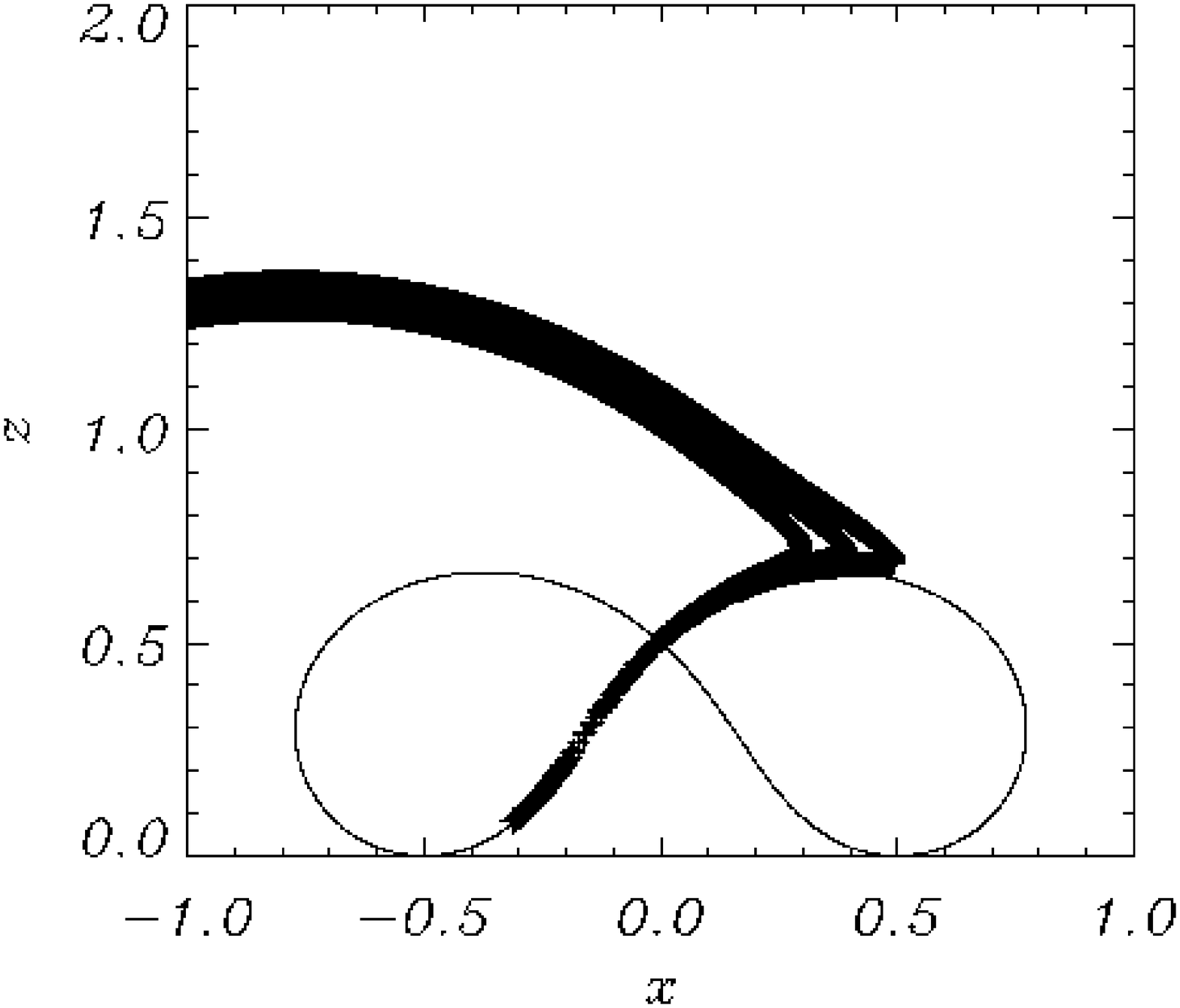}
\hspace{0.0in}
\includegraphics[width=1.2in]{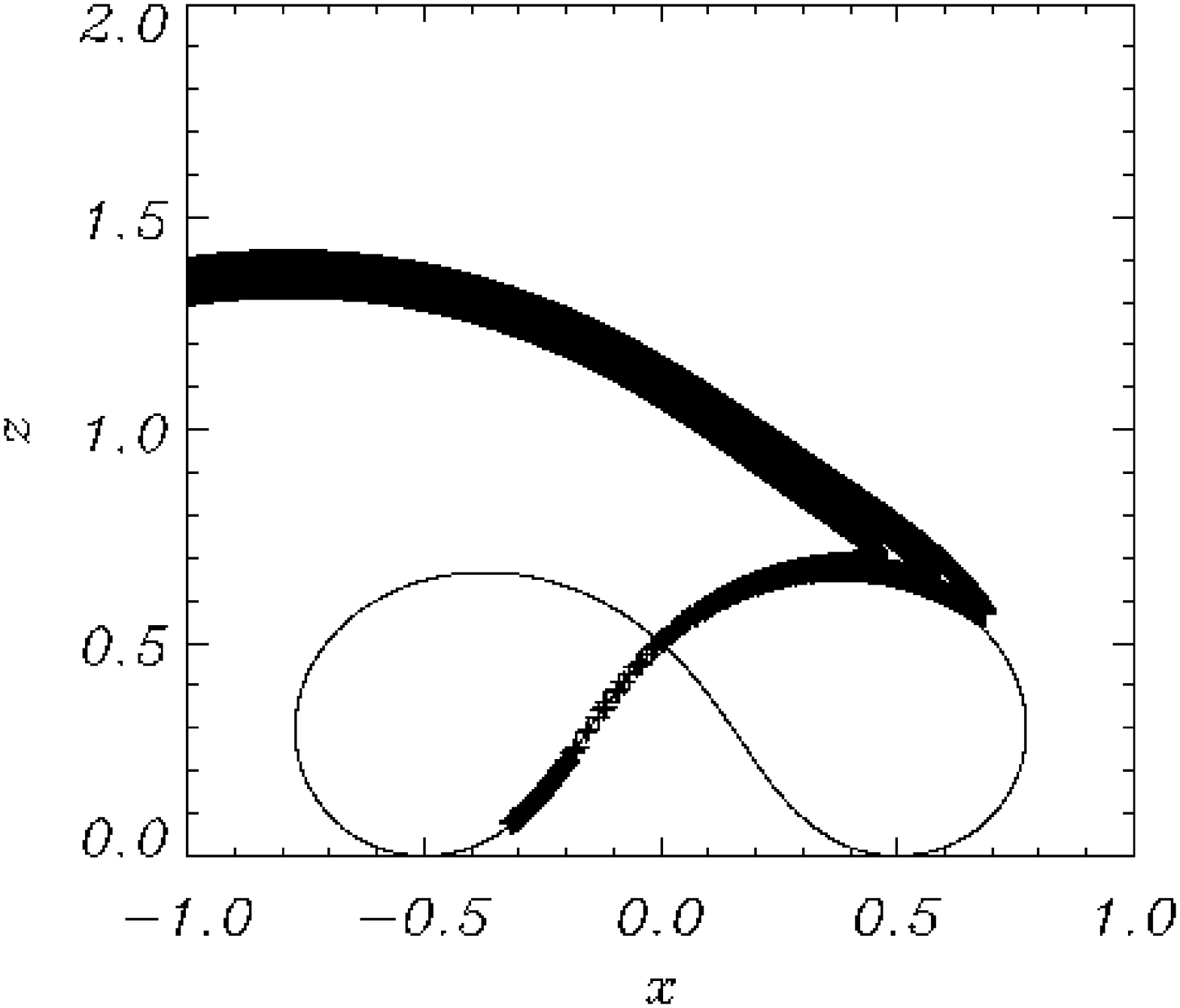}
\hspace{0.0in}
\includegraphics[width=1.2in]{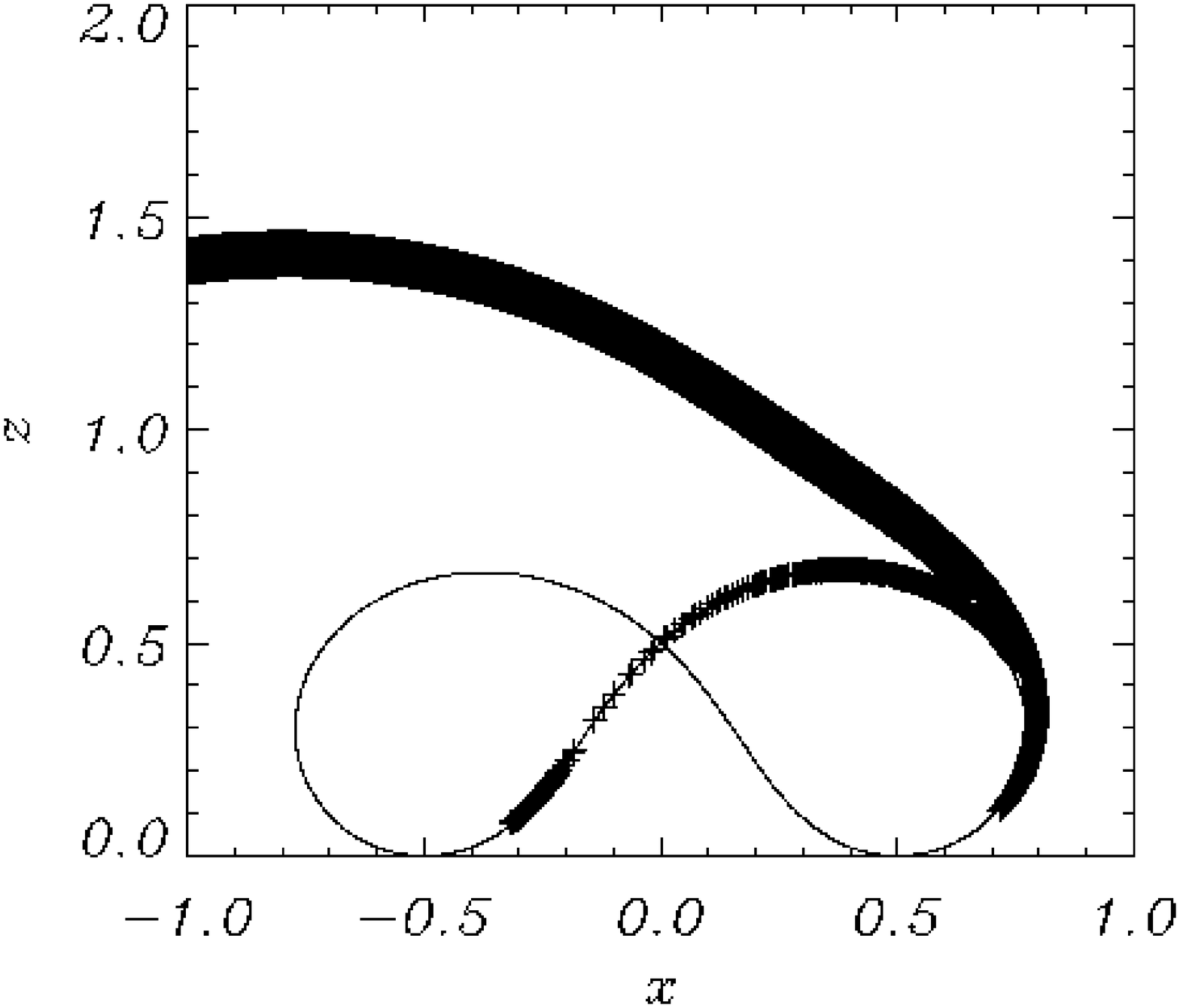}
\hspace{0.0in}
\includegraphics[width=1.2in]{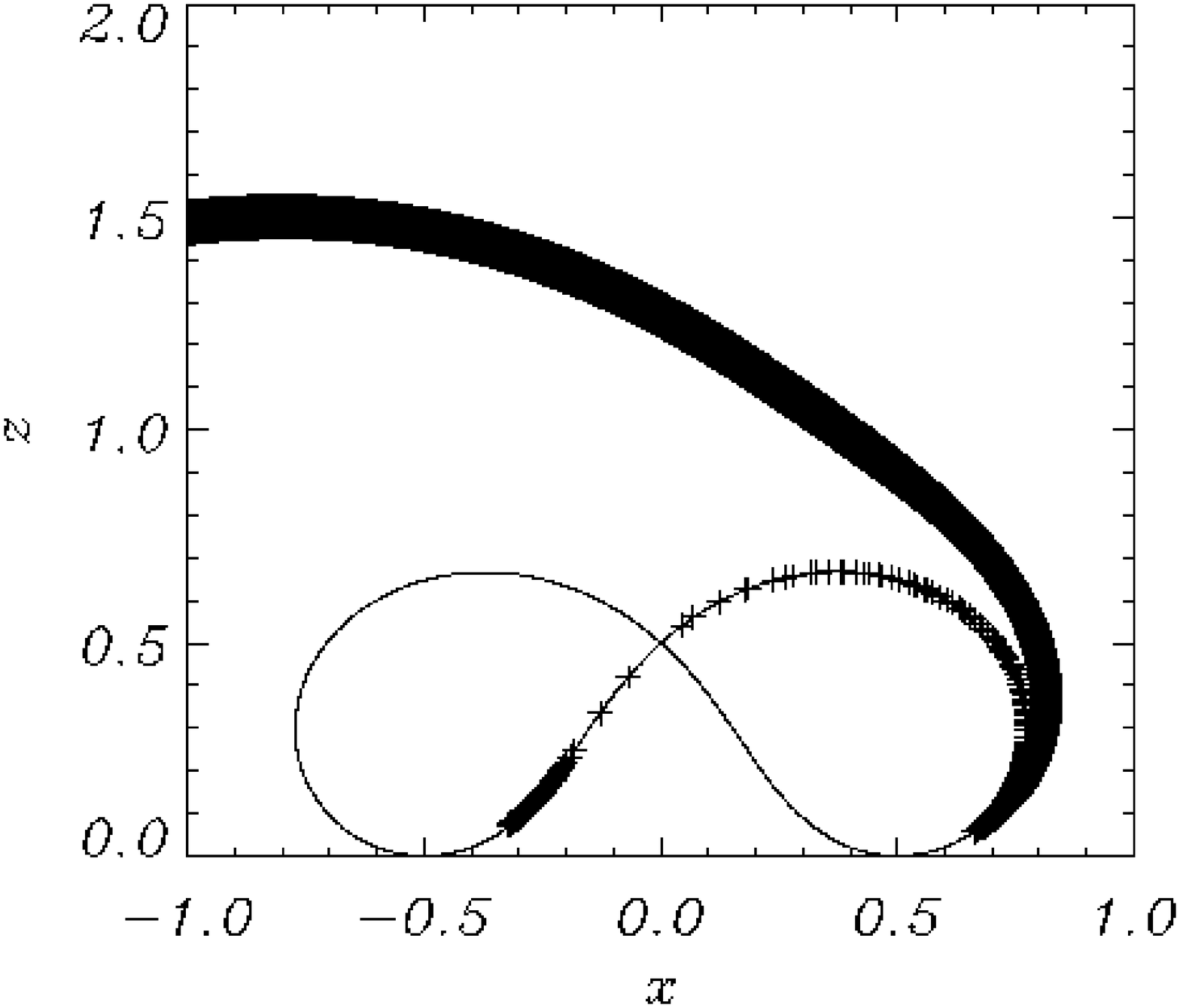}
\caption{Analytical solution for WKB approximation of an  Alfv\'en wave sent in from lower boundary for $-1 \leq x \leq -0.7 $, and its resultant propagation at times  $(a)$ $t$=0.05, $(b)$ $t$=0.1, $(c)$ $t$=0.15, $(d)$ $t=0.2$, $(e)$ $t$=0.4, $(f)$ $t$=0.6, $(g)$ $t=0.8$, $(h)$ $t$=0.9, $(i)$ $t$=1.0, $(j)$ $t=1.1$, $(k)$ $t$=1.2, $(l)$ $t$=1.4,  labelling from top left to bottom right. The lines represent the front, middle and back edges of the WKB wave solution.}
\label{fig:4.3.4.2.buffy}
\end{figure*}

By plotting positions of constant $\phi$ we can visualise the wavefront. This can be seen in Figure \ref{fig:4.3.4.2.buffy}. Here, we have plotted the wave fronts at the same times as in the numerical simulation (Figure \ref{fig:4.2.buffy}) to allow a direct comparison. The agreement is very good. Again, each wavefront is indicated by tiny crosses. Not only can we see part of the wave accumualting along the separatrices more clearly now, but we can also see how stretched the wave gets across the skeleton.

\begin{figure}
\begin{center}
\includegraphics[width=2.0in]{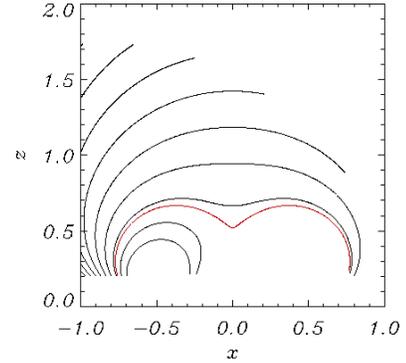}
\caption{Plots of WKB solution for an Alfv\'en wave sent in from the lower boundary and its resultant particle paths (thick lines). Starting points of $-1 \leq x_0 \leq -0.7$ at intervals of $\frac{1}{30}$ are plotted. Also shown in red is starting point $x_0=-0.758$} 
\label{fig:4.3.4.2.2}
\end{center}
\end{figure}

We can also use our WKB solution to plot the particle paths of individual elements from the initial wave. These can be seen in Figure \ref{fig:4.3.4.2.2}.

\section{Conclusions}\label{sec:4}

This paper describes an investigation into the nature of MHD waves in the  neighbourhood of two magnetic dipoles. From the work above, it has been seen that when a fast wave is generated on the lower boundary it propagates upwards and is influenced by the magnetic configuration. Firstly, it is influenced by the two regions of high Alfv\'en speed; this distorts the initally straight wave to form two peaks (with maxima located over the loci of the magnetic field). These localised areas of high speed is a new effect not seen previously in Papers I or II.

The wave continues to propagate upwards but travels slower as it approaches the null  (located at $(x,z)=(0,0.5)$) due to the decreasing Alfv\'en speed. Also, the linear fast wave cannot cross the null (as its speed drops to zero at that point). Meanwhile, the rest of the wave (the \emph{wings}) continue to propagate upwards and spread  out (the fast wave propagates roughly isotropically). Thus, this gives the visual impression that the wave is pivoting about the null point (which is acting as a fulcrum). As this pivoting continues, the two wings eventually cross each other and, due to the linear nature of the model, both wings pass each other (do not affect each other). A full non-linear treatment of the equations may reveal some very different behaviour.

The two wings continue to pivot about their fulcrum. Close to the null, part of the wave is wrapping  tightly around it. Near the null, a {\emph{refraction}} effect spirals the wave into the X-point, wrapping it around again and again. The rest of the wave (above the magnetic skeleton) appears to simply continue to rise and spread out. The wave is stretched between its two goals (part wrapping around the null and part rising away from the magnetic skeleton) and this leads to the wave splitting; this occurs near the regions of high Alfv\'en speed (the localised high speed thins the wave and forces the split). Thus, part of the (now split) wave spirals into the null and the other part propagates away from the magnetic skeleton; no longer heavily influenced by the magnetic null point. These parts form a distinctive cross shape. It was also demonstrated that  $40\%$ of the  wave is trapped by the null, where the fraction captured is of the  form $\frac{1}{L}{x_{\small{\rm{\textit{crit}}}}\left(a,z_0\right)}$.

%Hence, we see that for fast magnetoacoustic waves in the neighbourhood of two dipoles, a fraction of the wave (and thus wave energy) is trapped by the null with the rest  escaping. The fraction captured by the null will be of the form $L-x_{\rm{\textit{critical}}}\left(a,z_0\right)$, where $L$ is the length of our lower boundary $\left-L\leq x \leq L\right)$ and $x_{\rm{\textit{critical}}}$ is the critical starting point that divides the particle paths into paths that spiral into the null and those that do not (e.g. the starting point connected to the green lines in Figure \ref{WKBtwodipolesparticlepaths_2}). For the linear fast wave investigated in this paper; $L=1$, $z=z_0=0.1$, $a=0.5$ and  $x_{\rm{\textit{critical}}}=0.4$, which means  $40\%$ of the  wave is trapped by the null.  Hence, $40\%$ of the wave energy accumulates at the null and ohmic dissipation will extract the energy in the wave at this point.

Thus, it has been shown that the key result for the fast wave from Paper I (i.e. that the fast wave accumulates at null points) persists in  this more complication magnetic configuration,  but only if the wave is close enough to the null point. This makes intuitive sense; if the fast wave is too far away from the magnetic null, it will not feel its effect (until it is much closer).  The physical significance of this is that any fast magnetoacoustic disturbance near such a magnetic configuration will propagate and eventally split; the parts close enough to the null will be drawn towards it and focus all its energy at that point, the rest will propagate away from the system (taking its wave energy with it). The part trapped by the refraction effect of the null will lead to large gradients in the  perturbed magnetic field, and thus this is where current density build-up will occur. We believe ohmic dissipation will extract the energy in the wave at this point and  wave heating will naturally occur at the null point (as in Paper I). However,  this is only a claim as  the numerical experiments and analytical work described above were all conducted in an ideal plasma.

We also looked at the behaviour of the Alfv\'en wave near our two dipoles configuration. We saw that when a linear Alfv\'en wave was generated on  the lower boundary, its propagation is confined to the path of the field line it starts on (the Alfv\'en wave cannot cross field lines). This agrees with one of the key results found for Alfv\'en waves in Paper II;   {{the Alfv\'en wave follows the field lines and each fluid element is trapped on the field line it starts on}}.      Thus, part of the wave  accumulates along the separatrix (depositing parts of itself as it goes) and this behaviour  stretches the wave thinly along the separatrices. The other part of the wave  appears to be propagating away from the magnetic skeleton, though it is actually just following the field lines (and these are spreading out). This was apparent in the numerical simulation but more clearly seen in the WKB approximation. These two goals for different parts of the wave result in a sharp wavefront being formed.  Dissipation may extract the energy from the wave at this point.

%All the behaviour of the Alfv\'en wave agrees with what we learnt from Chapter 4 and 5; that {{the Alfv\'en wave follows the field lines and each fluid element is trapped on the field line it starts on}}. The (WKB) particle paths also shows this. 

Furthermore, since we have the Alfv\'en wave accumulating and stretched along the separatrices, we will have a build-up of current density at these locations. Therefore, since current is accumulating along the separatrices, this will be the location where ohmic dissipation will extract the energy in the wave. Thus, {{Alfv\'en wave heating will naturally occur along the separatrices}} due to dissipation. The rest of the Alfv\'en wave (the part travelling away from the magnetic skeleton) does not create large gradients and thus does not generate large currents. This part of the wave simply travels out of our system (taking its energy with it).

% This makes intuitive sense; if $d$ is very large then little of the wave energy will be trapped by the null (far away the effect of the null is negligible), or if $d$ is less than $x_{0{\textrm{critical}}}$ then all the wave will refract into the null (none of the wave can escape).

\section*{Acknowledgements}

James McLaughlin acknowledges financial assistance from the Particle Physics and Astronomy Research Council (PPARC).  He also wishes to thank Toni D\'iaz for helpful discussions.

%He also wishes to thank Tom Bogdan, Valery Nakariakov and Erwin for helpful discussions.

%*********************************************************************************************
%

%*********************************************************************************************

\end{document}